\documentclass[proof]{WileyASNA-v1}
\articletype{Article Type}%

\newcommand{\ind}[1]{_{\mathrm{#1}}}
\newcommand{\diff}{\mathrm{d}}

\newcommand\Dnu{\Delta\nu}

\newcommand\numax{\nu\ind{max}}

\newcommand\Teff{T\ind{eff}}
\newcommand\Msol{{M\ind{\odot}}}
\newcommand\Rsol{{R\ind{\odot}}}
\newcommand\Tsol{{T\ind{eff,\odot}}}

\newcommand\numaxsol{{\nu\ind{max,\odot}}}
\newcommand\Dnusol{\Dnu\ind{\odot}}

\begin{document}

\title{FRA -- A new Fast, Robust and Automated pipeline for the detection and measurement of solar-like oscillations in time-series photometry of red-giant stars}

\author[1,2]{C. Gehan*}
\author[2,3]{T. L. Campante}
\author[2,3]{M. S. Cunha}
\author[2,3]{F. Pereira}

\authormark{C. GEHAN \textsc{et al}}

\address[1]{\orgname{Max-Planck-Institut für Sonnensystemforschung}, \orgaddress{\state{Justus-von-Liebig-Weg 3, 37077 Göttingen}, \country{Germany}}}
\address[2]{\orgdiv{Instituto de Astrofísica e Ciências do Espaço}, \orgname{Universidade do Porto, CAUP}, \orgaddress{\state{Rua das Estrelas, PT4150-762 Porto}, \country{Portugal}}}
\address[3]{\orgdiv{Departamento de F\'{\i}sica e Astronomia}, \orgname{Faculdade de Ci\^{e}ncias da Universidade do Porto}, \orgaddress{\state{Rua do Campo Alegre, s/n, PT4169-007 Porto}, \country{Portugal}}}

\corres{*Charlotte Gehan, \email{gehan@mps.mpg.de}}

\abstract{We developed, tested and validated a new Fast, Robust and Automated (FRA) tool to detect solar-like pulsations. FRA is based on the detection and measurement of the frequency of maximum oscillation power $\numax$, independently from the large frequency separation $\Dnu$. We applied the FRA pipeline to 254 artificial power spectra representative of TESS red giants, as well as 1689 red giants observed by \textit{Kepler} and 2344 red giants observed by TESS. We obtain a consistency rate for $\numax$ compared with existing measurements above 99\% for \textit{Kepler} red giants and above 97\% for TESS red giants. We find that using $\numax$ as an input parameter to guide the search for $\Dnu$ through the existing Envelope AutoCorrelation Function (EACF) method significantly improves the consistency of the measured $\Dnu$ in the case of TESS stars, allowing to reach a consistency rate above 99 \%. Our analysis reveals that we can expect to get consistent $\numax$ and $\Dnu$ measurements while minimizing both the false positive measurements and the non-detections for stars with a minimum of four observed sectors and a maximum G magnitude of 9.5.}


\maketitle

\section{Introduction}\label{introduction}

Many fields of astrophysics rely on the precise knowledge of the stellar properties, in particular stellar masses, radii, and ages. The characterisation of exoplanets requires the characterisation of their host star \citep{Rauer}. Similarly, Galactic archaeology, which aims at tracing the formation and evolution of the Milky Way, is based on the study of the stellar populations of its various structural components \citep{Miglio}. The detailed understanding of stellar evolution and its physical mechanisms is therefore fundamental and urgent for these fields, which are undergoing an exponential development.

In this context, we need powerful tools for the observational diagnosis of stellar interiors. This is made possible by asteroseismology, which consists in studying stellar oscillations containing information on the physical conditions prevailing in the interior of stars. Pulsating evolved low-mass stars, such as red giants and subgiants, represent ideal laboratories to study the physical mechanisms governing stellar cores, as their oscillation spectrum exhibits so-called mixed modes probing the most inner radiative regions in addition to the external convective envelope \citep{Bedding, Beck_2012}.

Important breakthroughs in stellar physics have been made in the last 15 years with the advent of the ultra-high
precision photometry space missions CoRoT (ESA, 2006-2012) and \textit{Kepler}/K2 (NASA, 2009-2018), which led to the
detection of oscillations in tens of thousands of stars. The TESS space mission (NASA) has been launched in 2018, and its observation strategy allows an entire hemisphere to be mapped in a single year. Most of the sky will receive 27 days of continuous observations, while a small area around the poles will get more or less continuous coverage for the entire year. TESS completed its 2-year, all-sky nominal survey in July 2020 and is dramatically increasing the number of red giants with detected oscillations, which is expected to be on the order of $10^5$ \citep{Campante_2016, Aguirre, Mackereth}. TESS enables in particular the measurement of the two global asteroseismic properties $\numax$ and $\Dnu$ \citep{Mackereth}. The frequency of maximum oscillation power $\numax$ is proportional to the surface gravity $g$ and effective temperature $\Teff$ of the star such that \citep{Brown, Kjeldsen}
\begin{equation}\label{eqt-numax}
\numax \propto \frac{g}{\sqrt{\Teff}}.
\end{equation}
$\numax$ corresponds to the center of the Gaussian envelope characterising solar-like oscillations in the power spectrum.
The large frequency separation $\Dnu$ is related to the mean stellar density $\rho$ as \citep{Ulrich}
\begin{equation}\label{eqt-Dnu}
\Dnu \propto \sqrt{\rho}.
\end{equation}
$\Dnu$ characterises the regular frequency spacing of acoustic oscillations modes.
The asymptotic expression of the large separation represents the inverse of the time it takes for a pressure wave to travel back and forth from the centre to the surface of the star, such that \citep{Ulrich}
\begin{equation}\label{eqt-Dnu-as}
\Dnu = \left ( 2 \, \int^R_0 \frac{\diff r}{c\ind{s}} \right )^{-1},
\end{equation}
where $R$ is the stellar radius and $c\ind{s}$ is the internal sound-speed.

We can determine accurate seismic masses and radii using $\numax$ and $\Dnu$, independently from modelling, through the scaling relations \citep{Kjeldsen, Kallinger_2010a, Mosser_2013}
\begin{equation}\label{eqt-mass}
\frac{M}{\Msol} = \left(\frac{\numax}{\numaxsol}\right)^3 \left(\frac{\Dnu}{\Dnusol}\right)^{-4} \left(\frac{\Teff}{\Tsol}\right)^{3/2},
\end{equation}
and
\begin{equation}\label{eqt-radius}
\frac{R}{\Rsol} = \left(\frac{\numax}{\numaxsol}\right) \left(\frac{\Dnu}{\Dnusol}\right)^{-2} \left(\frac{\Teff}{\Tsol}\right)^{1/2},
\end{equation}
where $\numaxsol = 3050 \, \mu$Hz, $\Dnusol = 135.5 \, \mu$Hz and $\Tsol = 5777$ K are the solar values. When spectroscopic measurements of the effective temperature are not available, we can use a proxy for red giants as given by \cite{Mosser_2012b}
\begin{equation}\label{eqt-Teff}
\Teff = 4800 \, \left ( \frac{\numax}{40} \right )^{0.06},
\end{equation}
with $\numax$ in $\mu$Hz.

The precision on the exoplanet properties such as the radius, mass and age is directly related to the precision reached on the determination of stellar properties. TESS allows us to reach a relative precision of $\sim$ 4\% in $\Dnu$ and $\sim$ 3.5\% in $\numax$ \citep{Aguirre}. Radii, masses, and ages can be obtained with uncertainties at the 6\%, 14\%, and 50\% level, and decrease to 3\%, 6\%, and 20\% when parallax information from Gaia DR2 \citep{Gaia} is included. These precision levels are similar to those obtained with the 4-year long \textit{Kepler} datasets, which give rise to relative precisions of 2.9\% in radius, 7.8\% in mass and 25\% in age \citep{Wu, Yu}. Additionally, Galactic archaeology also requires precise stellar radii, masses and ages \citep{Miglio_2017}.

Several methods exist to measure $\numax$ and/or $\Dnu$, in particular:
\begin{itemize}
\item the COR \citep{Mosser_2009}, OCT \citep{Hekker_2010} and A2Z \citep{Mathur_2010} methods first search for the signature of $\Dnu$, using the Envelope AutoCorrelation Function (EACF) method in the case of the COR pipeline and by computing the power spectrum of the power spectrum for the other methods, for several windows inside the spectrum in the case of the OCT pipeline and for the whole spectrum for the A2Z pipeline. The three methods then rely on the $\Dnu$ measurement to target the search for $\numax$ through the local fit of a Gaussian envelope for the oscillations;
\item the CAN \citep{Kallinger_2010b}, BAM \citep{Zinn} and BHM \citep{Elsworth} pipelines use a Bayesian Markov-Chain Monte-Carlo (MCMC) algorithm to fit a global model to the observed power density spectrum, including a description of the background signal and a Gaussian envelope for the oscillations. The CAN and BAM methods then fit Lorentzian profiles for the radial, dipole and quadrupole modes, and finally derive $\Dnu$ from this mode identification. The BHM method then computes the power spectrum of the power spectrum to derive $\Dnu$;
\item the approach of \cite{Hon_2018} consists in using supervised deep learning to detect oscillations in red giants and estimate $\numax$.
\end{itemize}

However, it is crucial to be able to detect $\numax$ independently from $\Dnu$, which is not the case of COR, OCT and A2Z. Indeed, $\Dnu$ is not necessarily measurable. A short observation duration and/or a low signal-to-noise ratio results in a resolution of the oscillation modes that is insufficient to reveal a regular spacing of pressure modes, while the Gaussian bump of oscillations can still be visible, making it possible to measure $\numax$ and to detect solar-like oscillations. Additionally, CAN, BAM and BHM are not optimized to assess whether or not a given star exhibits oscillations, being rather aimed at deriving an accurate $\numax$ measurement for stars that are known to exhibit clear oscillations \citep{Hon_2018}. On the other side, the method of \cite{Hon_2018} aims at mimicking a visual detection of oscillations by an expert eye and does not rely on a statistical criteria. We thus need a pipeline allowing to statistically assess the existence of solar-like oscillations and to measure $\numax$ independently from $\Dnu$. Such a pipeline has to be fast, robust and automated in order to efficiently analyse the harvest of red giants observed by TESS and, in a near future, PLATO.

We here present a new Fast, Robust and Automated (FRA) pipeline to detect solar-like oscillations, which can be easily implemented. It relies on the statistical detection of $\numax$ to validate the presence of oscillations. FRA presents the advantage of giving a $\numax$ measurement which is independent from $\Dnu$. Additionally, FRA uses an innovative approach relying on a local search for $\numax$ over the power spectrum, while methods like CAN, BAM and BHM require a MCMC algorithm to fit the stellar background together with the oscillations. We deal with reduced computation times since our approach only requires 5 free parameters, for which we use realistic bounds to guide the search. Our pipeline has already been used in several studies \citep[][and Jiang et al. submitted]{Huber_2022, Gaulme_2022} as well as in several collaborations within the TESS Asteroseismic Science Operations Center (TASOC)\footnote{https://tasoc.dk/}. Section \ref{methods} is dedicated to the description of the FRA pipeline and to the measurement of $\numax$. In Sect.~\ref{test}, we test and validate the FRA pipeline on TESS-like synthetic oscillation spectra built with different values of $\numax$ typical of low-luminosity red-giant branch (LLRGB) stars and corresponding to different numbers of observed TESS sectors. We also discuss the performance of our pipeline as a function of the stellar magnitude of the number of observed TESS sectors. In Sect.~\ref{results}, we test and validate our FRA pipeline on \textit{Kepler} red giants benefiting from exquisite 4 year-long observations, as well as TESS red giants from the Southern Continuous Viewing Zone (SCVZ) having between 2 sectors (54.8 days) and 13 sectors (about 1 year) of observations. In Sect.~\ref{Dnu-measurement}, we use the EACF method \citep{Mosser_2009} to obtain $\Dnu$ measurements in addition to $\numax$ and consider its performance compared to the FRA pipeline. We provide an independent catalogue of $\numax$ and $\Dnu$ measurements for the \textit{Kepler} and SCVZ TESS targets analyzed in this study\footnote{Results are publicly available on ADS.}, which can be valuable for future studies. Sect.~\ref{conclusion} is devoted to conclusions.

\section{The FRA pipeline to detect solar-like oscillations: measurement of $\numax$}\label{methods}

We developed the Fast, Robust and Automated (FRA) method to measure $\numax$ based on the fit of the smoothed power spectrum, with the aim of detecting the bump of oscillations in the power spectrum associated to a Gaussian envelope. The novelty of FRA lies in its ability to assess whether or not a given star exhibits oscillations. To that matter, FRA performs a totally blind search for $\numax$, independent from the $\Dnu$ measurement. We describe in the following the different steps we implemented.

\begin{figure*}[h!]
\centering
\includegraphics[width=8.8cm]{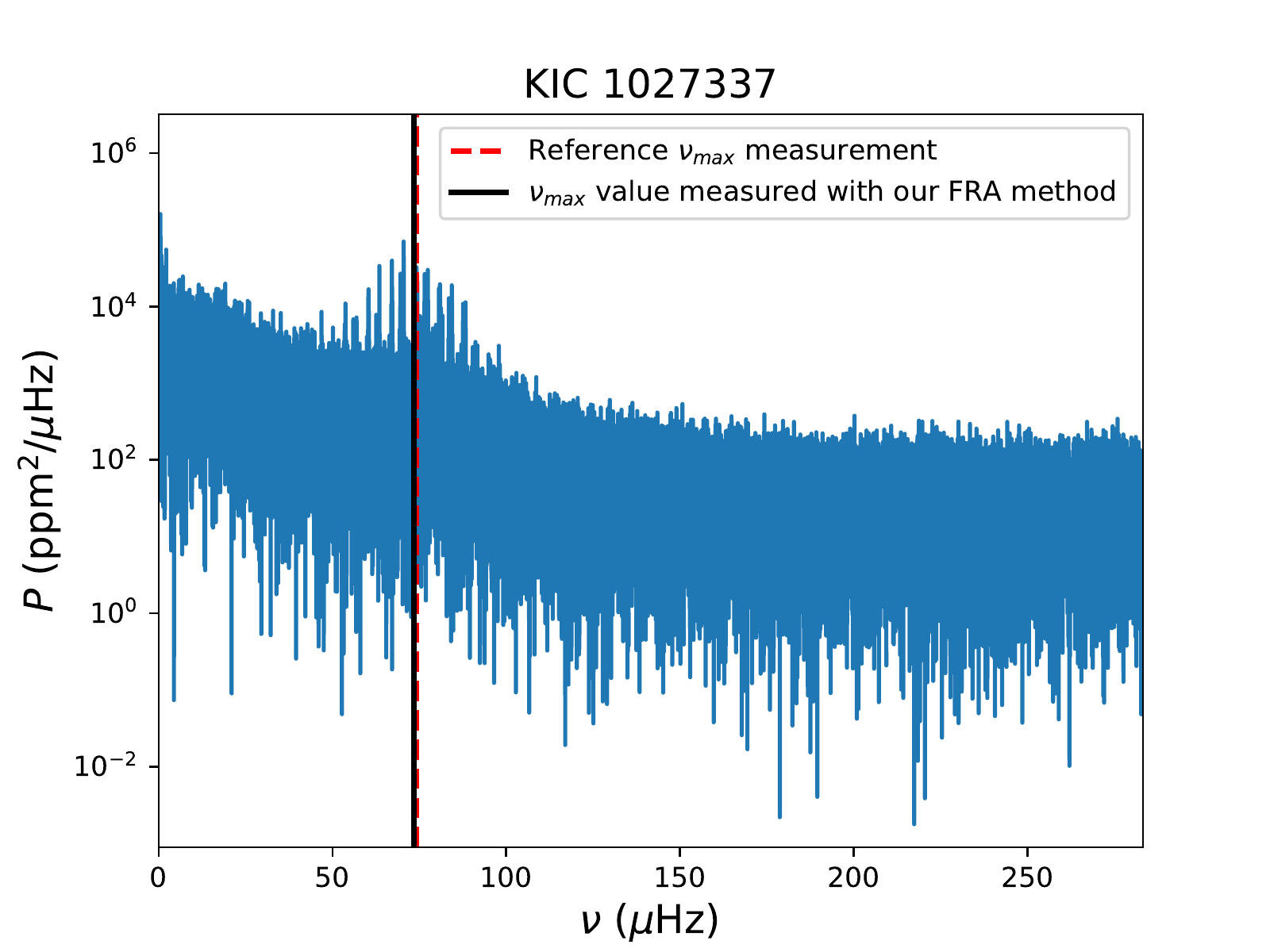}
\includegraphics[width=8.8cm]{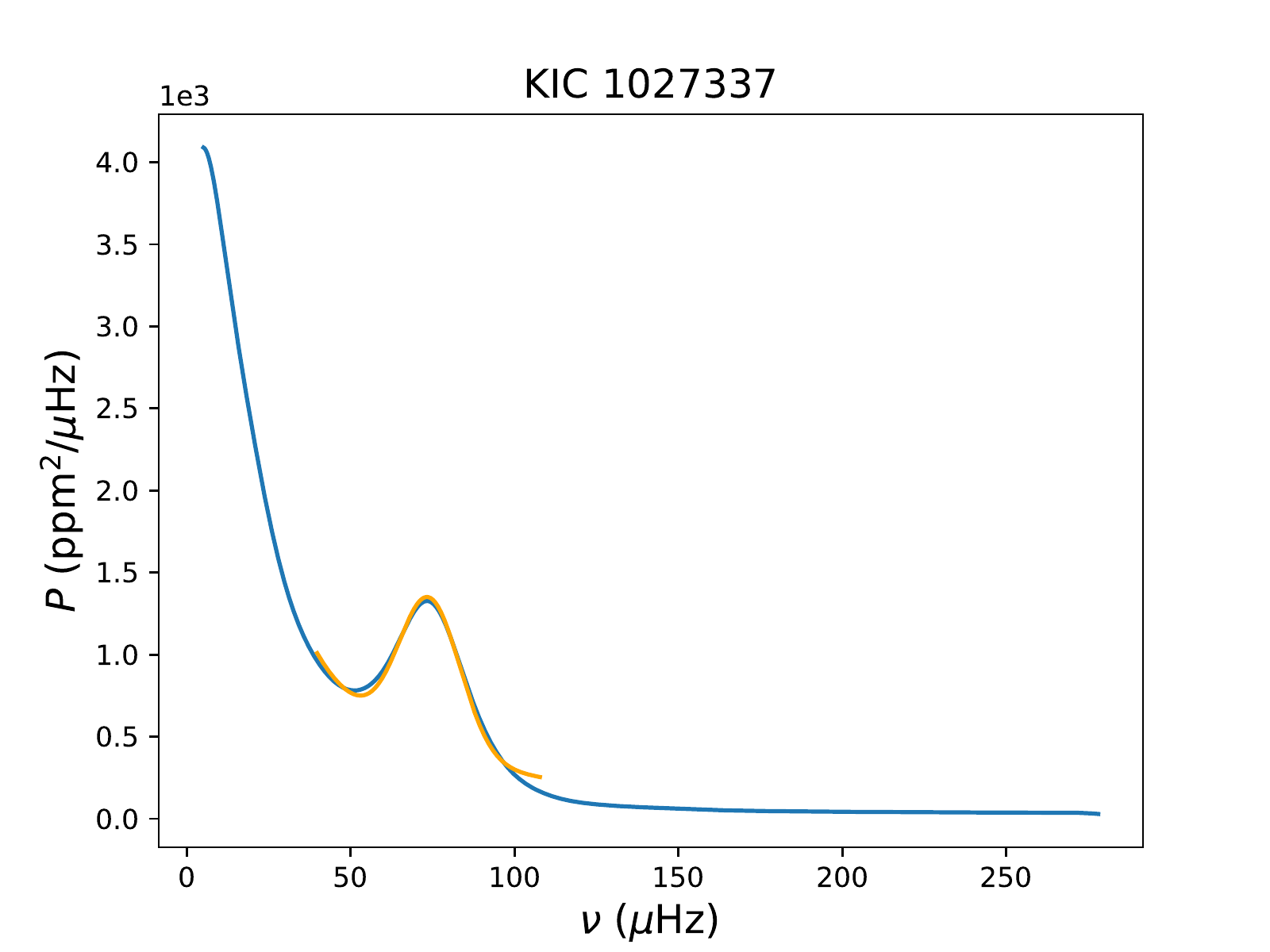}
\caption{Spectrum of the \textit{Kepler} red giant KIC 1027337. \textit{Left:} Raw spectrum. The reference $\numax$ measurement and our measurement with the FRA pipeline are represented by the vertical red and black lines, respectively, which are superimposed. \textit{Right:} Optimally smoothed spectrum. The orange line represents the local fit of oscillations from Eq. \ref{eqt-oscillations}, centered around $\numax$.}
\label{fig-spectrum}
\end{figure*}

\subsection{Smoothing the power spectrum}\label{smoothing}

The spectrum first needs to be smoothed (Fig.~\ref{fig-spectrum}). To that end, 20 input values $\nu\ind{c}$ are tested for $\numax$, regularly spaced between 30 and 270 $\mu$Hz, since the Nyquist frequency of 30-mins cadence TESS data is 278 $\mu$Hz. For each $\nu\ind{c}$ value, an input large separation $\Dnu\ind{c}$ is tested, derived through the scaling relation for red giants \citep{Mosser_2010}
\begin{equation}\label{eqt-Dnu-c}
\Dnu\ind{c} = 0.28 \, \nu\ind{c}^{0.75}.
\end{equation}
These $\Dnu\ind{c}$ values are used to derive 20 different smoothed spectra by convolving the power spectrum with a Gaussian of full width at half maximum (FWHM) \citep{Mosser_2009}
\begin{equation}\label{eqt-dnu-env}
\delta\nu\ind{env} = 3 \, \Dnu\ind{c}.
\end{equation}

We highlight here that we are exploring a range of input $\Dnu\ind{c}$ values for the large separation, hence we do not need to know $\Dnu$ to smooth the spectrum. The scaling relation from Eq. \ref{eqt-Dnu-c} then allows us to target a frequency range to search for $\numax$ for each smoothed spectrum. The detection of $\numax$ is thus performed without knowing $\Dnu$.

Border effects occur when convolving the power spectrum with a Gaussian. They prevent us from accessing the extreme frequencies and, thus, to measure very low and very high $\numax$. In practice, the FRA pipeline is limited to $\numax \gtrsim 10 \, \mu$Hz. Using 30-mins cadence data, FRA is in addition limited to $\numax \lesssim 270 \, \mu$Hz, but this limitation can be overcome using higher cadence data, for example in the 2-mins cadence.

We here distinguish between two slightly different configurations of the $\numax$ measurement within the FRA pipeline:
\begin{itemize}
\item FRA1 applies to each of the 20 smoothed spectra;
\item FRA2 applies to a unique smoothed spectrum constructed by taking the median of the power spectrum density among all the 20 smoothed spectra.
\end{itemize}
FRA2 presents the advantage to be much faster than FRA1. However, for a given $\numax$, the height of the Gaussian envelope of oscillations appears smaller with FRA2 compared to FRA1. This effect is particularly emphasized at high $\numax$, since the power spectral density at $\numax$ decreases with $\numax$. Using FRA2 thus requires data allowing the height of the Gaussian envelope to be large enough in order to ensure that oscillations associated to a high $\numax$ value are still detectable. This is the case for long observation durations such as the $\sim$ 4 years of \textit{Kepler}, which  allow the power spectral density of oscillations to remain significantly high even at high $\numax$ (see Sect. \ref{kepler-numax}). However, the Gaussian envelope of oscillations is too diluted at high $\numax$ to be detectable using FRA2 for TESS stars, with observation durations ranging from 1 month to 1 year, for which FRA1 is required (see Sect. \ref{TESS-numax}).



\subsection{Local fitting of oscillations}\label{local-fit}

For each of the 20 individual smoothings performed in Sect.~\ref{smoothing}, we fit a Gaussian envelope $G(\nu)$ for the oscillations with a local contribution for the background $B(\nu)$, such that
\begin{equation}\label{eqt-oscillations}
P\ind{G}(\nu) = G(\nu) + B(\nu),
\end{equation}
with \citep{Mosser_2012b}
\begin{equation}\label{eqt-background}
B(\nu) = \alpha \left (\frac{\nu}{\nu\ind{c}}\right )^\beta,
\end{equation}
where $\alpha$ and $\beta$ are free parameters.

The Gaussian $G(\nu)$ is centered around $\nu\ind{c}$, and its standard deviation $\sigma\ind{G}$ is computed as \citep{Mosser_2010}
\begin{equation}\label{eqt-sigma-G}
\sigma\ind{G} = \frac{\delta\nu\ind{G}}{2 \sqrt{2 \ln 2}},
\end{equation}
where $\delta\nu\ind{G}$ is the FWHM of the Gaussian estimated as \citep{Mosser_2010}
\begin{equation}\label{eqt-dnuG}
\delta\nu\ind{G} = 0.59 \, \nu\ind{c}^{0.90}.
\end{equation}

The shot noise in the raw power spectrum follows a two-degree-of-freedom chi-square statistics \citep[e.g.][]{Garcia_2019}. However, the smoothed power spectrum is approximately normally distributed as a consequence of the Central Limit Theorem. We therefore use Gaussian statistics for the minimization to fit the oscillations. We set bounds on the variables and use the Trust Region Reflective optimization algorithm \citep{Branch_1999}. The fit to the power spectrum by Eq. \ref{eqt-oscillations} is performed over a frequency range encompassing $\nu\ind{c} \, \pm 1.2 \, \delta\nu\ind{G}$, with the following boundaries for the free parameters:
\begin{itemize}
\item the center of the Gaussian, $\nu\ind{c}$, is explored over the frequency range $[\nu\ind{c} - \delta\nu\ind{env} + \delta\nu\ind{G}/3$, $\nu\ind{c} + \delta\nu\ind{env} - \delta\nu\ind{G}/3]$ with $\delta\nu\ind{env}$ defined by Eq. \ref{eqt-dnu-env} and $\delta\nu\ind{G}$ defined by Eq. \ref{eqt-dnuG};
\item the amplitude of the Gaussian is explored in the range $[ P\ind{c}/15$, $15 \, P\ind{c} ]$ where $P\ind{c}$ is the smoothed PSD at $\nu\ind{c}$;
\item the standard deviation of the Gaussian, $\sigma\ind{G}$, is allowed to vary by $\pm$ 10\% from Eq. \ref{eqt-sigma-G};
\item the background parameter $\alpha$ in Eq. \ref{eqt-background} is explored between 0 and the maximum value of the smoothed PSD present in the smoothed spectrum in the frequency range where we are fitting the bump of oscillations;
\item the background exponent $\beta$ in Eq. \ref{eqt-background} is explored in the range $[-5$, $0]$ \citep{Mosser_2012b}.
\end{itemize}

\begin{figure}
\centering
\includegraphics[width=9.1cm]{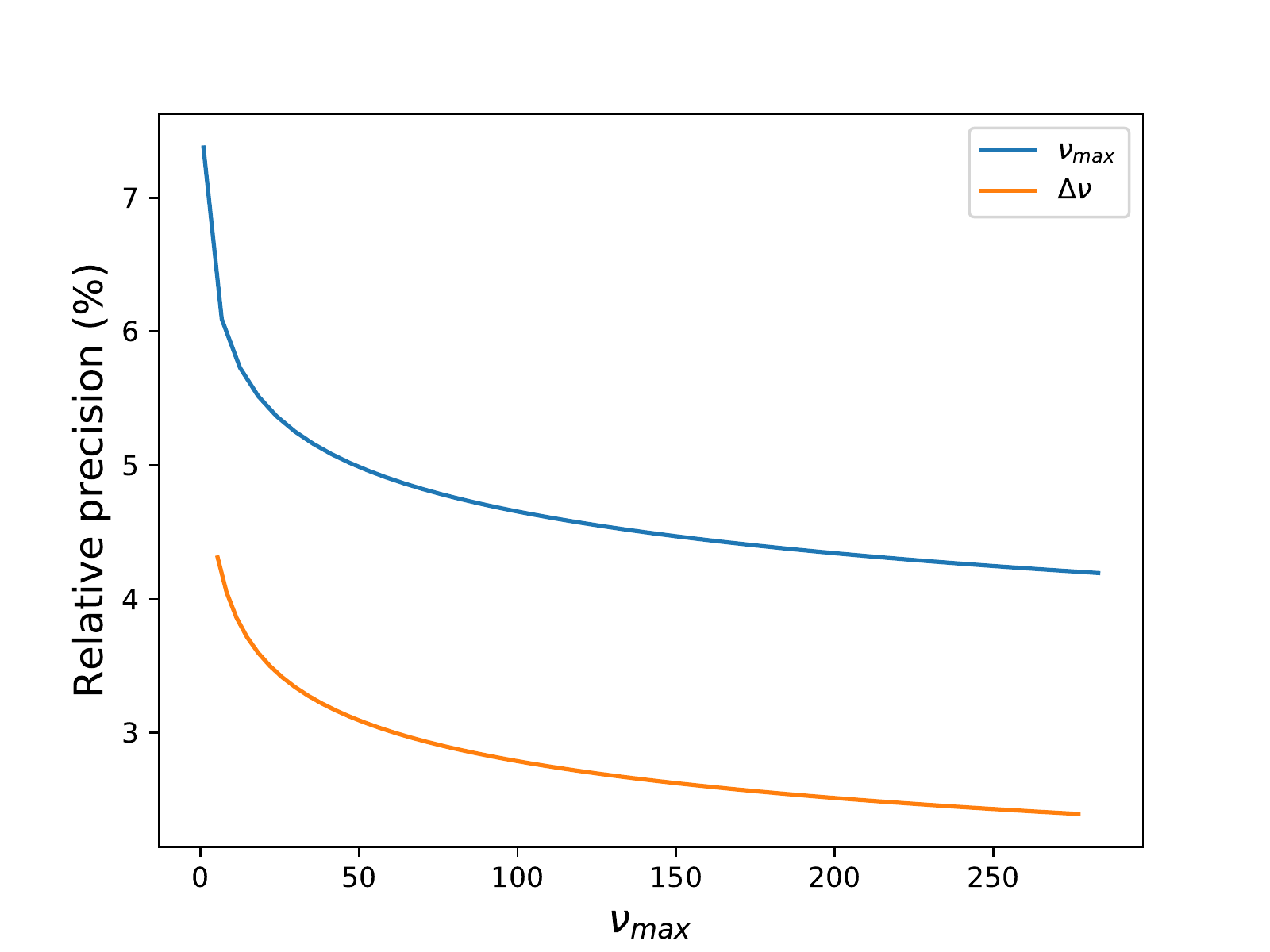}
\caption{Relative precision on the $\numax$ measurement (blue line) and the $\Dnu$ measurement (orange line) as a function of $\numax$. The relative precision on $\numax$ has been computed using Eqs. \ref{eqt-dnuG} and \ref{eqt-numax-uncertainty}. The relative precision on $\Dnu$ has been computed using Eq. \ref{eqt-precision-Dnu} with $A\ind{lim} = 10$ and $\delta\nu\ind{H}$ estimated with Eq. \ref{eqt-dnuH}.}
\label{fig-rel-precision}
\end{figure}

\begin{figure*}[h!]
\centering
\includegraphics[width=8.8cm]{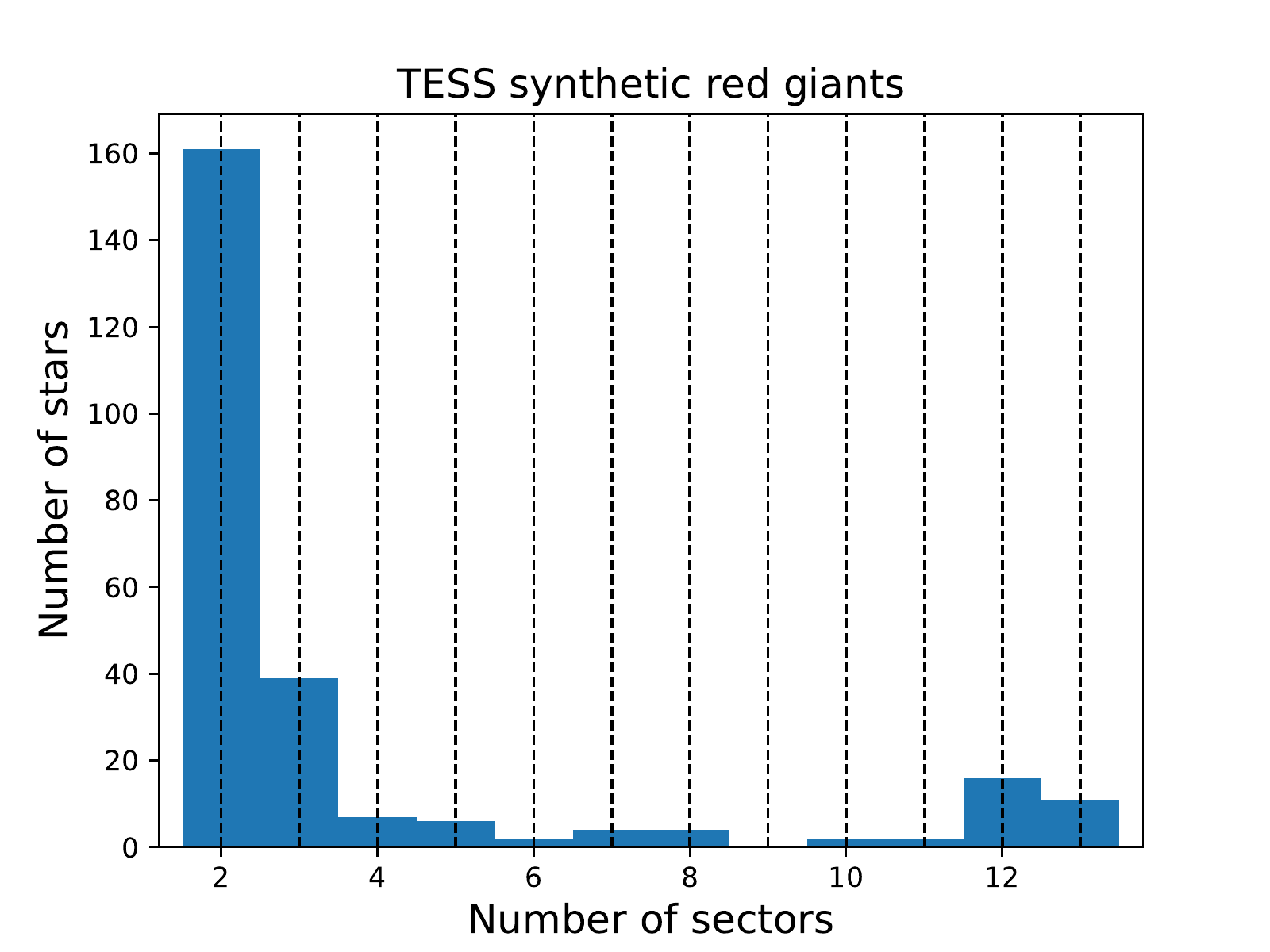}
\includegraphics[width=8.8cm]{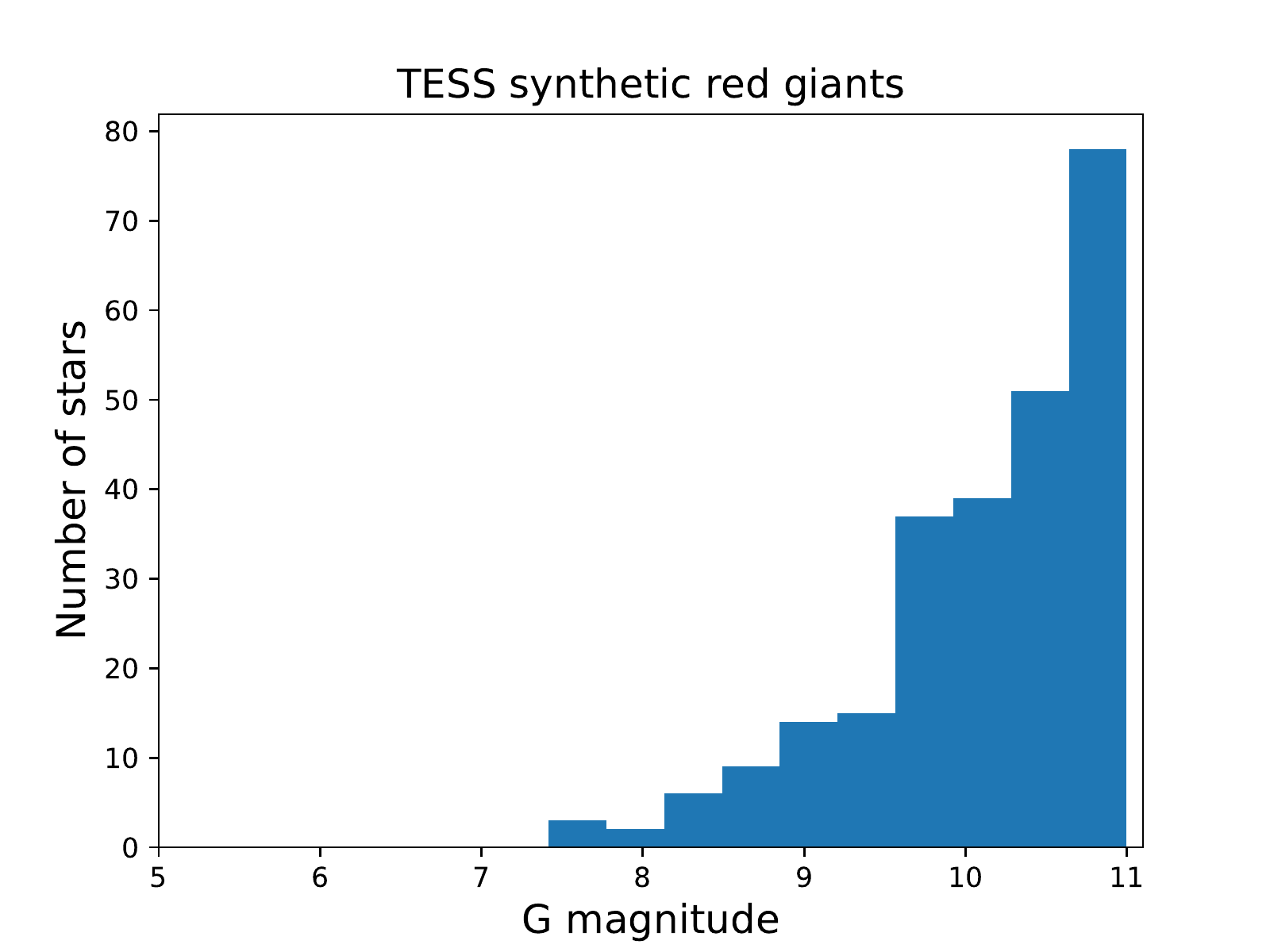}
\caption{Characteristics of the sample of 254 artificial red giants analyzed in this study. \textit{Left:} Distribution of the number of observed TESS sectors. Vertical dashed lines indicate sector numbers between 2 and 13. \textit{Left:} Distribution of the G magnitude.}
\label{fig-TESS-synthetic}
\end{figure*}

\subsection{Measuring $\numax$ and associated uncertainty}

Our FRA pipeline then keeps the configuration corresponding to the highest relative variation between the smoothed PSD at $\numax$ and the local background at $\numax$, ensuring that the bump of oscillations has a PSD at $\numax$ significantly above the local background. This step provides a first estimate of $\numax$. We then obtain an estimate of $\Dnu$ through Eq. \ref{eqt-Dnu-c}, and the optimal smoothing is performed with a Gaussian filter of FWHM computed through Eq. \ref{eqt-dnu-env}. The step presented in Sect.~\ref{local-fit} is then performed again, to refine the measurement of $\numax$ (Fig.~\ref{fig-spectrum}). We also fit a model without oscillations, including only a local background contribution (Eq. \ref{eqt-background}). We then compare the significance of these two models with respect to the data through the likelihood ratio test \citep{Appourchaux_1998}. This hypothesis test compares the goodness-of-fit of two models to determine which offers a better fit to the data. 
The second model includes the first one, with some additional parameters. The null hypothesis is that the data is better fitted by the first model with less parameters. The likelihood ratio test writes
\begin{equation}
R = -2 \, \ln \left ( \frac{\mathcal{L\ind{1}}}{\mathcal{L\ind{2}}} \right ),
\end{equation}
where $\mathcal{L\ind{1}}$ (resp. $\mathcal{L\ind{2}}$) is the likelihood of the model without oscillations (resp. with oscillations). 
The likelihood ratio approximately follows a chi-square distribution under the null hypothesis, where the number of degrees of freedom correspond to the number of additional parameters used in the model including oscillations compared to the model without oscillations \cite{Wilks_1938}. Since the model with oscillations including 2 additional parameters compared to the model without oscillations, we used used a chi-square statistics with 2 degrees of freedom.
We then compute the p-value, which is the probability of obtaining the observed results by assuming that the null hypothesis is true. We consider that the measured $\numax$ is significant and that oscillations are detected if the p-value is below 1\%.

\begin{figure*}[h!]
\centering
\includegraphics[width=8.8cm]{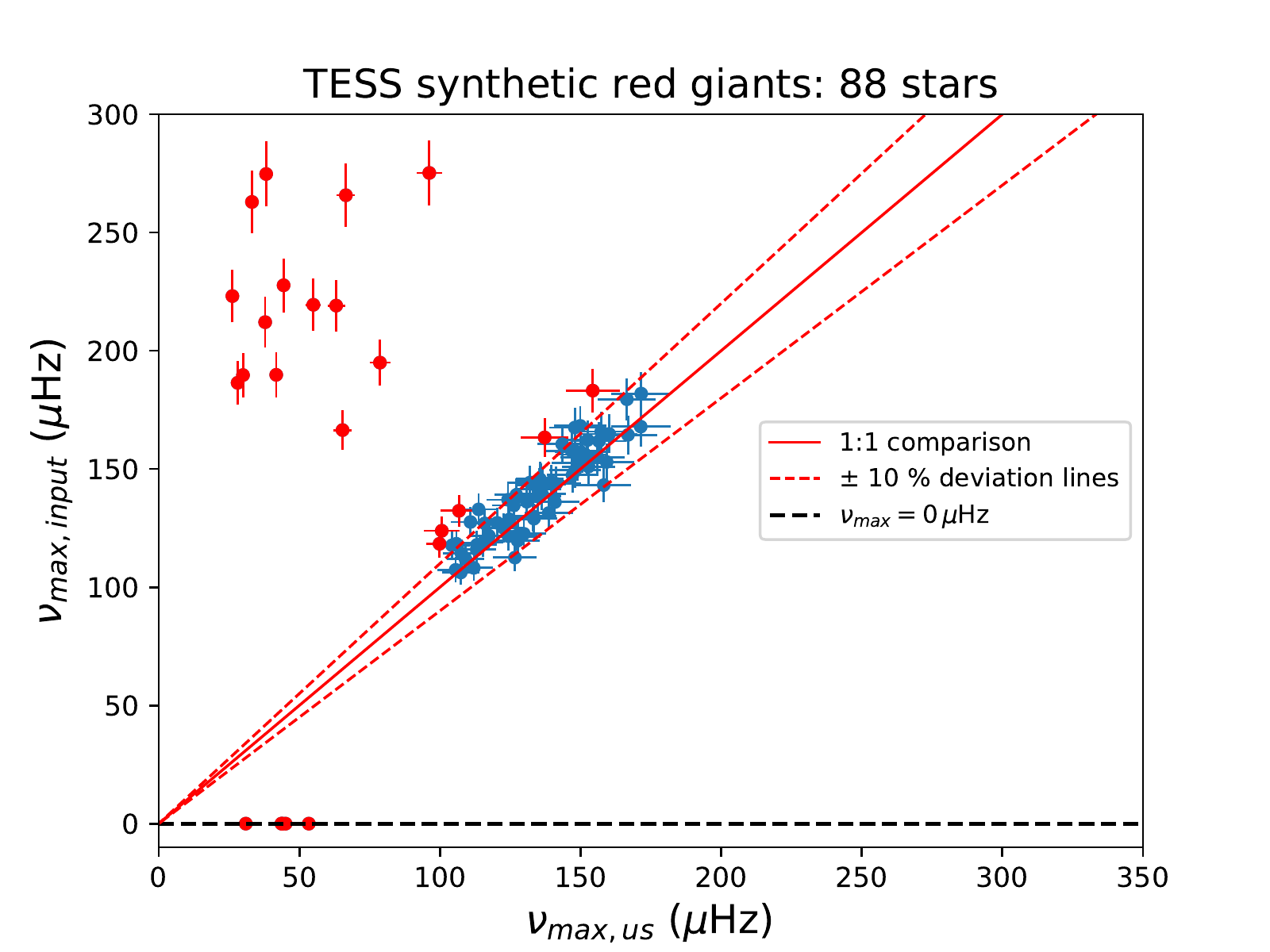}
\includegraphics[width=8.8cm]{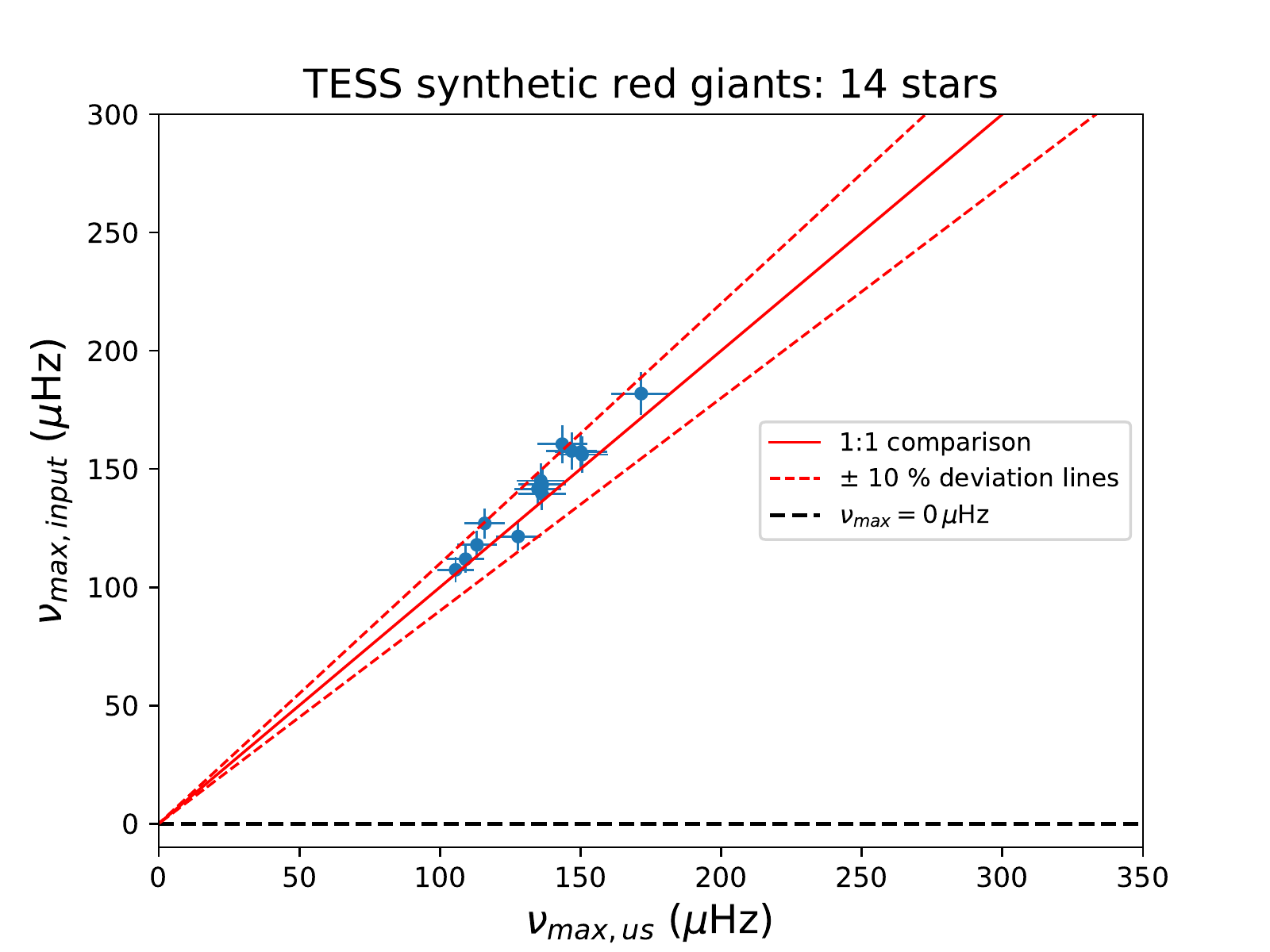}
\caption{\textit{Left:} comparison between our measurents and the input $\numax$ values for synthetic TESS targets. Red dots correspond to stars for which the relative deviation between measurements is at least 10\%. The red line represents a 1:1 comparison while the red dashed lines correspond to a deviation of $\pm$ 10 \% from the 1:1 comparison. The horizontal dashed lines indicate an input $\numax$ value of 0 $\mu$Hz, meaning that no oscillations were injected in the power spectrum. \textit{Right:} same as the left panel, but for $N\ind{sectors} > 3$ and ensuring that the associated $\Dnu$ measurement is consistent for G > 9.5.}
\label{fig-TESS-numax-synthetic}
\end{figure*}

\begin{table*}
\caption{Characteristics of the $\numax$ measurement for 254 synthetic power spectra representative of TESS red giants with a G magnitude below 11.}
\label{table:synthetic-numax}
\centering
\begin{tabular}{c c c c c c}
\hline\hline
Parameter & Detection & Consistency rate & Median relative & Median time & False positive\\
& rate & with input values & precision & per star & rate\\
\hline
$\numax$ & 43.2 \% & 77.1 \% & 6.2 \% & 1.7 s & 8.1 \%\\
$\numax$ extra step for mag > 9.5 & 28.0 \% & 100 \% & -- & 11.4 s & 0 \%\\
\hline
\end{tabular}
\begin{tablenotes}
\item Our measurements are considered as consistent when the relative deviation with the existing ones is below 10\%. 
\end{tablenotes}
\end{table*}

\begin{figure*}[h!]
\centering
\includegraphics[width=8.8cm]{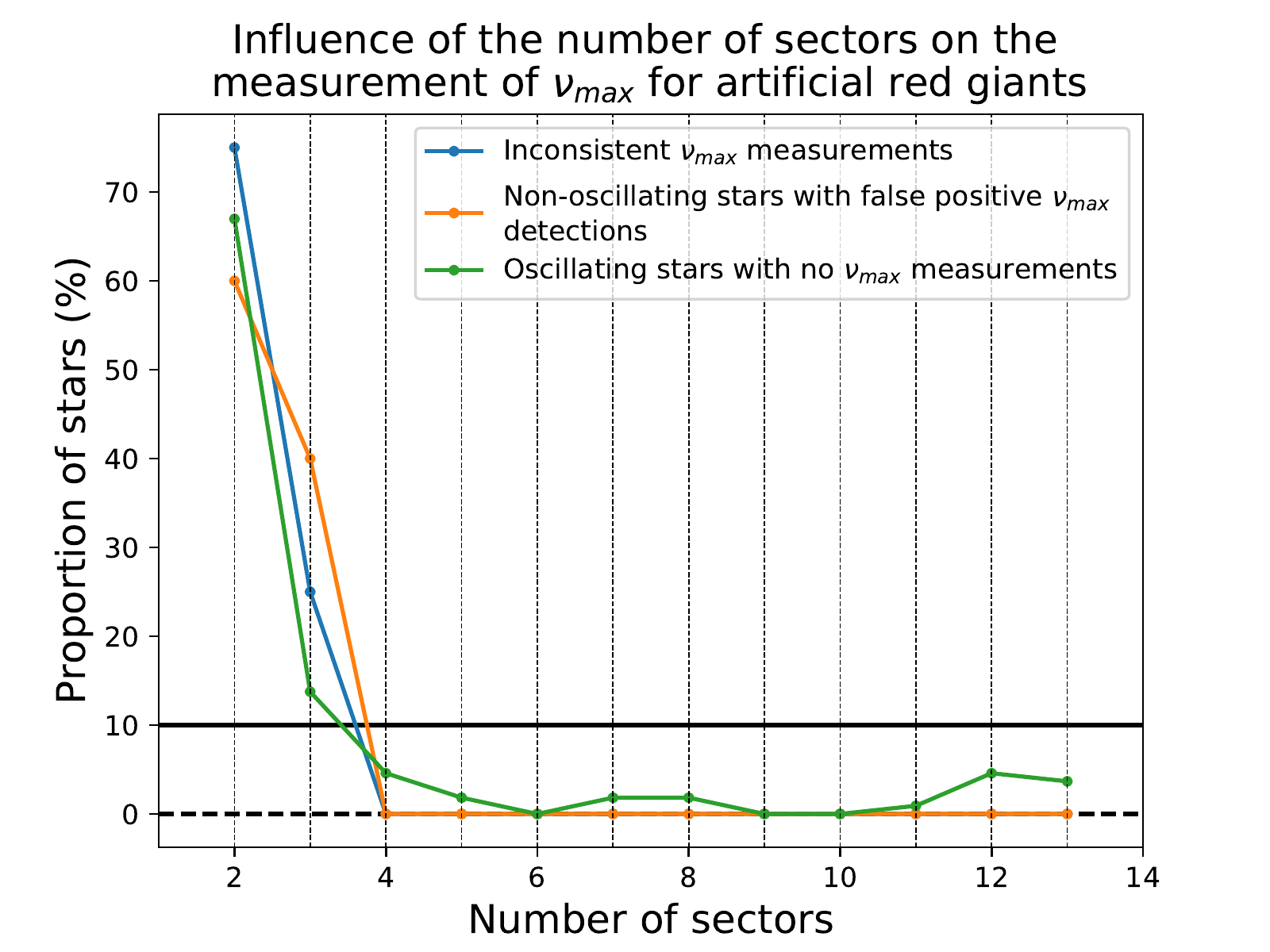}
\includegraphics[width=8.8cm]{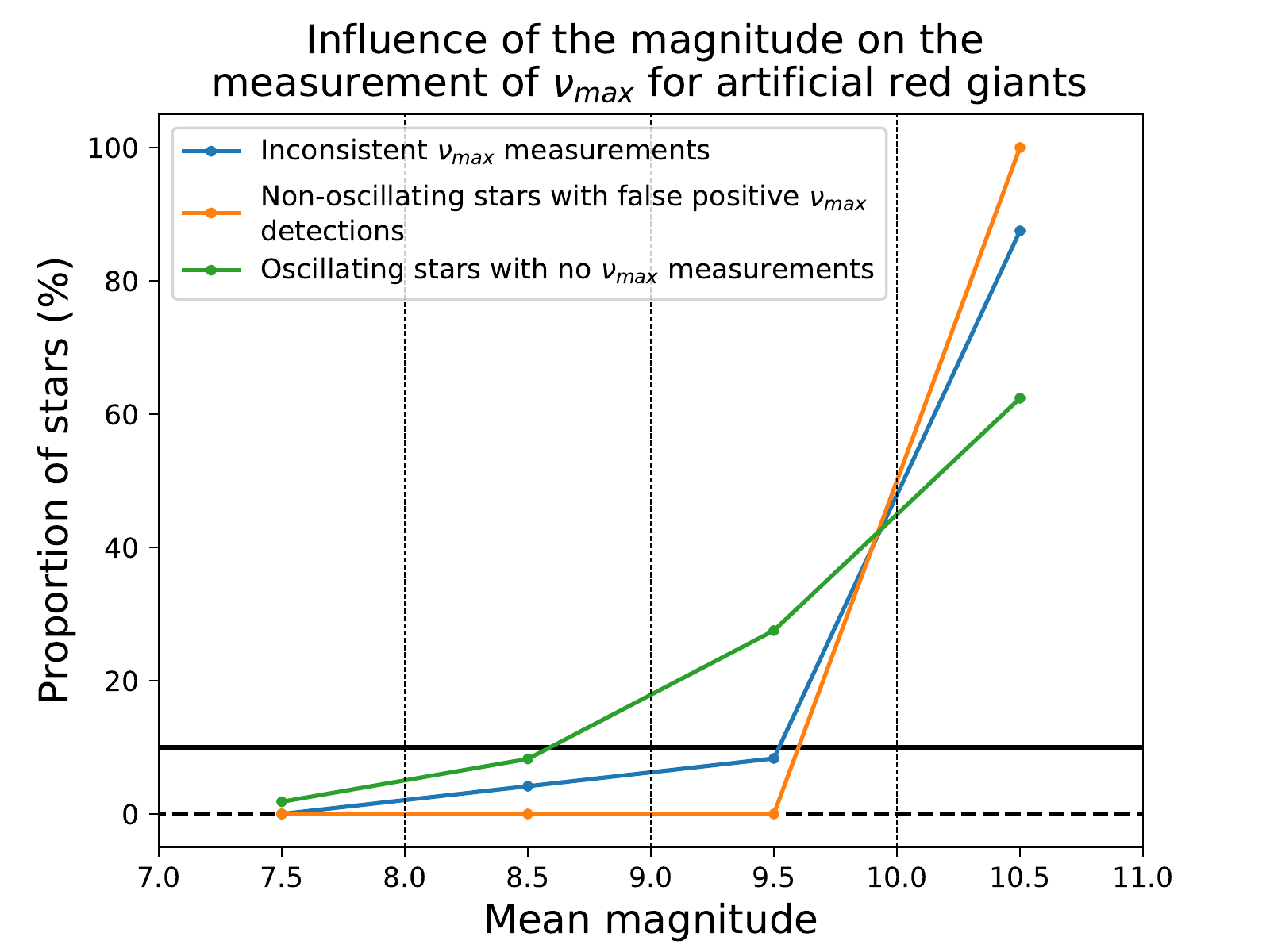}
\caption{Proportion of artificial red giants as a function of the number of TESS sectors (left panel) and the mean stellar magnitude (right panel). The blue curve corresponds to stars for which we derive inconsistent $\numax$ measurements, the orange curve to non-oscillating stars for which we obtain false positive $\numax$ measurements, and the green curve to oscillating stars for which we do not derive a $\numax$ measurement. Horizontal dashed lines indicate a proportion of stars of 0 \%, while the horizontal continuous line represents a proportion of stars of 10 \%.}
\label{fig-TESS-synthetic-impact-mag-Nsectors-numax}
\end{figure*}

We assume that the uncertainty on $\numax$ is a fraction of the width of the envelope of the oscillations, i.e a fraction of the FWHM of the fitted Gaussian. We determine the $1-\sigma$ uncertainty on $\numax$ such as
\begin{equation}\label{eqt-numax-uncertainty}
\sigma\ind{\numax} = \delta\nu\ind{G}/8,
\end{equation}
We checked that dividing the FWHM of the Gaussian by a factor of 8 results in uncertainties that are typical for TESS stars, i.e a relative precision of $\sim$ 5-6\% (Fig. \ref{fig-rel-precision}).

The relative precision is improved as $\numax$ increases and, therefore, as $\Dnu$ increases based on Eq. \ref{eqt-Dnu-c} (Fig.~\ref{fig-rel-precision}).





\section{Test and validation on artificial TESS oscillation spectra}\label{test}

In order to assess the performance of our methods, we applied them to a set of artificial power spectra representative of TESS. We used the artificial lightcurves produced by \cite{Campante_2018} for $\sim$ 30 000 LLRGB stars in 30-mins cadence and introduced gaps every 13.7 days to be consistent with TESS observations. We then derived artificial power spectra through a Lomb-Scargle periodogram, with an oversampling factor of 10 to match the power spectra analysed by \cite{Mackereth} and to maximise the detectability of oscillations. We selected the artificial power spectra which have the same number of observed TESS sectors, i.e. in the range $[2, 13]$, compared to the sample of TESS stars analysed by \cite{Mackereth}. The selected sample includes a much higher number of targets with a low number of observed sectors compared to TESS stars from \cite{Mackereth} (Fig.~\ref{fig-TESS-synthetic}). We additionally selected the artificial power spectra corresponding to stellar G magnitudes below 11, as the sample analysed by \cite{Mackereth}, since we do not expect to reliably detect oscillations for stars with larger stellar magnitudes. The selected sample has a G magnitude in the range $[7.4, 11]$ (Fig.~\ref{fig-TESS-synthetic}). In total, we used a set of 254 artificial power spectra, of which 192 (75.6 \%) have injected oscillations and 62 (24.4 \%) have oscillations with $\numax$ above the Nyquist frequency, and are thus considered as not having injected oscillations.

\subsection{Measurement of $\numax$}

We used the FRA1 method to analyse artificial power spectra representative of TESS red giants, as described in Sect.~\ref{methods}. We detect oscillations and measure $\numax$ for 88 stars (Fig.~\ref{fig-TESS-numax-synthetic}). These include 5 stars which have no injected oscillations, representing a false positive detection rate of 8.1 \% (among the 62 stars with no injected oscillations, Table \ref{table:synthetic-numax}). We also detect oscillations and derive a $\numax$ measurement for 83 stars with injected oscillations, representing a detection rate of 43.2 \% (among the 192 stars with injected oscillations). For 19 stars with injected oscillations, we derived a $\numax$ value with a relative deviation of at least 10\% compared to the input value, giving a consistency rate of 77.1 \%. We discuss in Sect.~\ref{mag-sectors-artificial-numax} the reasons for this rather high false positive rate and low consistency rate, and we demonstrate in Sect.~\ref{discussion-optimized-numax-synthetic} that we obtain a 100 \% consistency rate and a false positive detection rate of 0 \% by simply adding an extra validation step when analyzing stars with a G magnitude above 9.5 (Table \ref{table:synthetic-numax}).
We obtain a median relative precision of 6.2 \% and a median computation time spent for each star of 1.7 s.

\begin{figure*}[h!]
\centering
\includegraphics[width=9.1cm]{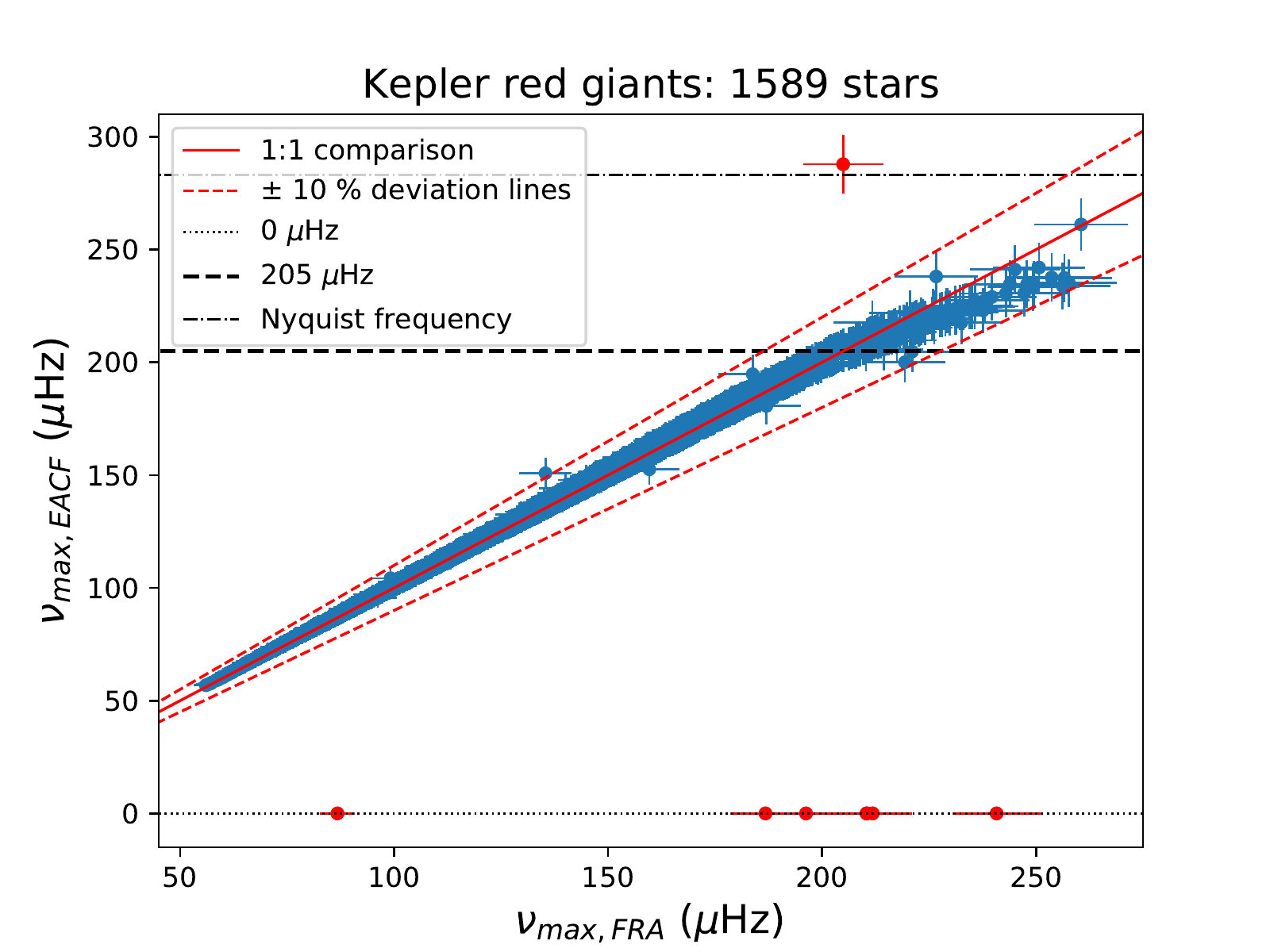}
\caption{Comparison between our $\numax$ measurements and existing ones for \textit{Kepler} stars. The color code is the same as in Fig.~\ref{fig-TESS-numax-synthetic}. The horizontal dashed line indicates $\numax = 205$ $\mu$Hz. Dotted lines and dot-dashed lines represent $\numax = 0 \, \mu$Hz and the Nyquist frequency of $283 \, \mu$Hz, respectively.}
\label{fig-Kepler-numax}
\end{figure*}

\subsection{Impact of the stellar magnitude and the number of TESS sectors on the detectability and consistency of $\numax$}\label{mag-sectors-artificial-numax}

We here discuss the impact of the stellar magnitude and the number of observed TESS sectors on the detectability of $\numax$ and as well as on the consistency of our measurements. We consider that inconsistent measurements for $\numax$ are associated to a relative deviation of at least 10\% compared to the input values. To that end, we computed the fraction of stars with inconsistent measurements falling in each stellar magnitude bin (resp. having a given number of observed TESS sectors). We then compared this distribution to the one obtained for stars for
which we get consistent measurements in order to highlight any trend difference between these two populations.

We find that the stellar magnitude and the number of observed TESS sectors impact the consistency, the detectability and the false detection rate of $\numax$ (Fig.~\ref{fig-TESS-synthetic-impact-mag-Nsectors-numax}). We note that the proportions of stars with inconsistent measurements, of non-oscillating stars with false positive measurements, and of oscillating stars with no measurement, exceed 10 \% for $N\ind{sectors} \leq 3$ and increase with stellar magnitude. We note that the proportions of stars with inconsistent $\numax$ measurements and of non-oscillating stars with false positive $\numax$ measurements exceed 10 \% for a stellar magnitude above 9, while the proportion of oscillating stars with no $\numax$ measurement exceeds 10 \% for a stellar magnitude above 9.


So far, \cite{Mackereth} only assessed the yield of seismic detection as a function of stellar magnitude and of the number of TESS sectors, while not distinguishing between the detection yield obtained from the $\numax$ measurement and the detection yield obtained from the $\Dnu$ measurement. Moreover, \cite{Mackereth} did not apply their pipeline to artificial power spectra. They thus did not address the false positive detection rate, neither as the consistency rate of their measurements. They also did not address the impact of the number of TESS sectors and of the stellar magnitude on these rates.

 \begin{figure*}[h!]
\centering
\includegraphics[width=8.8cm]{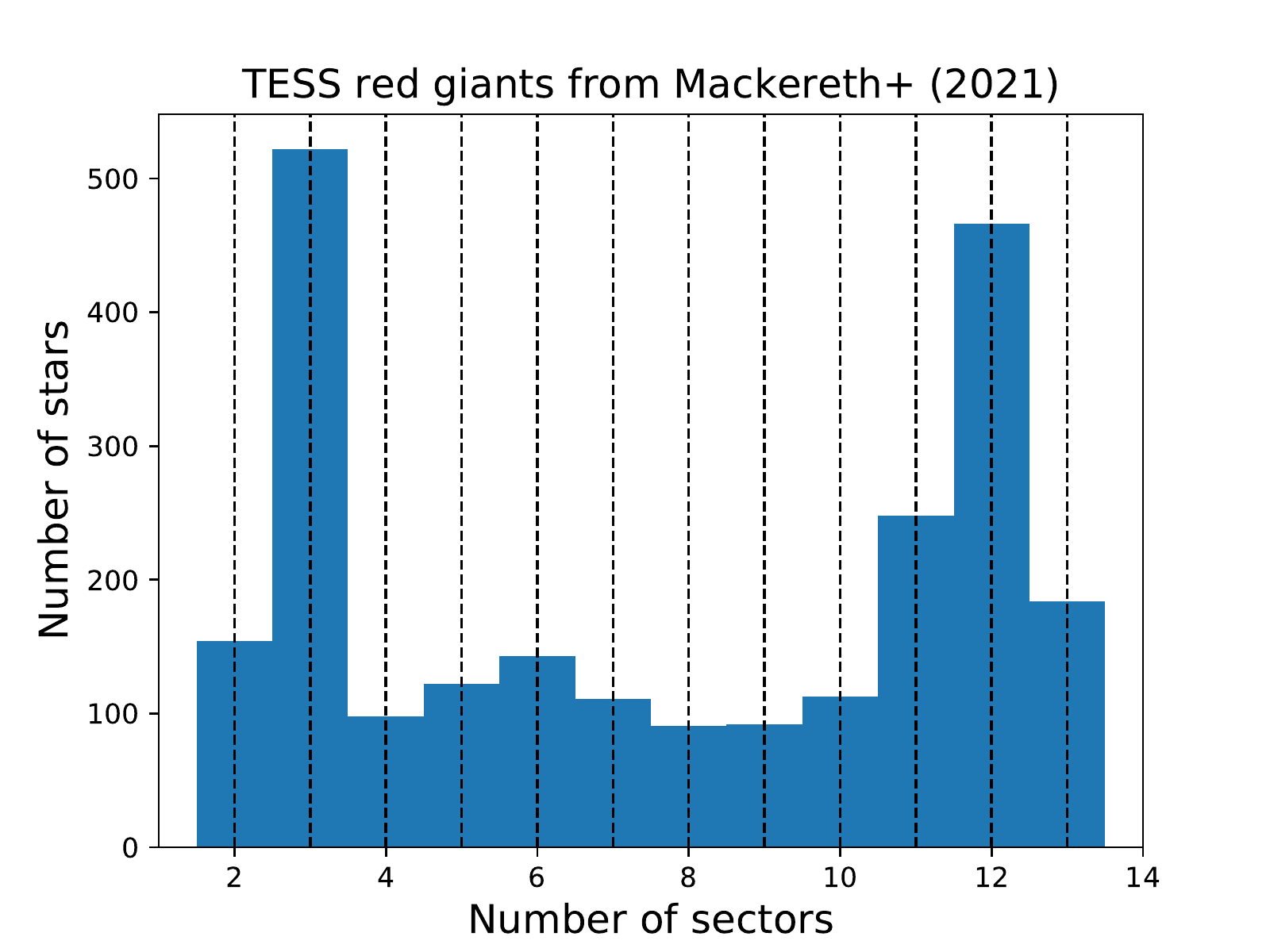}
\includegraphics[width=8.8cm]{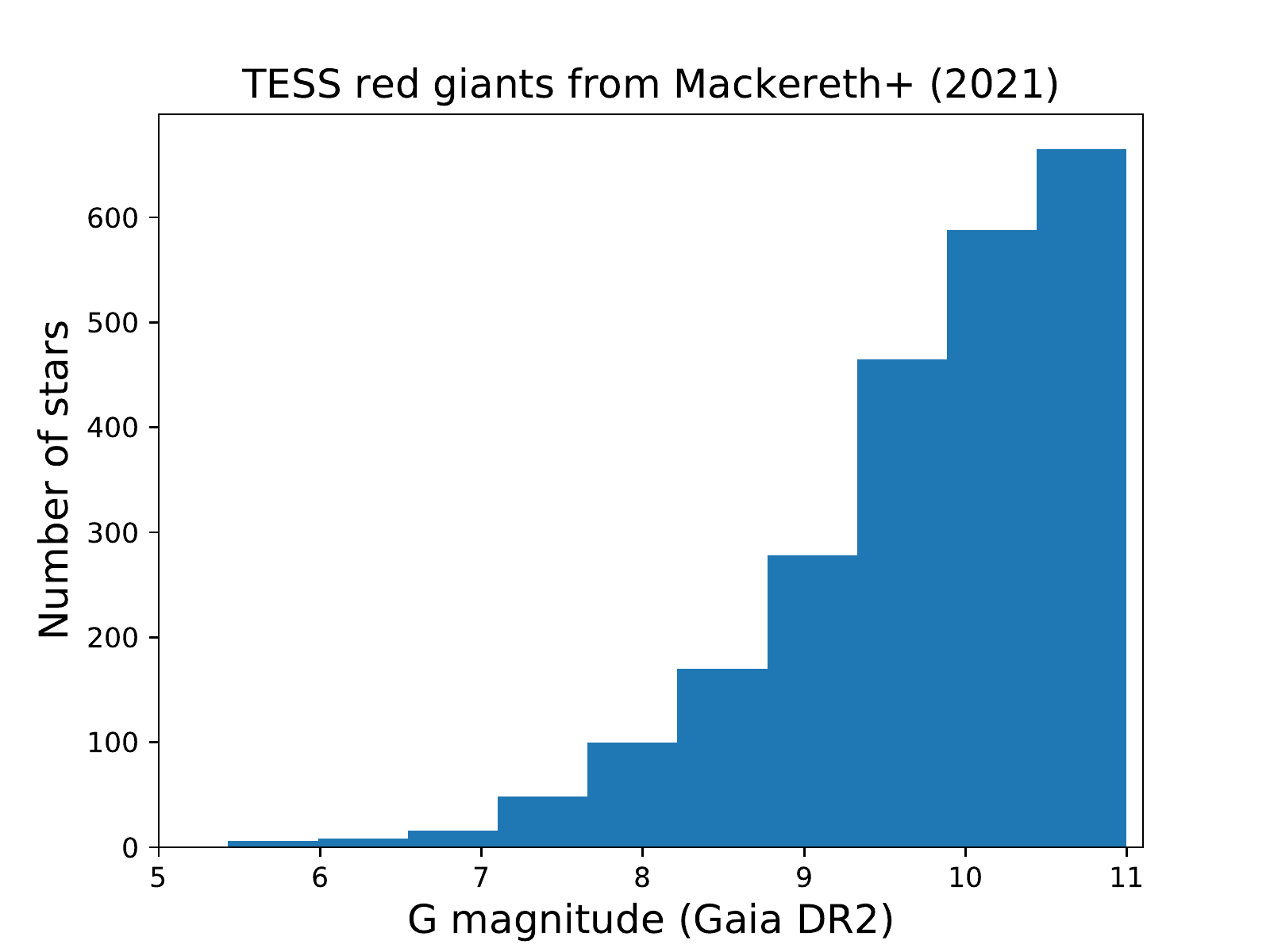}
\caption{Same Figure as Fig.~\ref{fig-TESS-synthetic} for 2344 SCVZ red giants from \cite{Mackereth}.}
\label{fig-TESS-sample}
\end{figure*}

\subsection{Optimized $\numax$ measurement}\label{discussion-optimized-numax-synthetic}

Based on the above, we selected only the artificial and observed power spectra associated to a number of observed TESS sectors $N\ind{sectors} > 3$. Additionally, we added another criterium to validate the $\numax$ values measured for a stellar magnitude above 9.5: our $\numax$ measurement for a given star is validated only if we also obtain a $\Dnu$ measurement, and this $\Dnu$ value has to be close to the $\Dnu$ estimated using $\numax$ from Eq.~\ref{eqt-numax-Dnu}, i.e. the relative deviation between these two $\Dnu$ values has to be below 10 \%. Otherwise, the $\numax$ we measure is not validated, and we consider that we do not have a $\numax$ detection. Our implementation of the EACF method to measure $\Dnu$ is described in Sect. \ref{EACF}. We now obtain a lower detection rate of 28.0 \% (i.e. for 14 stars out of 50 with injected oscillations), which is counterbalanced by a much higher consistency rate of 100 \% and a much lower false positive detection rate of 0 \% (Table~\ref{table:synthetic-numax} and right panel of Fig.~\ref{fig-TESS-numax-synthetic}). Hence, we infer that our pipeline provides consistent measurements while minimizing the false positive measurements and the non-detections for $\numax$, provided that the number of observed TESS sectors is $N\ind{sectors} > 3$. The median computation time per run is 11.4 s, which is much longer since we are now left with targets with higher number of observed TESS sectors.

\section{Results: analysing \textit{Kepler} and TESS red giants}\label{results}

We tested our FRA pipeline to measure $\numax$ for 3933 red giant stars observed by the space missions \textit{Kepler} and TESS.


\subsection{Analysis of 1589 \textit{Kepler} red giants}\label{kepler-numax}

We selected a sample of 1589 \textit{Kepler} stars with 4 year-long observations, without oversampling, belonging to the sample which \cite{Gehan_2018} and \cite{Gehan_2021} analysed to measure the mean core rotation rate and the inclination angle. We compare our results for these \textit{Kepler} red giants with $\numax$ and $\Dnu$ values measured using the COR method \citep{Mosser_2009}.

We used the FRA2 method to analyse \textit{Kepler} red giants, as described in Sect.~\ref{methods}. We were able to derive a $\numax$ measurement for all the 1589 Kepler red giants analysed in this study (Table \ref{table:numax}). We have a median relative precision of 4.6 \%, which is better than for artificial TESS targets (Table \ref{table:synthetic-numax}) because the \textit{Kepler} stars analyzed have higher $\numax$ values, which are more precise (Fig. \ref{fig-rel-precision}).

The computation time spent for each star ranges between 1.9 s and 8.3 s. This variation in the computation time comes from the fact that the number of local maxima to test in the oscillation spectrum to fit the oscillations envelope is star-dependent. The median computation time per run is 3.4 s, slightly longer compared to TESS artificial targets. Indeed, although TESS artificial targets have much shorter observation durations ranging between 1 month and 1 year, their power spectrum is oversampled with a factor of 10, increasing the computation time.

 For 8 stars, the relative deviation between $\numax$ obtained through the EACF method and our measurements is of at least 10\% (Fig.~\ref{fig-Kepler-numax}). We note that the EACF fails to propose a value for $\numax$ for 7 of these stars, appearing in the figure as having $\numax = 0$ $\mu$Hz. The EACF gives a $\numax$ value above the Nyquist frequency (for the 30-mins cadence used) for the remaining star. Our FRA pipeline successfully measures $\numax$ in these 8 cases (see \ref{appendix-1}). We note that the $\numax$ measurements obtained through the EACF method tend to be underestimated above 205 $\mu$Hz when compared to our $\numax$ values derived through the FRA pipeline, however the relative deviation remains below 10\%. This deviation seems to be due to a sharper cutoff of the edges of the spectrum when smoothing the spectrum with the EACF method compared to our method, Indeed, our $\numax$ actually targets the maximum of the PSD within the Gaussian envelope of oscillations, which is not the case for $\numax$ measured with the EACF method (see \ref{appendix-2}).

\subsection{Analysis of 2344 TESS red giants}\label{TESS-numax}

We also considered a sample of 2344 Southern Continuous Viewing Zone (SCVZ) TESS red giants for which \cite{Mackereth} indicated that COR, A2Z and BHM obtain consistent results accross the three methods using 30-mins cadence TESS data. Their power spectra are obtained using an oversampling factor of 10. This TESS sample has between 2 and 13 TESS sectors-long lightcurves and a G magnitude below 11 in order to select only stars bright enough to maximise the probability of oscillations being detectable (Fig.~\ref{fig-TESS-sample}). We compare our results for these TESS red giants with $\numax$ and $\Dnu$ values published in \cite{Mackereth}, which result from the mean of the values obtained by applying the COR \citep{Mosser_2009}, A2Z \citep{Mathur_2010} and BHM \citep{Elsworth} pipelines.

\begin{figure*}[h!]
\centering
\includegraphics[width=8.8cm]{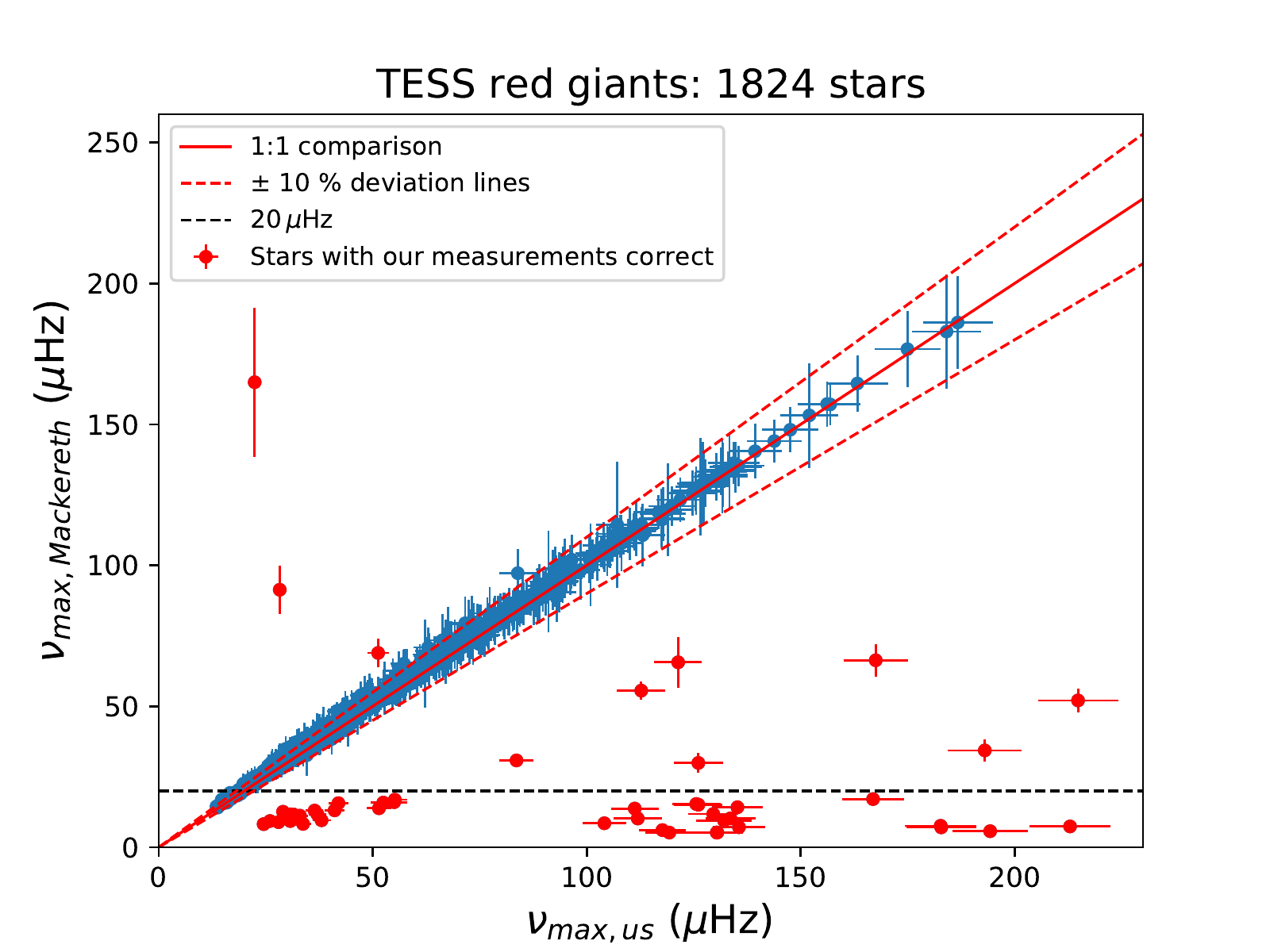}
\includegraphics[width=8.8cm]{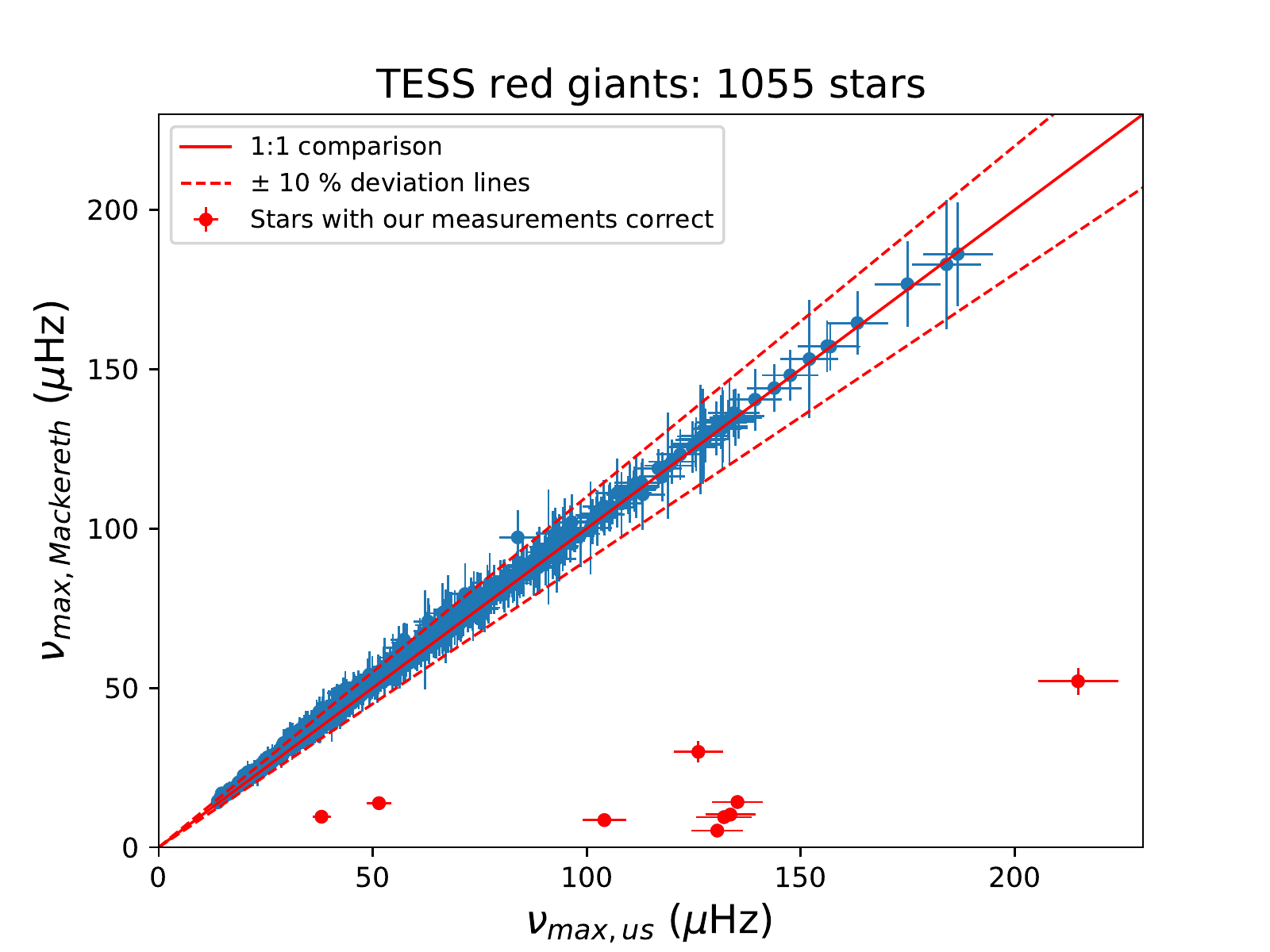}
\includegraphics[width=8.8cm]{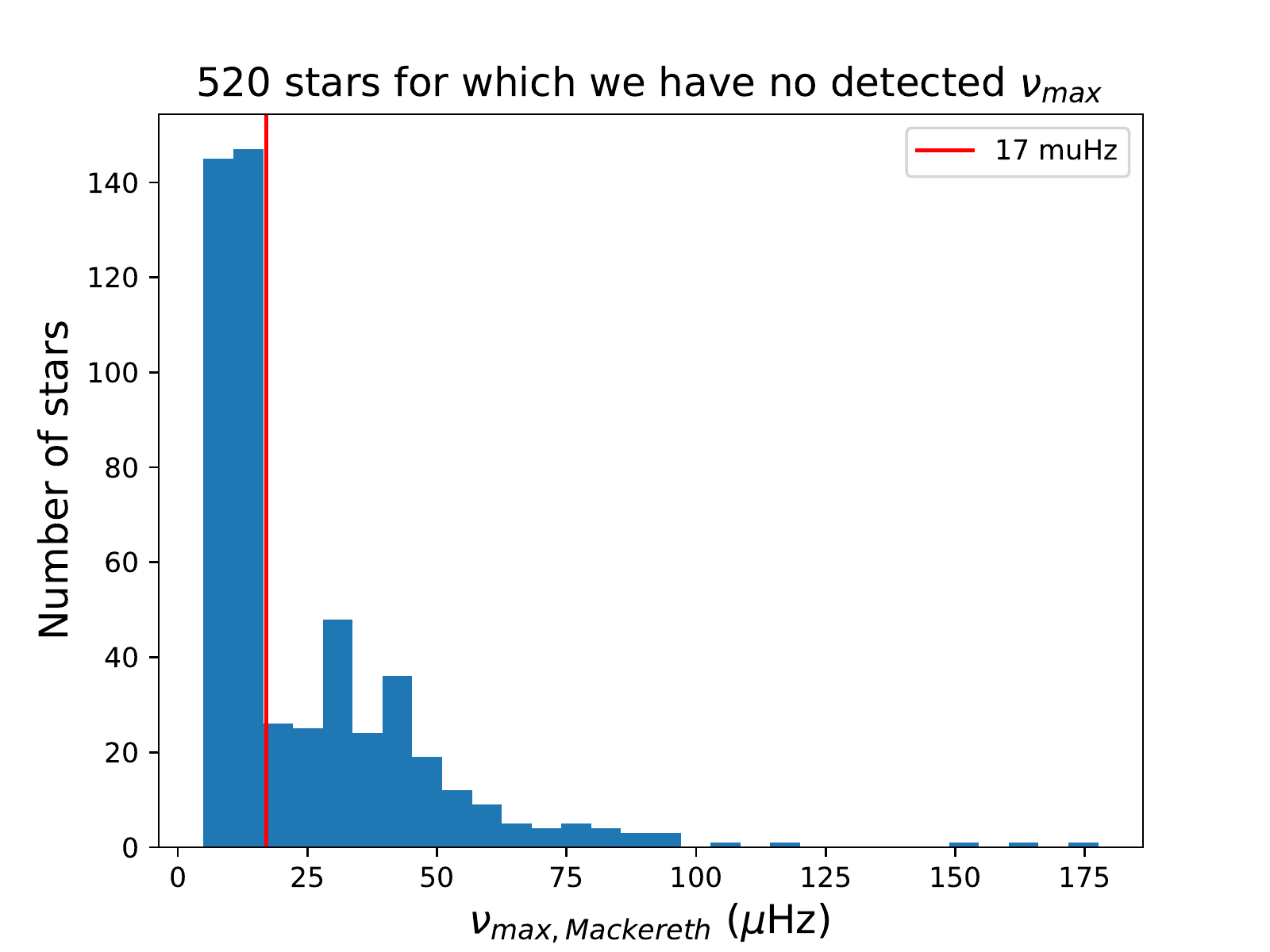}
\includegraphics[width=8.8cm]{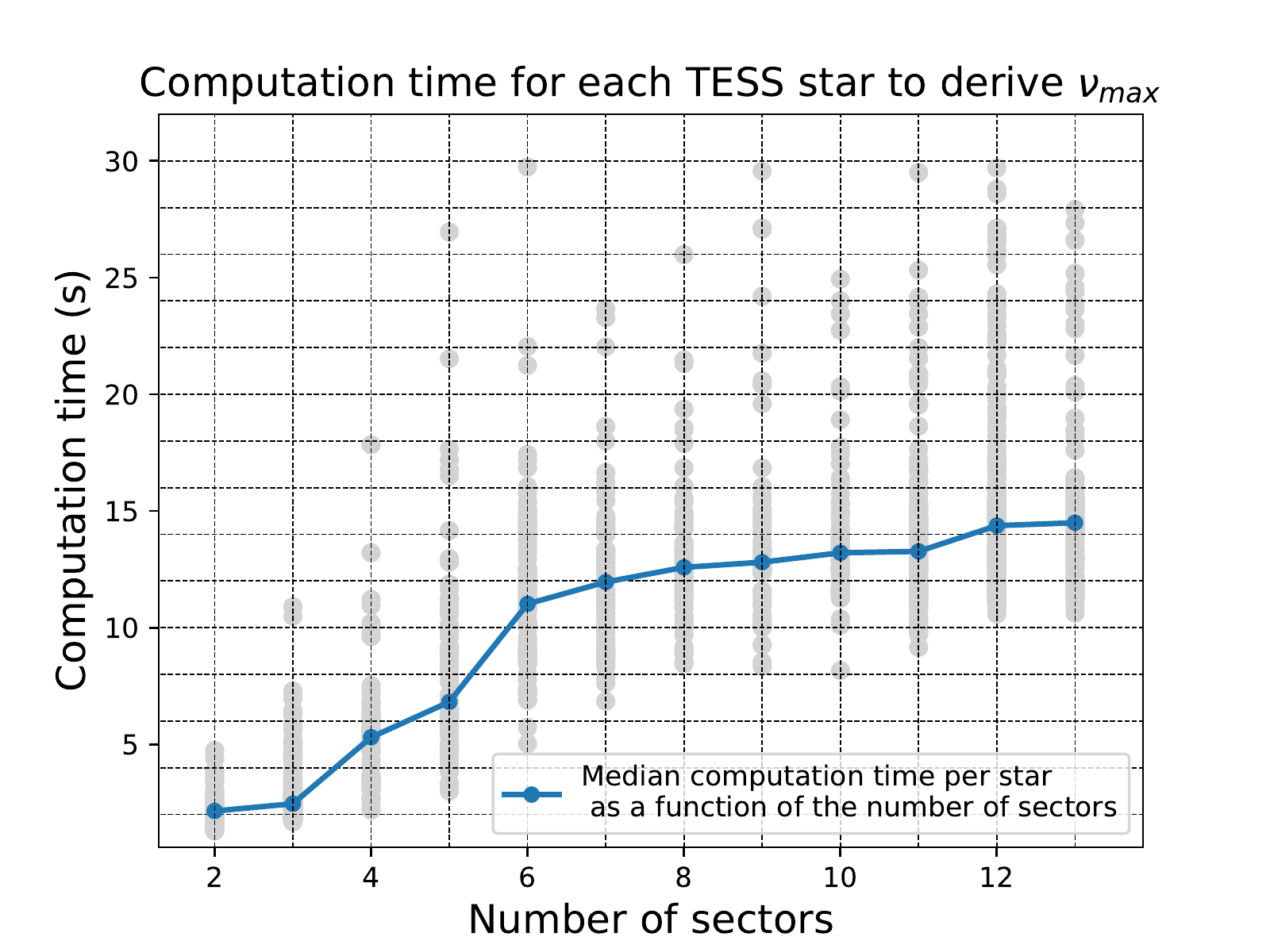}
\caption{Characteristics of the $\numax$ measurement for TESS stars. \textit{Upper left:} Same as the left panel of Fig.~\ref{fig-TESS-numax-synthetic}, but for TESS red giants. \textit{Upper right:} Same as the right panel of Fig.~\ref{fig-TESS-numax-synthetic}, but for TESS red giants. \textit{Upper right:} $\numax$ histogram of the non-detections in the case of the upper left panel. The vertical red line represents $\numax = 17 \, \mu$Hz. \textit{Bottom:} Computation time spent for each star as a function of the number of observed TESS sectors in the case of the upper left panel. The blue dots represent the median computation time. Horizontal lines indicate times with 2 s spacings.}
\label{fig-TESS-numax}
\end{figure*}

We used the FRA1 method to analyse TESS red giants, as described in Sect.~\ref{methods}. We checked that the FRA1 method maximises the consistent detection of $\numax$ for TESS stars. We derived a $\numax$ measurement for 1824 stars, i.e. 77.8 \% of the analysed sample (upper left panel of Fig.~\ref{fig-TESS-numax}). The majority of the 520 stars for which we did not detect oscillations based on the measurement of $\numax$ have low reference $\numax$ measurements, below 17 $\mu$Hz (upper right panel of Fig.~\ref{fig-TESS-numax}). We additionally note that the minimum $\numax$ we detect is 13 $\mu$Hz, while the minimum $\numax$ measured by \cite{Mackereth} is 7 $\mu$Hz. Indeed, as stated in Sect. \ref{smoothing}, we cannot use the extreme frequency edges of the smoothed spectrum that are affected by border effects, which makes it not possible to measure low $\numax$ with our method.

 \begin{table*}
\caption{Characteristics of the $\numax$ measurement for the 1589 \textit{Kepler} and 2344 TESS stars analysed.}
\label{table:numax}
\centering
\begin{tabular}{c c c c}
\hline\hline
Sample & Consistency rate with & Median relative & Median time\\
 & existing measurement & precision & per star\\
\hline
\textit{Kepler} & 99.5 \% & 4.6 \% & 3.4 s\\
\hline
TESS & 97.3 \% & 5.3 \% & 11.6 s \\
TESS extra step for mag > 9.5 & 99.1 \% & -- & 12.3 s \\
\hline
\end{tabular}
\begin{tablenotes}
\item Our measurements are considered as consistent when the relative deviation with the existing ones is below 10\%. 
\end{tablenotes}
\end{table*}

\begin{figure*}[h!]
\centering
\includegraphics[width=8.8cm]{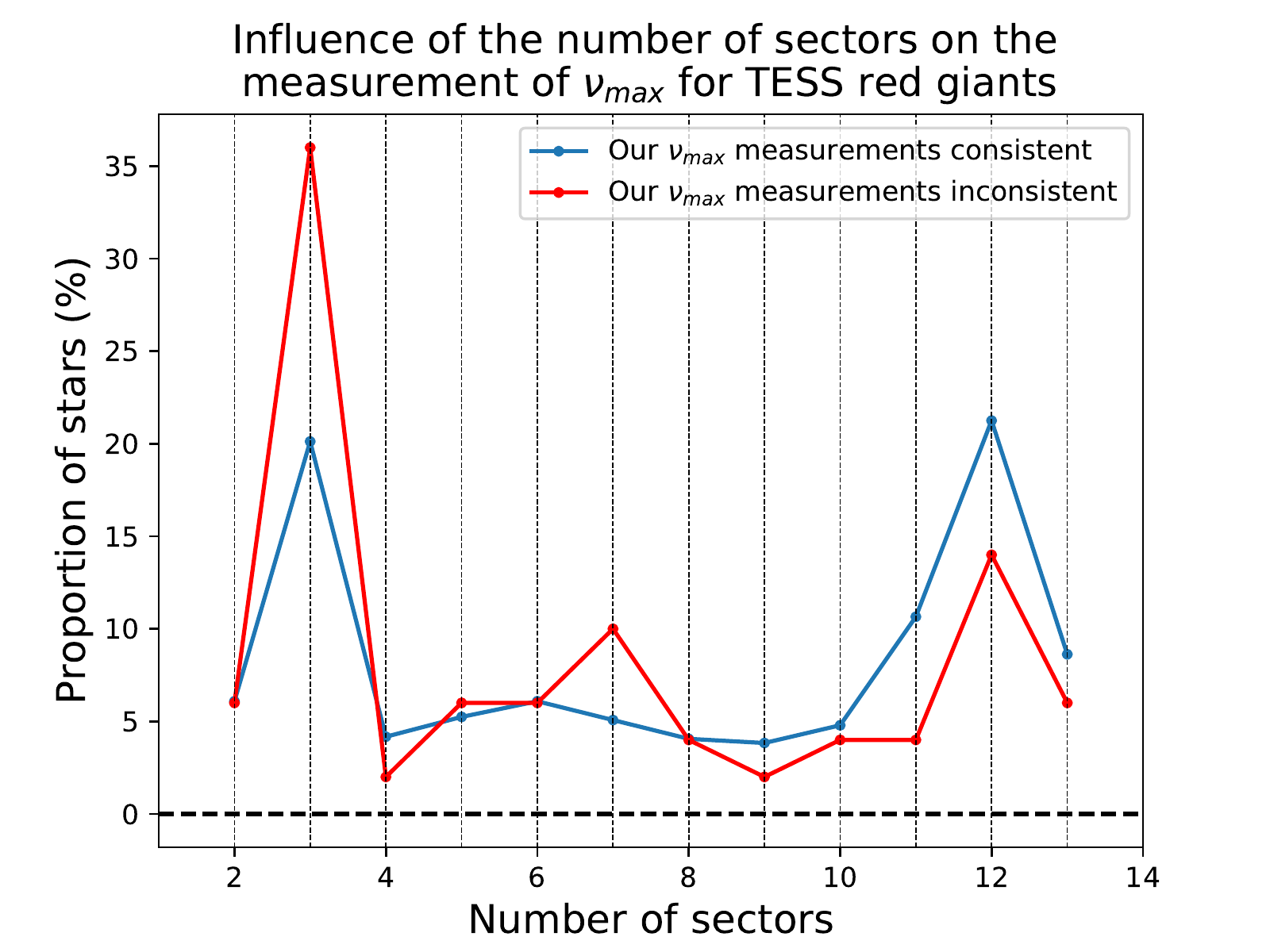}
\includegraphics[width=8.8cm]{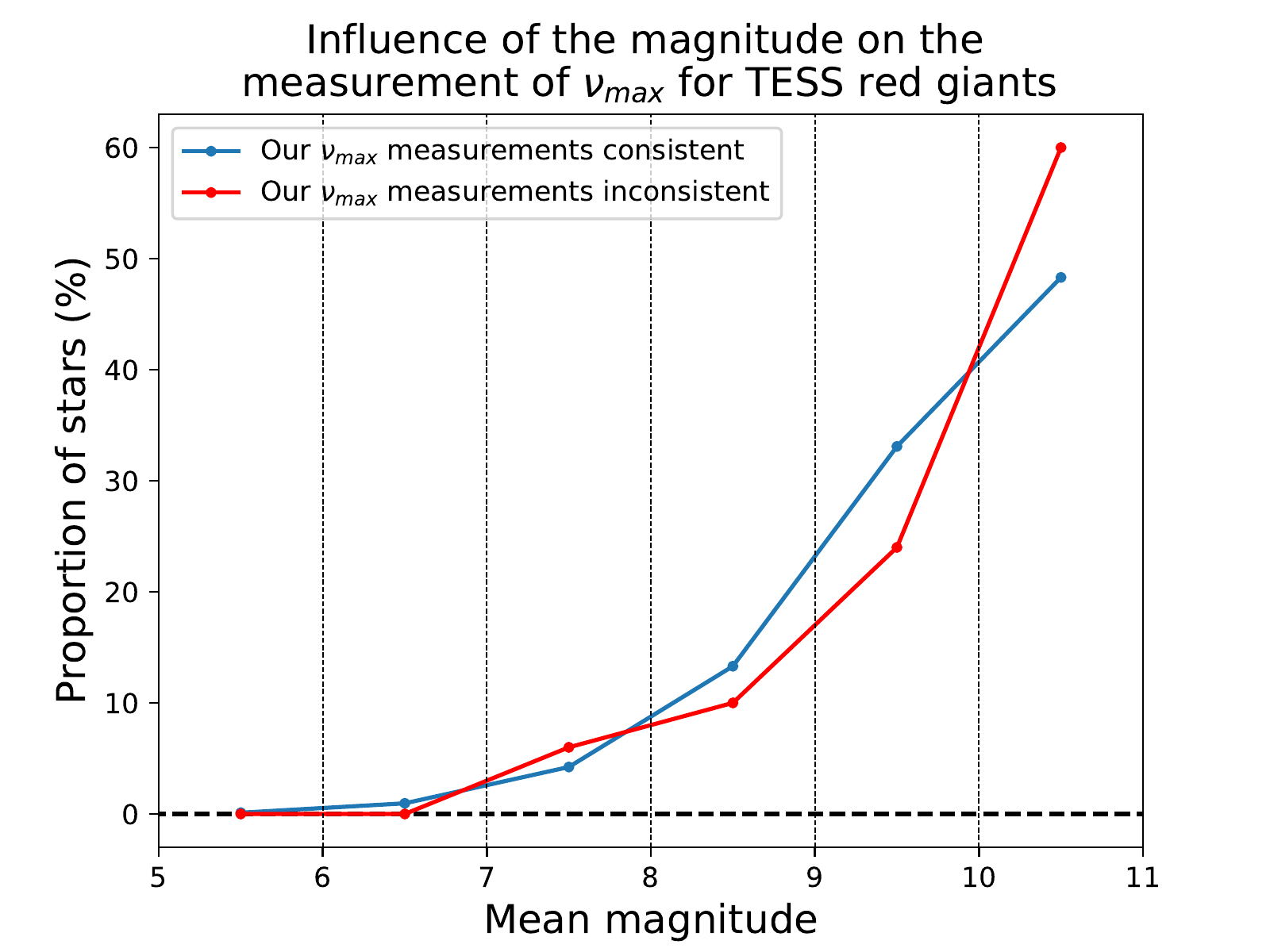}
\caption{Proportion of TESS red giants for which we derive consistent $\numax$ measurements (blue curve) and inconsistent measurements (red curve) as a function of the number of TESS sectors (left panel) and the mean stellar magnitude (right panel). Horizontal dashed lines indicate a proportion of stars of 0 \%.}
\label{fig-TESS-impact-mag-Nsectors-numax}
\end{figure*}

For 50 stars, the relative deviation between $\numax$ from \cite{Mackereth} and our measurements is of at least 10\%, among which 40 have $\numax < 20 \, \mu$Hz as measured by \cite{Mackereth} (upper left panel of Fig.~\ref{fig-TESS-numax}). Since our method cannot measure low $\numax$ in many cases, it is much more probable to obtain inaccurate measurements for low $\numax$ values, mainly when the signal is noisy and mimicks bumps in the power spectrum that are compatible with a fit with Eq. \ref{eqt-oscillations} including oscillations (see Sect. \ref{discussion-optimized-numax-TESS}). We provide such examples in \ref{appendix-3}. We obtain a consistency rate of 97.3 \% with existing measurements (Table \ref{table:numax}). It is not surprising that we obtain a higher consistency rate compared to artificial stars whithout requiring an extra validation for stellar magnitudes G > 9.5, since the TESS red giants we analyzed here are bona fide oscillating stars, for which oscillations have been detected through three different methods, which is not the case of the artificial sample analyzed.

We have a median relative precision of 5.3 \% (Table \ref{table:numax}). This is worse compared to \textit{Kepler} red giants, which is consistent with the overall lower $\numax$ of the analysed TESS sample (Fig.~\ref{fig-rel-precision}). The median relative precision is however better than what we obtain for artificial stars (Table \ref{table:synthetic-numax}), because we inconsistently retrieve many low $\numax$ values in the case of artificial stars, which are less precise (Fig.~\ref{fig-rel-precision}).

The computation time spent for each star is between 1.3 s and 29.9 s, with a median time of 11.6 s per run. The computation time is highly dependent on the number of observed TESS sectors (bottom panel of Fig.~\ref{fig-TESS-numax}). The median computation time is significantly higher for TESS red giants compared to \textit{Kepler} ones with 4-year long datasets (Table \ref{table:numax}); this comes from the fact that TESS power spectra were obtained with an oversampling factor of 10, while no oversampling was applied on \textit{Kepler} lightcurves. The median computation time is also much longer compared to artificial TESS targets (Table \ref{table:synthetic-numax}), since there is a greater proportion of real TESS stars with a higher number of observed TESS sectors.


\begin{figure*}[h!]
\centering
\includegraphics[width=8.8cm]{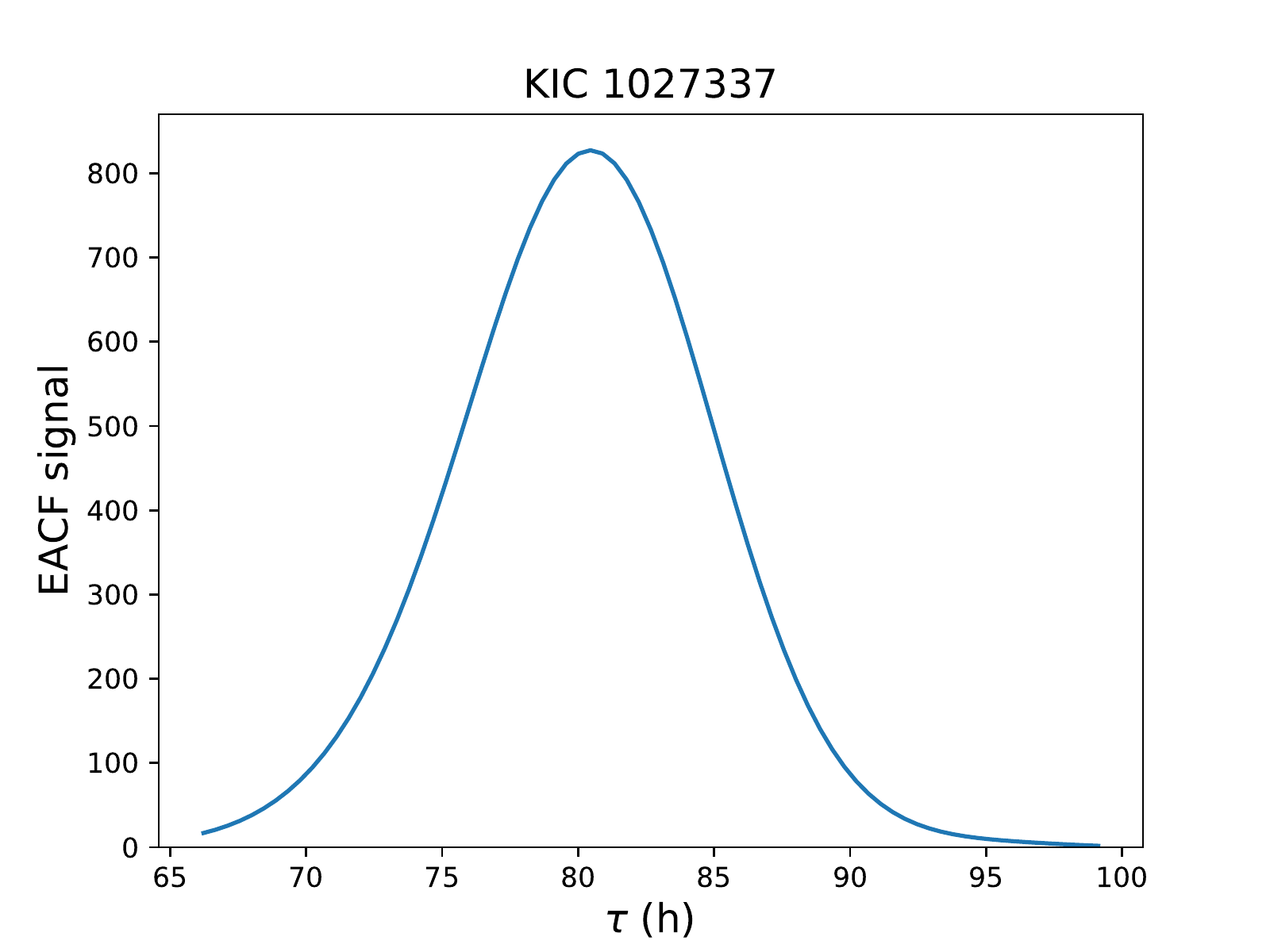}
\includegraphics[width=8.8cm]{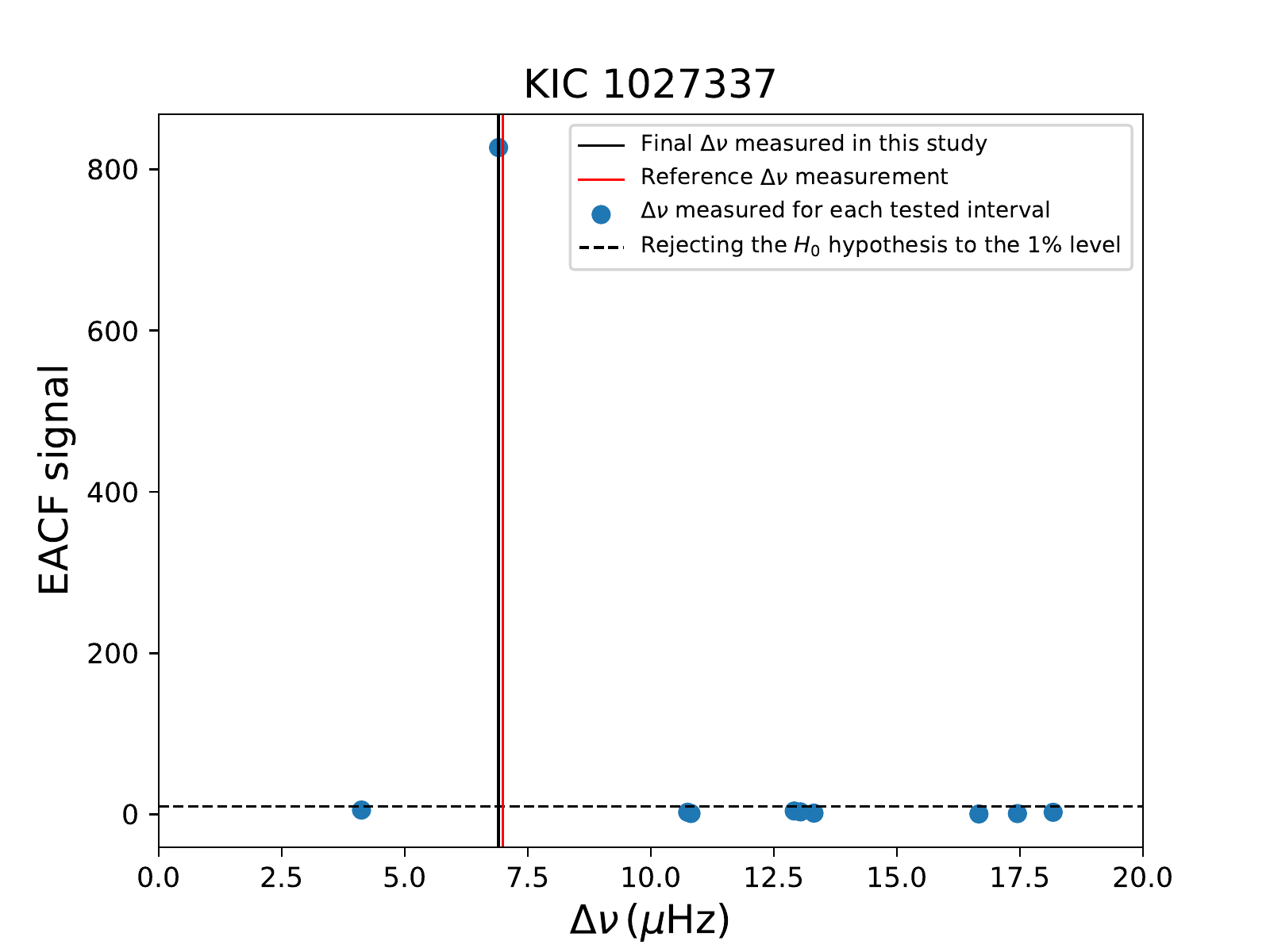}
\caption{Application of the EACF method to the \textit{Kepler} red giant KIC 1027337. \textit{Left:} Optimal local EACF signal as a function of the time lag in the autocorrelation space. \textit{Right:} EACF signals for the different $\Dnu$ measured for each tested interval. The reference $\Dnu$ measurement and our final measurement are represented by the vertical red and black lines, respectively, which are superimposed. The horizontal dashed line represents the limit above which the null hypothesis is rejected to the 0.4\% level.}
\label{fig-EACF}
\end{figure*}

\subsection{Impact of the stellar magnitude and the number of TESS sectors on the consistency of $\numax$}\label{discussion}

We led the same analysis as for the artificial sample in Sect. \ref{mag-sectors-artificial-numax} to assess the impact of the stellar magnitude and the number of observed TESS sectors on the consistency of our $\numax$ measurements compared either to \cite{Mackereth} measurements. As for the artificial targets analyzed in this study, we find that both the stellar magnitude and the number of observed TESS sectors $N\ind{sectors}$ impact the consistency of our $\numax$ measurements for the TESS stars analyzed by \cite{Mackereth} (Fig.~\ref{fig-TESS-impact-mag-Nsectors-numax}). Indeed, we note that the proportion of stars with inconsistent measurements is above the proportion of stars with consistent measurements for $N\ind{sectors} \leq 3$, representing 42 \% of the stars with inconsistent measurements. We note that 6 \% (resp. 10 \%) of the stars with inconsistent measurements have $N\ind{sectors}$ = 5 (resp. $N\ind{sectors}$ = 7), and this proportion is higher than for stars with consistent measurements. We checked that there are only 3 stars (resp. 5 stars) with inconsistent measurements and $N\ind{sectors}$ = 5 (resp. $N\ind{sectors}$ = 7), which all have stellar magnitudes G > 9.5, except one star with $N\ind{sectors}$ = 5 and G = 8.85. We additionally note that the proportion of stars with inconsistent measurements is above the proportion of stars with consistent measurements for a magnitude $G \geq 10$, representing 60 \% of stars with inconsistent measurements. These trends are not surprising since a low number of TESS sectors and/or a large stellar magnitude result in a lower signal-to-noise ratio in the power spectrum, making more challenging to unambiguously identify the bump of oscillations.

\subsection{Optimized $\numax$ measurement}\label{discussion-optimized-numax-TESS}

As for artificial targets (Sect. \ref{discussion-optimized-numax-synthetic}), we optimized the measurement of $\numax$ for TESS stars. We did not make a cut for $N\ind{sectors} > 3$ as for the artificial sample, since these are bona fide oscillating stars, for which $\numax$ has been consistently detected through three different methods. We only applied the same extra validation step as for the artificial sample for G magnitudes above 9.5. Similarly as for the artificial sample, we now obtain a $\numax$ detection for only 45.1 \% of the sample (i.e. for 1055 stars out of 2344, against 77.8 \% initially), together with a higher consistency rate of 99.1 \% (Table~\ref{table:numax} and upper right panel of Fig.~\ref{fig-TESS-numax}). We provide examples for 5 of the 9 stars for which we still measure an inconsistent $\numax$ in \ref{appendix-4}. The median computation time per run is 12.3 s, which is slightly longer since this extra validation step favors stars with a larger number of observed TESS sectors for G > 9.5.

\begin{figure*}[h!]
\centering
\includegraphics[width=8.8cm]{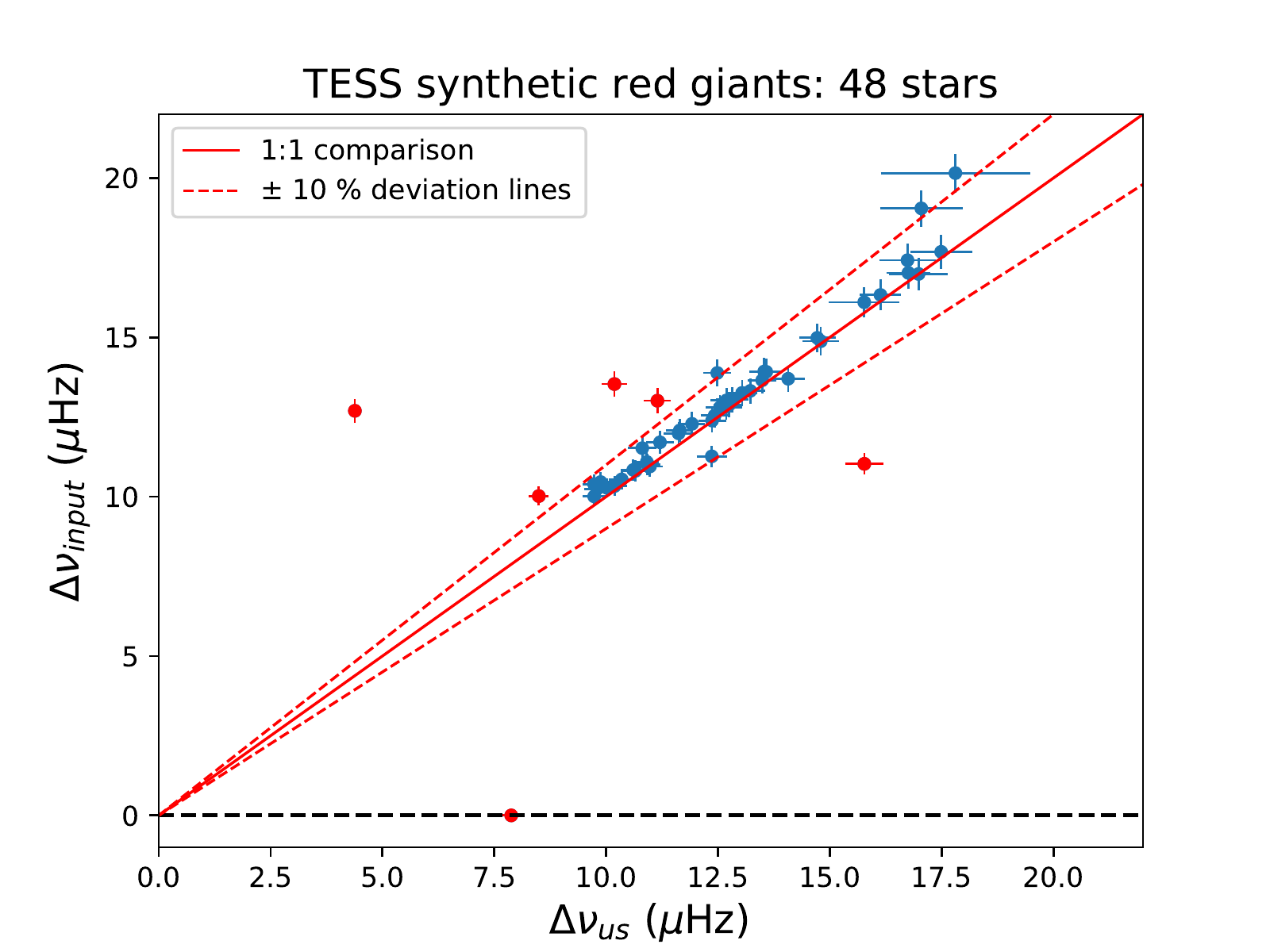}
\includegraphics[width=8.8cm]{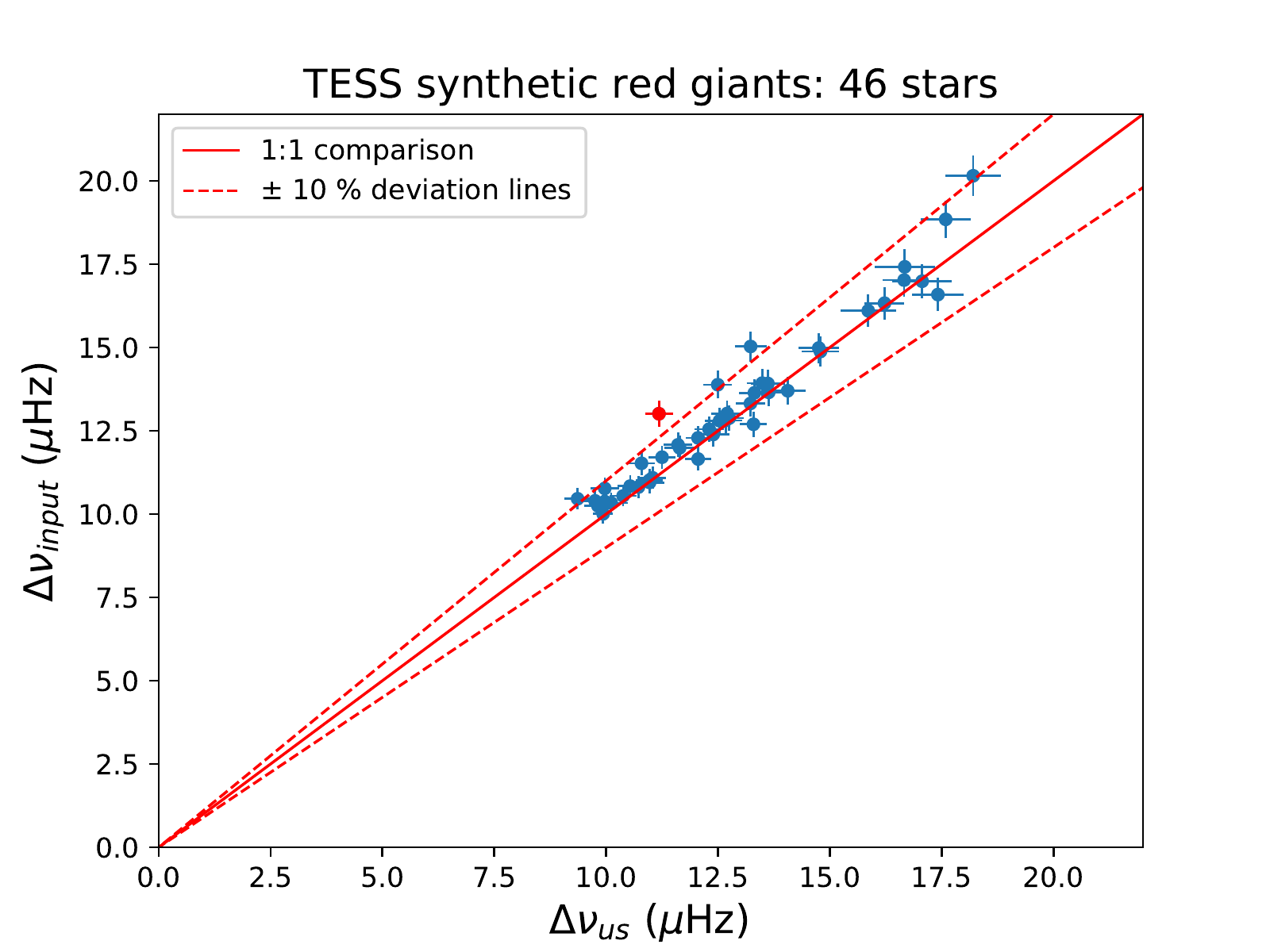}
\caption{Same as Fig. \ref{fig-TESS-numax-synthetic}, but for $\Dnu$. \textit{Left:} Comparison for $\Dnu$ measured through a blind search. The horizontal dashed lines indicate input $\Dnu$ values of 0 $\mu$Hz, meaning that no oscillations were injected in the oscillation spectrum. \textit{Bottom:} Comparison for $\Dnu$ measured through a guided search using $\numax$ as an input parameter.}
\label{fig-TESS-Dnu-synthetic}
\end{figure*}

\begin{table*}
\caption{Same as Table \ref{table:synthetic-numax}, but for $\Dnu$.}
\label{table:synthetic-Dnu}
\centering
\begin{tabular}{c c c c c c}
\hline\hline
Parameter & Detection & Consistency rate & Median relative & Median time & False positive\\
& rate & with input values & precision & per star & rate\\
\hline
$\Dnu$ blind & 18.5 \% & 89.6 \% & 2.7 \% & 2.3 s & 1.6 \%\\
$\Dnu$ guided & 18.1 \% & 97.8 \% & 2.8 \% & 1.5 s & 0.0 \%\\
\hline
\end{tabular}
\begin{tablenotes}
\item Our measurements are considered as consistent when the relative deviation with the existing ones is below 10\%. 
\end{tablenotes}
\end{table*}


\section{Measurement of $\Dnu$}\label{Dnu-measurement}

We implemented our own version of the EACF method to measure $\Dnu$, based on the Fourier transform of the raw power spectrum windowed by a Hanning filter \citep{Mosser_2009}. The EACF methods presents the advantage of not requiring the knowledge of $\numax$, allowing us to derive $\Dnu$ measurements that are completely independent from the $\numax$ ones derived with our FRA pipeline.

\subsection{The EACF method}\label{EACF}

We describe in the following the different steps we implemented, which include some adjustments compared to the procedure of \cite{Mosser_2009} in order to optimize the EACF method for red giant stars.

\subsubsection{Windowing the spectrum by a Hanning filter}\label{Hanning}

We first select a portion of the raw power spectrum using a Hanning filter. The EACF signal is the cleanest and the highest when the windowed spectrum is centered around $\numax$. In order to perform a blind search, independent from the measured $\numax$ value, we test 18 different frequencies $\nu\ind{c}$ to center the Hanning filter, which are regularly spaced between 10 and 270 $\mu$Hz. The FWHM of the Hanning filter is set to \citep{Mosser_2009}
\begin{equation}\label{eqt-dnuH}
\delta\nu\ind{H} = \alpha \, \gamma \, \Dnu\ind{c},
\end{equation}
with $\alpha$ = 1.05 \citep{Mosser_2009} and, for red giants \citep{Mosser_2010},
\begin{equation}\label{eqt-gamma}
\gamma = 2.08 \, \nu\ind{c}^{0.15}.
\end{equation}
In cases where $\delta\nu\ind{H}$ is too large and one edge of the Hanning filter falls outside the observed frequency range, $\delta\nu\ind{H}$ in Eq. \ref{eqt-dnuH} is reduced accordingly.

The resulting windowed spectrum is then zero-padded at high frequencies so that the resolution of the EACF signal is high enough to give a clean signal. We chose to set the number of data points in the zero-padded spectrum to
\begin{equation}\label{eqt-NH}
N\ind{H} = 10  \, N,
\end{equation}
where $N$ is the number of data points initially existing in the windowed spectrum.

\subsubsection{Measuring $\Dnu$}

The next step consists in performing the Fourier transform of the zero-padded windowed spectrum. The signature of the large separation corresponds to the first peak in the autocorrelation signal (Fig.~\ref{fig-EACF}).

For each of the 18 zero-padded spectra windowed around a given $\nu\ind{c}$ value, an input large separation $\Dnu\ind{c}$ is tested, derived through Eq. \ref{eqt-Dnu-c}. In the Fourier space, we explore possible values for $\Dnu$ in the range $\left [\Dnu\ind{c} / G, \, G \Dnu\ind{c} \right ]$, with $G=1.1$. 
For a given value of the large separation, the corresponding time shift in the autocorrelation space is 
\begin{equation}\label{eqt-tau-Dnu}
\tau\ind{\Dnu} = \frac{2}{\Dnu}.
\end{equation}
Given Eq. \ref{eqt-tau-Dnu}, we explore the range $\left [\frac{2}{G \Dnu\ind{c}}, \frac{2 G}{\Dnu\ind{c}} \right ]$ in the autocorrelation space to search for $\Dnu$ (Fig.~\ref{fig-EACF}).

\begin{figure*}[h!]
\centering
\includegraphics[width=8.8cm]{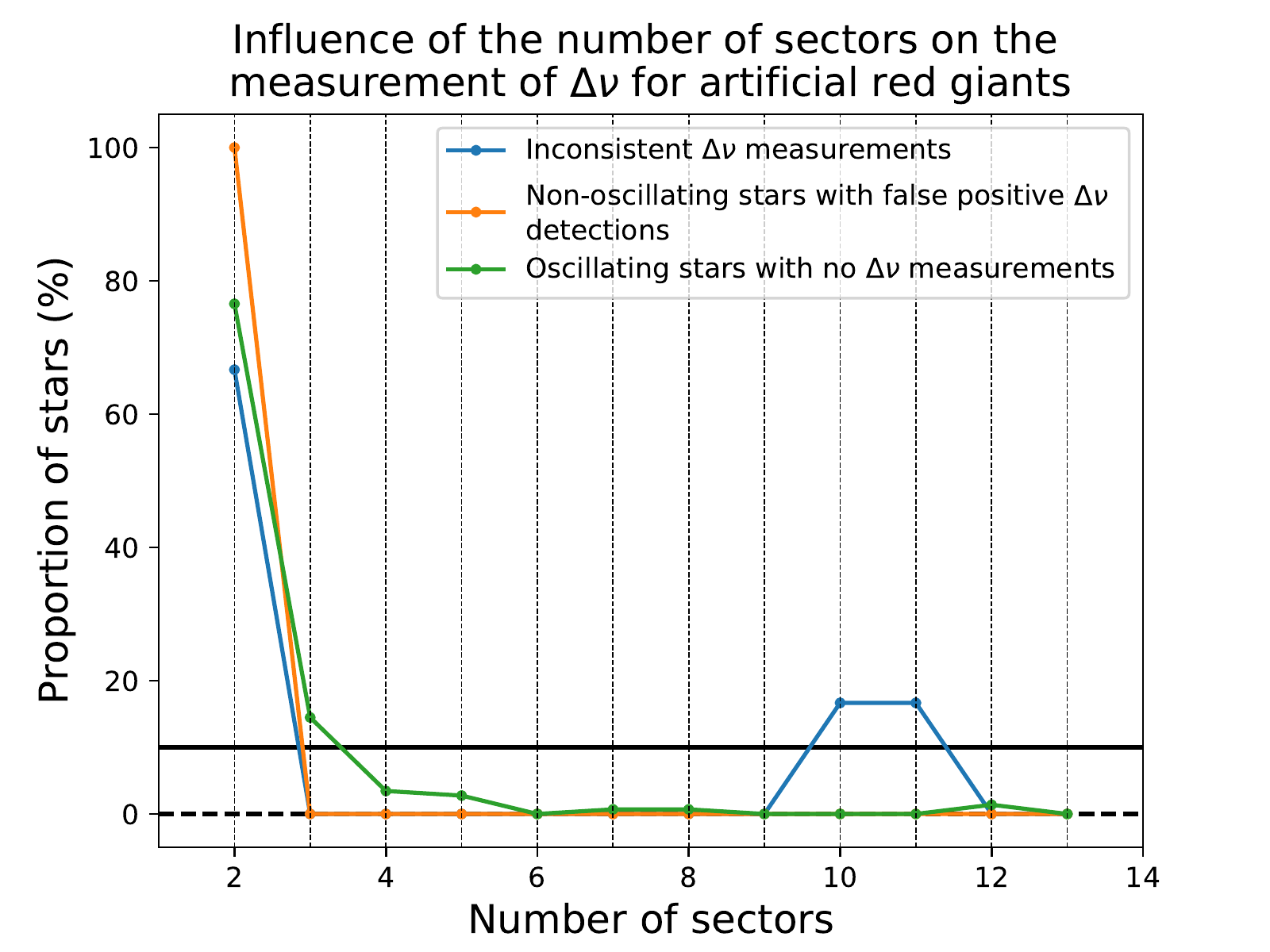}
\includegraphics[width=8.8cm]{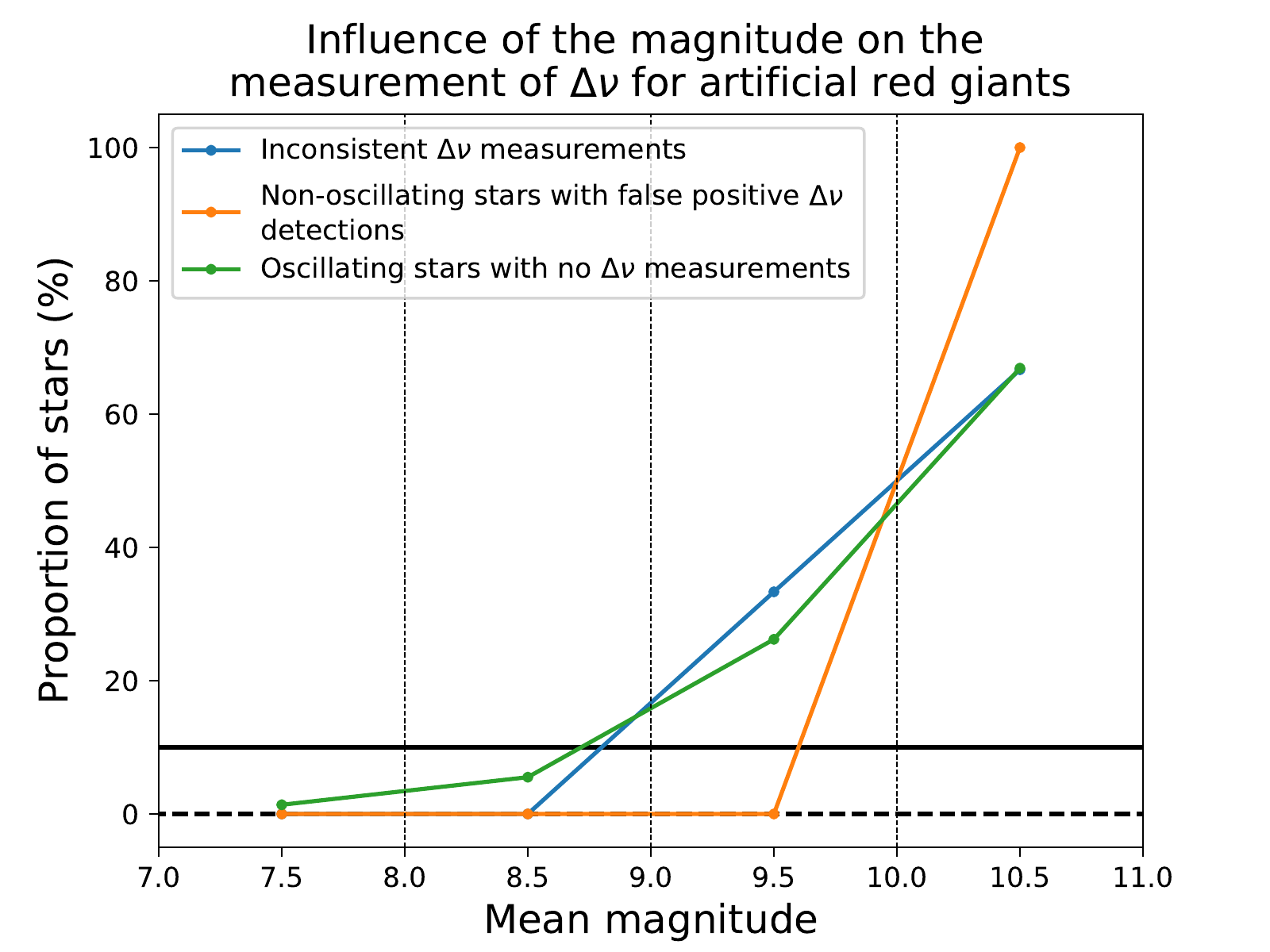}
\caption{Same as Fig. \ref{fig-TESS-synthetic-impact-mag-Nsectors-numax}, but for $\Dnu$ measured through a blind search.}
\label{fig-TESS-synthetic-impact-mag-Nsectors-Dnu}
\end{figure*}

We then normalize the autocorrelation signal $C(\tau)$ such that 
\begin{equation}\label{eqt-A-star}
A^\star = \frac{ \left | C(\tau)^2 \right |}{\left | C(0)^2 \right |}.
\end{equation}
The EACF signal $A$ is finally obtained by normalizing the autocorrelation signal to the mean noise level in the autocorrelation, $\sigma\ind{H}$, in order to accurately compare the strength of the EACF signal for each windowed spectrum (Fig.~\ref{fig-EACF}), such that
\begin{equation}\label{eqt-EACF}
A = \frac{A^\star}{\sigma\ind{H}},
\end{equation}
with
\begin{equation}\label{eqt-sigma-H}
\sigma\ind{H} = \frac{3}{2 (\frac{N\ind{H}}{N\ind{OS}} - 1)},
\end{equation}
where $N\ind{OS}$ is the oversampling factor, which equates to $2 \, N\ind{f} / N\ind{t}$, with $N\ind{f}$ the number of data points in the power spectrum and $N\ind{t}$ the number of data points in the time series.
\newline
The final $\Dnu$ value corresponds to the $\Dnu$ value with highest EACF signal (Fig.~\ref{fig-EACF}). 
The measurement of $\Dnu$ is considered as significant if its associated EACF signal is above or equal to a threshold value,  $A\ind{lim}$, defined below. Otherwise, no detection of $\Dnu$ is achieved.

\subsubsection{Reliability of the $\Dnu$ measurement}

For each of the 18 zero-padded spectra, the $\Dnu$ measurement is validated if the null hypothesis, i.e. that the detected signal can be explained by noise only, is rejected to a low level $p$. Eqs. (10) and (11) of \cite{Mosser_2009} define the threshold value above which the EACF signal rejects the null hypothesis to a given level $p$, as a function of the number of independent bins in the time interval over which the large separation is searched for. In the case of a zero-padded spectrum, \cite{Gabriel_2002} show that the number of independent bins has to be multiplied by a correction factor that depends on the padding factor. In this study, we are using a padding factor of 10 (Eq. \ref{eqt-NH}), which corresponds to a correction factor equal to 3 \citep{Gabriel_2002}. We thus revisit Eqs. (10) and (11) of \cite{Mosser_2009} in the case of a zero-padded spectrum such as
\begin{equation}\label{eqt-A-lim}
A\ind{lim} \simeq - \ln{p} + \ln{\left( 3\, \frac{\Delta \tau}{\delta \tau} \right)},
\end{equation}
where $\Delta \tau = 2/\Dnu$ is the time interval over which the large separation is searched for, and $\delta \tau$ is the FWHM of the autocorrelation peak. We chose to set $p$ to 0.4\%, giving $A\ind{lim} = 10$ (Fig.~\ref{fig-EACF}). We checked that we obtain the same threshold value based on Eq. (12) of \cite{Gabriel_2002}.

\subsubsection{Precision on the $\Dnu$ measurement}

The precision on $\Dnu$ is derived through \citep{Mosser_2009}
\begin{equation}\label{eqt-precision-Dnu}
\frac{\delta \Dnu}{\Dnu} = \frac{\beta}{2 \pi} \frac{b}{A\ind{lim}} \frac{\Dnu}{\delta \nu\ind{H}},
\end{equation}
where $\beta \simeq 0.763$ \citep{Mosser_2009} and the noise contribution is set to $b = A\ind{lim}$. The relative precision increases with $\Dnu$ and $\numax$ (Fig.~\ref{fig-rel-precision}), as we can infer from Eqs. \ref{eqt-dnuH} and \ref{eqt-gamma} that
\begin{equation}\label{eqt-precision-Dnu-with-numax}
\frac{\delta \Dnu}{\Dnu} \propto \numax^{-0.15}. 
\end{equation}

\begin{figure*}[h!]
\centering
\includegraphics[width=9.1cm]{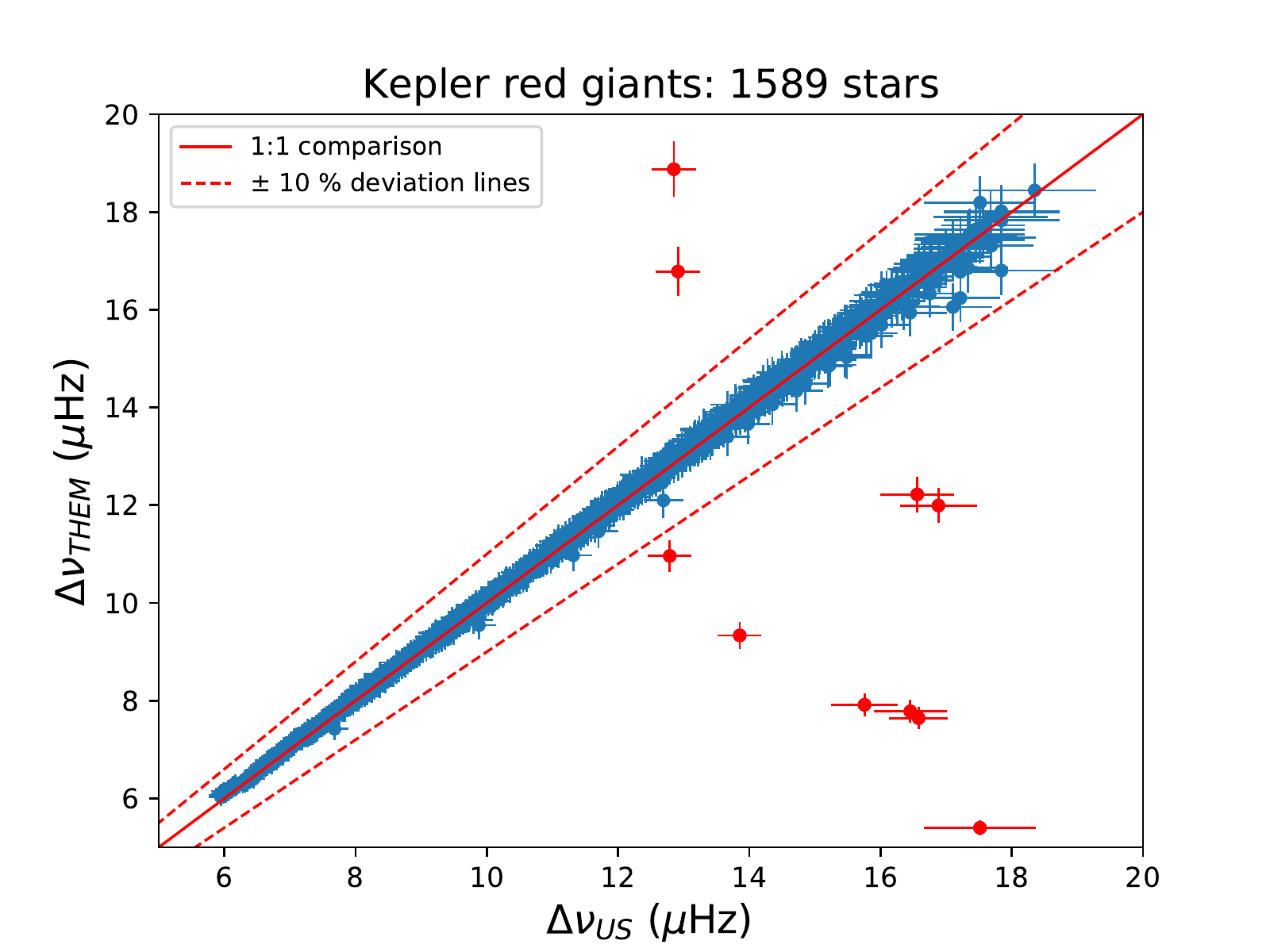}
\caption{Same as Fig. \ref{fig-Kepler-numax}, but for $\Dnu$.}
\label{fig-Kepler-Dnu}
\end{figure*}

\subsection{Test and validation on artificial TESS oscillation spectra}

We applied our implementation of the EACF method to the sample of artificial stars analyzed in Sect. \ref{test}.

\subsubsection{Blind search}

When measuring $\Dnu$ with a blind search, we detect oscillations for 48 stars (left panel of Fig.~\ref{fig-TESS-Dnu-synthetic}). These include 1 star which have no injected oscillations, representing a false positive detection rate of 1.6 \% (Table \ref{table:synthetic-Dnu}). We also derive a $\Dnu$ measurement for 47 stars with injected oscillations, representing a detection rate of 18.5 \%. 5 stars with injected oscillations have a relative deviation of at least 10\% compared to the input value, giving a consistency rate of 89.6 \%. 
We obtain a median relative precision of 2.7 \% and a median computation time spent for each star of 2.3 s.

\subsubsection{Guided search}

Given the rather low consistency rate obtained with a blind search, we also analysed TESS artificial red giants with a guided search that uses the measured $\numax$ as an input parameter to center the Hanning filter windowing the spectrum. We detect oscillations for 46 stars, all having injected oscillations (right panel of Fig.~\ref{fig-TESS-Dnu-synthetic}), representing a false positive detection rate of 0 \% and a a detection rate of 18.1 \% (Table \ref{table:synthetic-Dnu}). Among them, 1 has a relative deviation of at least 10\% compared to the input value, giving a consistency rate of 97.8 \%. 
We obtain a median relative precision of 2.8 \% (Table \ref{table:synthetic-Dnu}), which is similar to the median relative precision obtained for $\Dnu$ measured through a blind search.
Relying on the measured $\numax$ to derive $\Dnu$ appears to provide much more consistent results than a blind search for $\Dnu$. Given the high consistency rate of our measured $\numax$ compared to reference measurements, this is a sensible approach.

\subsubsection{Impact of the stellar magnitude and the number of TESS sectors on the detectability and consistency of $\Dnu$}

We note in Fig. \ref{fig-TESS-synthetic-impact-mag-Nsectors-Dnu} that both the proportion of stars with inconsistent measurements and of non-oscillating stars with false positive measurements exceeds 10 \% for $N\ind{sectors} = 2$, while the proportion of oscillating stars with no measurement exceeds 10 \% for $N\ind{sectors} \leq 3$. Additionally, we note that both the proportion of stars with inconsistent $\Dnu$ measurements and of oscillating stars with no $\Dnu$ measurement exceeds 10 \% for a stellar magnitude above 9, while the proportion of non-oscillating stars with false positive $\Dnu$ measurements exceeds 10 \% for a stellar magnitude above 10. This is globally consistent with what we obtain for the $\numax$ measurement in Sect. \ref{test}.

\begin{figure*}[h!]
\centering
\includegraphics[width=8.8cm]{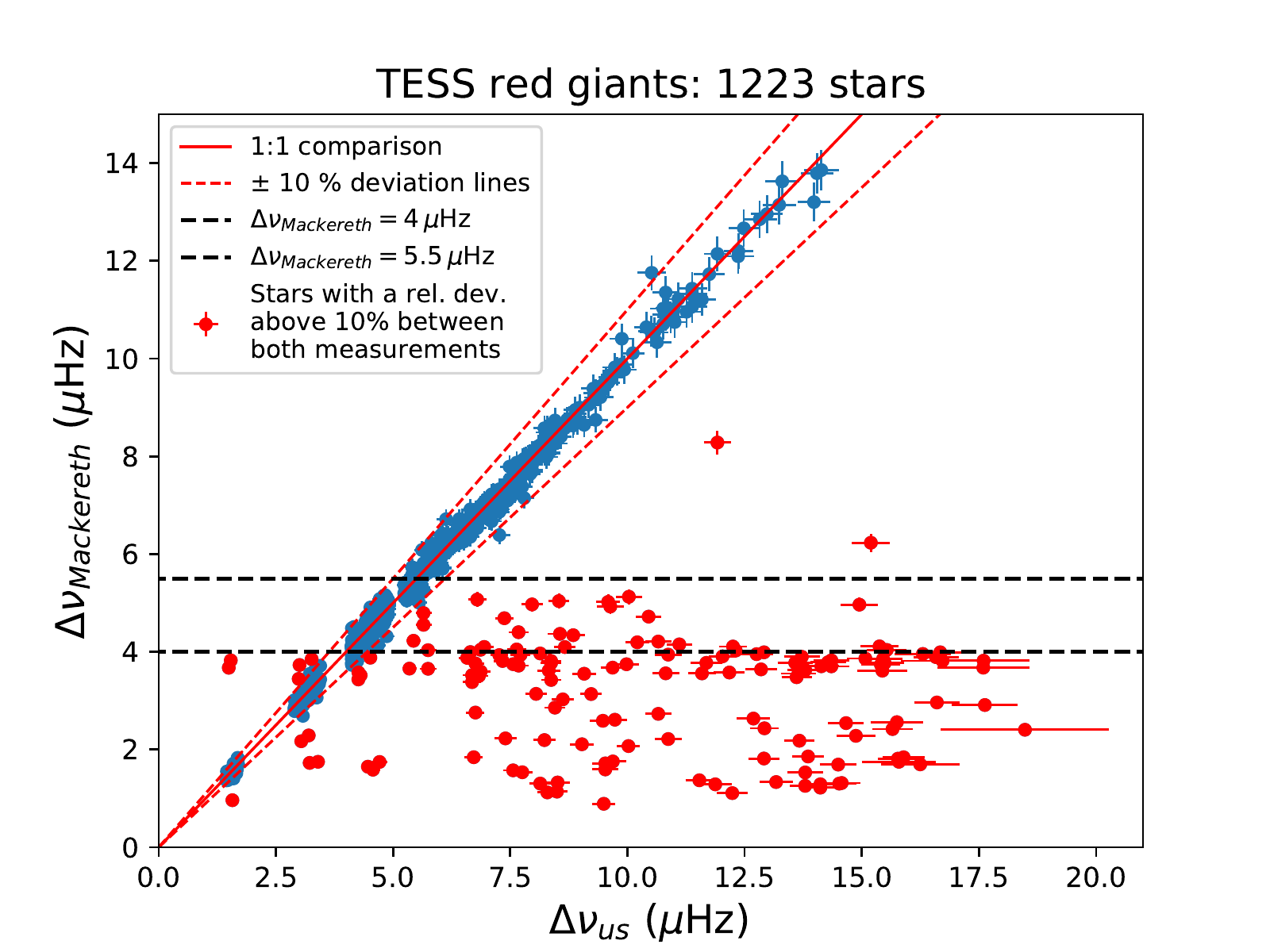}
\includegraphics[width=8.8cm]{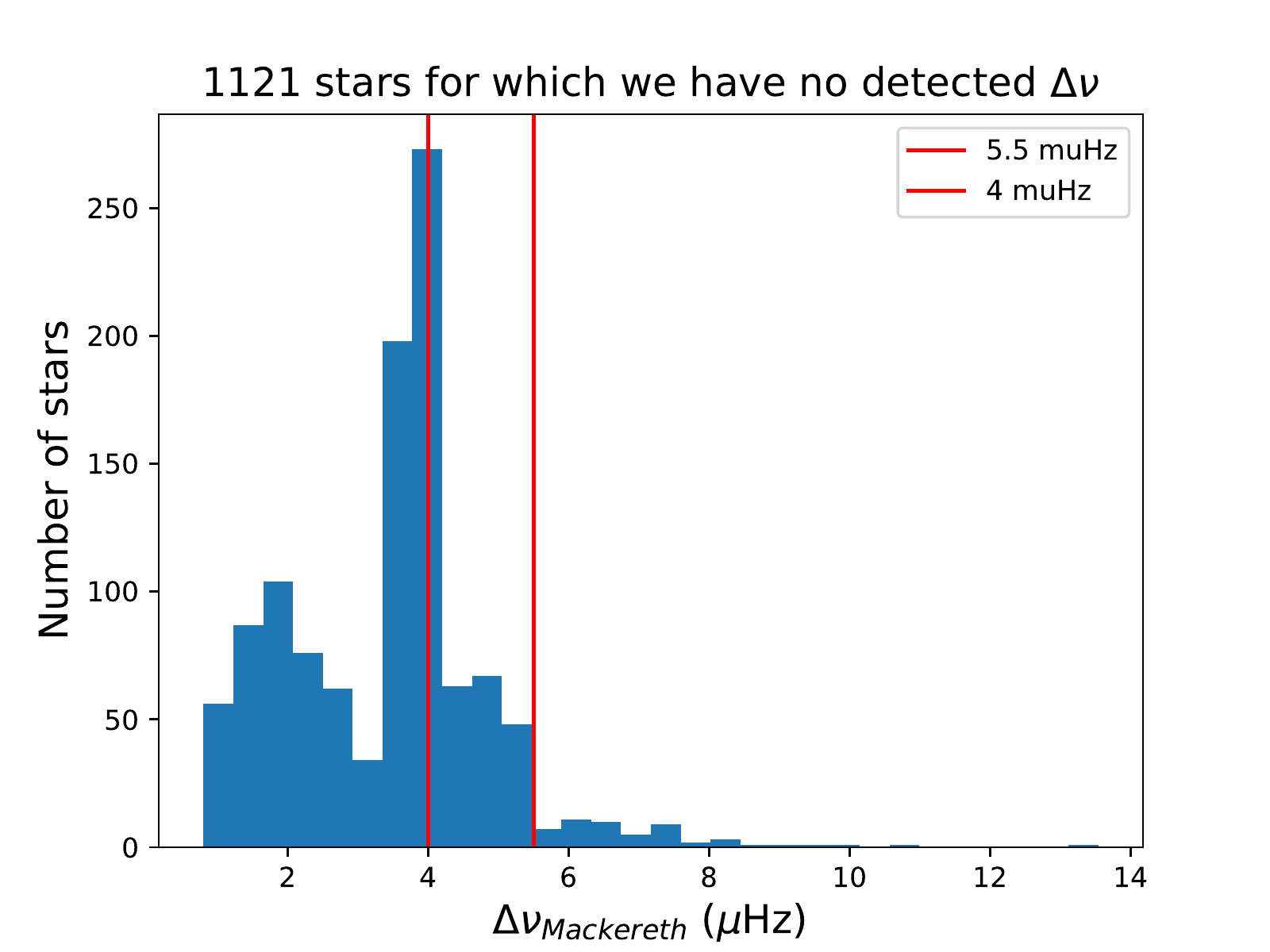}
\includegraphics[width=8.8cm]{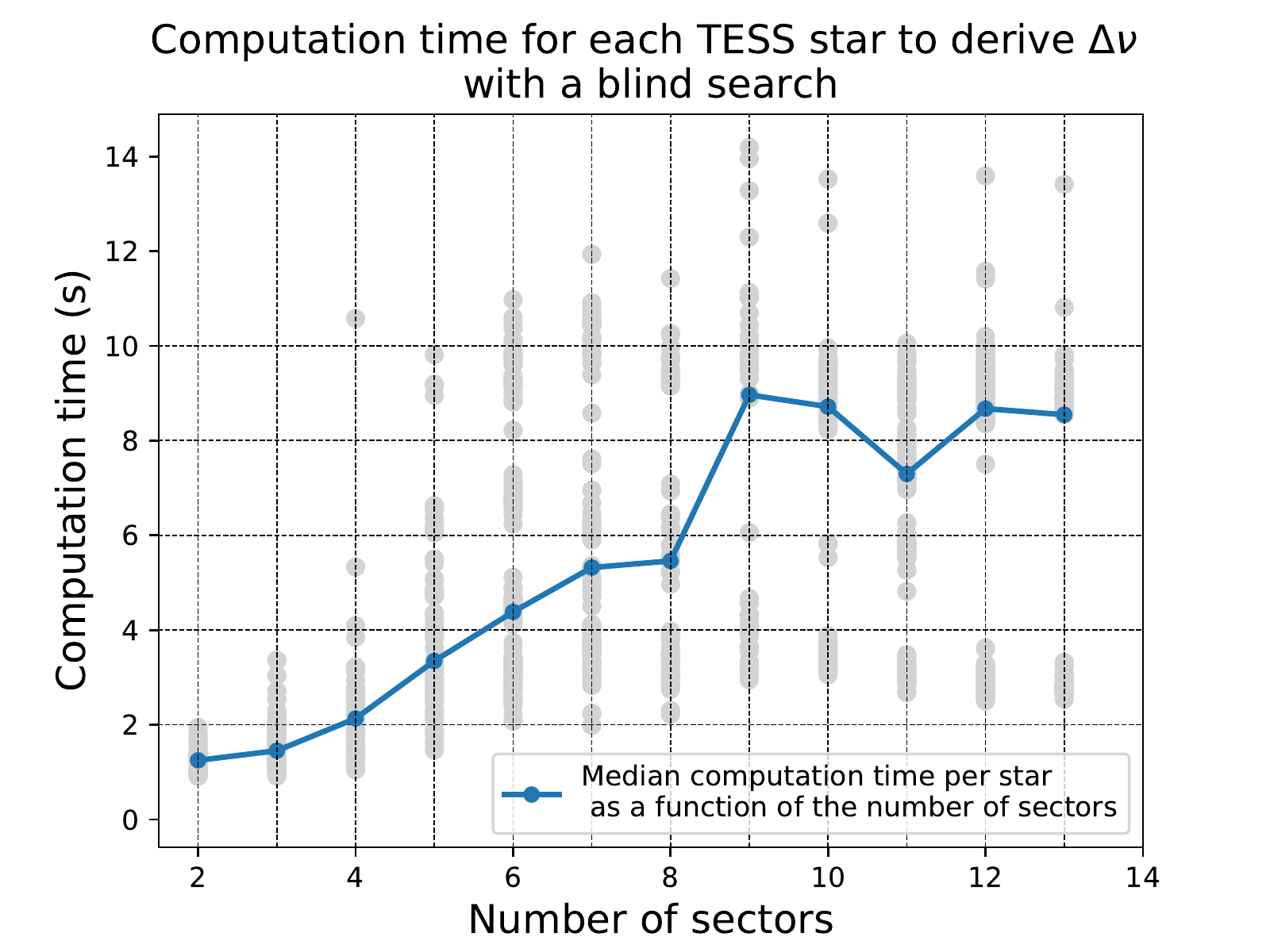}
\caption{Same as Fig.~\ref{fig-TESS-numax}, but for $\Dnu$ measured through a blind search. \textit{Upper left:} The horizontal dashed line represents $\Dnu\ind{Mackereth} = 4 \, \mu$Hz and $\Dnu = 5.5 \, \mu$Hz. \textit{Upper right:} Vertical red lines represent $\Dnu = 4 \, \mu$Hz and $\Dnu = 5.5 \, \mu$Hz.}
\label{fig-TESS-Dnu-blind}
\end{figure*}

\subsection{Analysis of 1589 \textit{Kepler} red giants}\label{kepler-Dnu}

We analysed \textit{Kepler} red giants with our own implementation of the EACF method \citep{Mosser_2009} presented in Sect.~\ref{EACF}. We obtained a $\Dnu$ measurement for all \textit{Kepler} stars analysed in this study (Table \ref{table:Dnu}). We have a median relative precision of 2.7 \%, as for TESS artificial targets (Table \ref{table:synthetic-Dnu}). The computation time spent for each star is between 1.1 s and 7.2 s, with a median time of 2.6 s per run, which is similar to what we obtain for TESS artificial targets (Table \ref{table:synthetic-Dnu}). For 10 stars, the relative deviation between our $\Dnu$ values and the existing ones is of at least 10\% (Fig.~\ref{fig-Kepler-Dnu}). In order to discriminate between the two measurements in these cases, we computed the $\numax$ estimate derived from the reference $\Dnu$ measurement using the scaling relation \citep{Mosser_2010}
\begin{equation}\label{eqt-numax-Dnu}
\nu\ind{max} = \left (\frac{\Dnu}{0.28}\right )^{4/3}.
\end{equation}
We then compared through a visual inspection this estimate with ours, which corresponds to the center of the Hanning filter used to window the spectrum (Sect.~\ref{Hanning}). We find that our $\numax$ estimates are consistent with the observed bump of oscillations for these 10 stars, contrary to the $\numax$ estimates obtained from Eq. \ref{eqt-numax-Dnu} using the reference $\Dnu$ measurements (see \ref{appendix-5}). This indicates that our $\Dnu$ measurement is self-consistent.

 \begin{figure*}[h!]
\centering
\includegraphics[width=8.8cm]{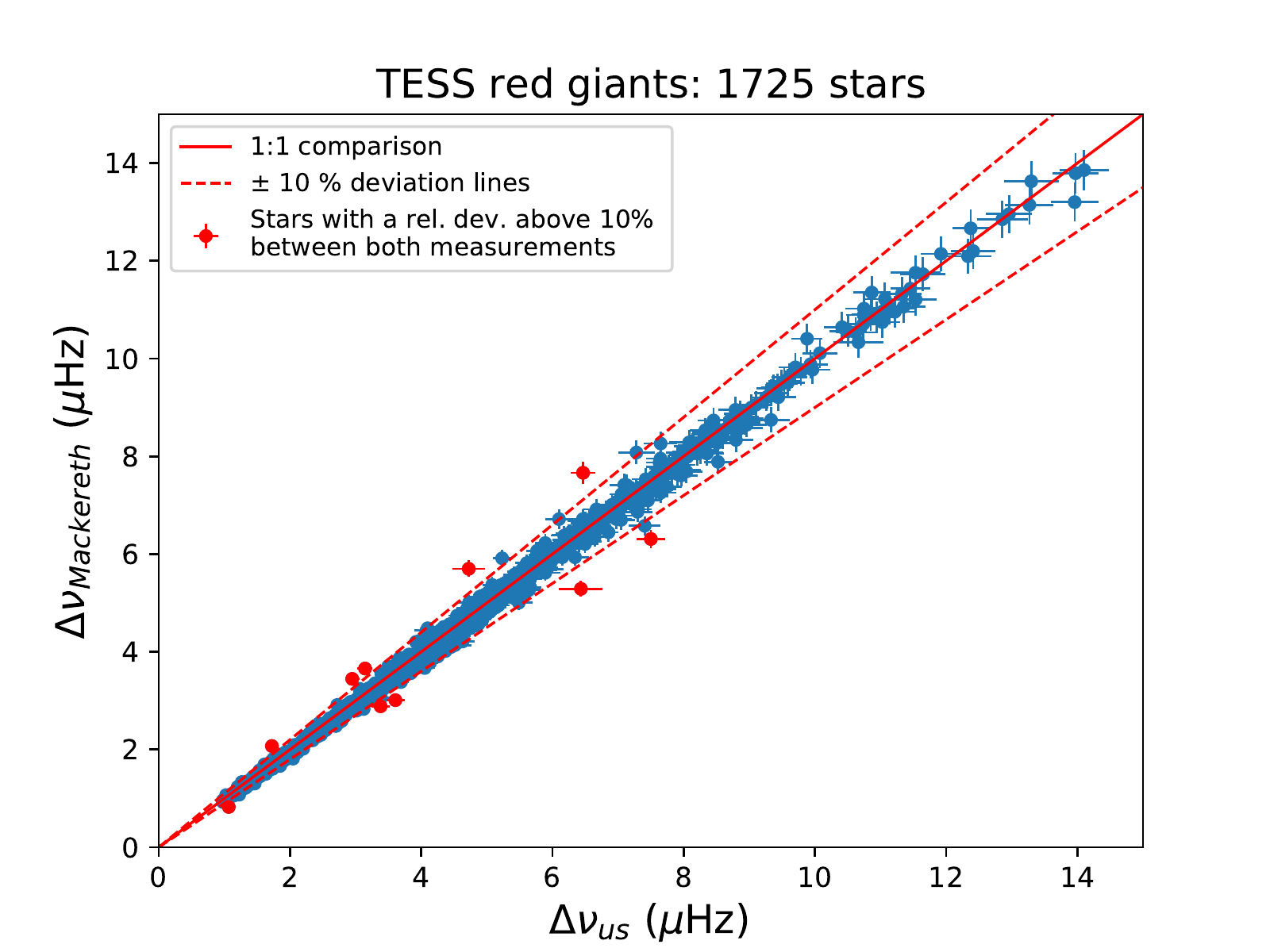}
\includegraphics[width=8.8cm]{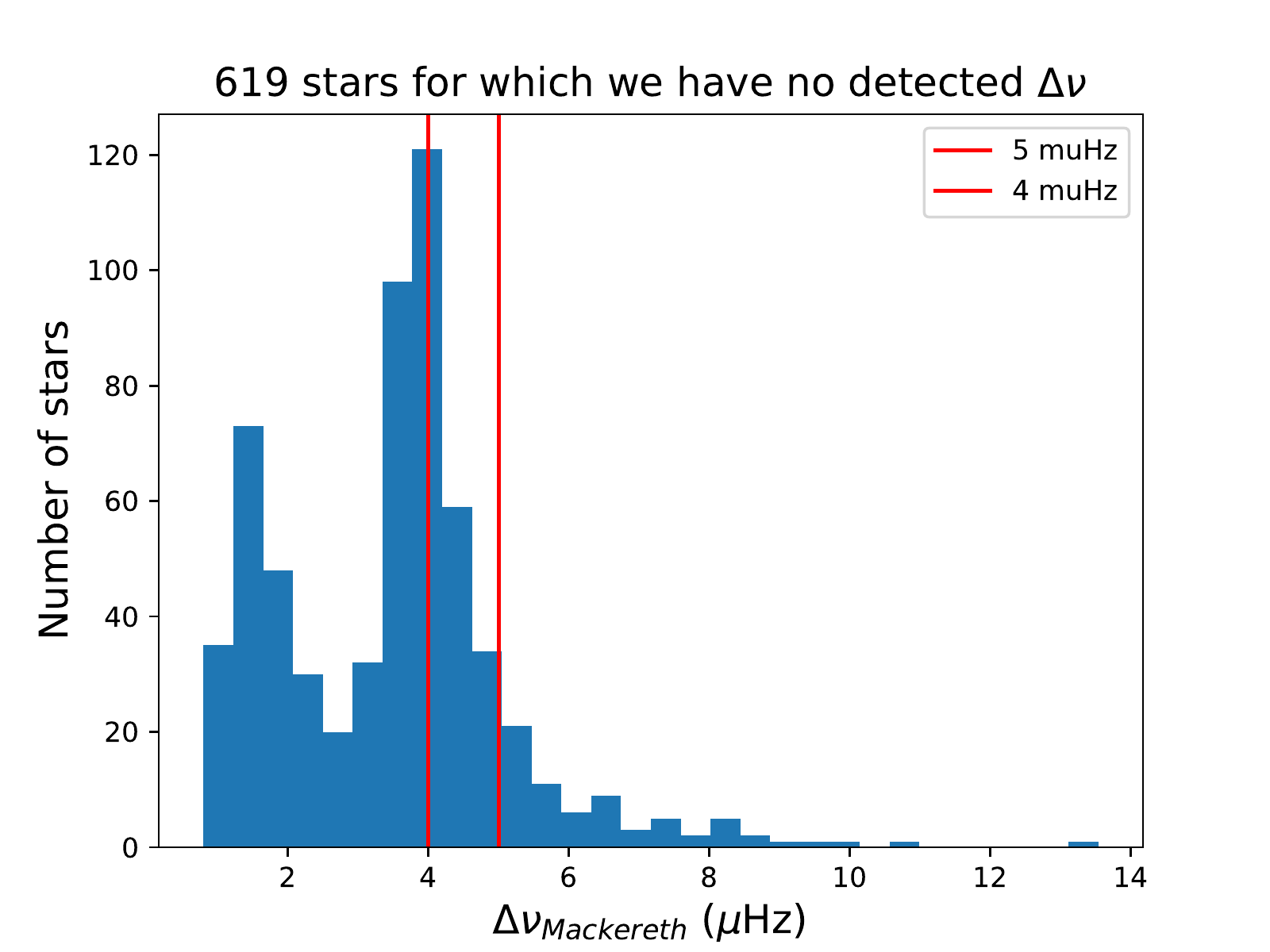}
\includegraphics[width=8.8cm]{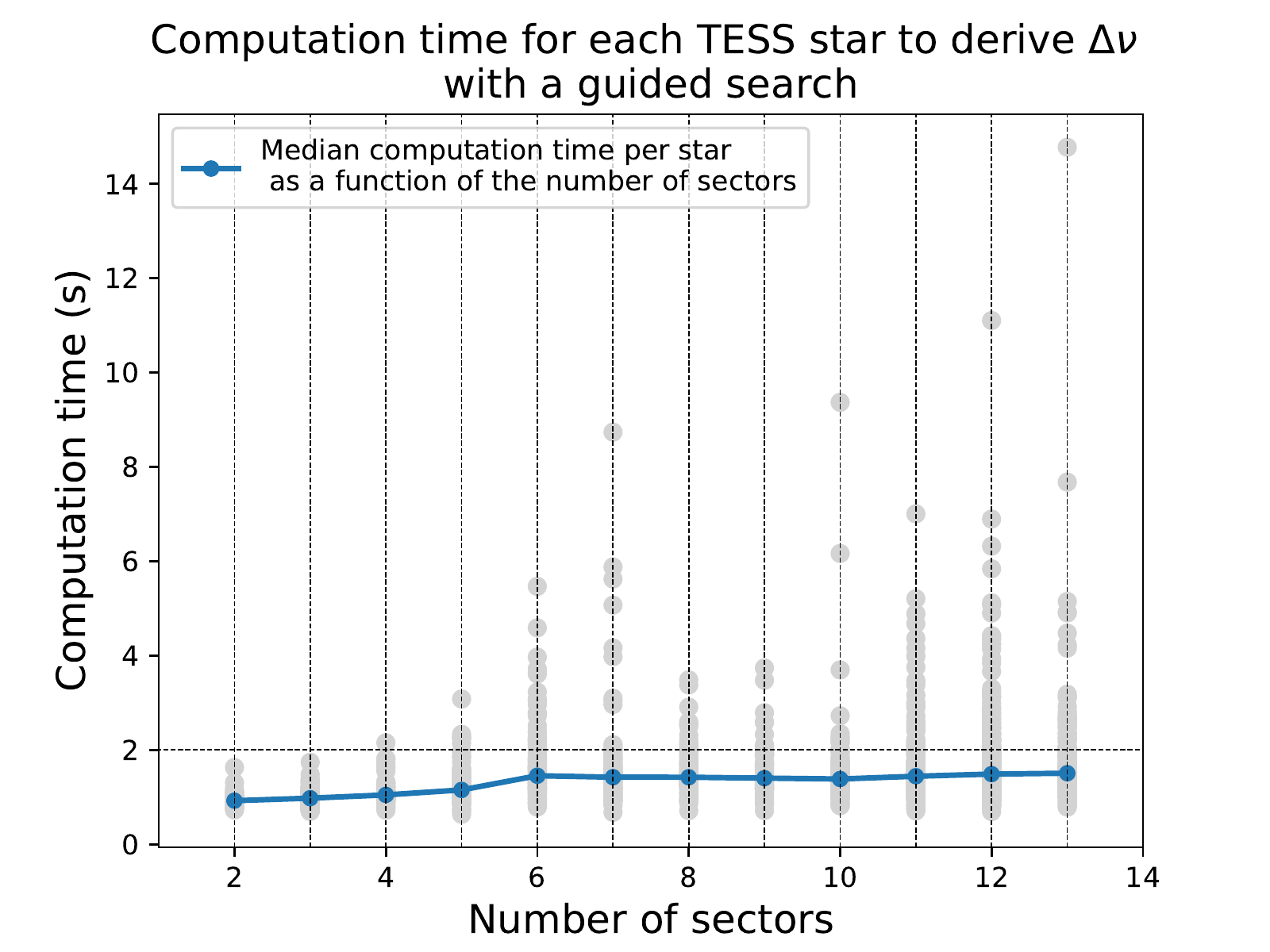}
\caption{Same as Fig.~\ref{fig-TESS-numax}, but for $\Dnu$ measured through a guided search using the measured $\numax$ as an input parameter. \textit{Upper right:} Vertical red lines represent $\Dnu = 4 \, \mu$Hz and $\Dnu = 5 \, \mu$Hz.}
\label{fig-TESS-Dnu-guided}
\end{figure*}

\subsection{Analysis of 2344 TESS red giants}\label{TESS-Dnu}

We analysed TESS red giants with our own implementation of the EACF method \citep{Mosser_2009} presented in Sect.~\ref{EACF}.

\subsubsection{Blind search}

We derived a $\Dnu$ measurement through a blind search for 1223 stars, i.e. 52.2 \% of the analyzed sample (upper left panel of Fig.~\ref{fig-TESS-Dnu-blind}). The majority of the 1121 stars for which we did not detect oscillations based on the measurement of $\Dnu$ have low reference $\Dnu$ measurements, below 5.5 $\mu$Hz with a peak around 4 $\mu$Hz (upper right panel of Fig.~\ref{fig-TESS-Dnu-blind}). We additionally note that the minimum $\Dnu$ we detect is 1.484 $\mu$Hz, while the minimum $\Dnu$ measured by \cite{Mackereth} is 0.890 $\mu$Hz. We discuss the difficulty we encounter in measuring low $\Dnu$ values in Sect.~\ref{discussion-Dnu}.

For 150 stars, the relative deviation between $\Dnu$ from \cite{Mackereth} and our measurements is of at least 10\% (upper left panel of Fig.~\ref{fig-TESS-Dnu-blind}). We provide examples in \ref{appendix-6}. This gives a consistency rate of 87.7 \% (Table \ref{table:Dnu}). This is lower than for TESS artificial stars (Table \ref{table:synthetic-Dnu}) because inconsistent measurements for real TESS stars occur for $\Dnu \leq 5.5 \,\mu$Hz, while the artificial sample have $\Dnu \geq 10 \,\mu$Hz. The vast majority of these discrepant stars are characterised by a greatly overestimated $\Dnu$ using our method. In general, $\Dnu < 5.5$ $\mu$Hz as measured by \cite{Mackereth} for these stars, of which 120 (i.e 80 \%) have $\Dnu \leq 4$ $\mu$Hz. We do not encounter such disagreement with regard to our \textit{Kepler} sample because the observation duration is much longer, thus the EACF signal is much higher. In the case of TESS stars with much shorter observation durations, the signal-to-noise ratio is lower even when using oversampling, and this significantly impacts the quality of the autocorrelation signal. The maximal EACF signal then actually rejects the null hypothesis in some cases, leading to spurious detections when carrying a blind search for $\Dnu$ with the EACF method (see \ref{appendix-6}). These incorrect detections are more frequent for low $\Dnu$ because such low values are complicated to detect with the EACF method, as we already mentioned. In many of these cases, the reference $\Dnu$ measurement from \cite{Mackereth} does not even lead to a corresponding acceptable EACF signal because the Hanning filter is too narrow and the peak in the autocorrelation appears too close to the edges of the scanned autocorrelation space to be considered as relevant (see \ref{appendix-6}). In such cases, the EACF signal corresponding to the accurate $\Dnu$ is not kept and thus cannot overpass the maximal EACF signal for another $\Dnu$ value.

We have a median relative precision of 3 \% (Table \ref{table:Dnu}). This is slightly worse than for \textit{Kepler} red giants and artificial TESS targets (Table \ref{table:synthetic-Dnu}). This is consistent with the overall lower $\Dnu$ values of the analysed TESS sample, which are less precise (Fig.~\ref{fig-rel-precision}). The median computation time spent for each star is 2.3 s, which is higher to what we obtain for artificial TESS targets (Table \ref{table:synthetic-Dnu}) as there is a lower number of stars in our real TESS sample with a low number of observed TESS sectors.

 \begin{table*}
\caption{Characteristics of the $\Dnu$ measurement for the 1589 \textit{Kepler} and 2344 TESS stars analysed.}
\label{table:Dnu}
\centering
\begin{tabular}{c c c c}
\hline\hline
Sample & Consistency rate with & Median relative & Median time\\
& existing measurement & precision & per star\\
\hline
\textit{Kepler} & 99.4 \% & 2.7 \% & 2.6 s\\
\hline
TESS blind & 87.7 \% & 3.0 \% & 3.7 s\\
TESS guided & 99.4 \% & 3.2 \% & 1.3 s\\
\hline
\end{tabular}
\begin{tablenotes}
\item \textit{blind} stands for a blind search of $\Dnu$, \textit{guided} stands for a guided search of $\Dnu$ using $\numax$. Our measurements are considered as consistent when the relative deviation with the existing ones is below 10\%. 
\end{tablenotes}
\end{table*}

 The computation time spent for each star is between 0.9 s and 14.2 s, with a median time of 3.7 s per run. The computation time is highly dependent on the number of observed TESS sectors (bottom panel of Fig.~\ref{fig-TESS-Dnu-blind}). As for the $\numax$ measurement, the median computation time is higher for TESS red giants compared to \textit{Kepler} ones with 4-year long datasets (Table \ref{table:Dnu}), due to the oversampling factor applied to obtain TESS power spectra. It is also slightly higher compared to artificial TESS targets, as there is a greater proportion of longer numbers of sectors among the real TESS red giants.
 
  \begin{figure*}[h!]
\centering
\includegraphics[width=8.8cm]{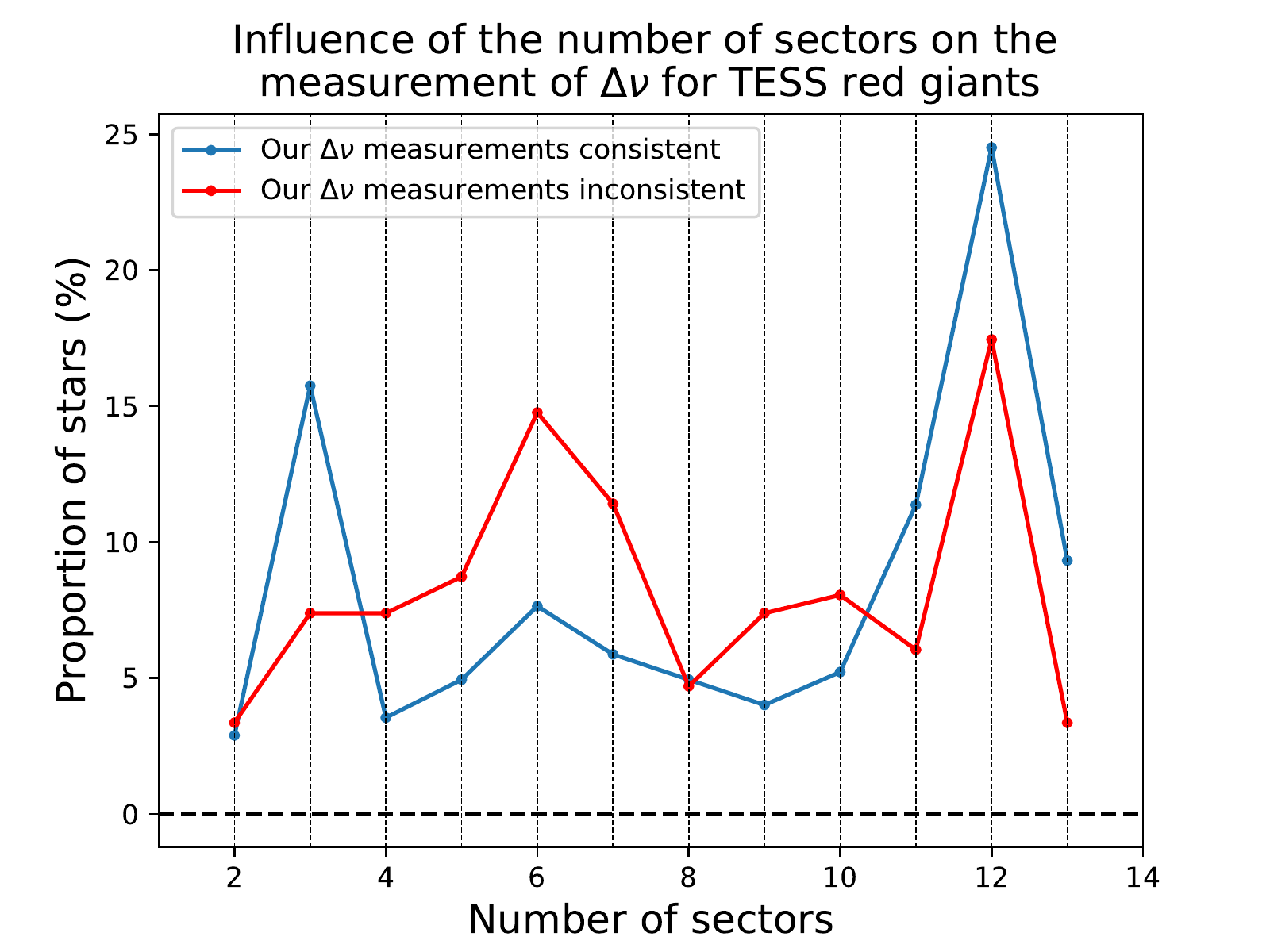}
\includegraphics[width=8.8cm]{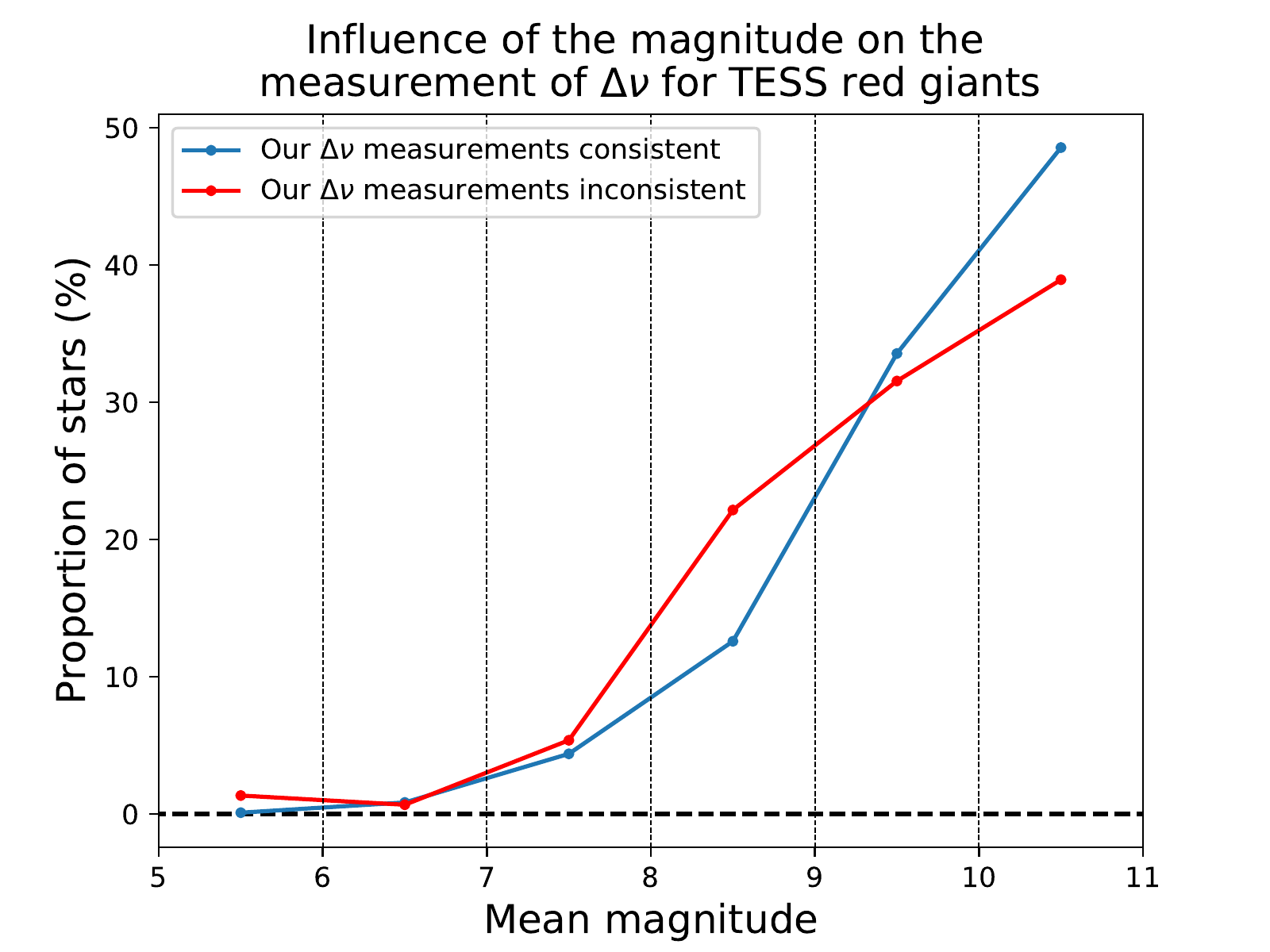}
\caption{Same as Fig. \ref{fig-TESS-impact-mag-Nsectors-numax}, but for $\Dnu$.}
\label{fig-TESS-impact-mag-Nsectors-Dnu}
\end{figure*}

\subsubsection{Guided search using $\numax$}

When measuring $\Dnu$ through a guided search using $\numax$ as an input parameter, we derived a $\Dnu$ measurement for 1725 stars, i.e. 73.6 \% of the analyzed sample (upper left panel of Fig.~\ref{fig-TESS-Dnu-guided}). As for the blinded search, the majority of the 619 stars for which we did not detect oscillations based on the measurement of $\Dnu$ have low reference $\Dnu$ measurements, below 5 $\mu$Hz with a peak around 4 $\mu$Hz (upper right panel of Fig.~\ref{fig-TESS-Dnu-guided}). These non-detections are discussed in Section \ref{discussion-Dnu}.

We have a median relative precision of 3.2 \% (Table \ref{table:Dnu}). This is slightly lower than for the blind search as we could retrieve a larger number of low $\Dnu$ values, which are more challenging to measure and are therefore less precise (Fig.~\ref{fig-rel-precision}). The computation time spent for each star is between 0.6 s and 14.8 s, with a median time of 1.3 s per run (Table \ref{table:Dnu}). This is similar to what we obtain for artificial TESS targets (Table \ref{table:synthetic-Dnu}).
 The median computation time is no longer significantly dependent on the number of observed TESS sectors (bottom panel of Fig.~\ref{fig-TESS-Dnu-guided}). Indeed, our guided search uses only one position of the Hanning filter using $\numax$ as an input parameter, so the search for $\Dnu$ is performed only for one targeted portion of the power spectrum instead of 18 different portions as with the blind search, minimising the significant increase in the number of data points in the spectrum as a result of the larger number of observed TESS sectors.

We now have only 10 stars for which the relative deviation between $\Dnu$ from \cite{Mackereth} and our measurements is of at least 10\% (upper left panel of Fig.~\ref{fig-TESS-Dnu-guided}). This gives a consistency rate of 99.4 \% (Table \ref{table:Dnu}). This consistency rate is higher than for TESS artificial stars for the same reason than for $\numax$.

 \begin{figure*}[h!]
\centering
\includegraphics[width=8.8cm]{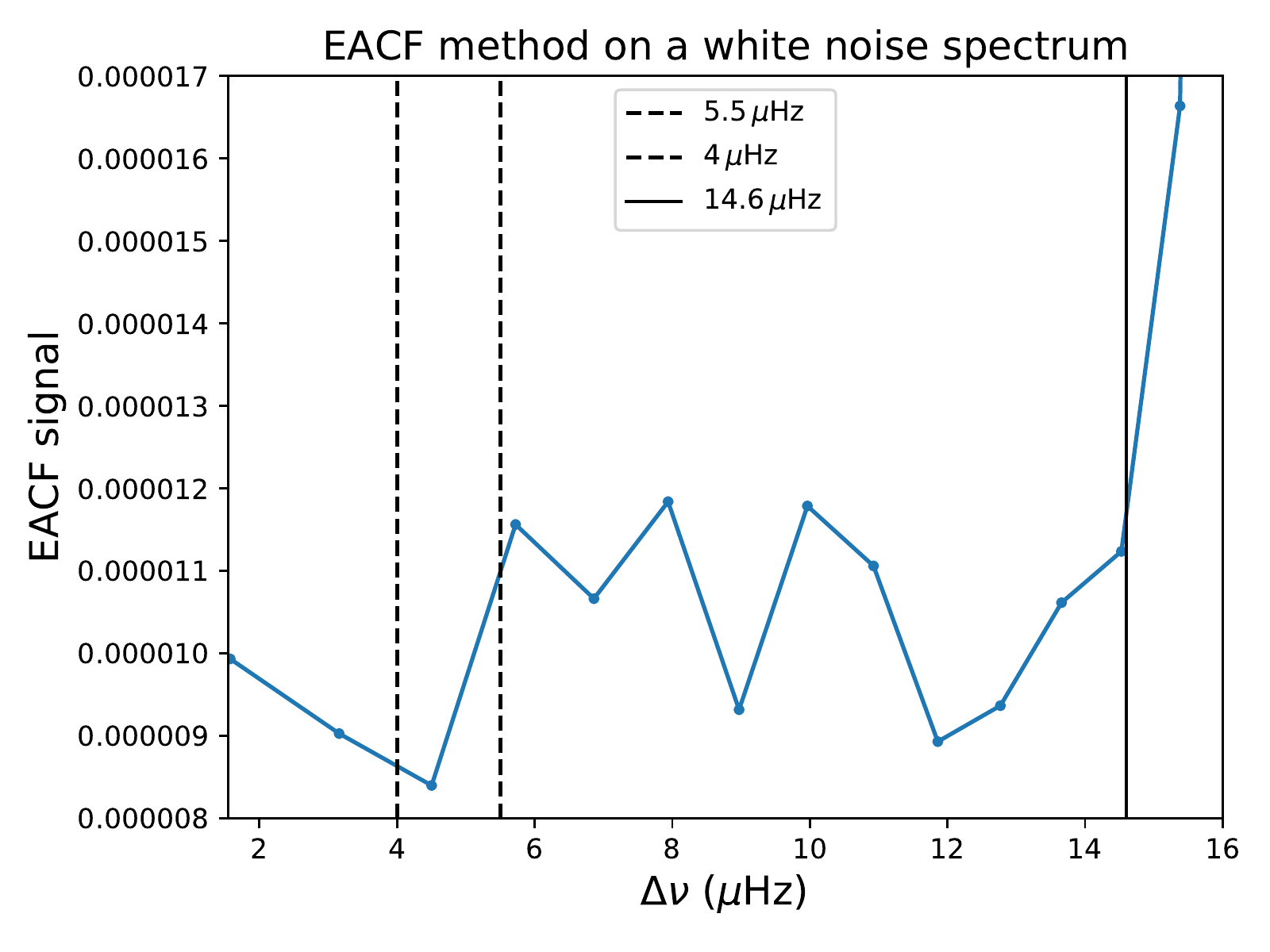}
\includegraphics[width=8.8cm]{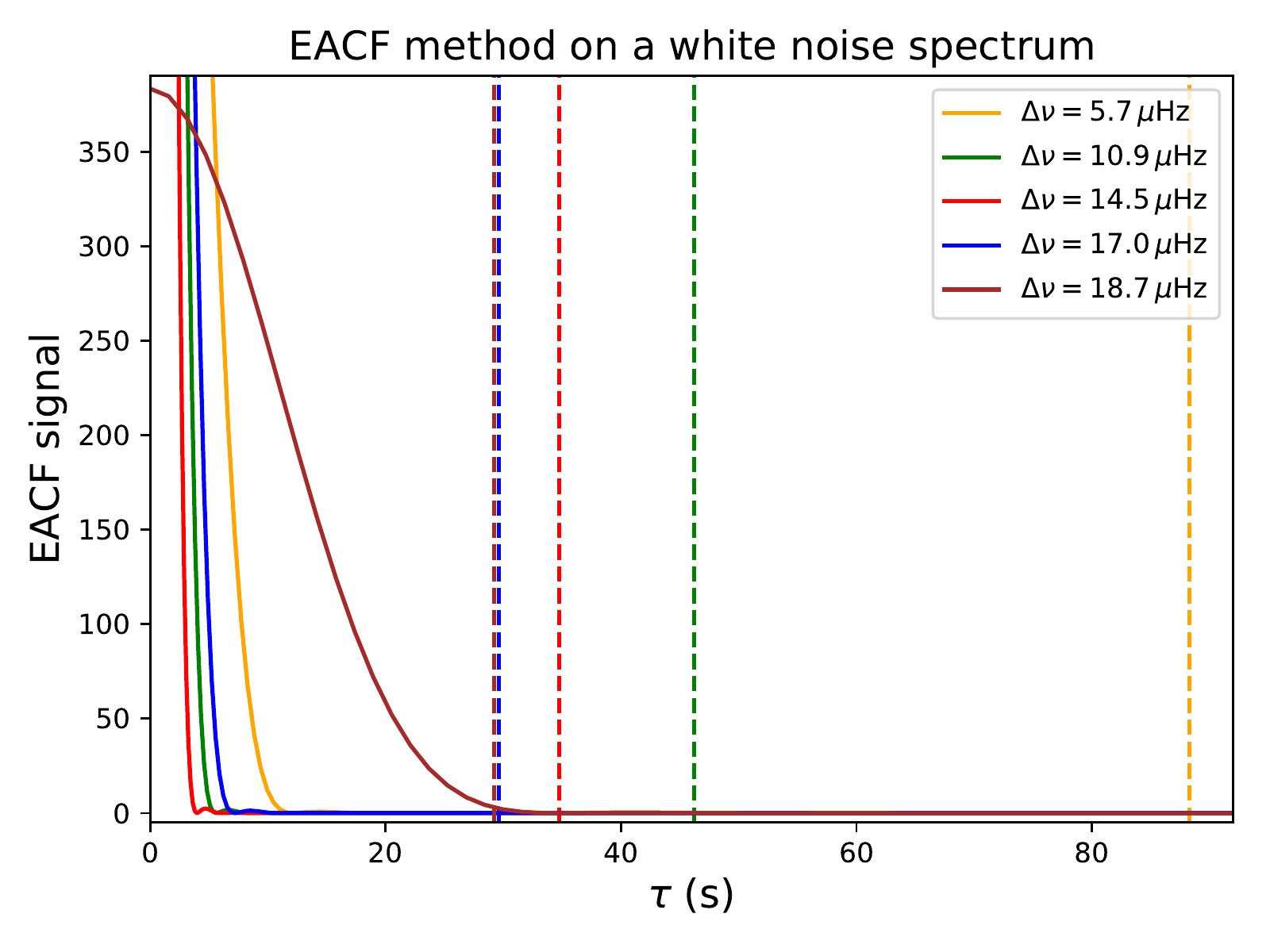}
\includegraphics[width=8.8cm]{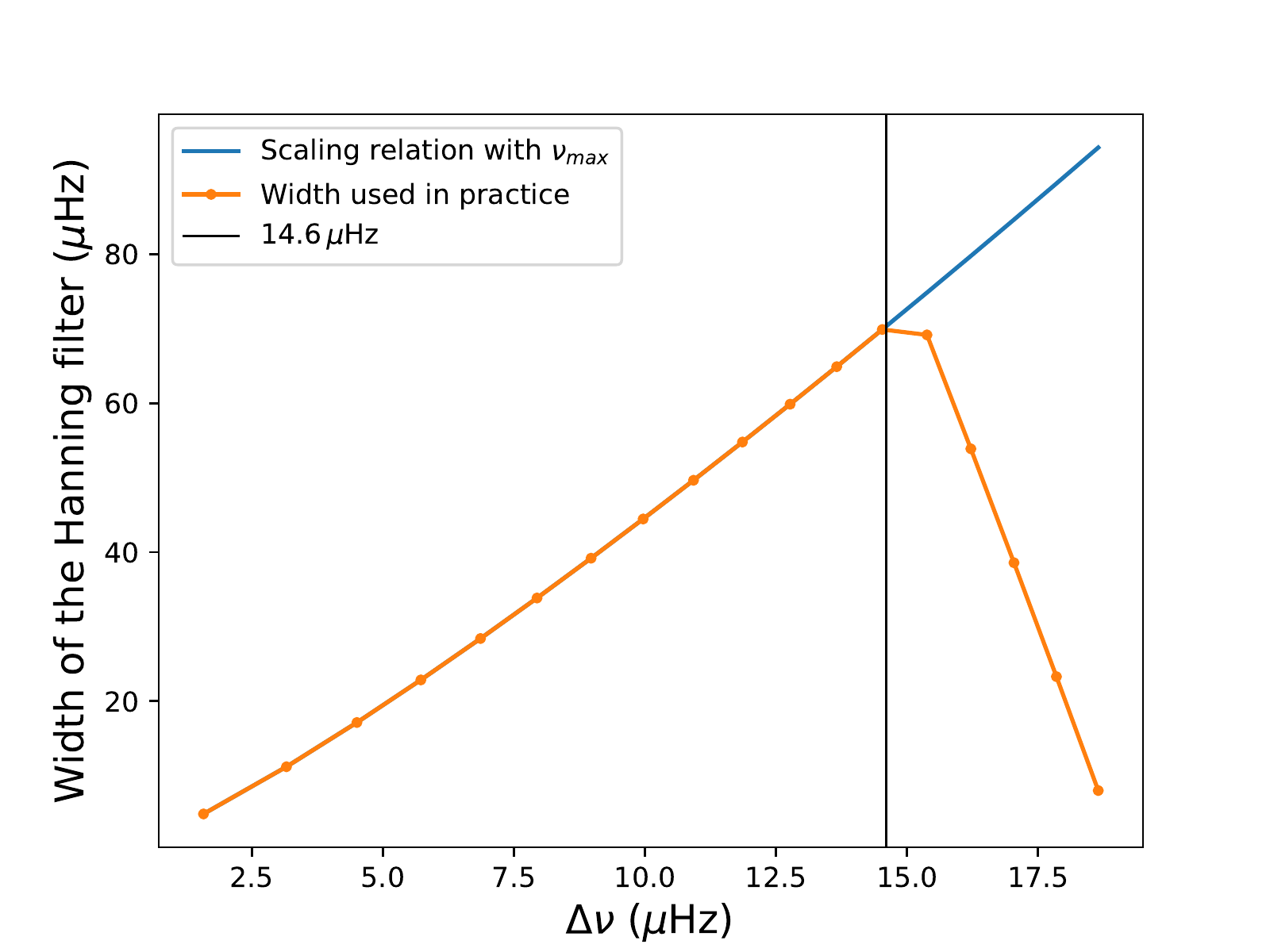}
\caption{\textit{Upper left:} Median value of the maximum EACF signal computed using 500 white noise power spectra, as a function of the tested $\Dnu$ value. Vertical dashed lines represent $\Dnu = 4 \, \mu$Hz and $\Dnu = 5.5 \, \mu$Hz, while the vertical continuous line represent $\Dnu = 14.6 \, \mu$Hz. \textit{Upper right:} EACF signal of a white noise power spectrum for 5 tested $\Dnu$ values, as a function of the time lag in the autocorrelation space. Vertical dashed lines represent the lower limit of the $\tau$ interval used to look for $\Dnu$ in the autocorrelation space, with same color code as the tested $\Dnu$ values. The dashed lines for $\Dnu = 17.0 \, \mu$Hz and $\Dnu = 18.7 \, \mu$Hz are almost superimposed. \textit{Bottom:} FWHM of the Hanning filter used to search for $\Dnu$ as a function of $\Dnu$. The blue curve corresponds to a width proportional to $\numax$ (Eq. \ref{eqt-dnuH}) while the orange curve corresponds to the actual width we use in this study.}
\label{fig-TESS-Hanning-impact-EACF}
\end{figure*}

\subsection{Impact of the stellar magnitude and the number of TESS sectors on the consistency of $\Dnu$}\label{discussion-Dnu}

Contrary to what we observe for $\numax$ measured for TESS red giants, we do not see such clear impact of the number of observed TESS sectors and the stellar magnitude on the consistency of our $\Dnu$ measurements obtained through a blind search for the TESS stars analyzed by \cite{Mackereth} (Fig.~\ref{fig-TESS-impact-mag-Nsectors-Dnu}). Indeed, the proportion of stars with inconsistent measurements appears to be above the proportion of stars with consistent measurements at intermediate $N\ind{sectors}$, between 4 and 10, and at intermediate stellar magnitude, between 7 and 9. The consistency of our $\Dnu$ measurements obtained through a blind search mostly depends on $\Dnu$ itself. Indeed, as stated earlier almost all our inconsistent measurements correspond to $\Dnu$ < 5.5 $\mu$Hz, of which 80 \% have $\Dnu \leq 4$ $\mu$Hz (upper left panel of Fig.~\ref{fig-TESS-Dnu-blind}). This is strikingly similar to the distribution of $\Dnu$ for stars for which we do not have a $\Dnu$ detection (upper right panel of Fig.~\ref{fig-TESS-Dnu-blind}). Hence, both the non-detections of low $\Dnu$ values and the inconsistent measurements of many low $\Dnu$ values have a similar cause, which has to do with the strength of the EACF signal.

We thus tested the strength of the EACF signal against the tested $\Dnu$ values by applying the EACF method to 500 artificial power spectra made of pure white noise. For each $\Dnu$ tested, we then compared the median value of the maximum EACF signal computed accross the 500 power spectra. We found that the EACF signal tends to be particularly low for low $\Dnu$ values, significantly decreasing around $\Dnu = 5.5 \, \mu$Hz and reaching a minimum close to $\Dnu = 4 \, \mu$Hz (upper left panel of Fig.~\ref{fig-TESS-Hanning-impact-EACF}). This is completely consistent with the $\Dnu$ values for which we do not find a detection and for which we obtain inconsistent measurements. This observed trend of the EACF signal being slightly higher for higher $\Dnu$ values leads to spurious detections in some cases, in particular for values around $\Dnu = 4 \, \mu$Hz for which the EACF signal is particularly low. We explored the reasons behind this trend. The higher the tested $\Dnu$ value, the smaller the lower limit of the $\tau$ interval used to look for $\Dnu$ in the autocorrelation space (upper right panel of Fig.~\ref{fig-TESS-Hanning-impact-EACF}). Hence, we are exploring an interval closer to the main lobe as $\Dnu$ increases, leading to a slightly higher EACF signal. This trend is even more pronounced for $\Dnu > 14.6 \, \mu$Hz. Indeed, the center frequency of the Hanning filter becomes too close to the Nyquist frequency to allow us to use the optimal width for the Hanning filter, forcing us to reduce this width (lower panel of Fig.~\ref{fig-TESS-Hanning-impact-EACF}). However, the smaller the FWHM of the Hanning filter, the larger the main lobe in the autocorrelation space. We are thus testing a $\tau$ interval which lower limit becomes particularly close to the edge of the main lobe for $\Dnu > 14.6 \, \mu$Hz (upper right panel of Fig.~\ref{fig-TESS-Hanning-impact-EACF}). We note that such cases do not lead to too many spurious detections since in many cases, the maximum of the EACF signal appears too close to the lower edge of the tested $\tau$ interval to allow this configuration to be considered as valid (see some examples in \ref{appendix-6}). We do not encounter this problem for \textit{Kepler} targets for two reasons. First, there are no $\Dnu < 6 \, \mu$Hz in the sample we analyzed (Fig.~\ref{fig-Kepler-Dnu}). Second, the increasing EACF signal with increasing $\Dnu$ should not be a problem when the signal-to-noise ratio is high, as it is the case for \textit{Kepler} targets with 4-year long observations, since the EACF signal is then high enough for low $\Dnu$ values to deliver a consistent measurement. However, this observed trend between the EACF signal and $\Dnu$ becomes a limiting factor for noisier data, making it challenging to derive consistent measurements when $\Dnu \leq$ 5.5 $\mu$Hz. Since our $\numax$ measurements proved to be highly consistent for TESS targets (Table \ref{table:numax}), using $\numax$ as an input parameter to guide the search for $\Dnu$ represents a viable alternative for these power spectra and largely reduces the spurious detections for stars with low $\Dnu$ values (Table \ref{table:Dnu}).

\section{Conclusions}\label{conclusion}

We developed a new pipeline to detect solar-like oscillations, which we named FRA, based on the detection and the measurement of $\numax$ through a local fit of the envelope of oscillations. FRA is entirely automated, fast (few seconds) and relies on statistical criteria to assess the presence of oscillations. It operates blindly, without any needed a priori information on the presence and the location of oscillations. It can detect solar-like oscillations and provide $\numax$ measurements for $\numax \gtrsim 10 \, \mu$Hz. We also used the Envelope AutoCorrelation Function (EACF) method \citep{Mosser_2009} to measure $\Dnu$ in addition to $\numax$, since both parameters are crucial to derive precise and accurate stellar masses and radii.

We applied our pipeline to a set of 1589 red giants observed by \textit{Kepler} which have 4-year long lightcurves \citep{Gehan_2018, Gehan_2021}, as well as a set of 2344 TESS red giants having between 2 and 13 observed sectors \citep{Mackereth}. All these are bona fide stars, for which the presence of oscillations is already established. We obtain consistent $\numax$ and $\Dnu$ measurements for all \textit{Kepler} stars. For TESS stars, we obtain consistent $\numax$ measurements in more than 97 \% of the cases, and consistent $\Dnu$ measurements in almost 88 \% of the cases. The inconsistent $\numax$ measurements we obtain majoritarily correspond to a low number of observed TESS sectors, $N\ind{sectors} \leq 3$, and/or a large G magnitude, above 10. Regarding $\Dnu$, we majoritarily get inconsistent measurements for low values, i.e. $\Dnu \leq 5.5 \, \mu$Hz, independently of the stellar magnitude and the number of observed TESS sectors. We tested our implementation of the EACF method on artificial power spectra made of pure white noise and found that the strength of the EACF signal tends to increase with the tested $\Dnu$ value, resulting in many spurious measurements for low $\Dnu$ values. This behaviour is significant for TESS targets with much shorter observation durations than \textit{Kepler} and, therefore, lower signal-to-noise ratios. We could overcome this limitation by using the measured $\numax$ as an input parameter to guide the search for $\Dnu$, which leads to a consistency rate above 99 \% with existing measurements for $\Dnu$ for TESS stars. Given the high consistency of our $\numax$ measurements, this approach appears sensible to optimize the $\Dnu$ measurement for TESS targets.

We additionally analyzed a set of 254 artificial power spectra representative of TESS red giants, of which 76 \% have injected oscillations and 24 \% have no injected oscillations. Analyzing this artificial data set provides limits in stellar magnitude and in the number of observed TESS sectors to obtain consistent measurements, to maximise the detectability of oscillations and to minimise the false positive detections. Our analysis reveals that we can expect to get consistent $\numax$ and $\Dnu$ measurements while minimizing both the false positive measurements and the non-detections for a number of observed TESS sectors $N\ind{sectors} > 3$. This is in agreement with the limit we obtained to derive consistent $\numax$ measurements for the TESS targets from \cite{Mackereth}. For a G magnitude above 9.5, one extra step has to be performed to discard spurious $\numax$ measurements by assessing that the obtained $\numax$ and $\Dnu$ are consistent with each other.

\section*{Acknowledgments}
CG thanks B. Mosser and J. T. Mackereth for providing the power spectra of the \textit{Kepler} and the TESS red giants analyzed in this study. The authors acknowledge the support by FCT/MCTES through the research grants UIDB/04434/2020, UIDP/04434/2020 and PTDC/FIS-AST/30389/2017, and by FEDER - Fundo Europeu de Desenvolvimento Regional through COMPETE2020 - Programa Operacional Competitividade e Internacionalização (grant: POCI-01-0145-FEDER-030389). CG was also supported by Max Planck Society (Max Planck Gesellschaft) grant “Preparations for PLATO Science” M.FE.A.Aero 0011. MSC and TLC are supported by national funds through FCT in the form of work contracts (CEECIND/02619/2017 and CEECIND/00476/2018, respectively).

\bibliography{./bibtex/biblio}


\appendix

\section{\textit{Kepler} red giants for which we have a relative deviation of at least 10\% compared to existing $\numax$ measurements}\label{appendix-1}

\begin{figure*}
\centering
\includegraphics[width=8.8cm]{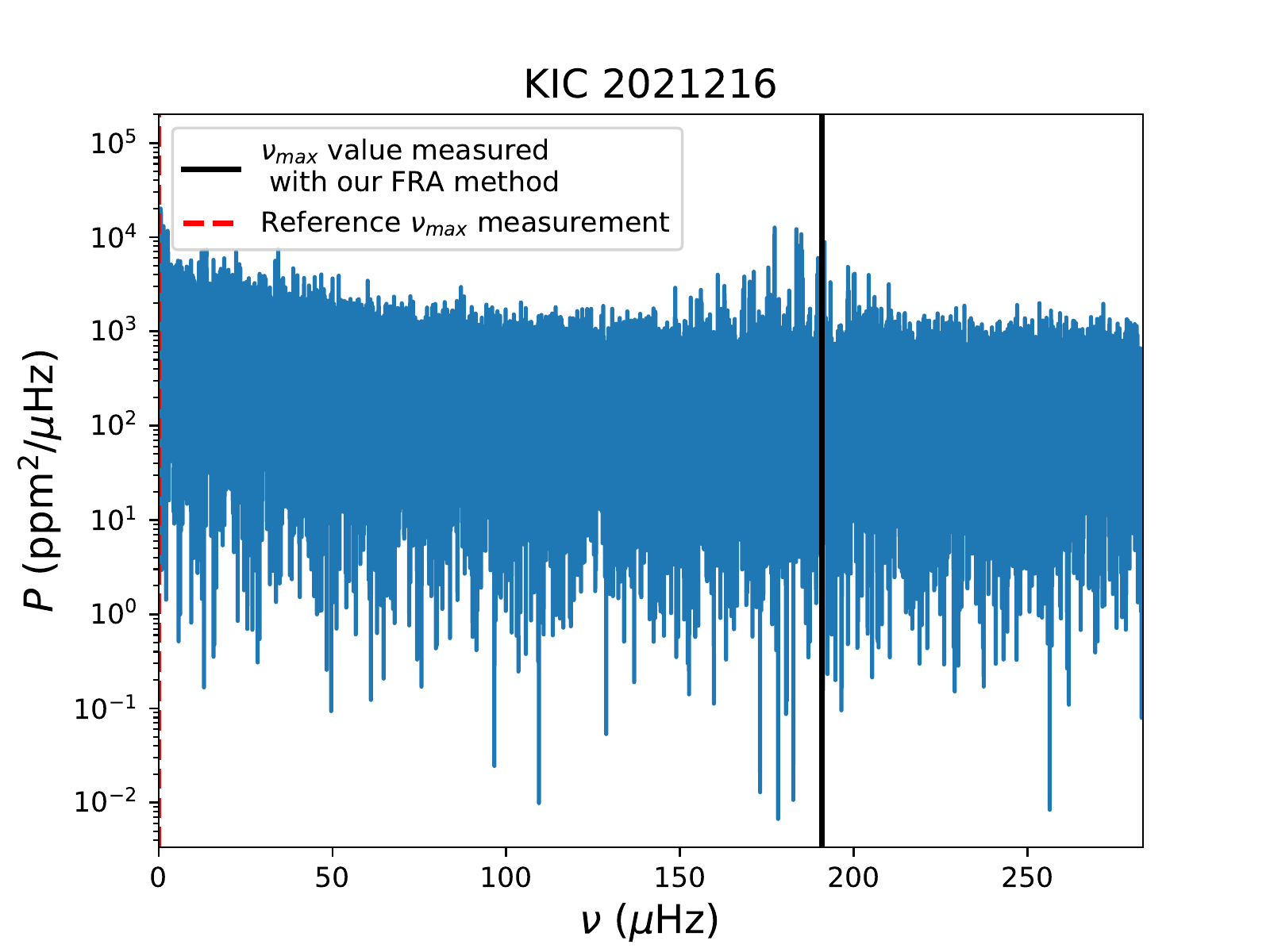}
\includegraphics[width=8.8cm]{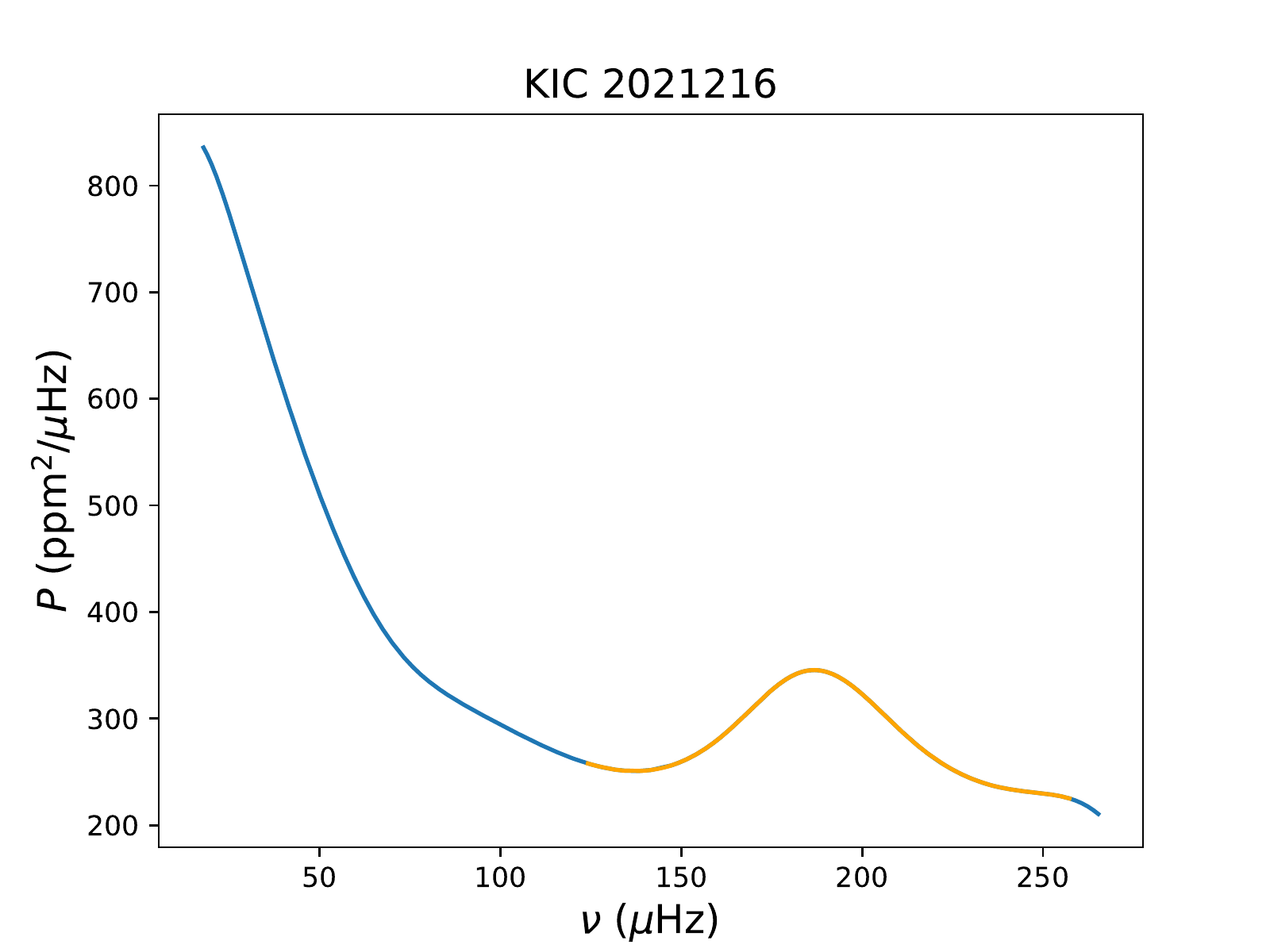}
\caption{Same as Fig.~\ref{fig-spectrum} for KIC 2021216. The reference $\numax$ value is 0 $\mu$Hz.}
\label{fig-Kepler-high-rel-dev-1}
\end{figure*}

\begin{figure*}
\centering
\includegraphics[width=8.8cm]{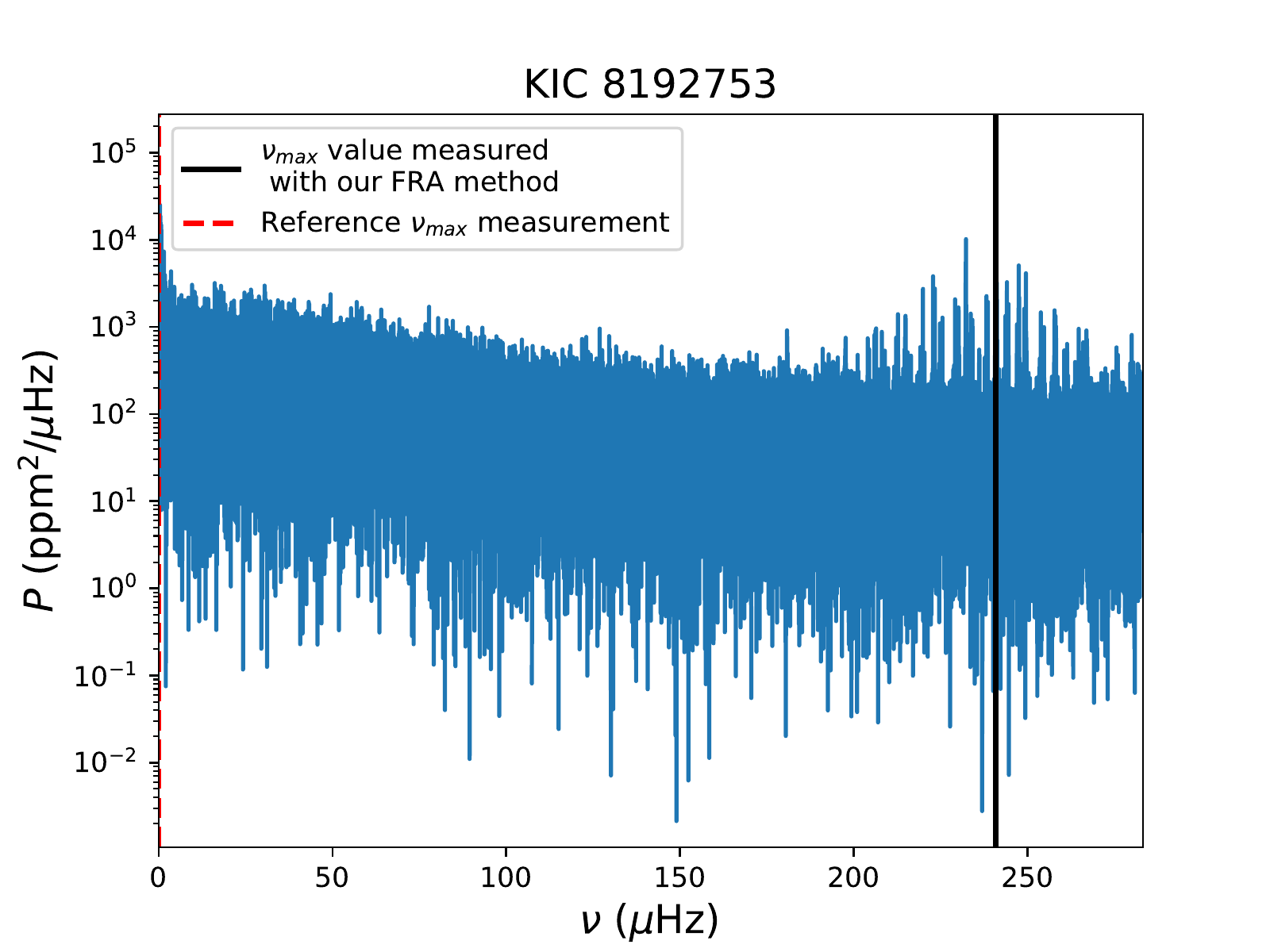}
\includegraphics[width=8.8cm]{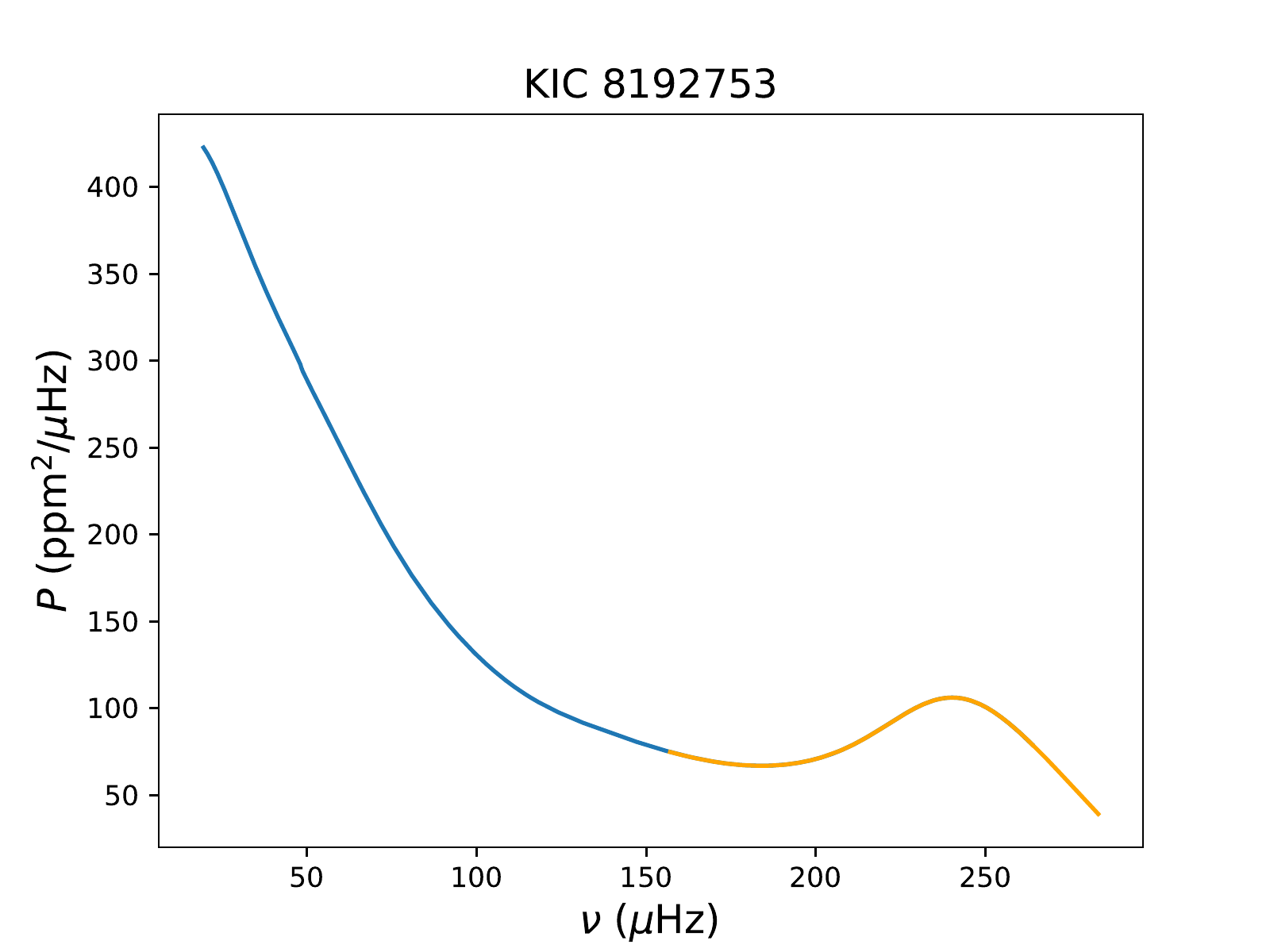}
\caption{Same as Fig.~\ref{fig-Kepler-high-rel-dev-1} for KIC 8192753. The reference $\numax$ value is 0 $\mu$Hz.}
\label{fig-Kepler-high-rel-dev-2}
\end{figure*}

\begin{figure*}
\centering
\includegraphics[width=8.8cm]{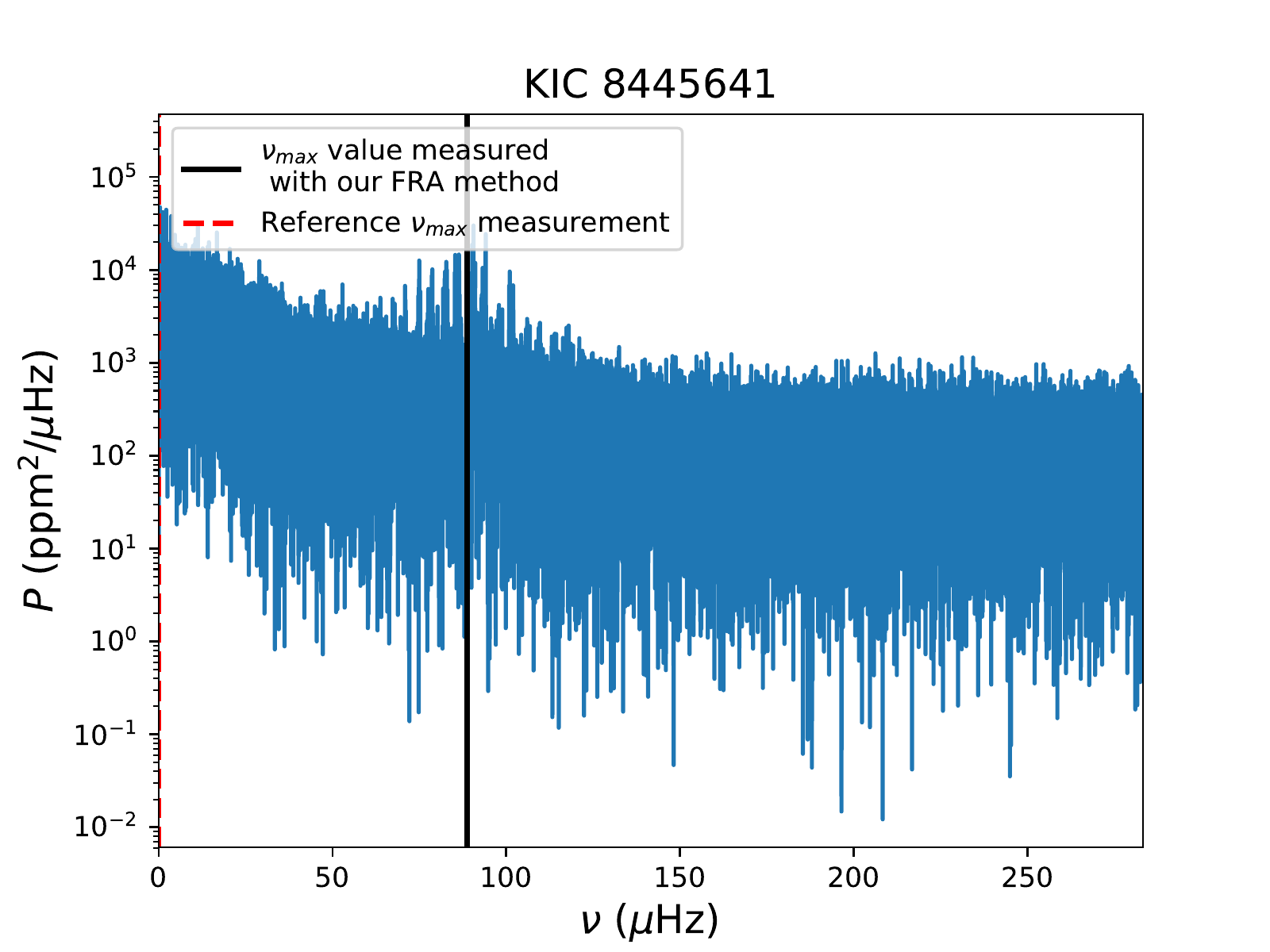}
\includegraphics[width=8.8cm]{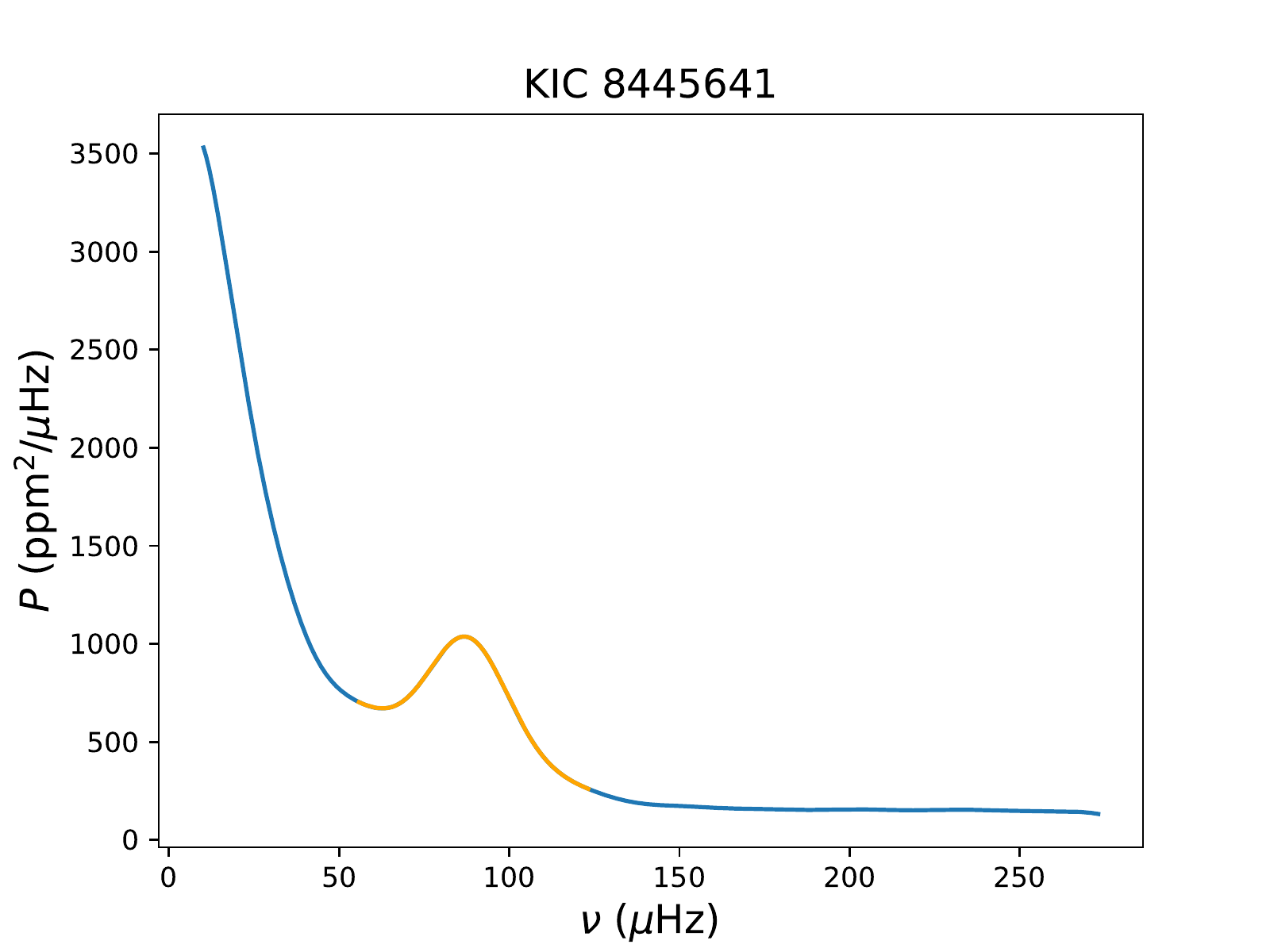}
\caption{Same as Fig.~\ref{fig-Kepler-high-rel-dev-1} for KIC 8445641. The reference $\numax$ value is 0 $\mu$Hz.}
\label{fig-Kepler-high-rel-dev-3}
\end{figure*}

\begin{figure*}
\centering
\includegraphics[width=8.8cm]{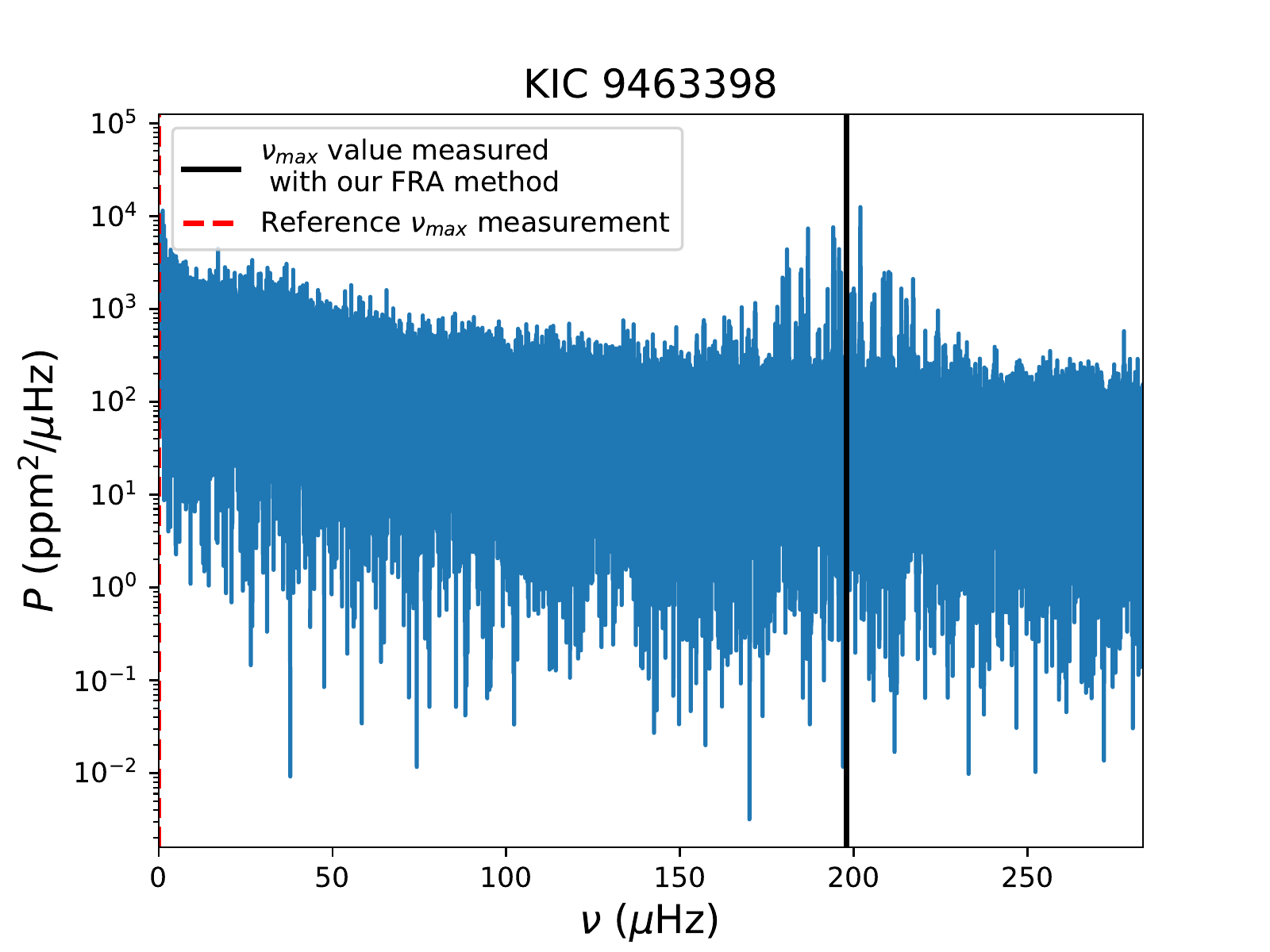}
\includegraphics[width=8.8cm]{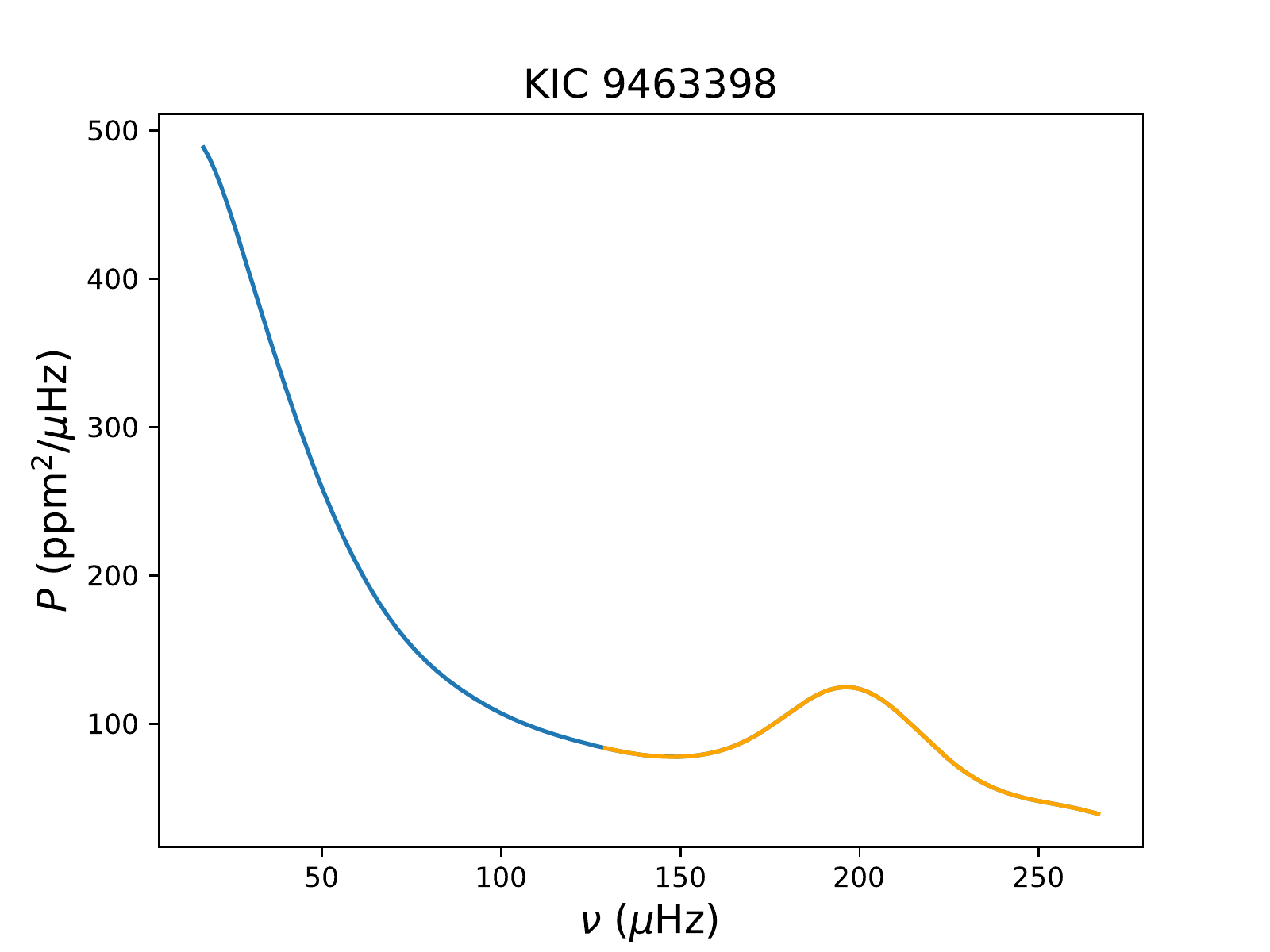}
\caption{Same as Fig.~\ref{fig-Kepler-high-rel-dev-1} for KIC 9463398. The reference $\numax$ value is 0 $\mu$Hz.}
\label{fig-Kepler-high-rel-dev-4}
\end{figure*}

\begin{figure*}
\centering
\includegraphics[width=8.8cm]{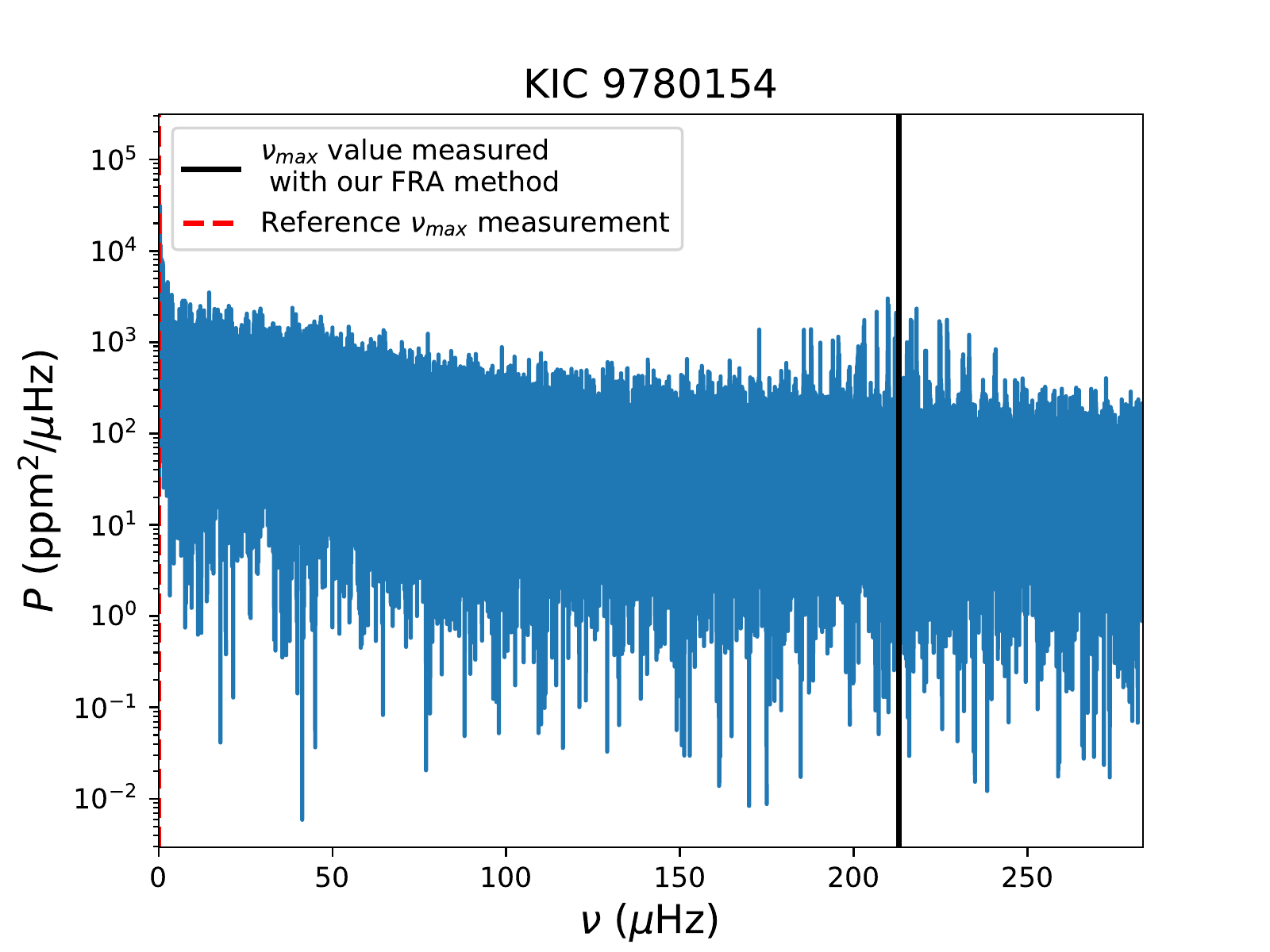}
\includegraphics[width=8.8cm]{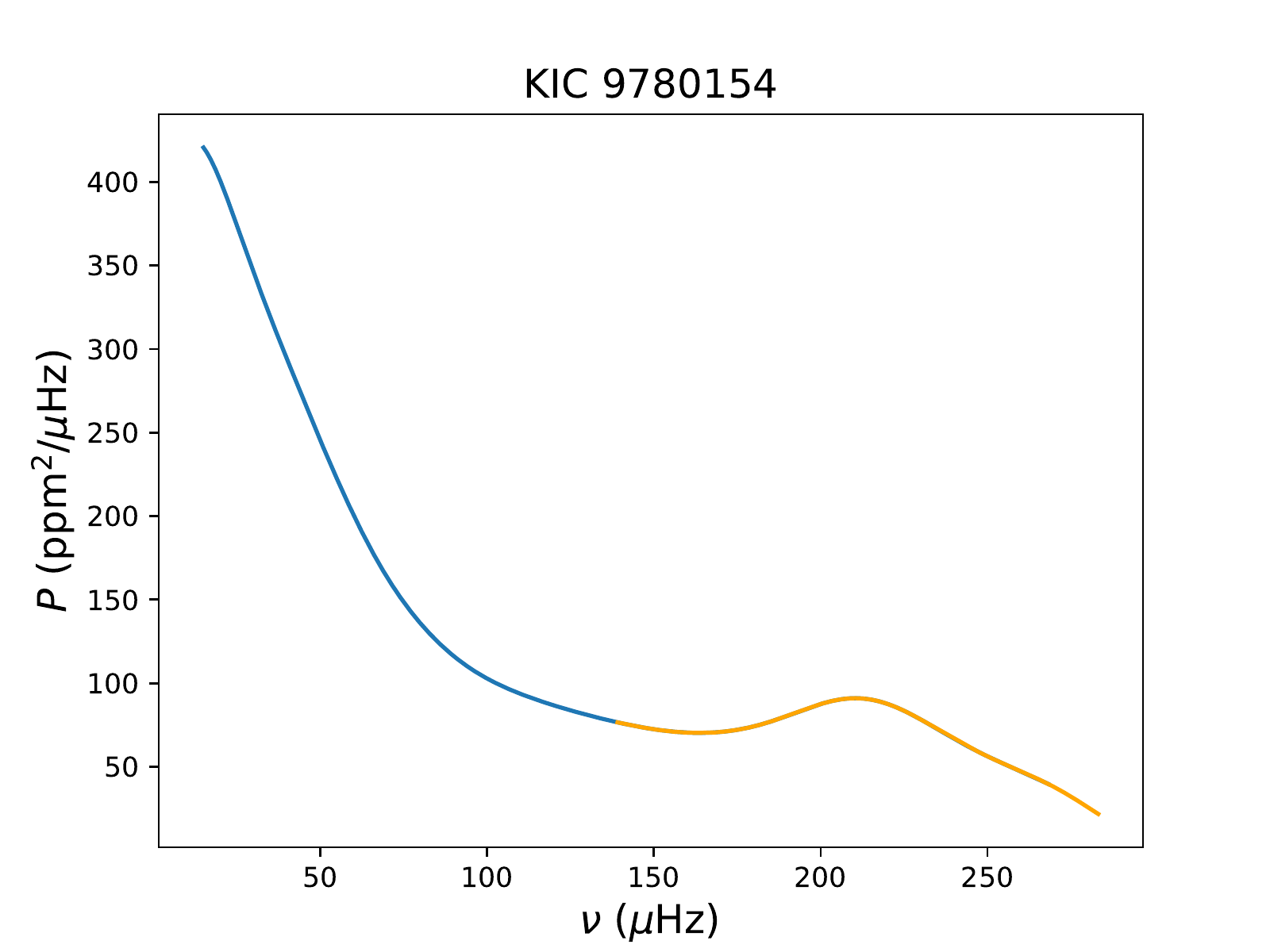}
\caption{Same as Fig.~\ref{fig-Kepler-high-rel-dev-1} for KIC 9780154. The reference $\numax$ value is 0 $\mu$Hz.}
\label{fig-Kepler-high-rel-dev-5}
\end{figure*}

\begin{figure*}
\centering
\includegraphics[width=8.8cm]{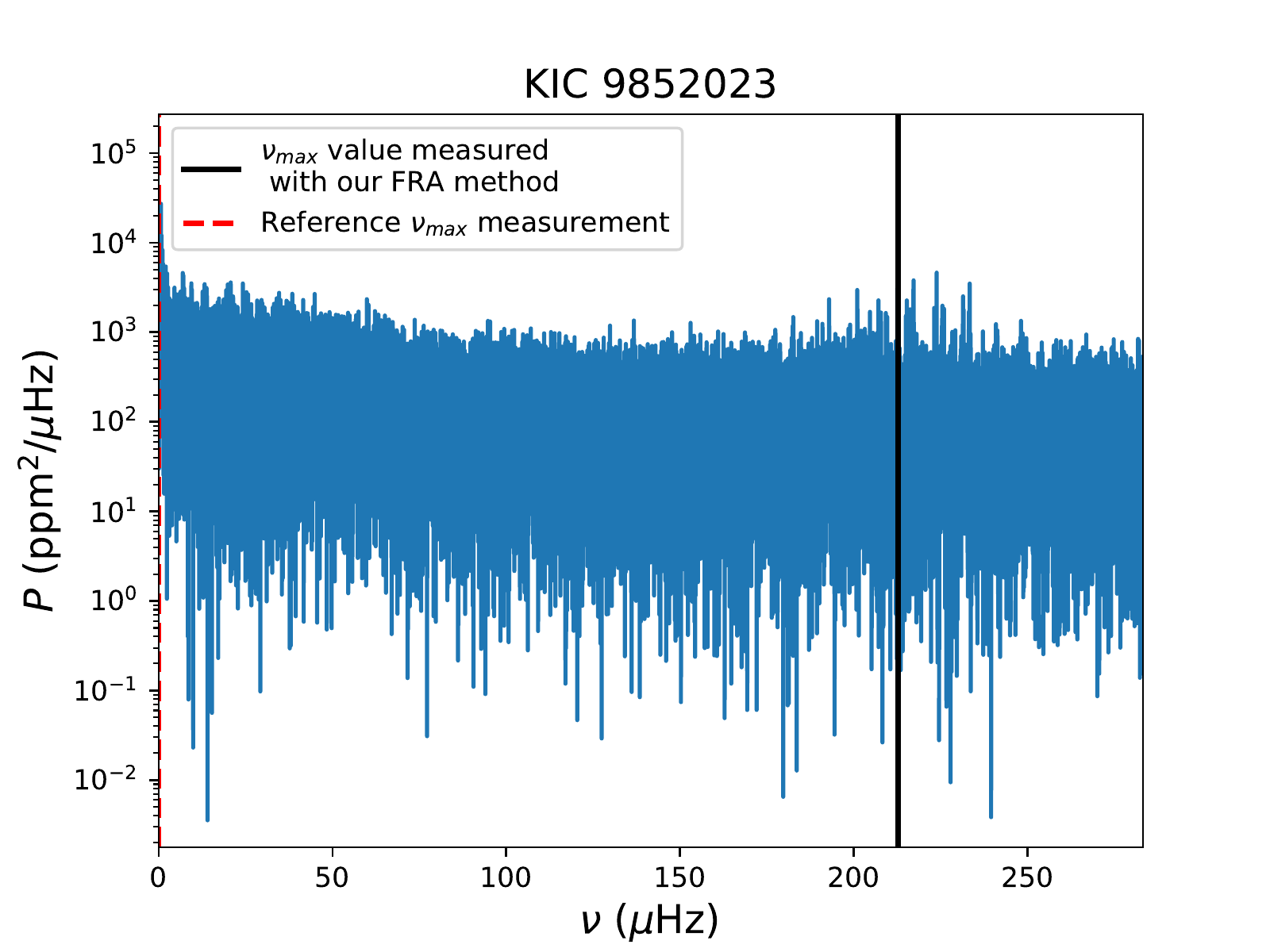}
\includegraphics[width=8.8cm]{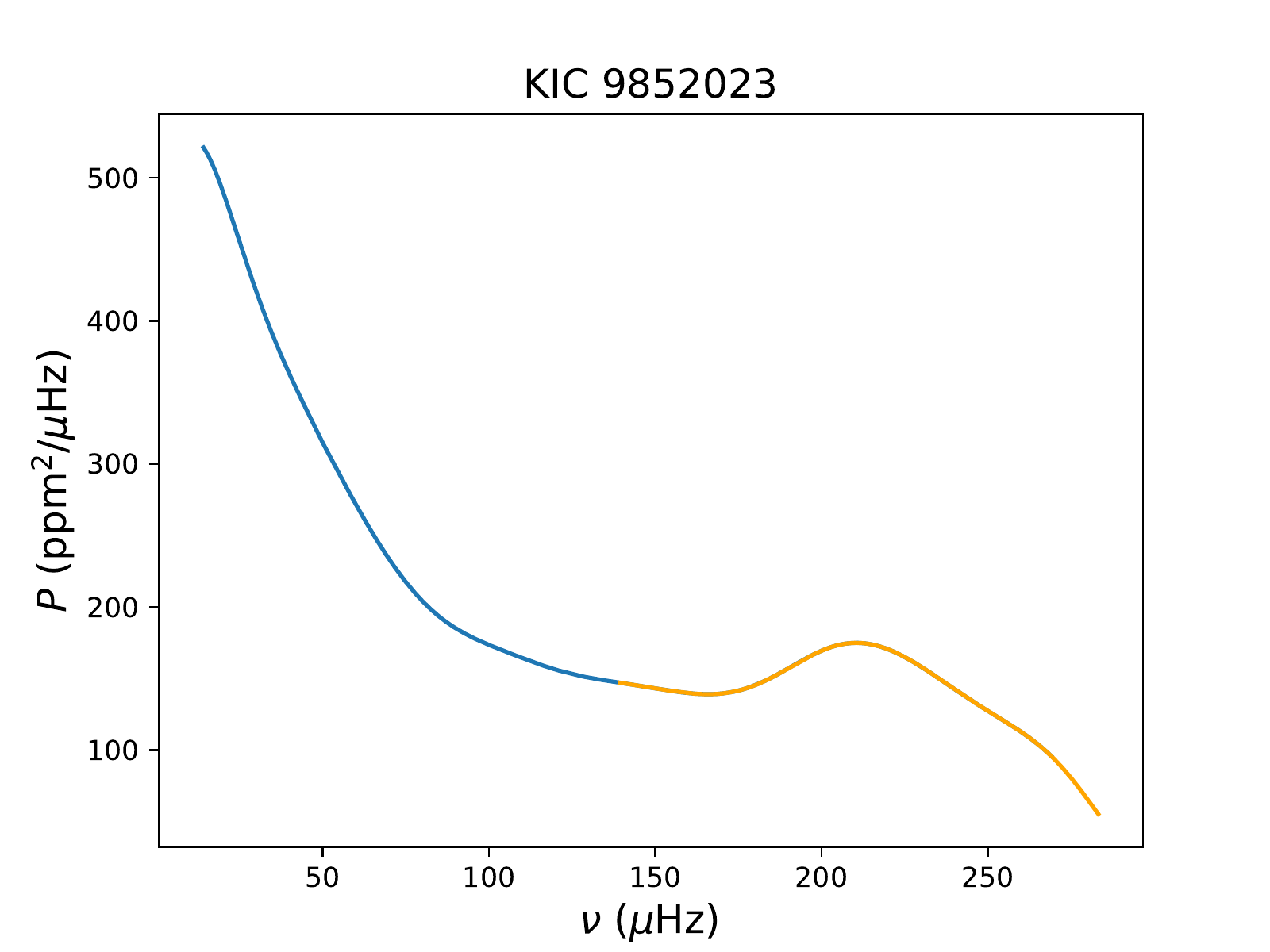}
\caption{Same as Fig.~\ref{fig-Kepler-high-rel-dev-1} for KIC 9852023. The reference $\numax$ value is 0 $\mu$Hz.}
\label{fig-Kepler-high-rel-dev-6}
\end{figure*}

\begin{figure*}
\centering
\includegraphics[width=8.8cm]{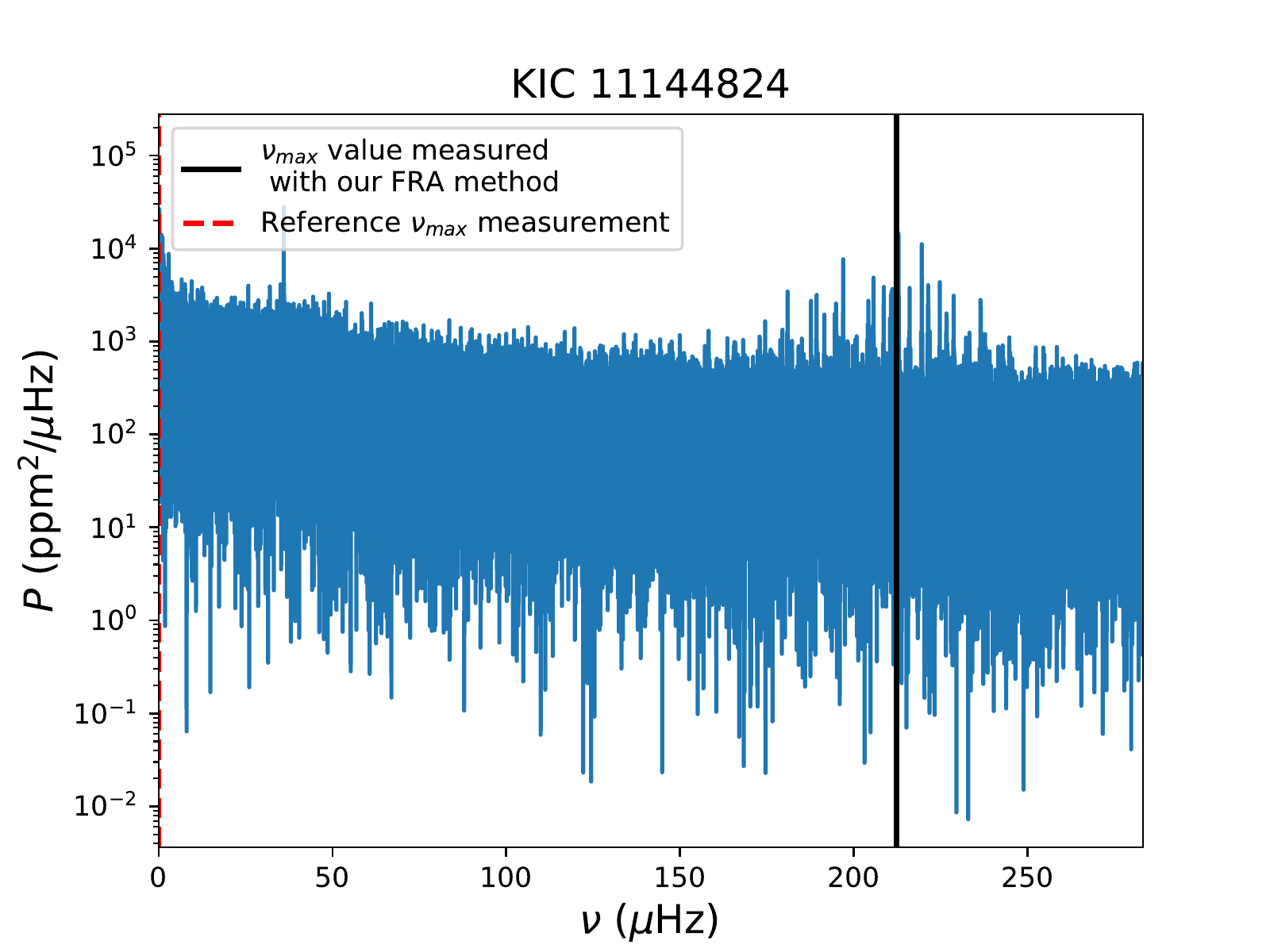}
\includegraphics[width=8.8cm]{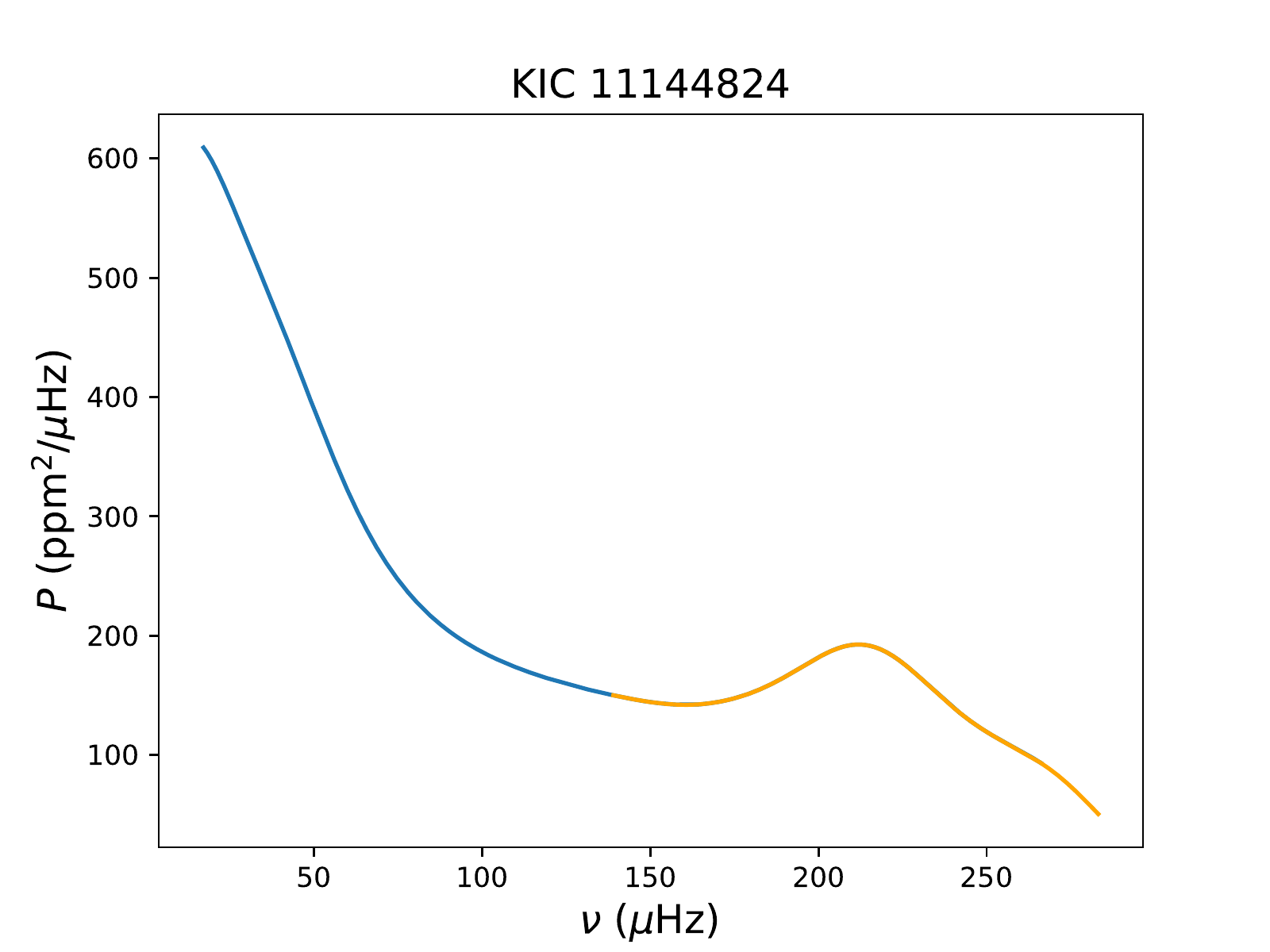}
\caption{Same as Fig.~\ref{fig-Kepler-high-rel-dev-1} for KIC 11144824. The reference $\numax$ value is 0 $\mu$Hz.}
\label{fig-Kepler-high-rel-dev-7}
\end{figure*}

\begin{figure*}
\centering
\includegraphics[width=8.8cm]{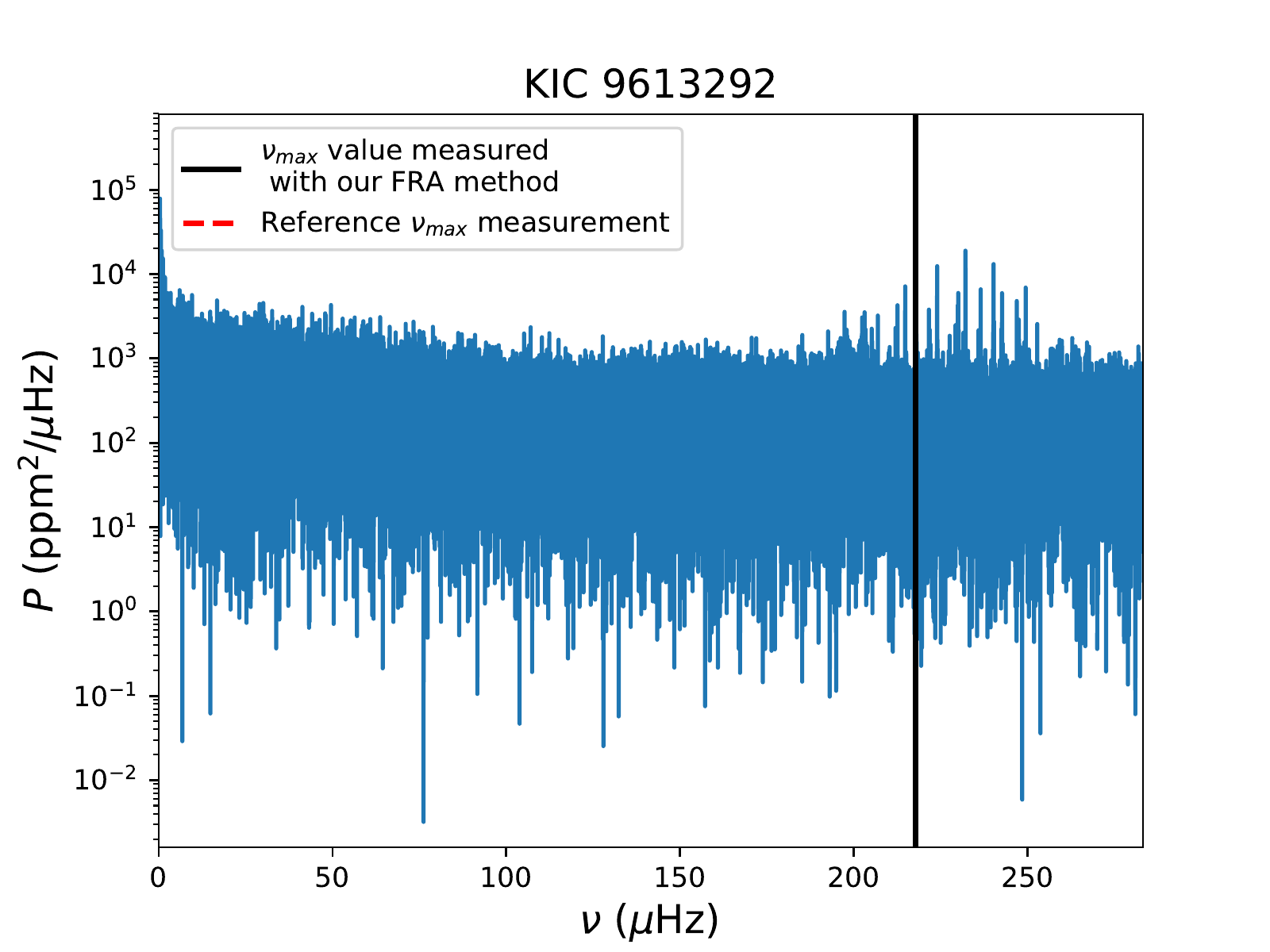}
\includegraphics[width=8.8cm]{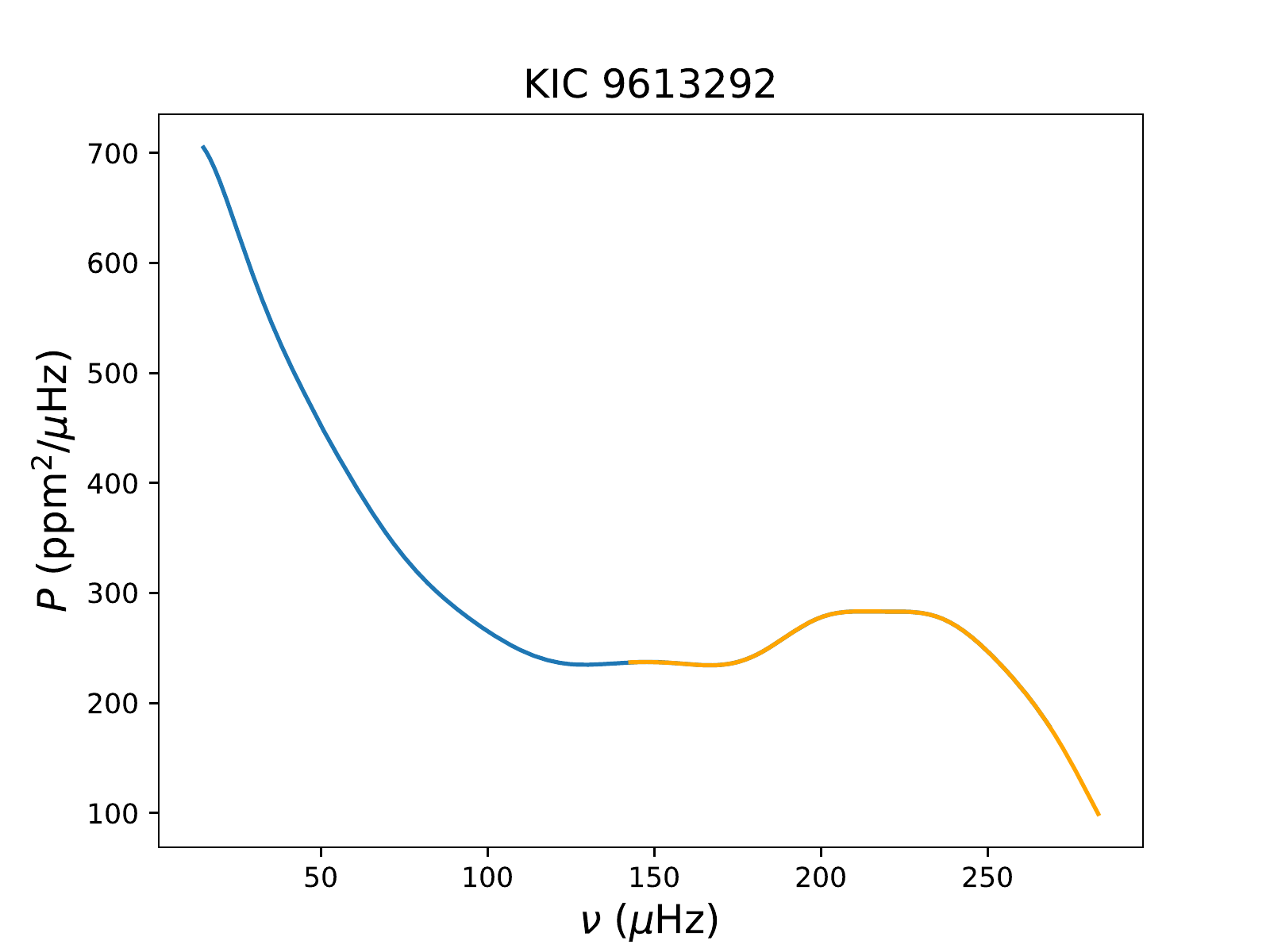}
\caption{Same as Fig.~\ref{fig-Kepler-high-rel-dev-1} for KIC 9613292. The reference $\numax$ value is 287.84 $\mu$Hz.}
\label{fig-Kepler-high-rel-dev-8}
\end{figure*}

There are 7 \textit{Kepler} red giants for which the EACF method from \cite{Mosser_2009} did not succeed in returning a $\numax$ value, which appear as having $\numax = 0$ $\mu$Hz, while our FRA pipeline gives an accurate measurement:
\begin{itemize}
\item KIC 2021216 (Fig.~\ref{fig-Kepler-high-rel-dev-1});
\item KIC 8192753 (Fig.~\ref{fig-Kepler-high-rel-dev-2});
\item KIC 8445641 (Fig.~\ref{fig-Kepler-high-rel-dev-3});
\item KIC 9463398 (Fig.~\ref{fig-Kepler-high-rel-dev-4});
\item KIC 9780154 (Fig.~\ref{fig-Kepler-high-rel-dev-5});
\item KIC 9852023 (Fig.~\ref{fig-Kepler-high-rel-dev-6});
\item KIC 11144824 (Fig.~\ref{fig-Kepler-high-rel-dev-7}).
\end{itemize}
There is additionally KIC 9613292 for which the EACF method from \cite{Mosser_2009} gave a $\numax$ value above the Nyquist frequency of 283 $\mu$Hz for \textit{Kepler} 30 minutes-cadence data, for which our FRA pipeline gives an accurate measurement (Fig.~\ref{fig-Kepler-high-rel-dev-8}).

\section{Examples of \textit{Kepler} red giants wih $\numax > 205$ $\mu$Hz}\label{appendix-2}

\begin{figure*}
\centering
\includegraphics[width=8.8cm]{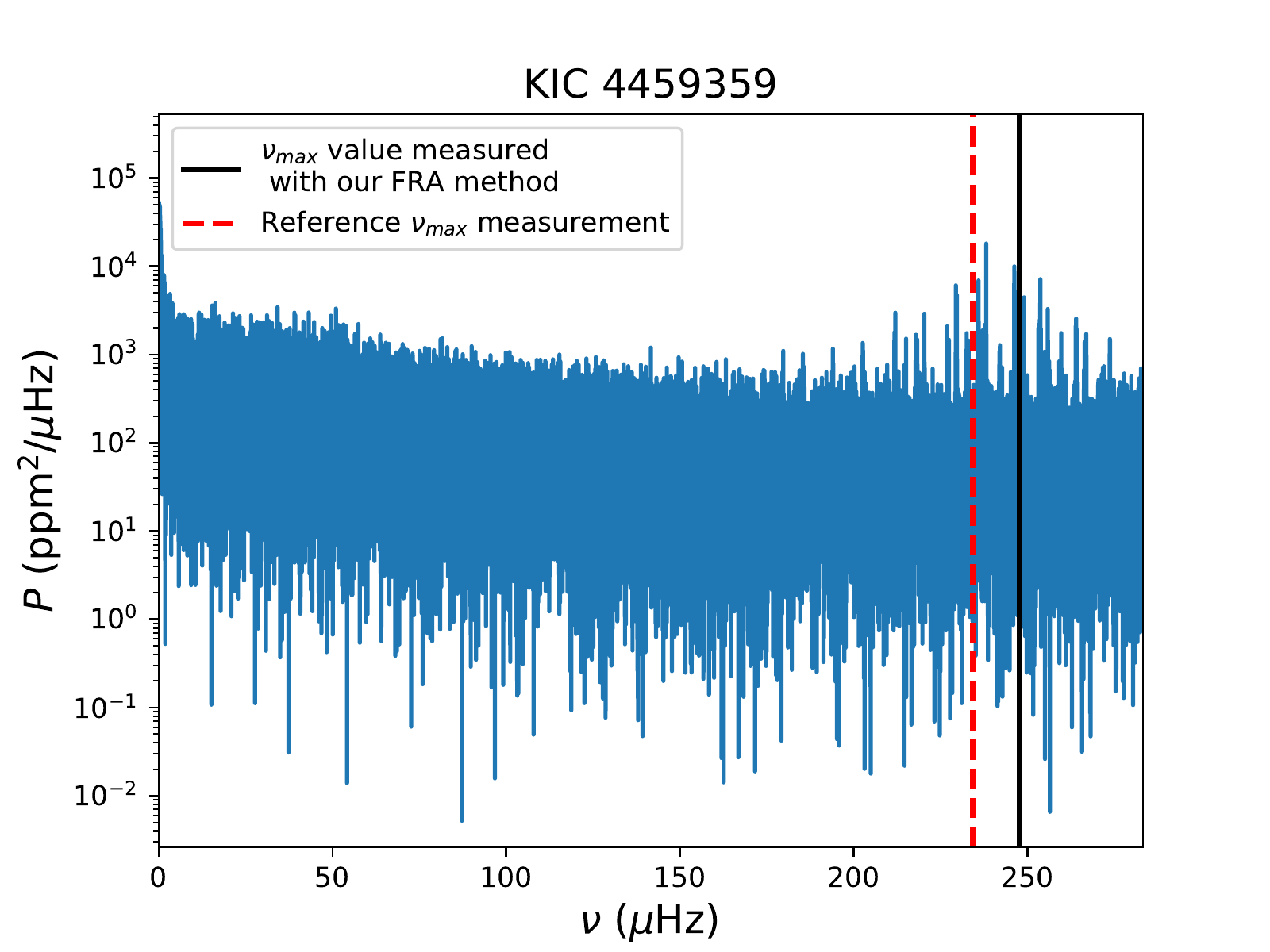}
\includegraphics[width=8.8cm]{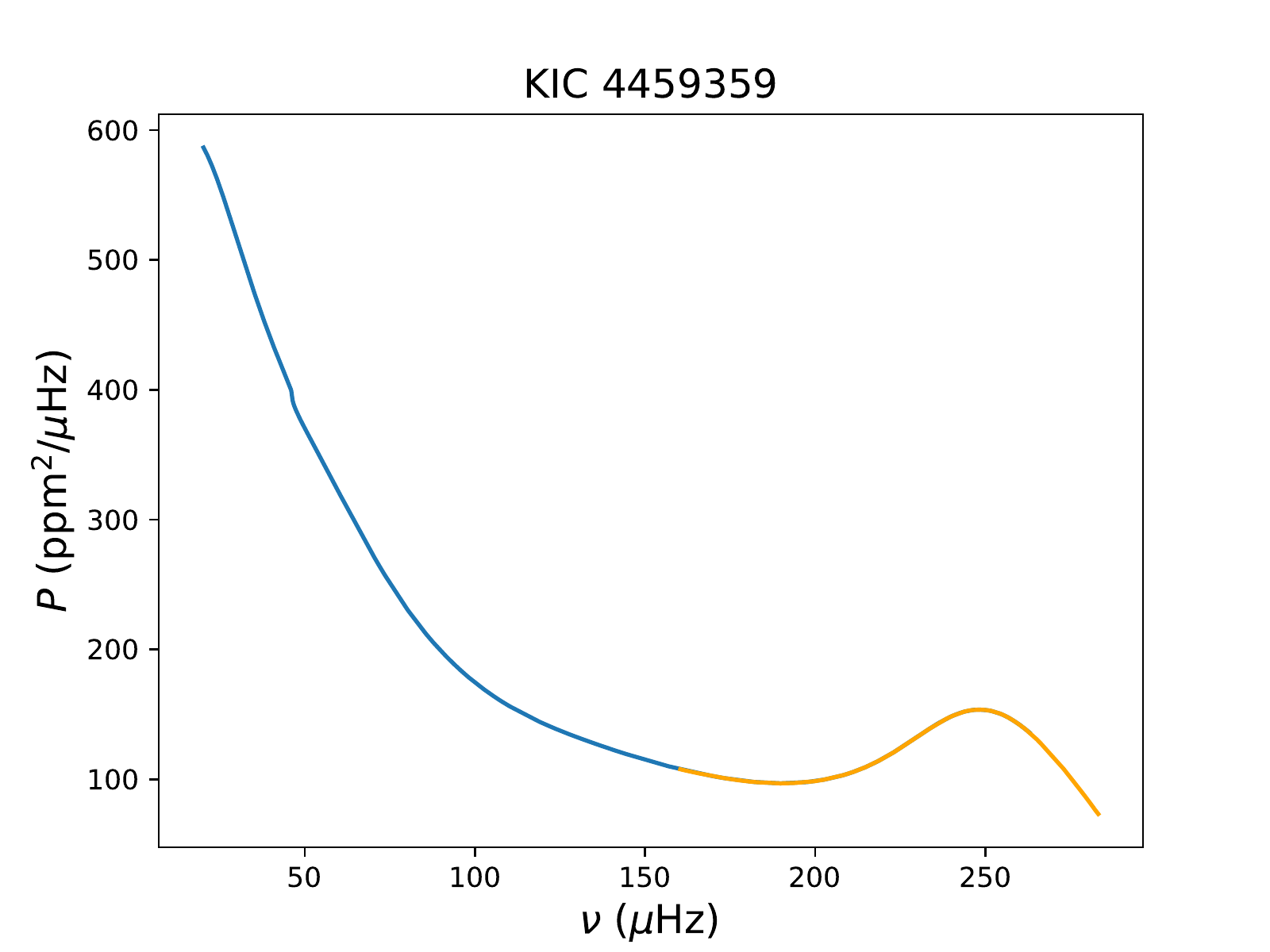}
\caption{Same as Fig.~\ref{fig-Kepler-high-rel-dev-1} for KIC 4459359. }
\label{fig-Kepler-high-numax-1}
\end{figure*}

\begin{figure*}
\centering
\includegraphics[width=8.8cm]{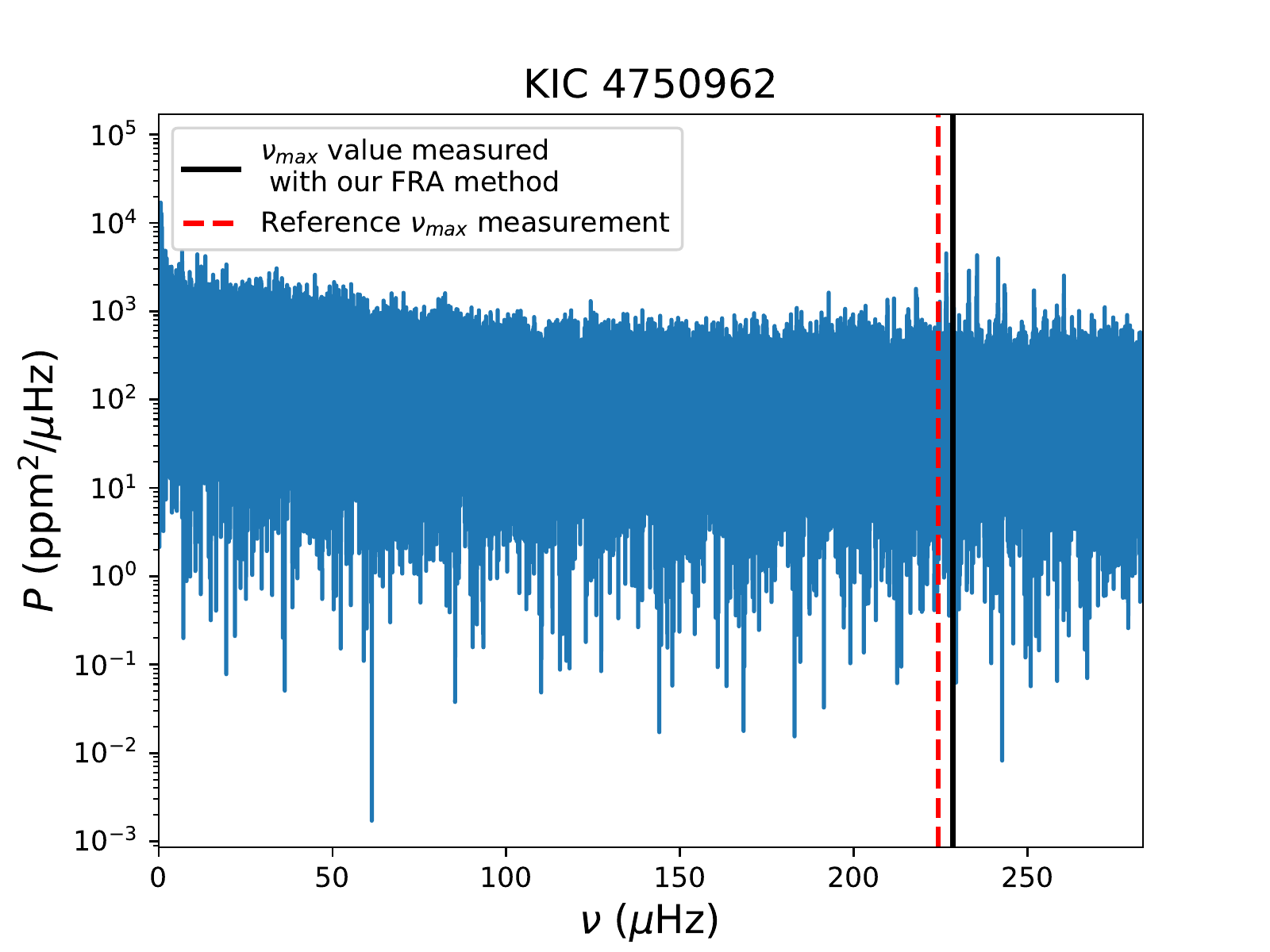}
\includegraphics[width=8.8cm]{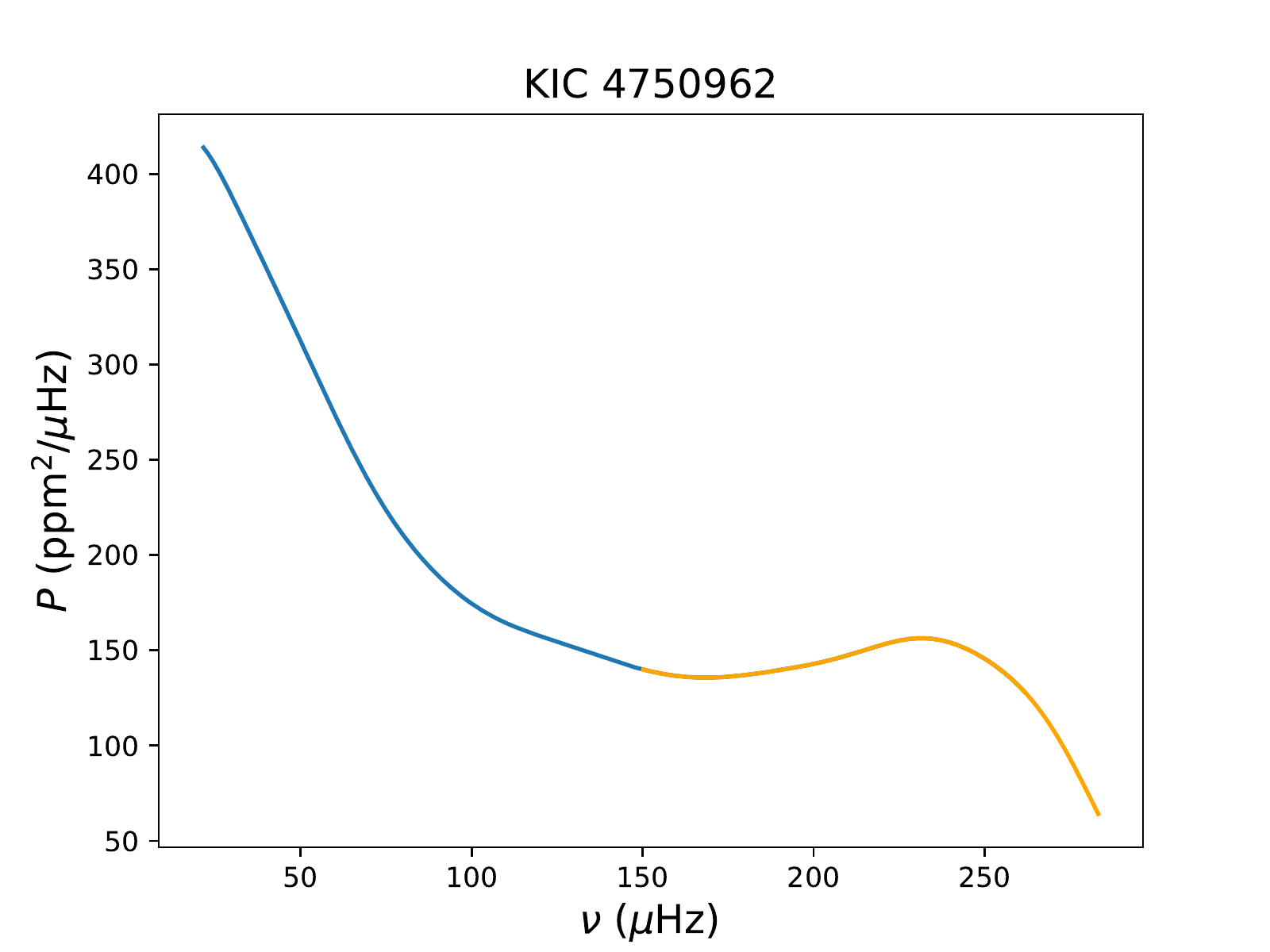}
\caption{Same as Fig.~\ref{fig-Kepler-high-rel-dev-1} for KIC 4750962. }
\label{fig-Kepler-high-numax-2}
\end{figure*}

\begin{figure*}
\centering
\includegraphics[width=8.8cm]{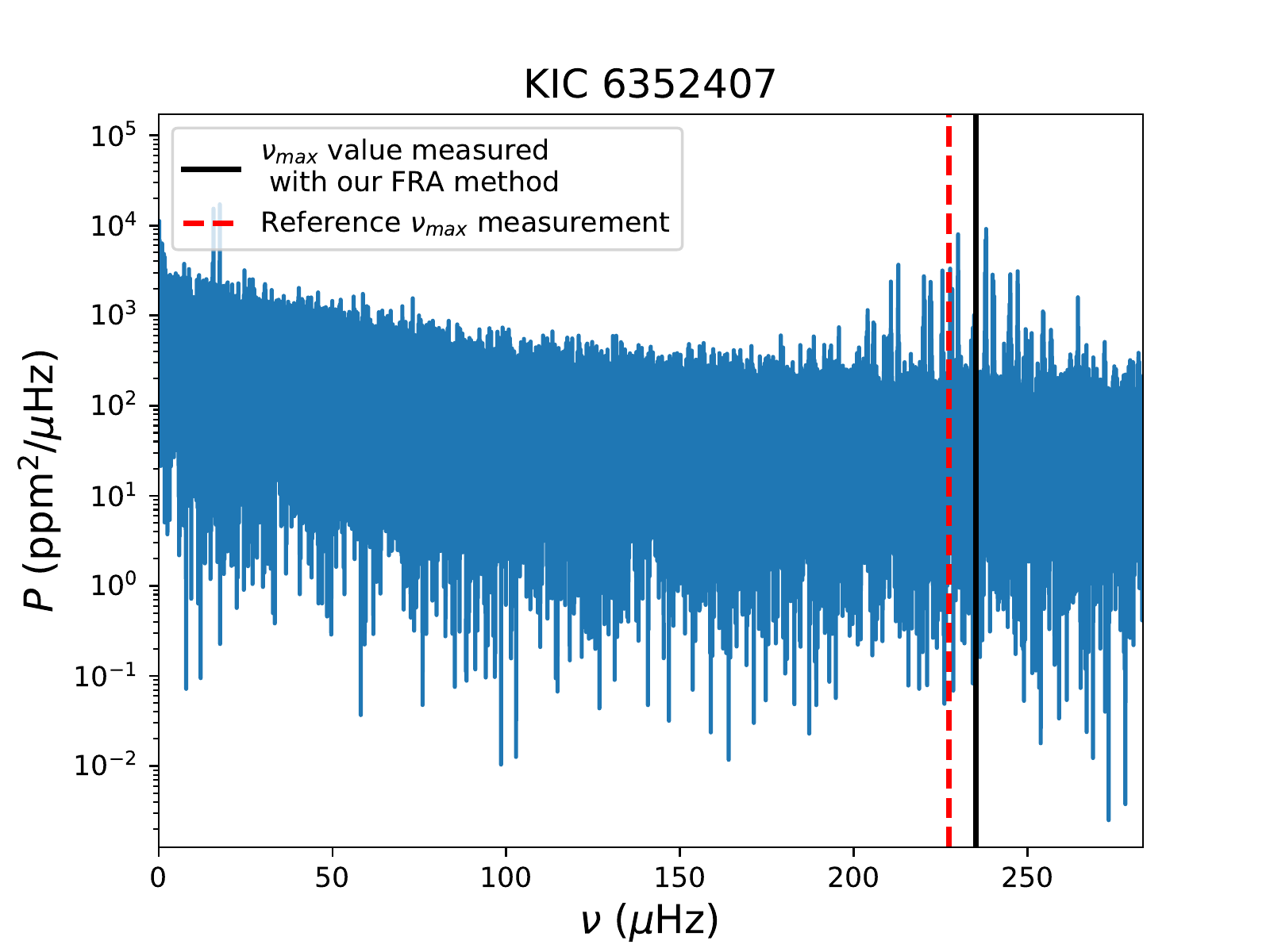}
\includegraphics[width=8.8cm]{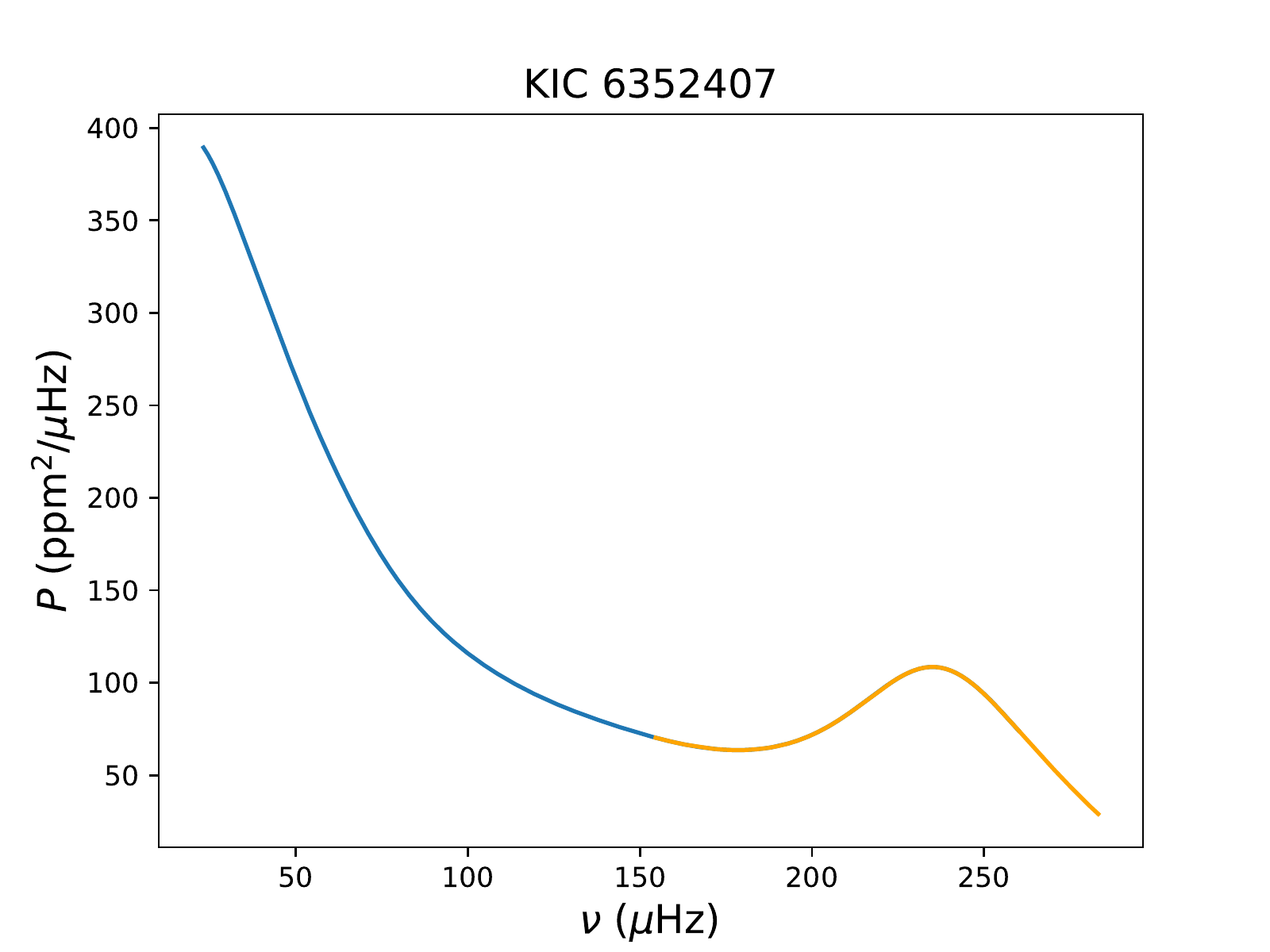}
\caption{Same as Fig.~\ref{fig-Kepler-high-rel-dev-1} for KIC 6352407. }
\label{fig-Kepler-high-numax-3}
\end{figure*}

\begin{figure*}
\centering
\includegraphics[width=8.8cm]{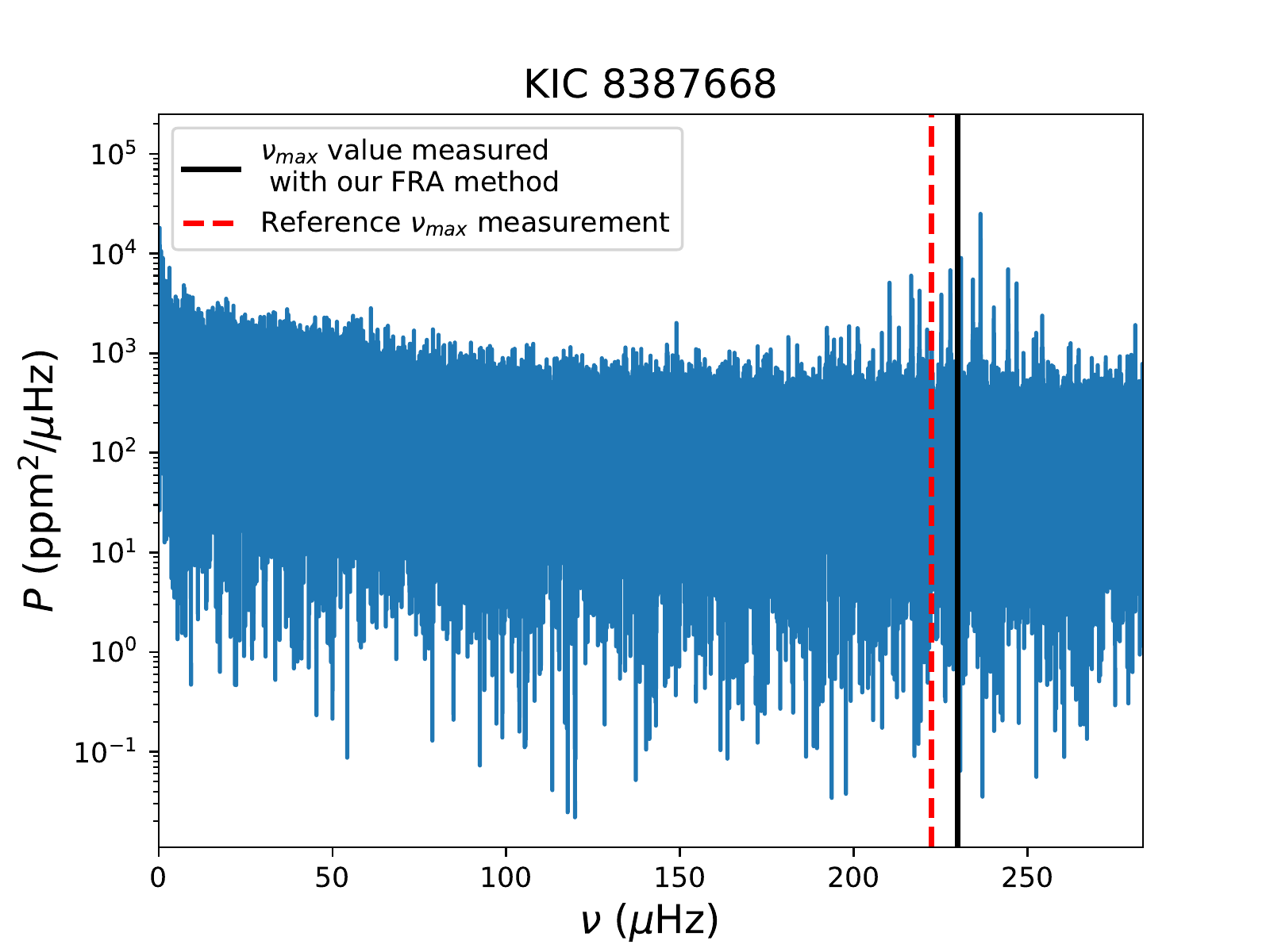}
\includegraphics[width=8.8cm]{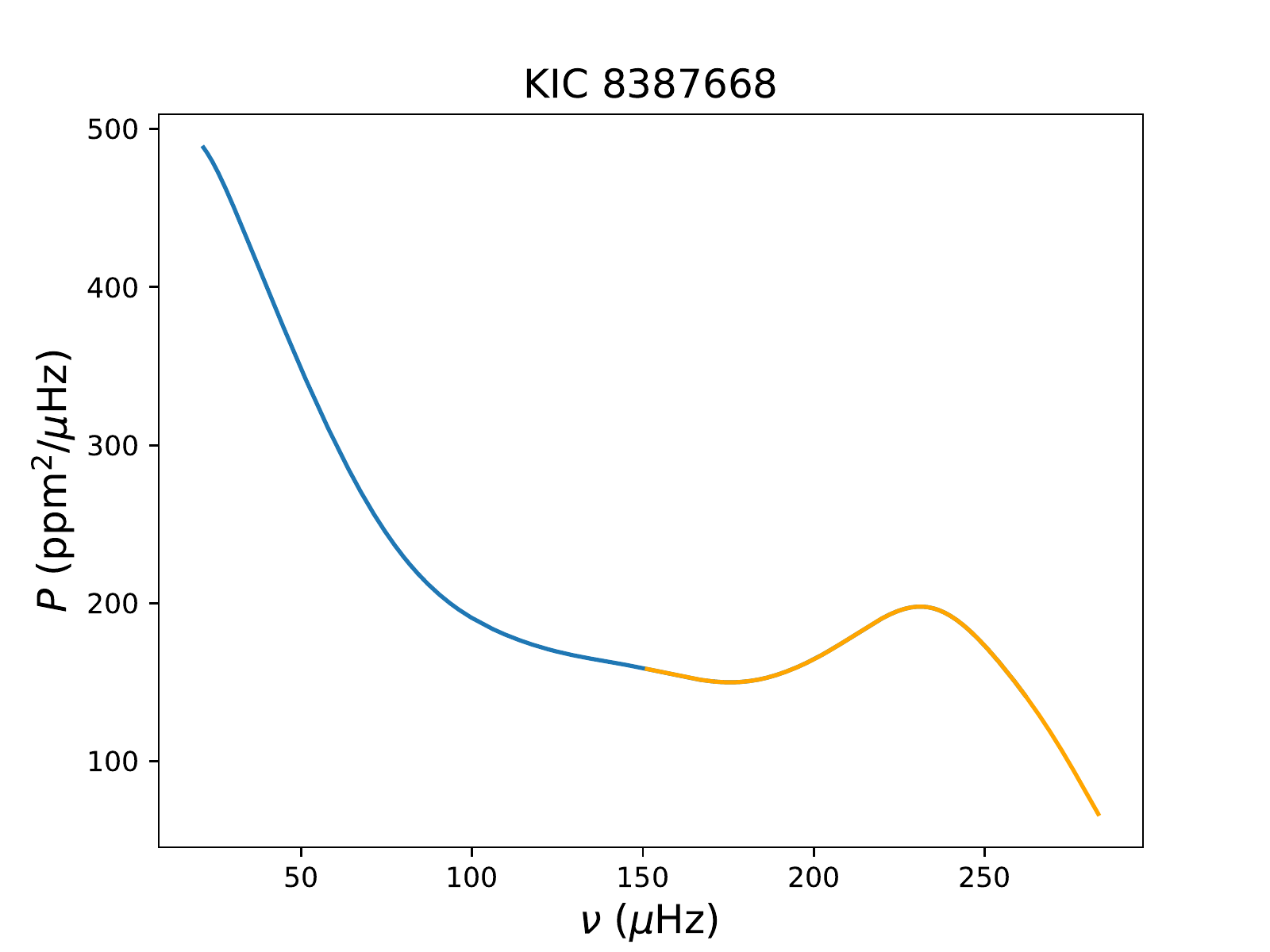}
\caption{Same as Fig.~\ref{fig-Kepler-high-rel-dev-1} for KIC 8387668. }
\label{fig-Kepler-high-numax-4}
\end{figure*}

\begin{figure*}
\centering
\includegraphics[width=8.8cm]{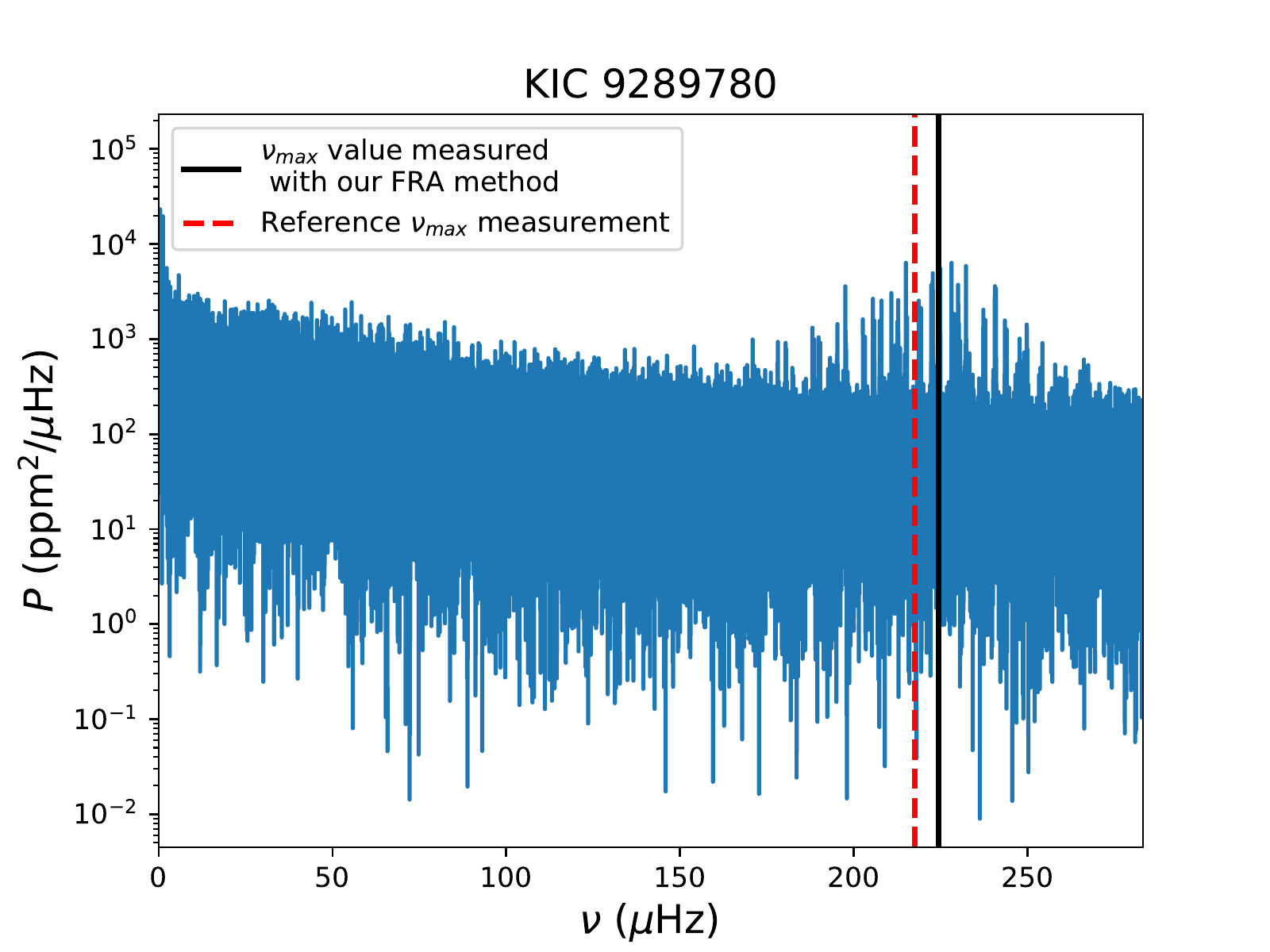}
\includegraphics[width=8.8cm]{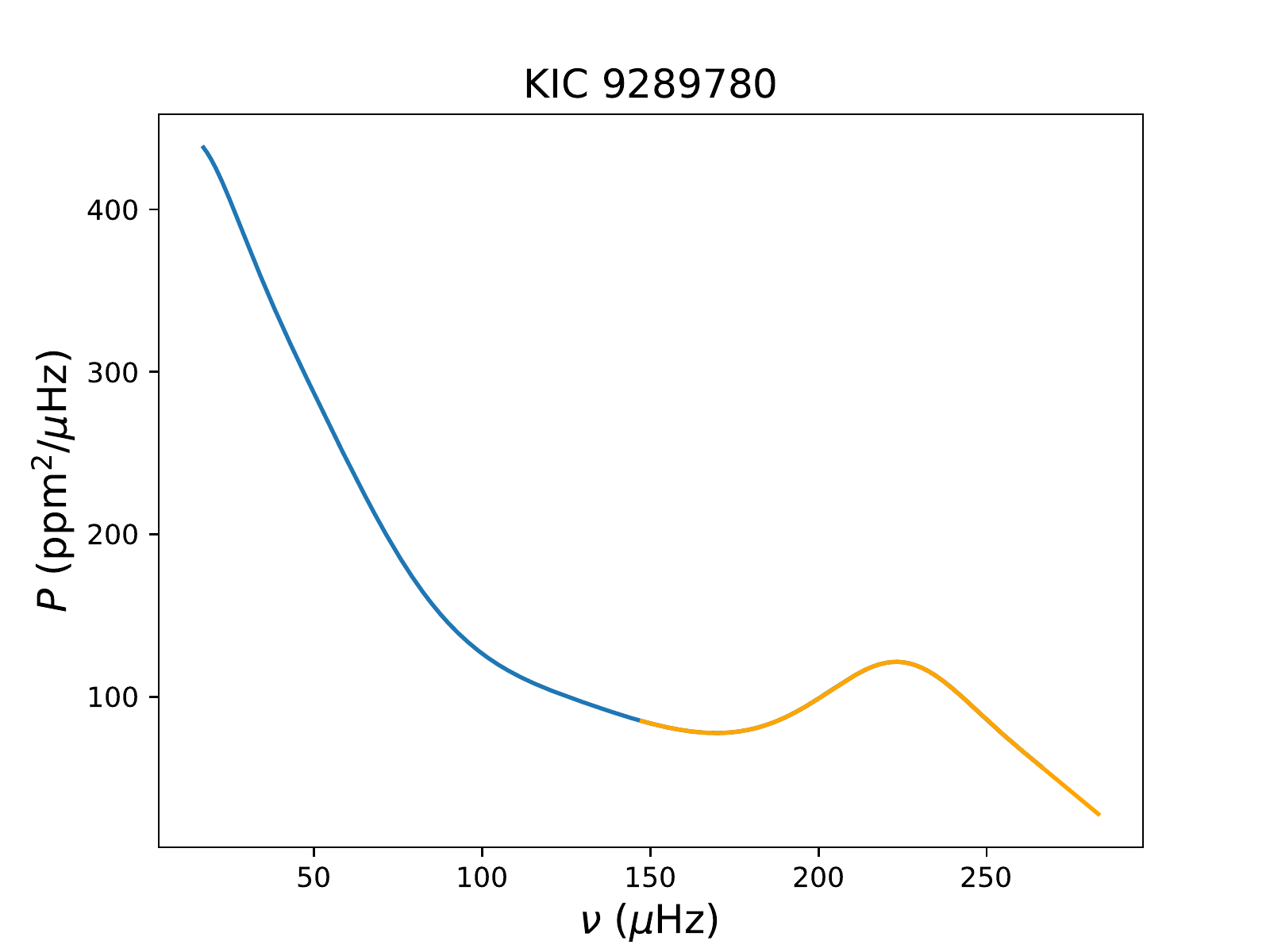}
\caption{Same as Fig.~\ref{fig-Kepler-high-rel-dev-1} for KIC 9289780. }
\label{fig-Kepler-high-numax-5}
\end{figure*}

We provide here examples of \textit{Kepler} red giants wih $\numax > 205$ $\mu$Hz:
\begin{itemize}
\item KIC 4459359 (Fig.~\ref{fig-Kepler-high-numax-1});
\item KIC 4750962 (Fig.~\ref{fig-Kepler-high-numax-2});
\item KIC 6352407 (Fig.~\ref{fig-Kepler-high-numax-3});
\item KIC 8387668 (Fig.~\ref{fig-Kepler-high-numax-4});
\item KIC 9289780 (Fig.~\ref{fig-Kepler-high-numax-5}).
\end{itemize}
These stars have $\numax$ close to the Nyquist frequency of 283 $\mu$Hz for \textit{Kepler} long cadence data. The COR method systematically underestimate $\numax$ for these stars while our FRA pipeline provides an accurate measurement.

\section{\textit{TESS} red giants for which we have a relative deviation of at least 10\% compared to existing $\nu\ind{max}$ measurements}\label{appendix-3}

\begin{figure*}
\centering
\includegraphics[width=8.8cm]{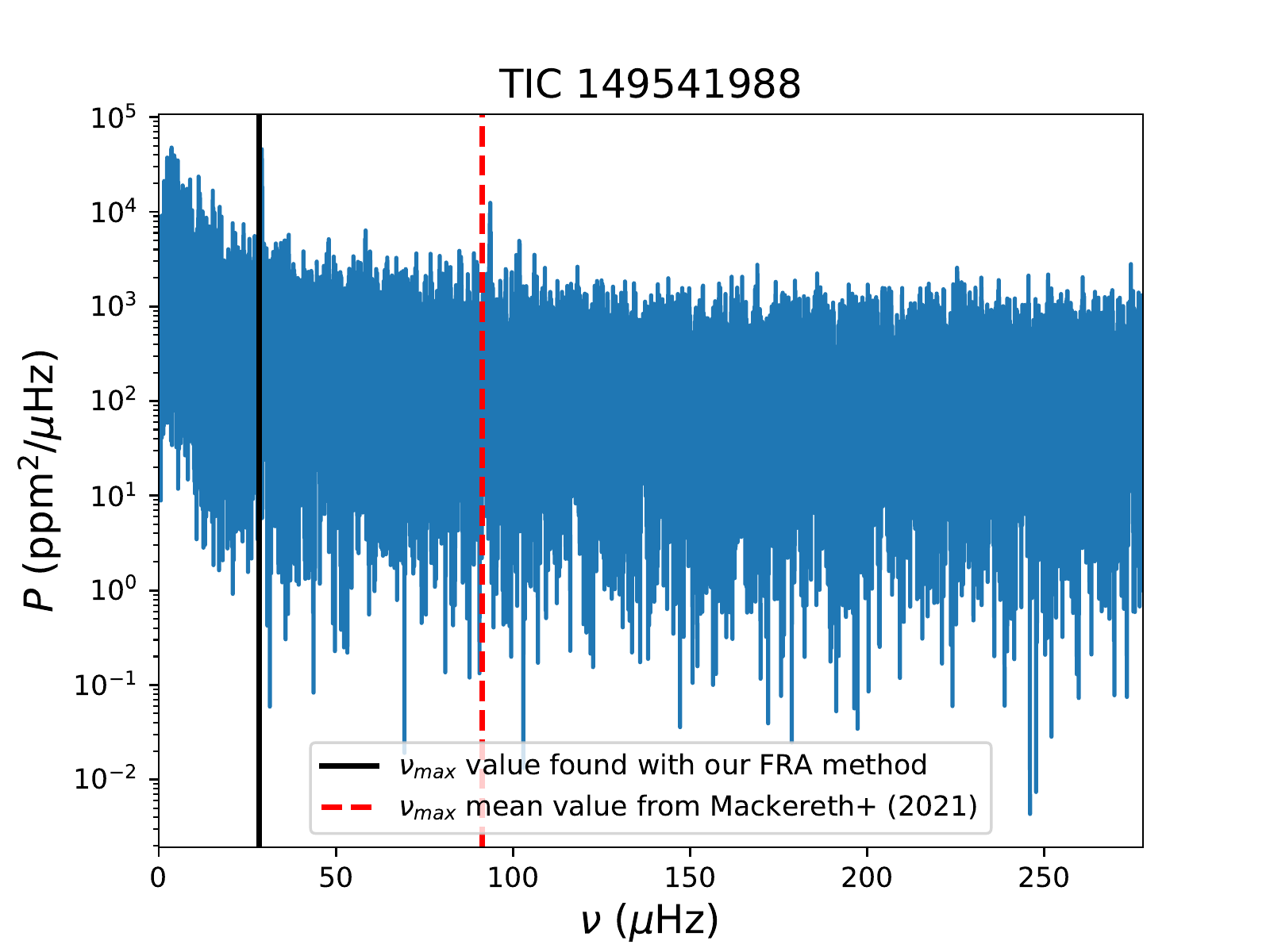}
\includegraphics[width=8.8cm]{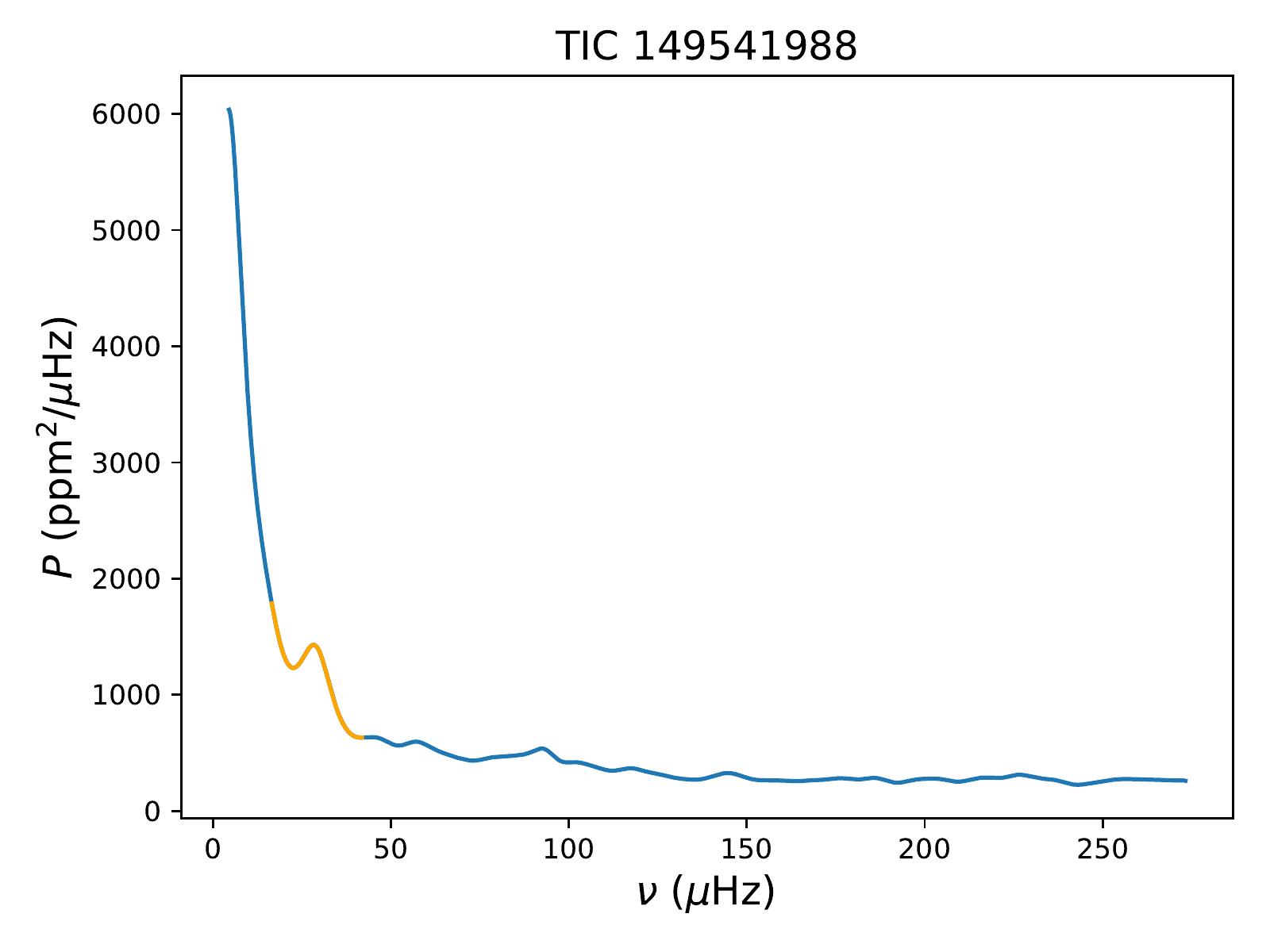}
\caption{Same as Fig.~\ref{fig-Kepler-high-rel-dev-1} for TIC 149541988.}
\label{fig-TESS-numax-1}
\end{figure*}

\begin{figure*}
\centering
\includegraphics[width=8.8cm]{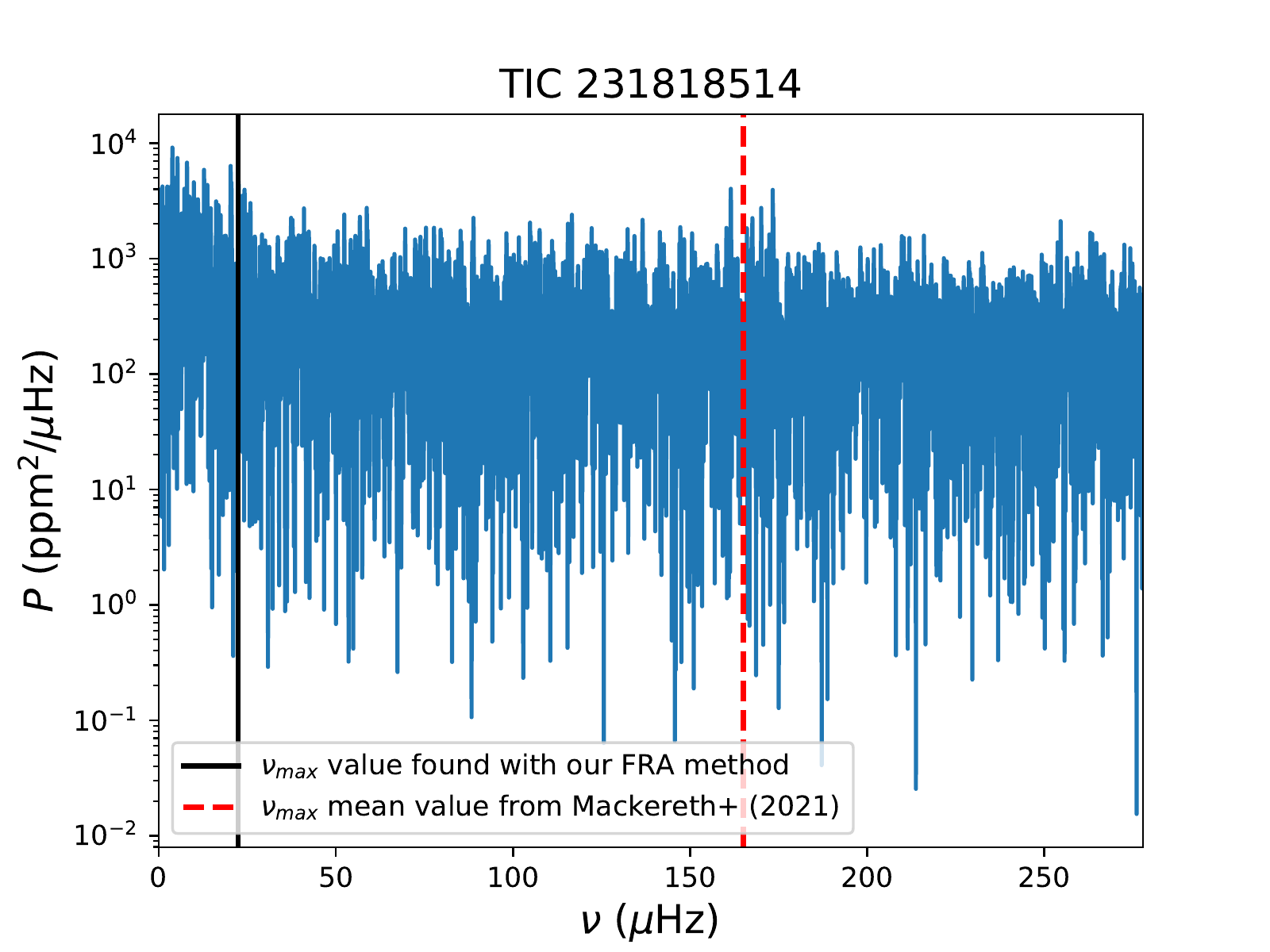}
\includegraphics[width=8.8cm]{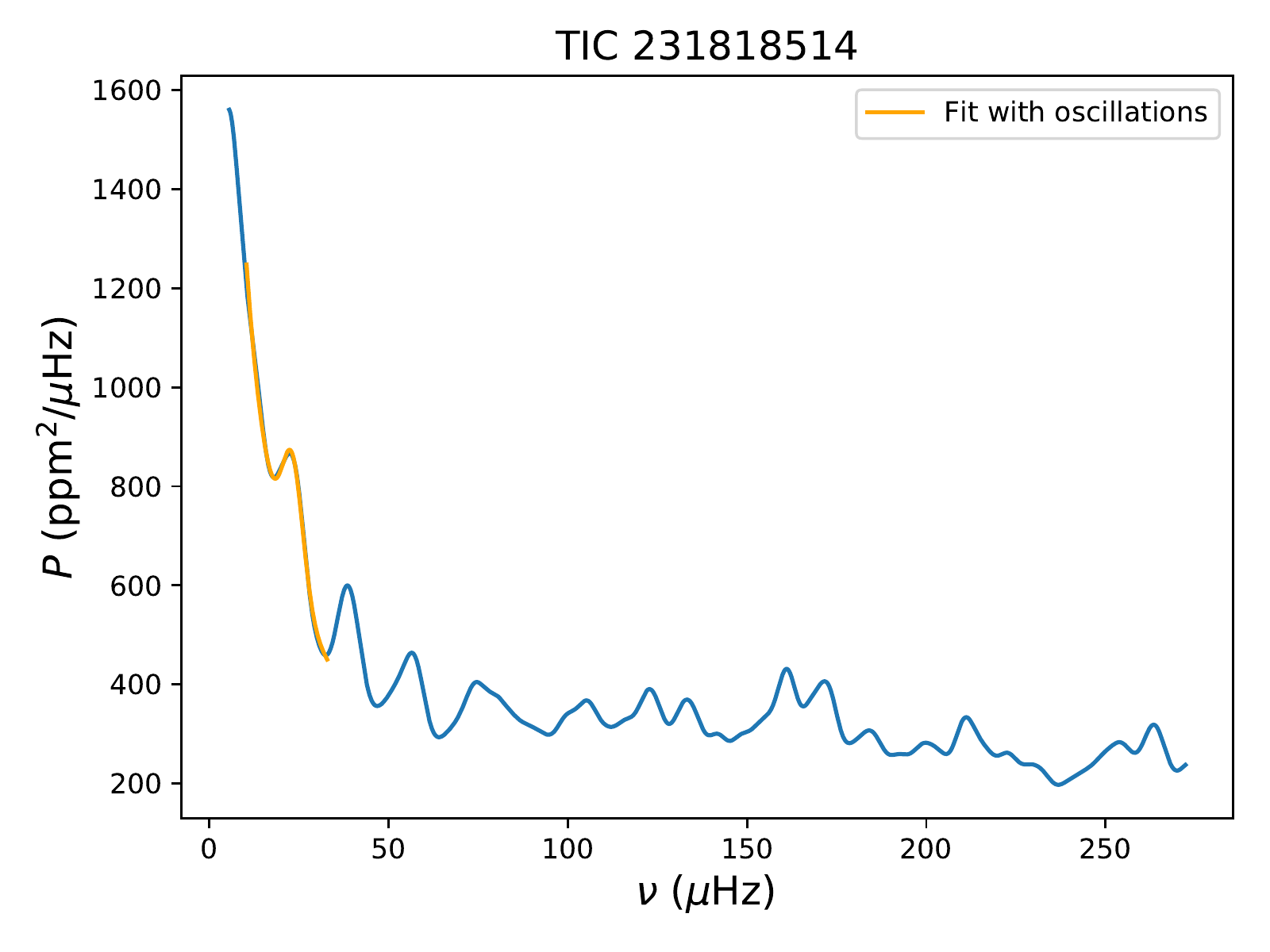}
\caption{Same as Fig.~\ref{fig-Kepler-high-rel-dev-1} for TIC 231818514.}
\label{fig-TESS-numax-2}
\end{figure*}

\begin{figure*}
\centering
\includegraphics[width=8.8cm]{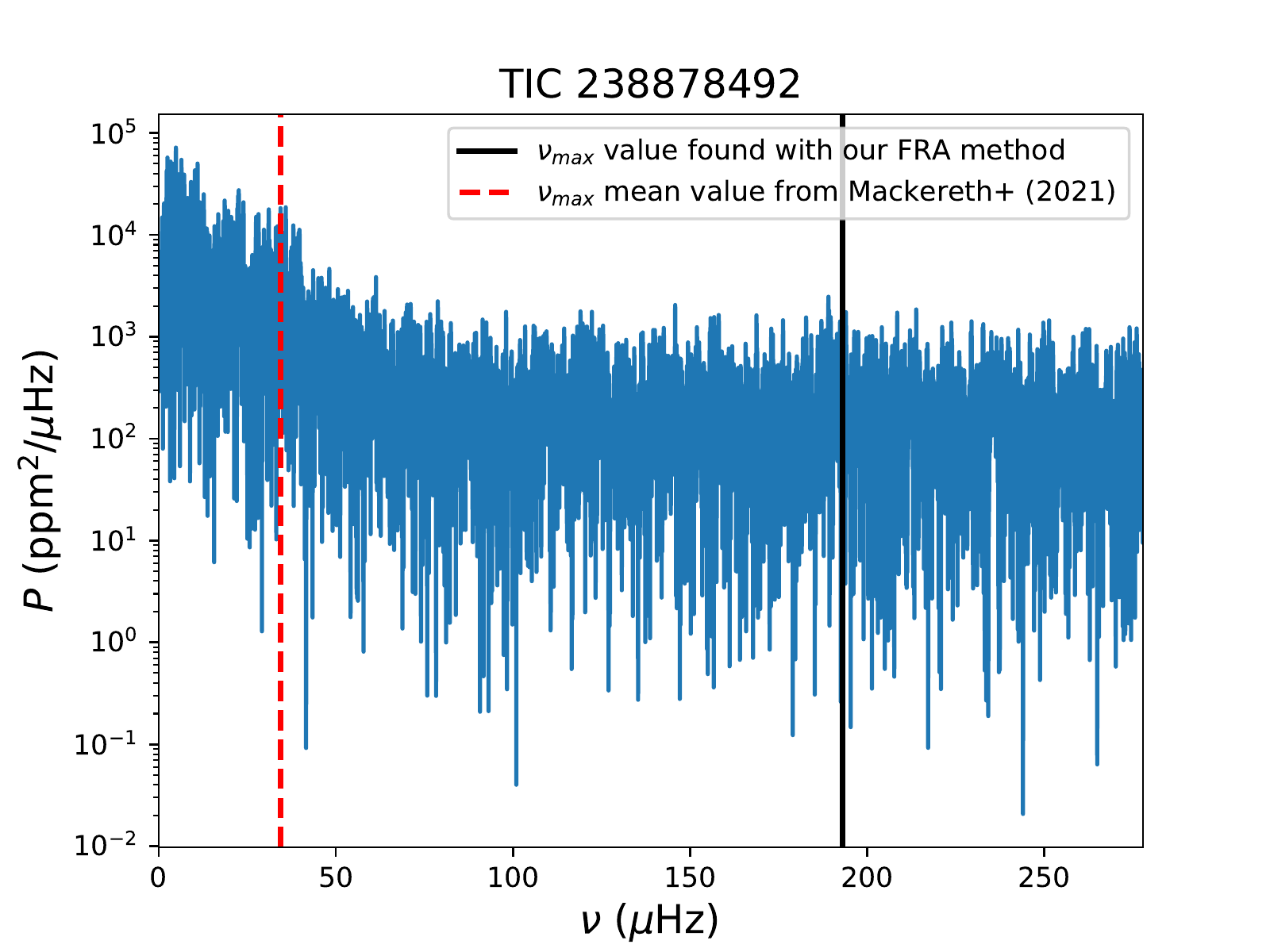}
\includegraphics[width=8.8cm]{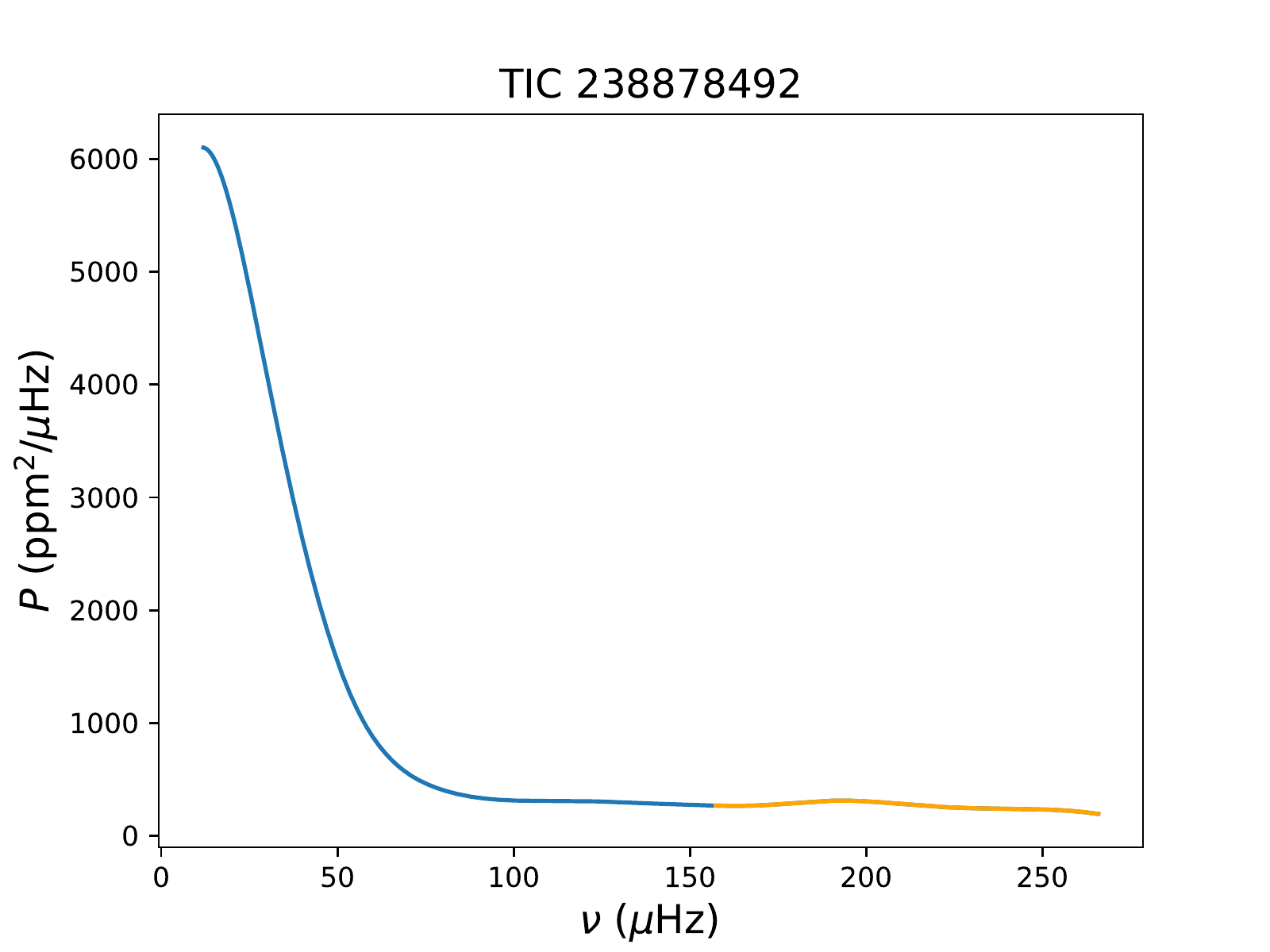}
\caption{Same as Fig.~\ref{fig-Kepler-high-rel-dev-1} for TIC 238878492.}
\label{fig-TESS-numax-3}
\end{figure*}

\begin{figure*}
\centering
\includegraphics[width=8.8cm]{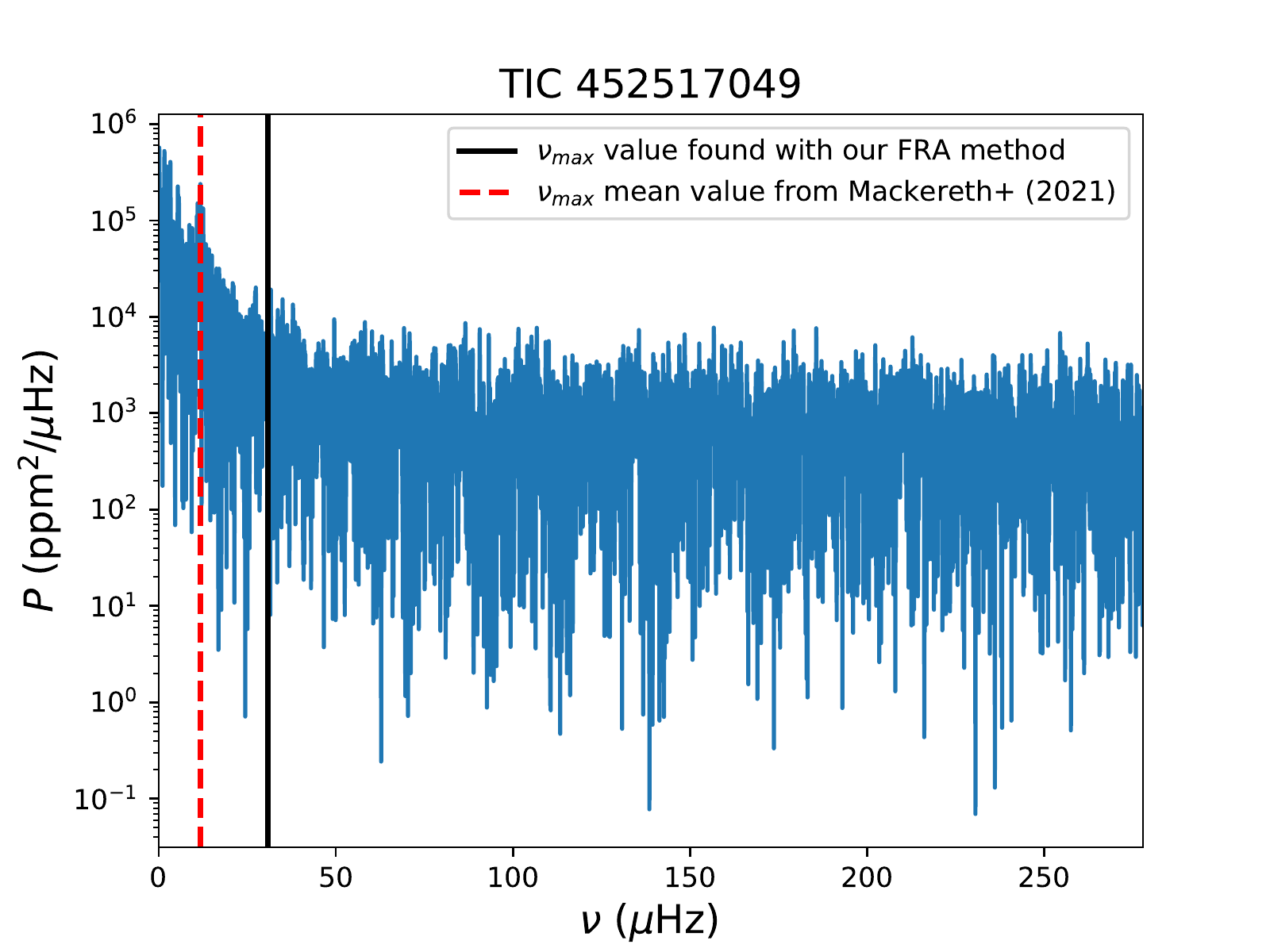}
\includegraphics[width=8.8cm]{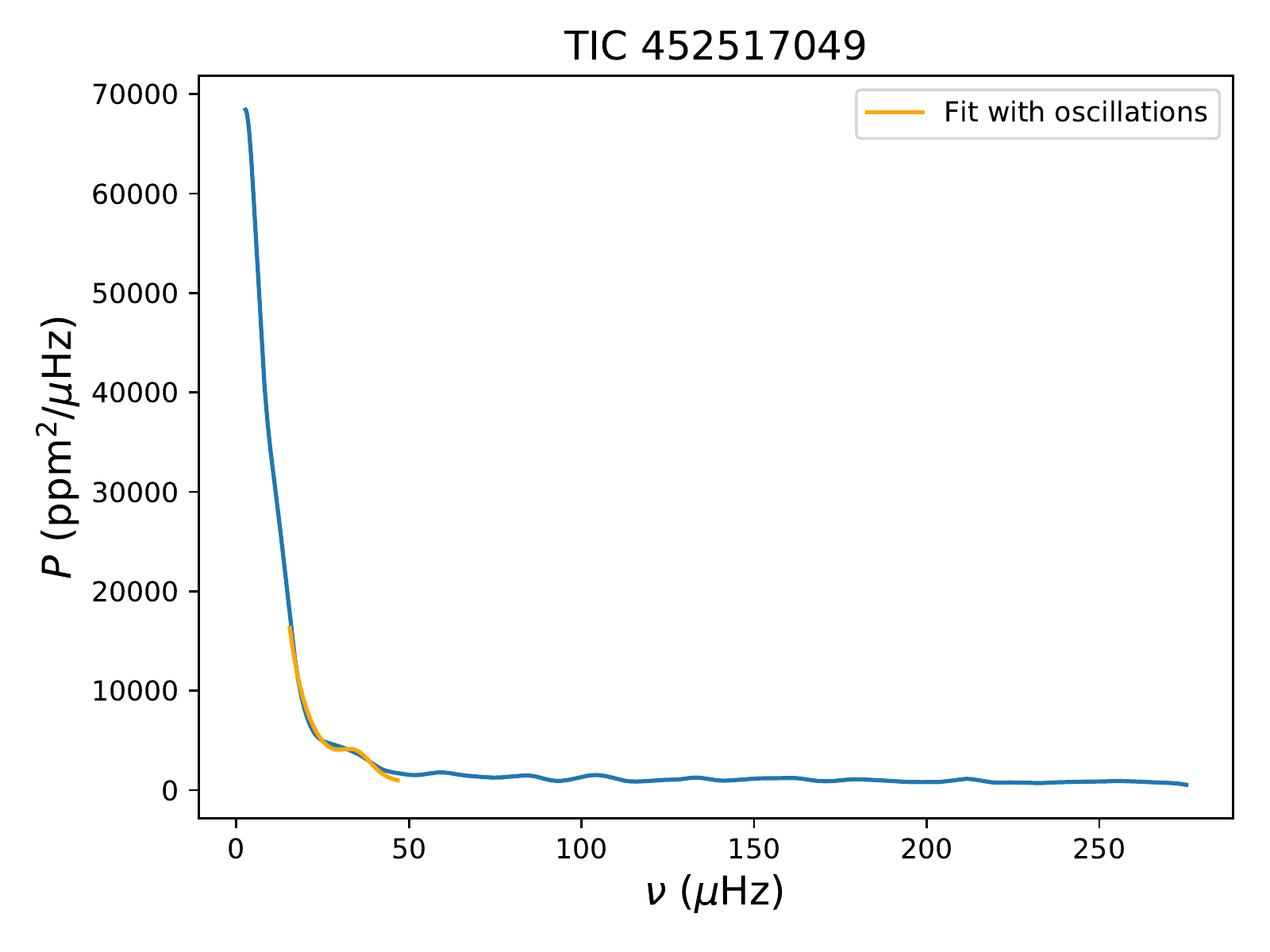}
\caption{Same as Fig.~\ref{fig-Kepler-high-rel-dev-1} for TIC 452517049.}
\label{fig-TESS-numax-4}
\end{figure*}

\begin{figure*}
\centering
\includegraphics[width=8.8cm]{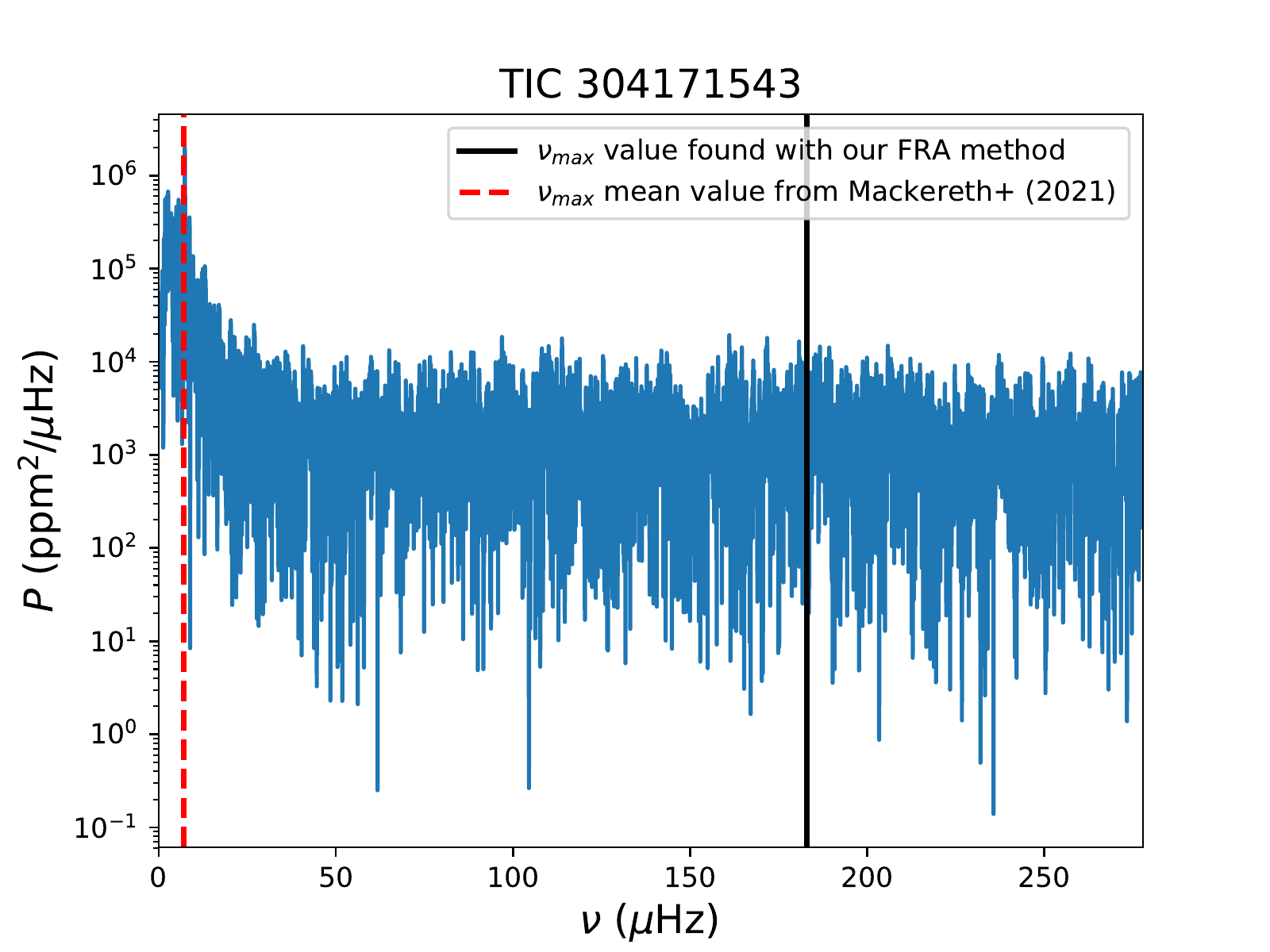}
\includegraphics[width=8.8cm]{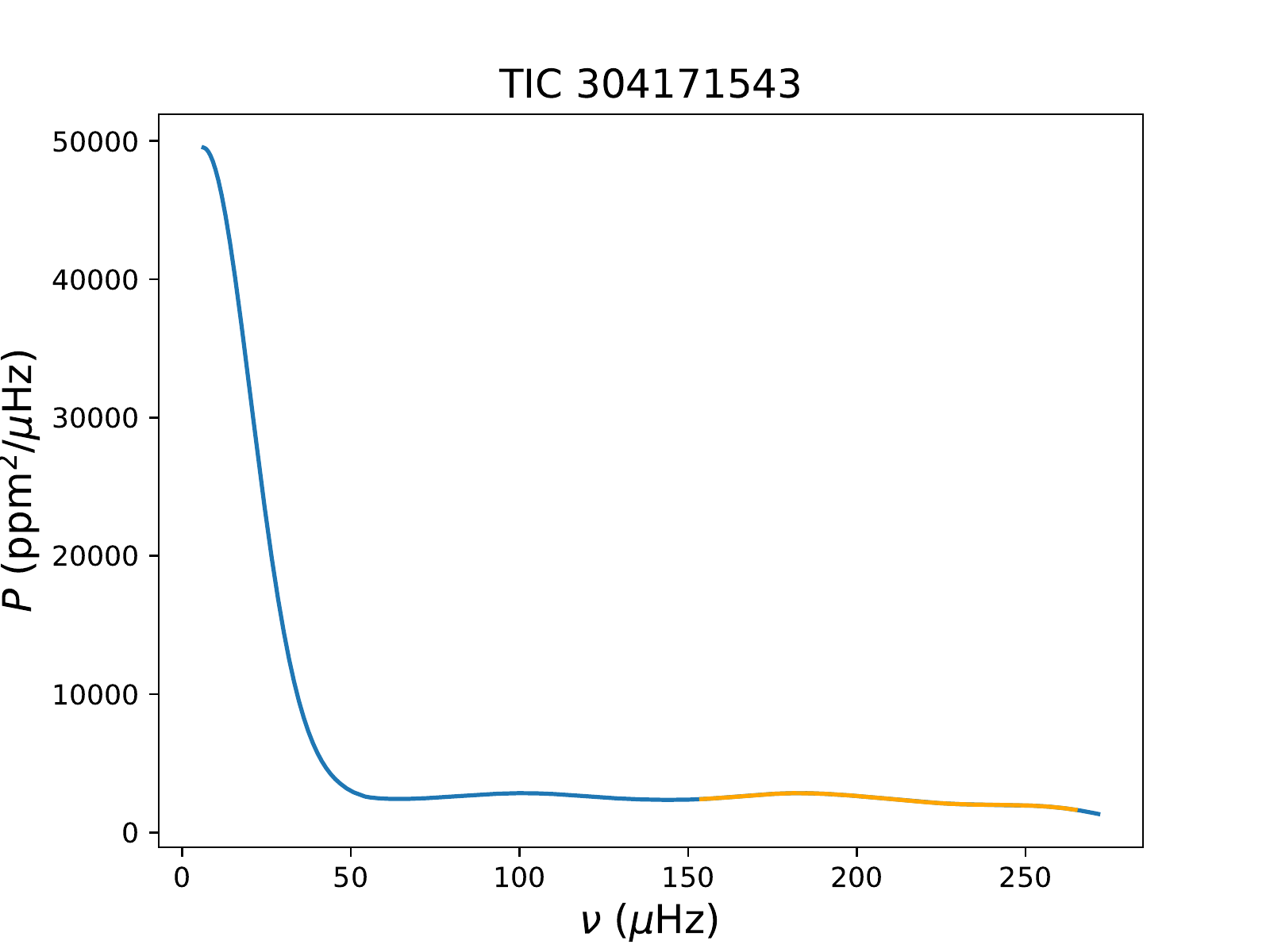}
\caption{Same as Fig.~\ref{fig-Kepler-high-rel-dev-1} for TIC 304171543.}
\label{fig-TESS-numax-5}
\end{figure*}

There are 50 TESS red giants for which we have a relative deviation of at least 10\% between our measurements obtained with our FRA pipeline and the measurements from \cite{Mackereth}. Here are some examples for which our analysis fails to provide an accurate $\numax$:
\begin{itemize}
\item TIC 149541988 (Fig.~\ref{fig-TESS-numax-1}) for which we underestimate $\numax$;
\item TIC 231818514 (Fig.~\ref{fig-TESS-numax-2}) for which we underestimate $\numax$;
\item TIC 238878492 (Fig.~\ref{fig-TESS-numax-3}) for which we overestimate $\numax$;
\item TIC 452517049 (Fig.~\ref{fig-TESS-numax-4}) for which we overestimate $\numax$;
\item TIC 304171543 (Fig.~\ref{fig-TESS-numax-5}) for which we overestimate $\numax$.
\end{itemize}

\clearpage


\section{\textit{TESS} red giants for which we have a relative deviation of at least 10\% compared to existing $\nu\ind{max}$ measurements with an extra validation step for $G > 9.5$}\label{appendix-4}

\begin{figure*}
\centering
\includegraphics[width=8.8cm]{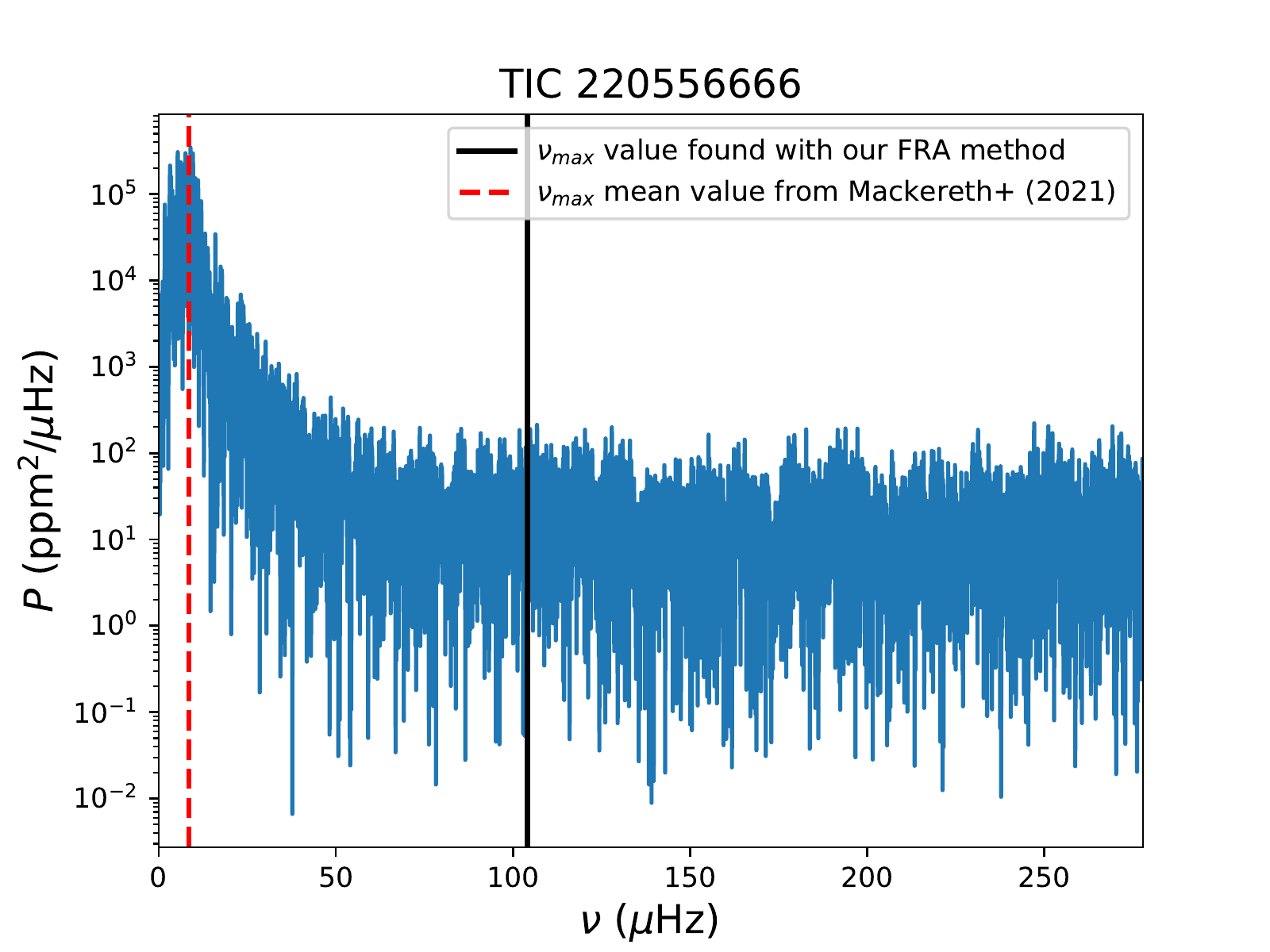}
\includegraphics[width=8.8cm]{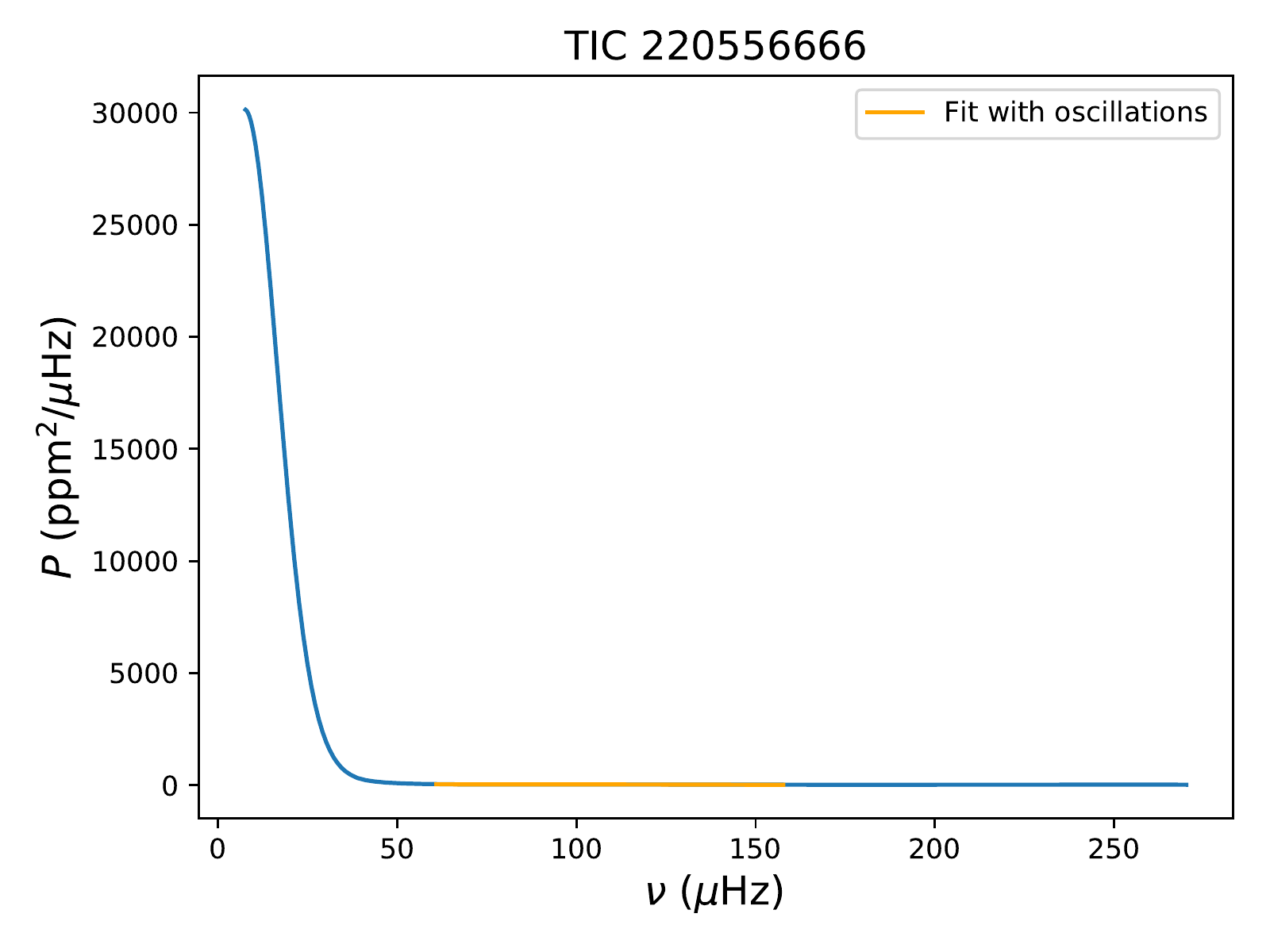}
\caption{Same as Fig.~\ref{fig-Kepler-high-rel-dev-1} for TIC 220556666.}
\label{fig-TESS-numax-6}
\end{figure*}

\begin{figure*}
\centering
\includegraphics[width=8.8cm]{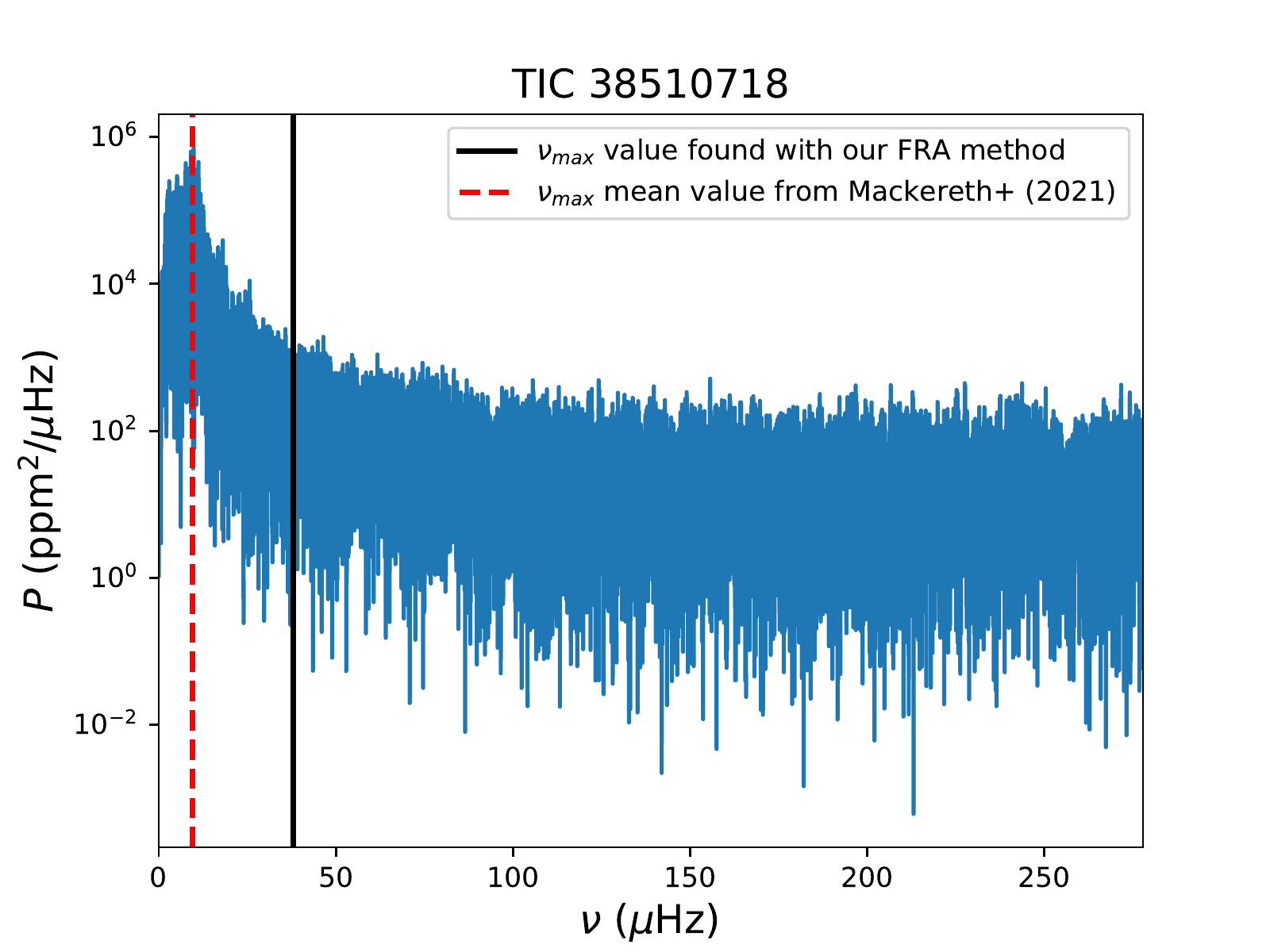}
\includegraphics[width=8.8cm]{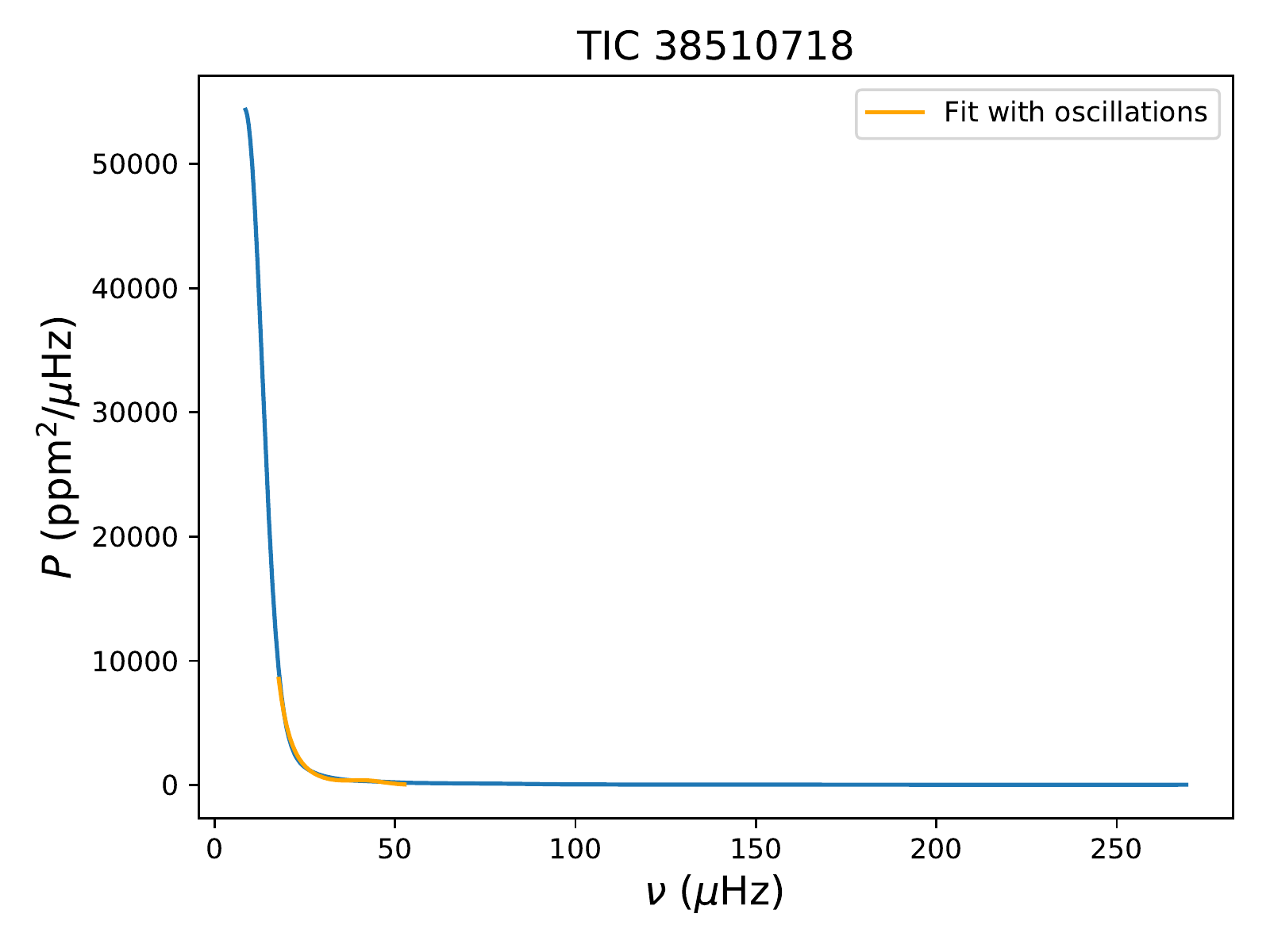}
\caption{Same as Fig.~\ref{fig-Kepler-high-rel-dev-1} for TIC 38510718.}
\label{fig-TESS-numax-7}
\end{figure*}

\begin{figure*}
\centering
\includegraphics[width=8.8cm]{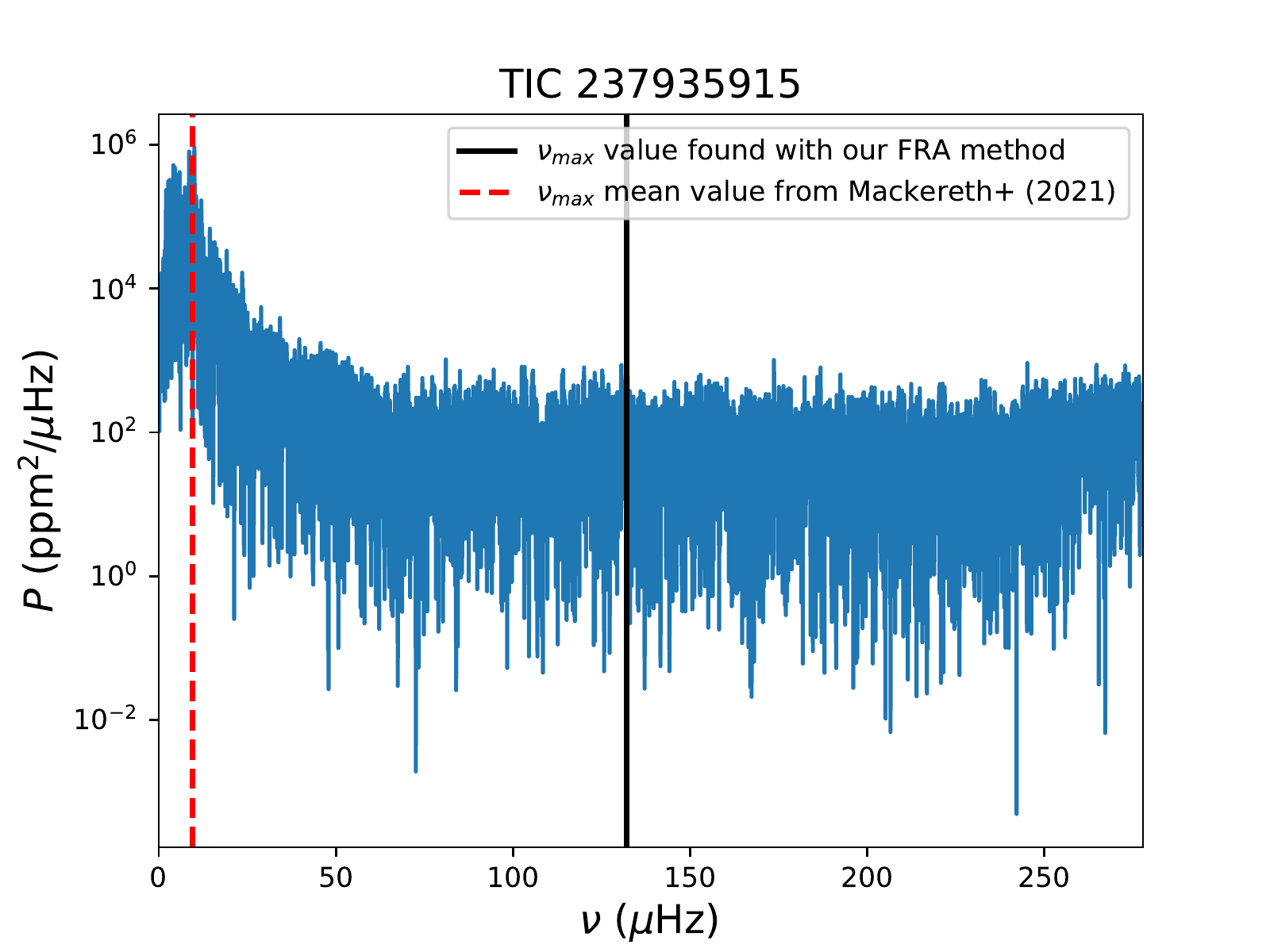}
\includegraphics[width=8.8cm]{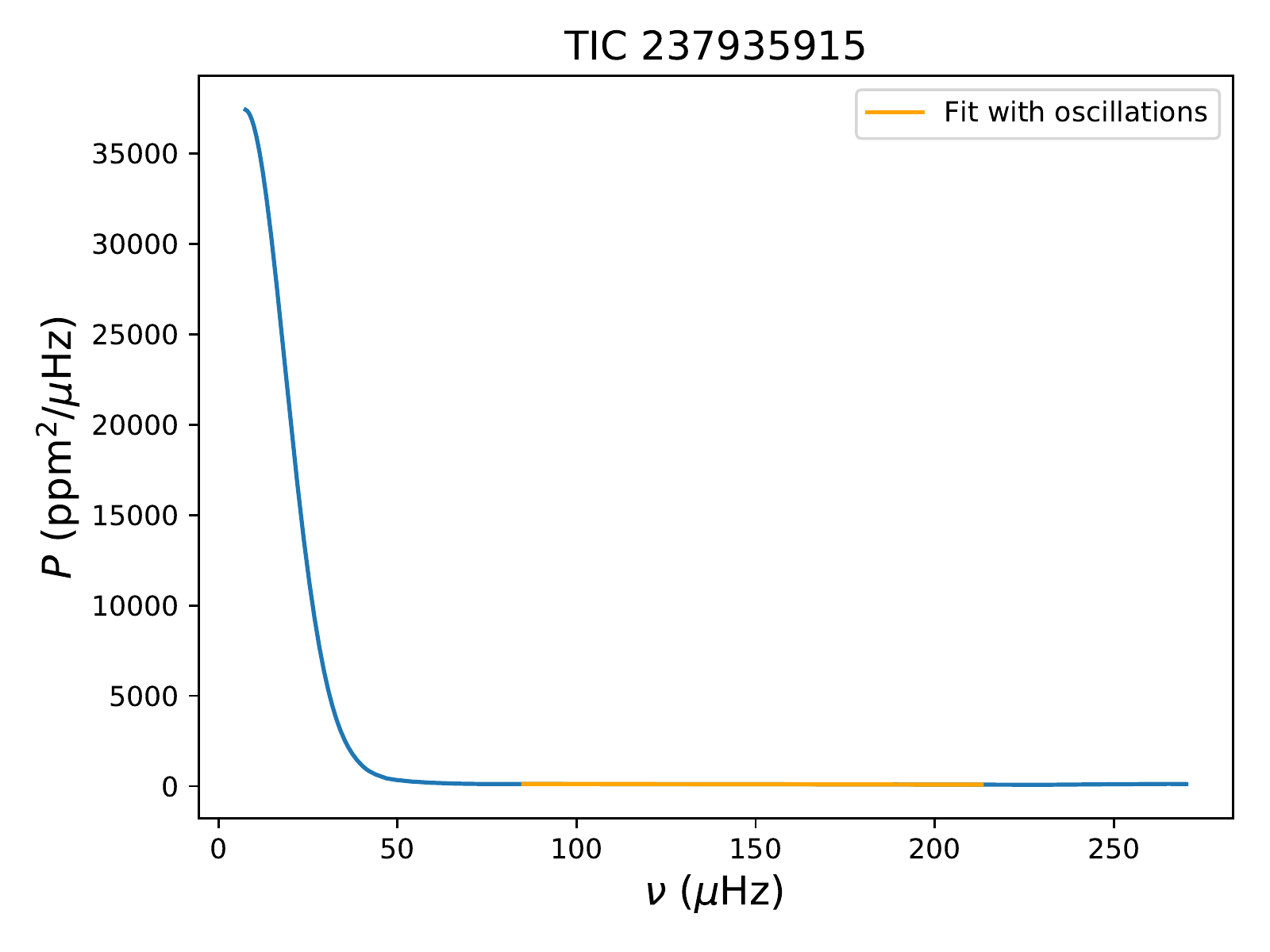}
\caption{Same as Fig.~\ref{fig-Kepler-high-rel-dev-1} for TIC 237935915.}
\label{fig-TESS-numax-8}
\end{figure*}

\begin{figure*}
\centering
\includegraphics[width=8.8cm]{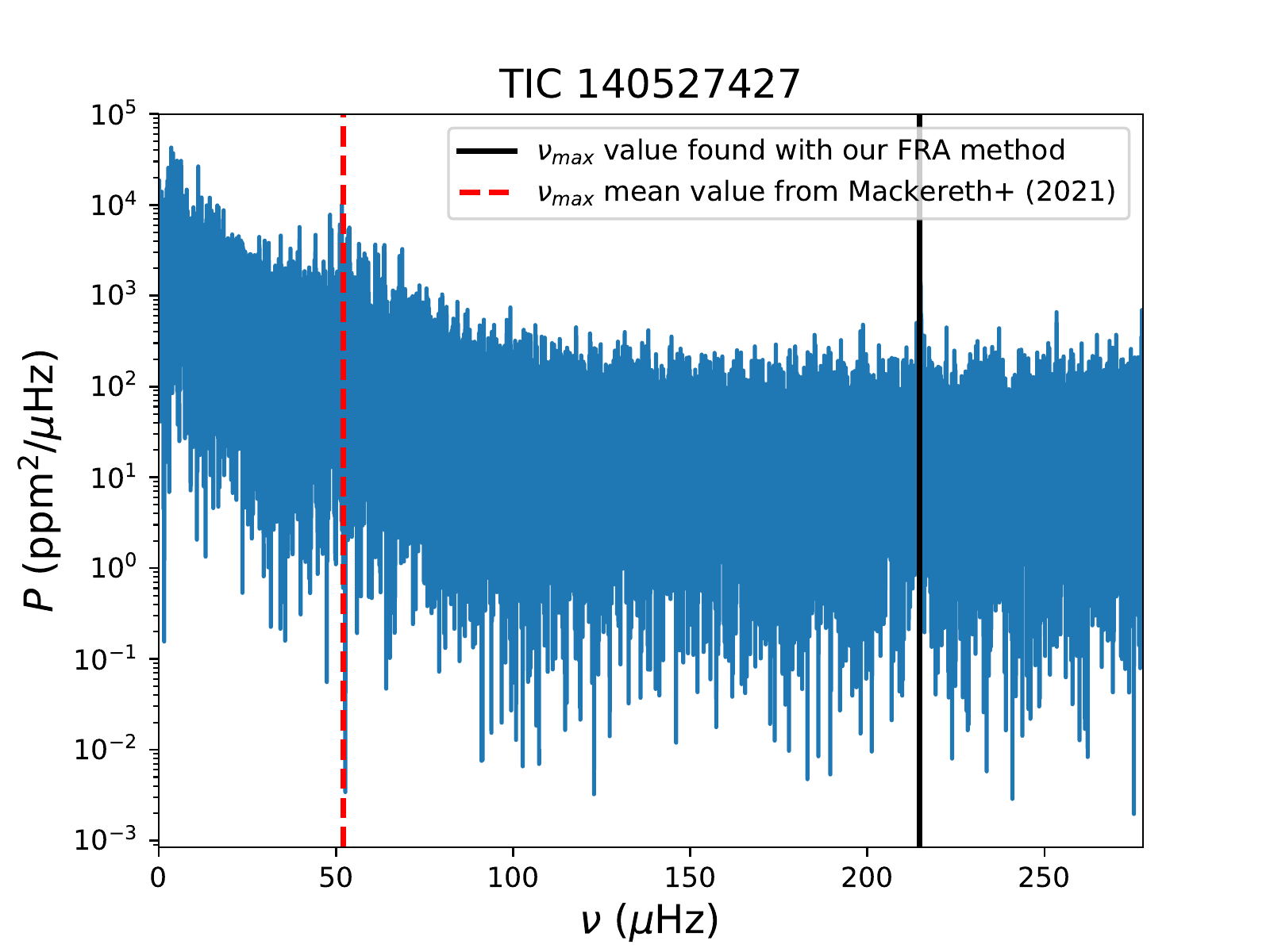}
\includegraphics[width=8.8cm]{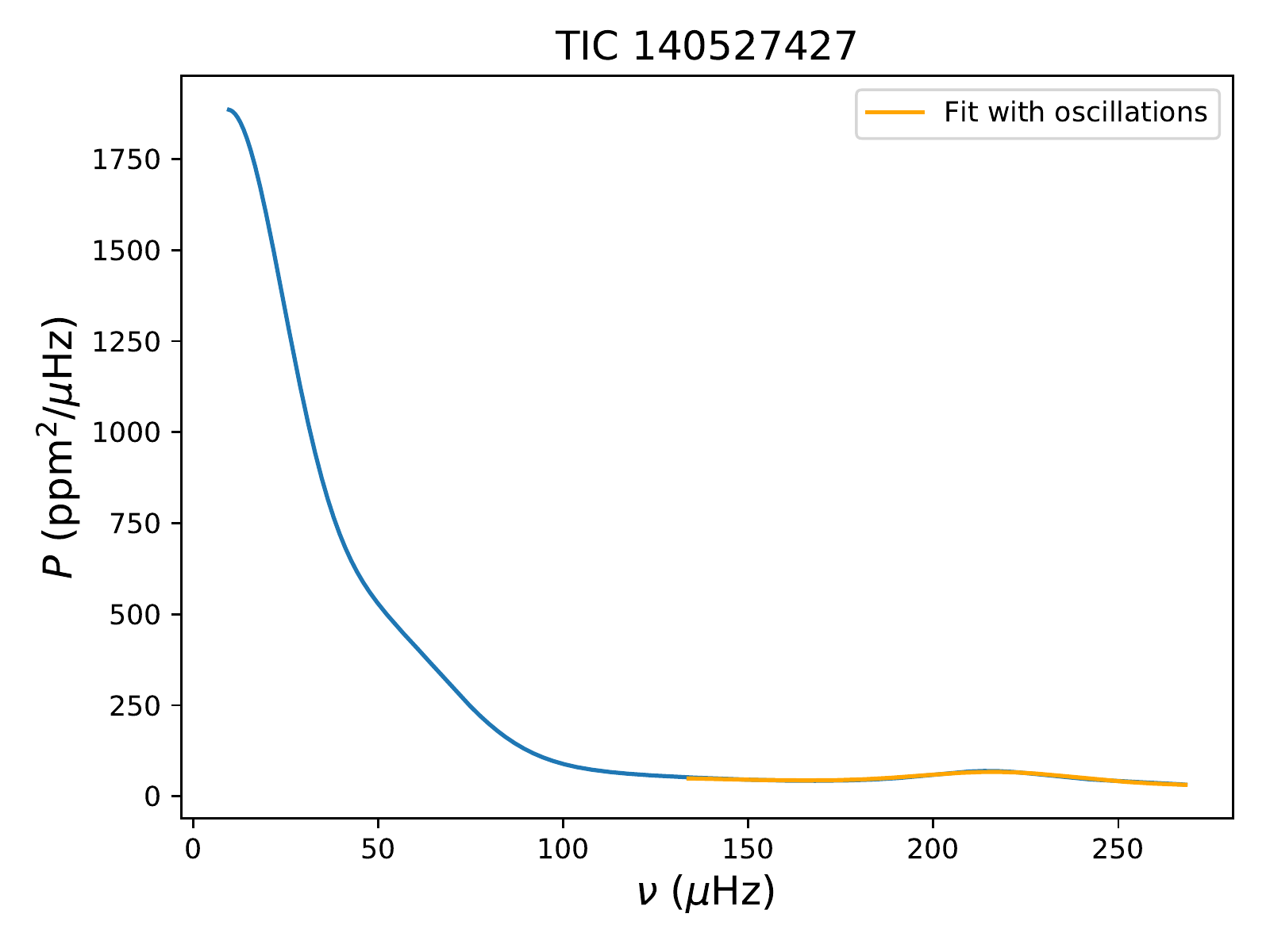}
\caption{Same as Fig.~\ref{fig-Kepler-high-rel-dev-1} for TIC 140527427.}
\label{fig-TESS-numax-9}
\end{figure*}

\begin{figure*}
\centering
\includegraphics[width=8.8cm]{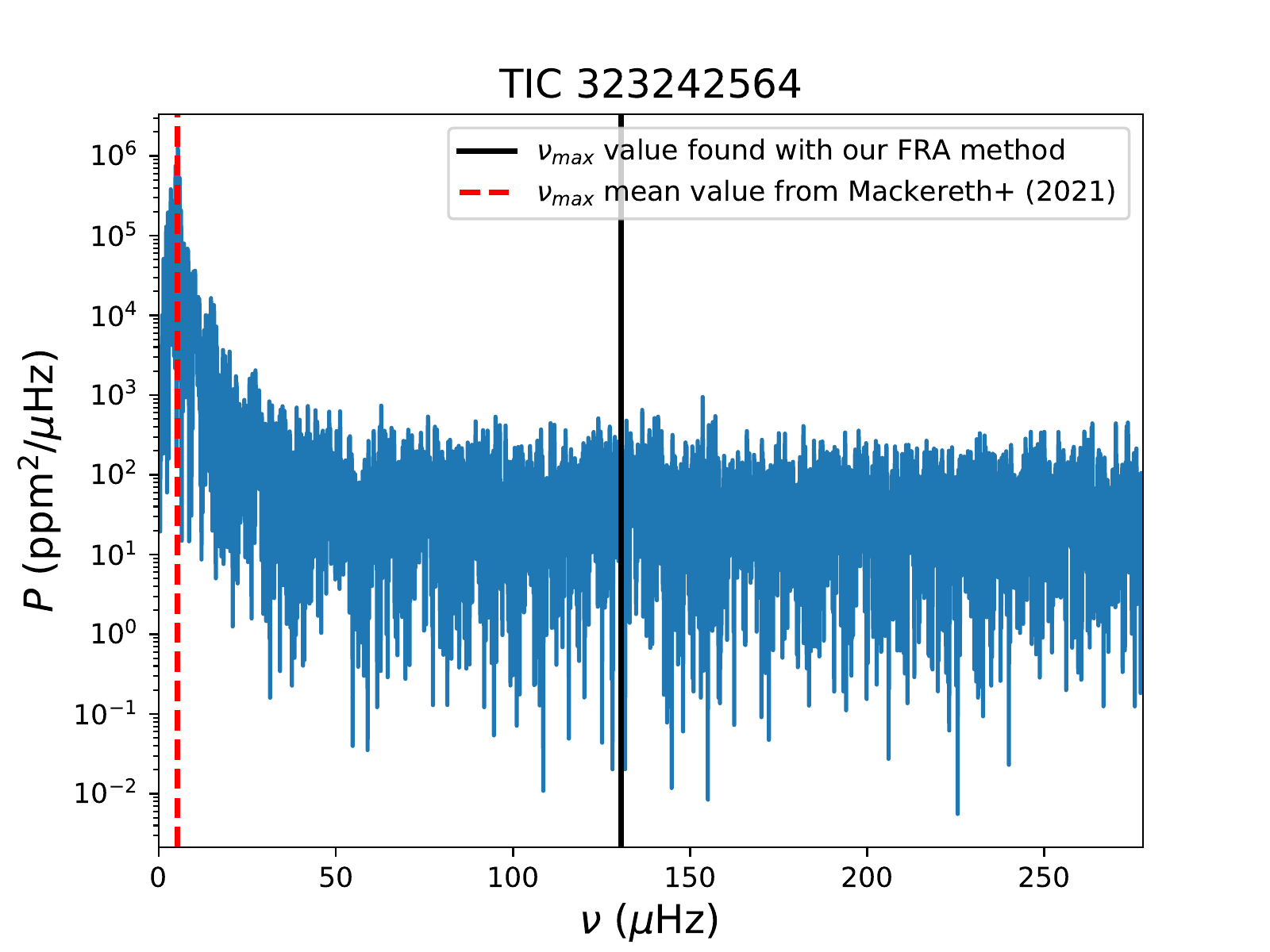}
\includegraphics[width=8.8cm]{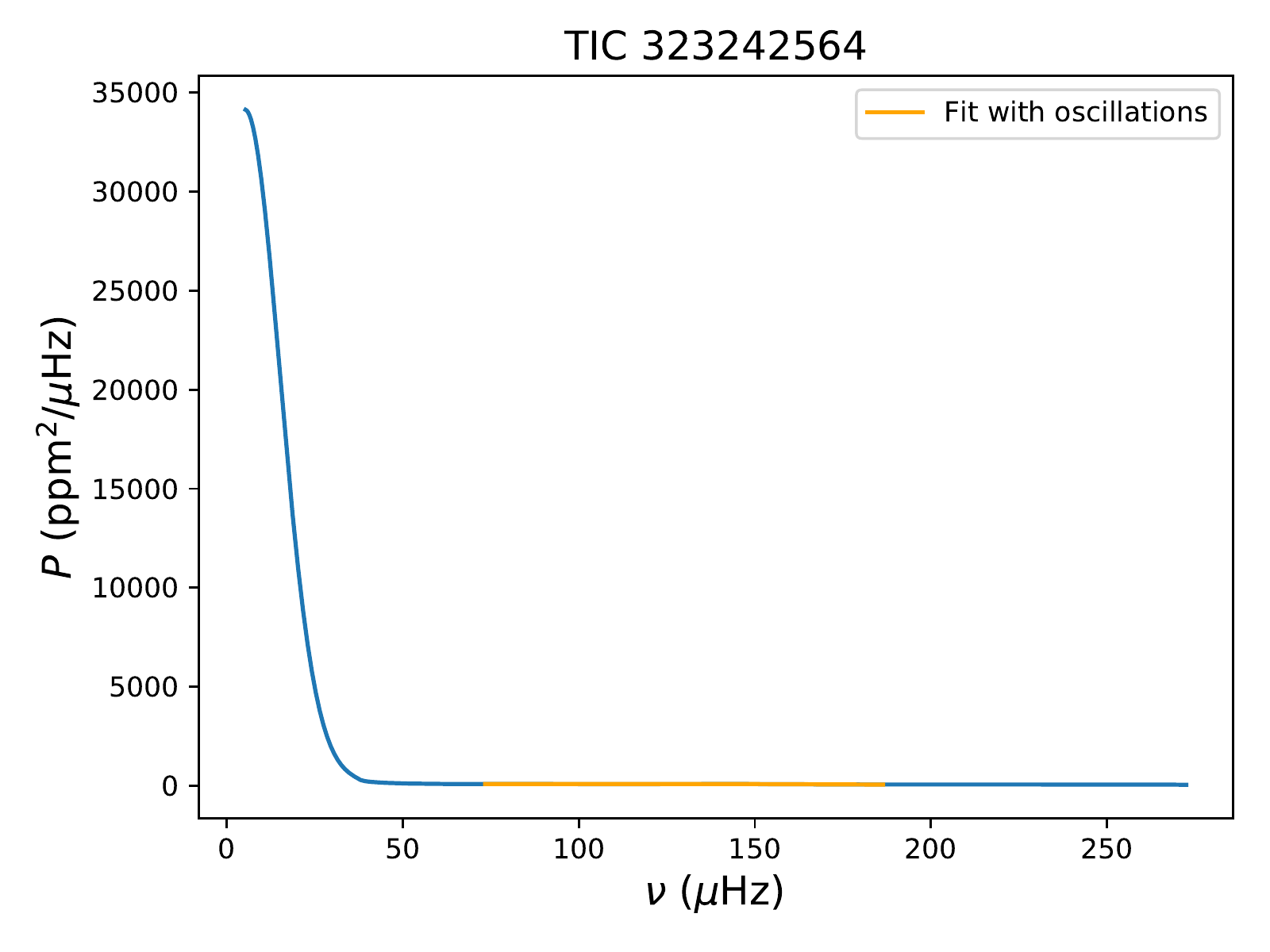}
\caption{Same as Fig.~\ref{fig-Kepler-high-rel-dev-1} for TIC 323242564.}
\label{fig-TESS-numax-10}
\end{figure*}

\begin{figure*}
\centering
\includegraphics[width=8.8cm]{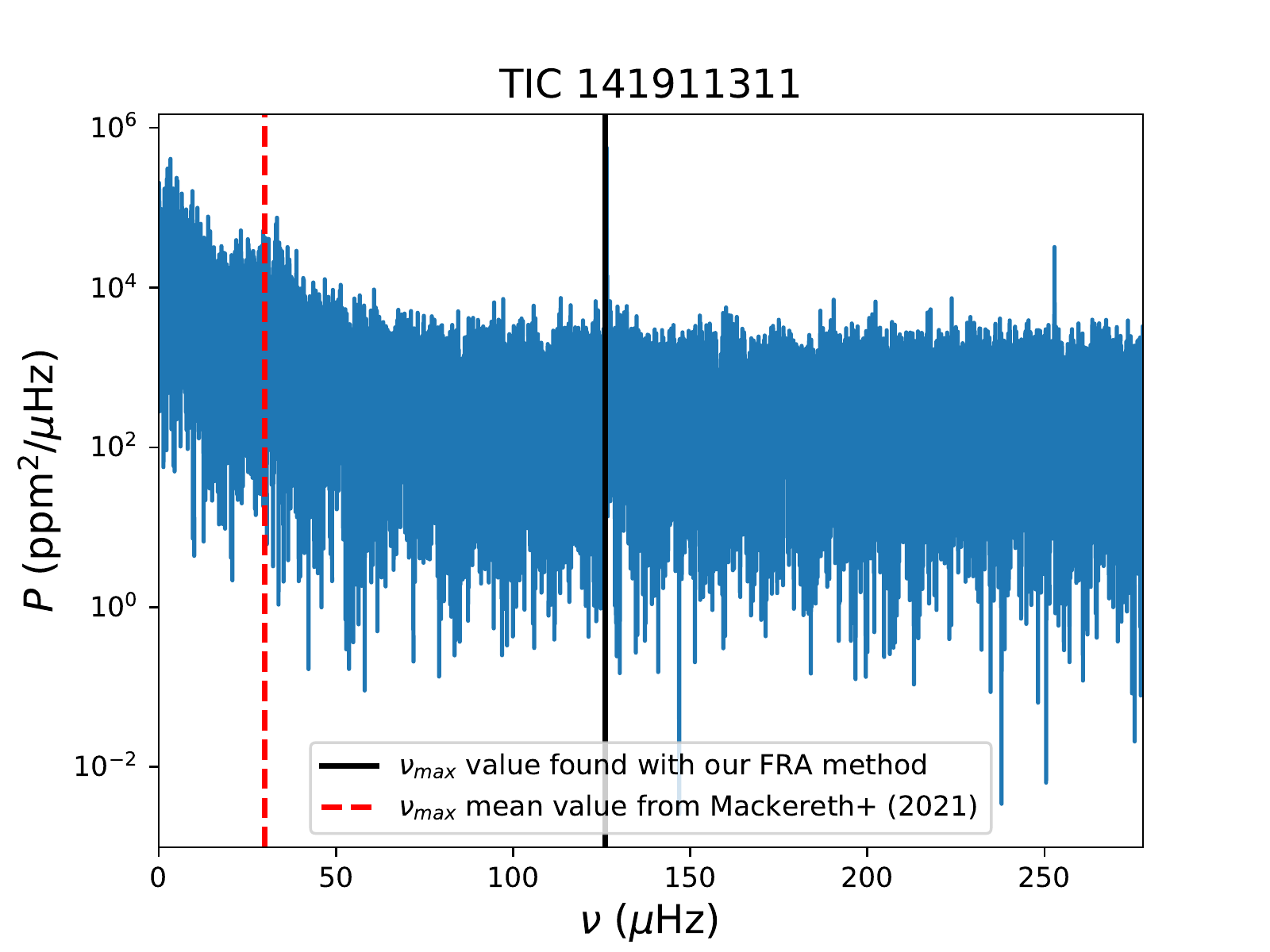}
\includegraphics[width=8.8cm]{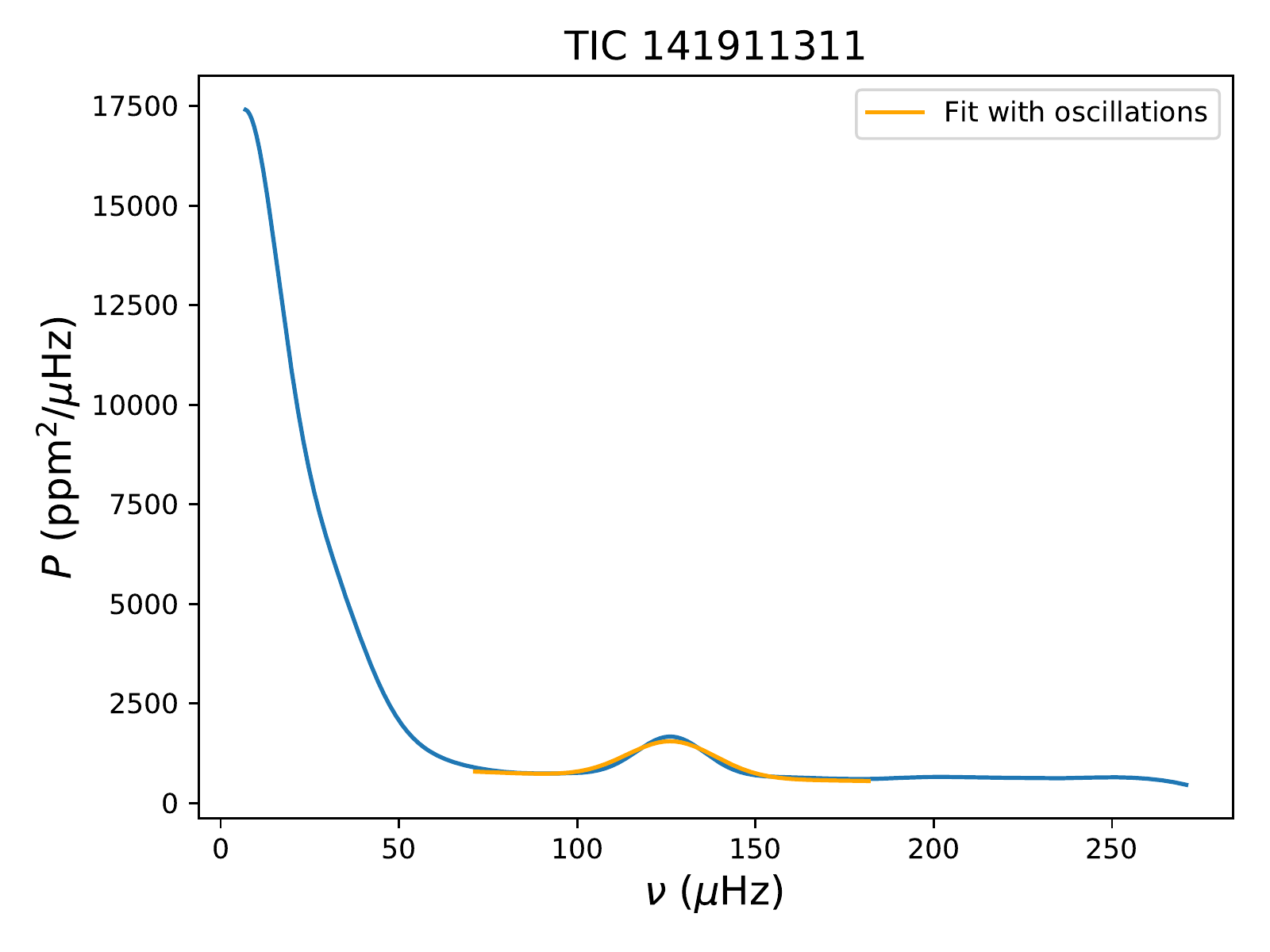}
\caption{Same as Fig.~\ref{fig-Kepler-high-rel-dev-1} for TIC 141911311.}
\label{fig-TESS-numax-11}
\end{figure*}

\begin{figure*}
\centering
\includegraphics[width=8.8cm]{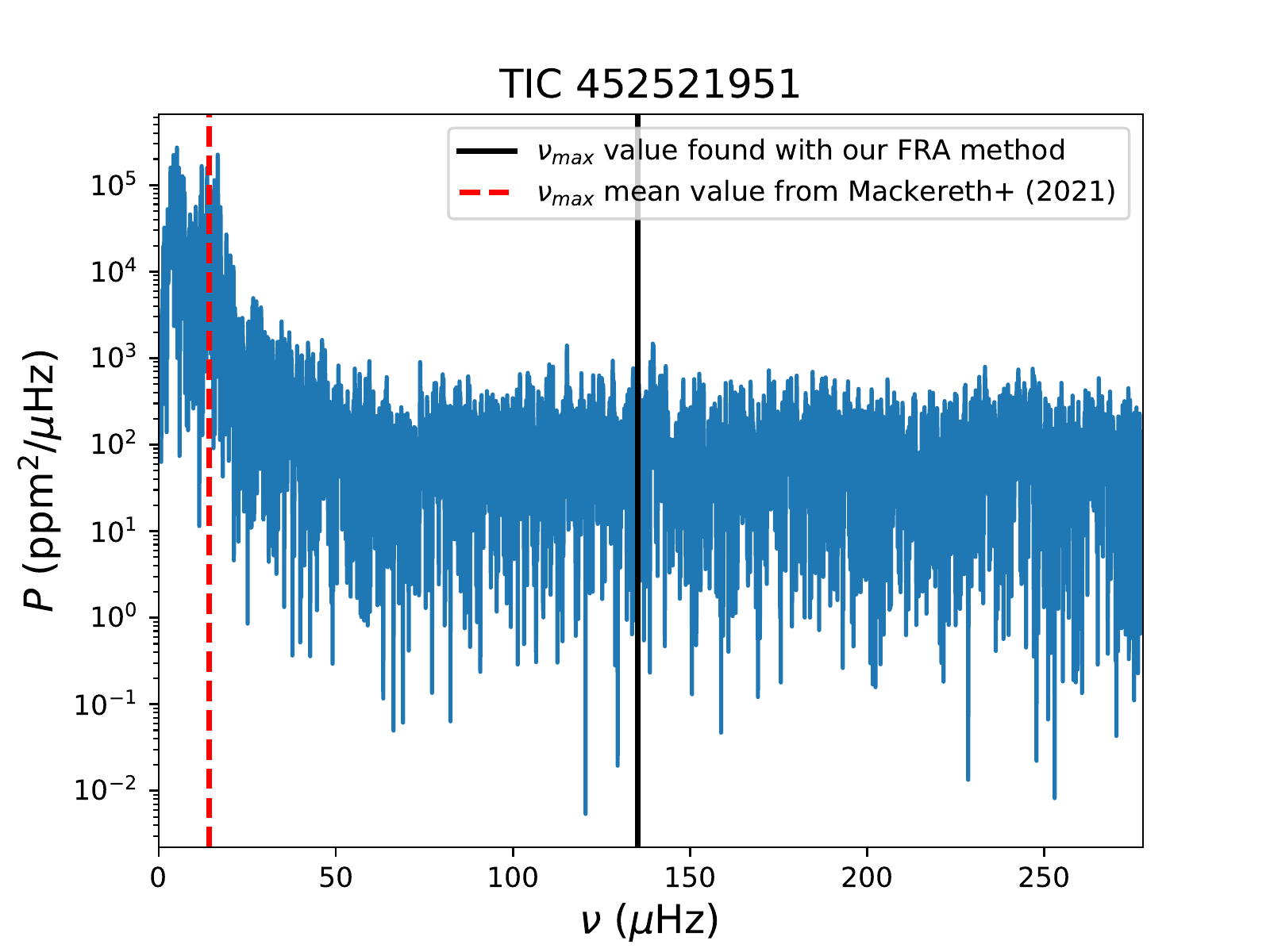}
\includegraphics[width=8.8cm]{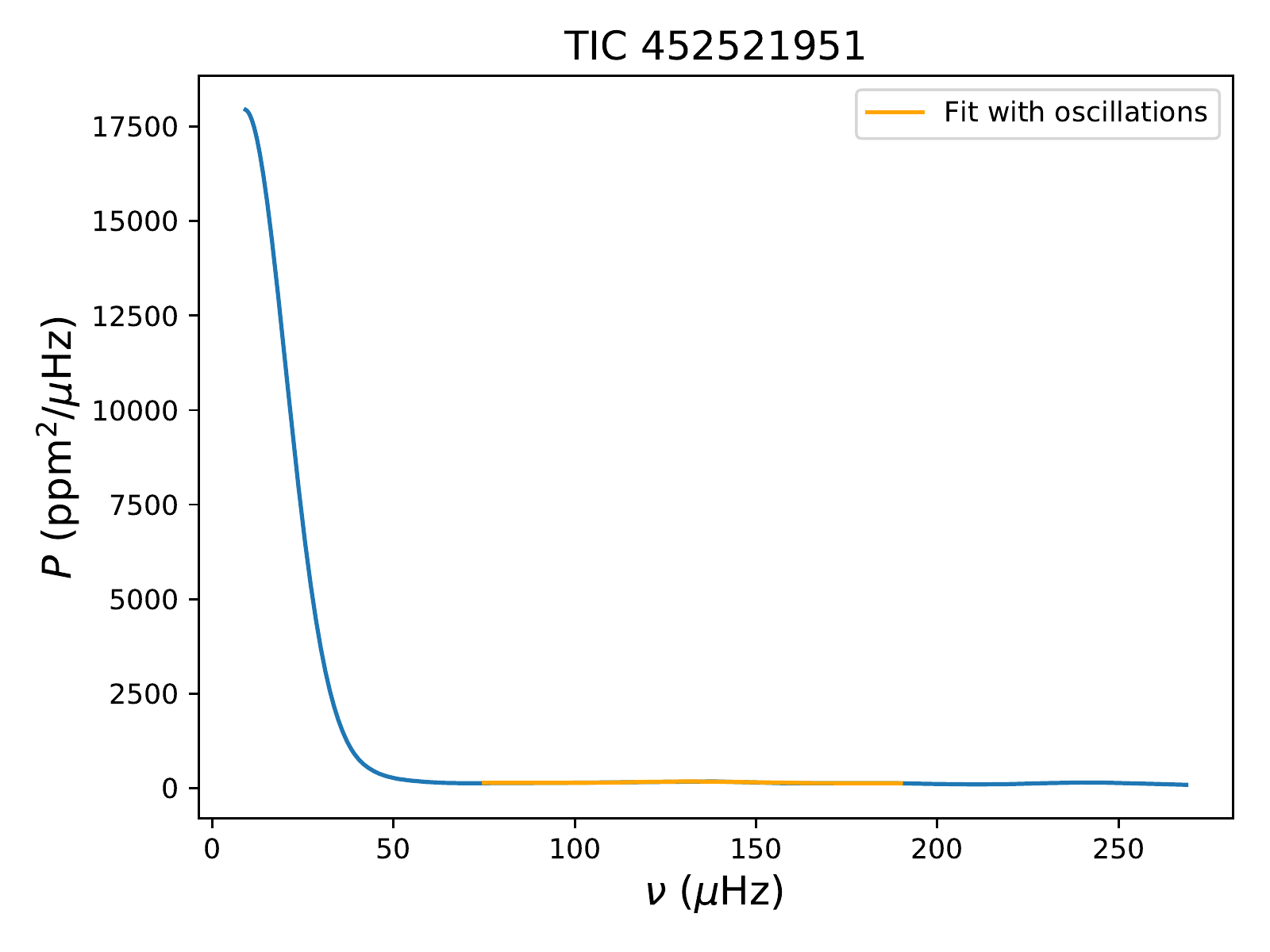}
\caption{Same as Fig.~\ref{fig-Kepler-high-rel-dev-1} for TIC 452521951.}
\label{fig-TESS-numax-12}
\end{figure*}

\begin{figure*}
\centering
\includegraphics[width=8.8cm]{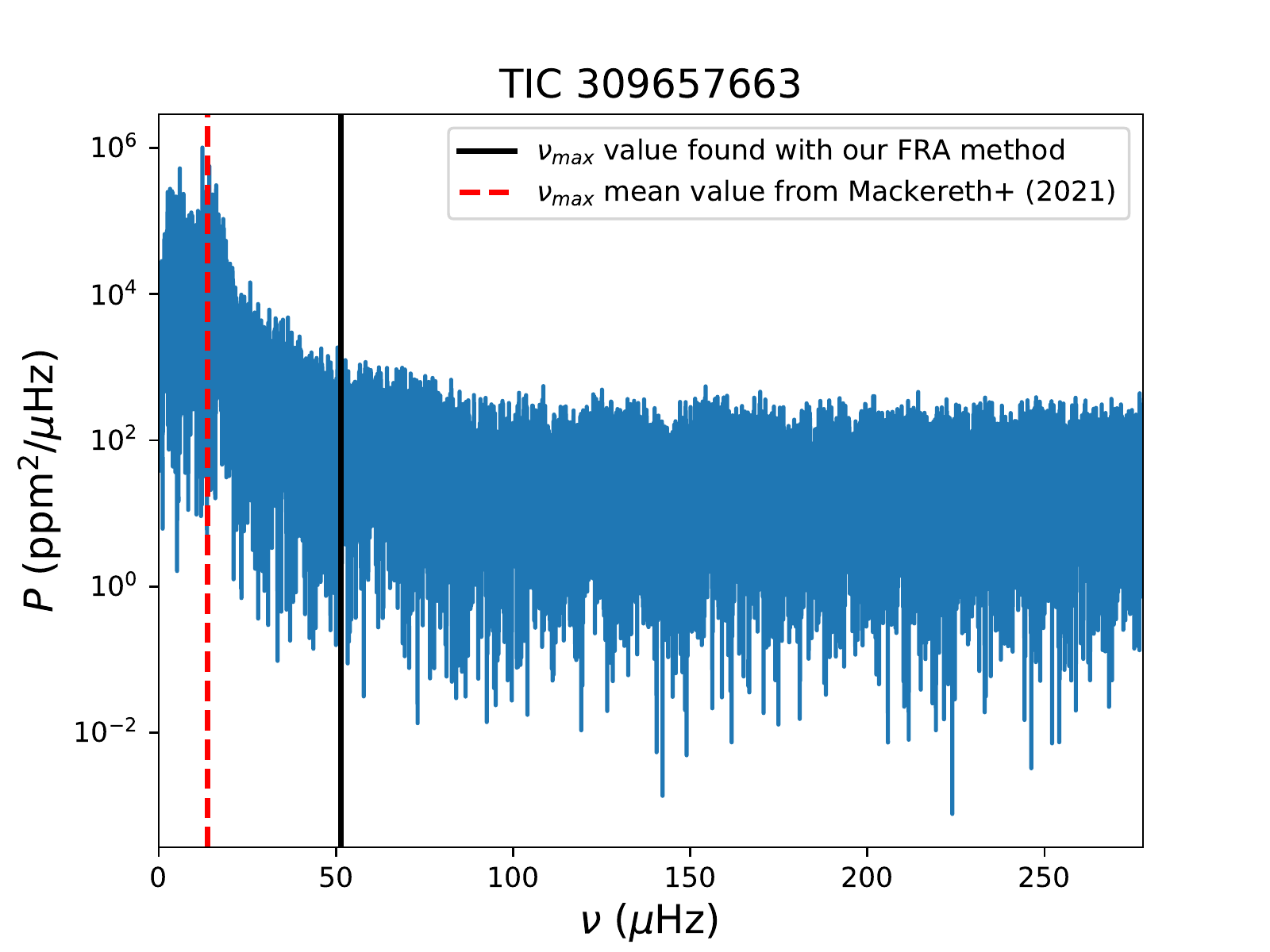}
\includegraphics[width=8.8cm]{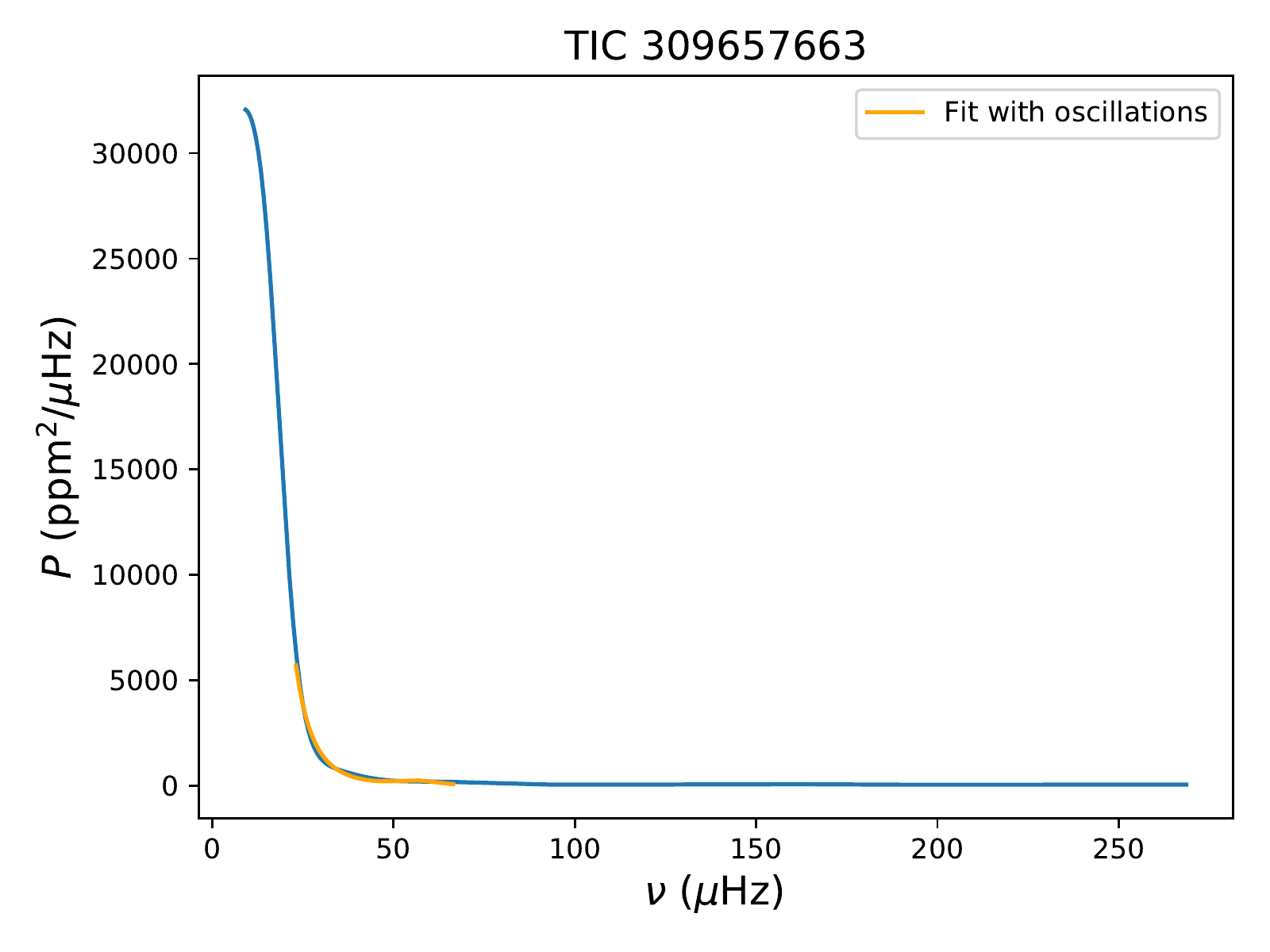}
\caption{Same as Fig.~\ref{fig-Kepler-high-rel-dev-1} for TIC 309657663.}
\label{fig-TESS-numax-13}
\end{figure*}

\begin{figure*}
\centering
\includegraphics[width=8.8cm]{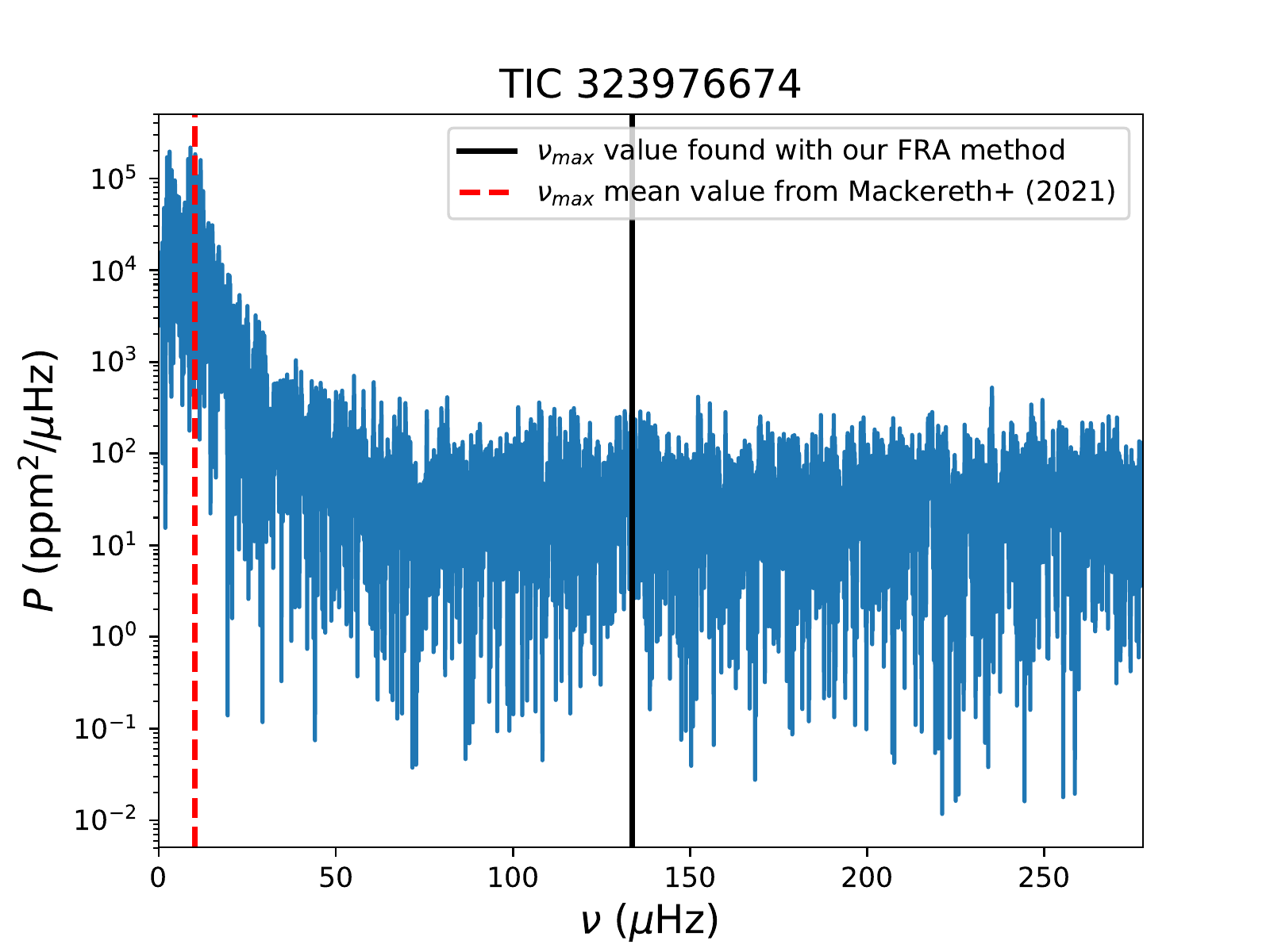}
\includegraphics[width=8.8cm]{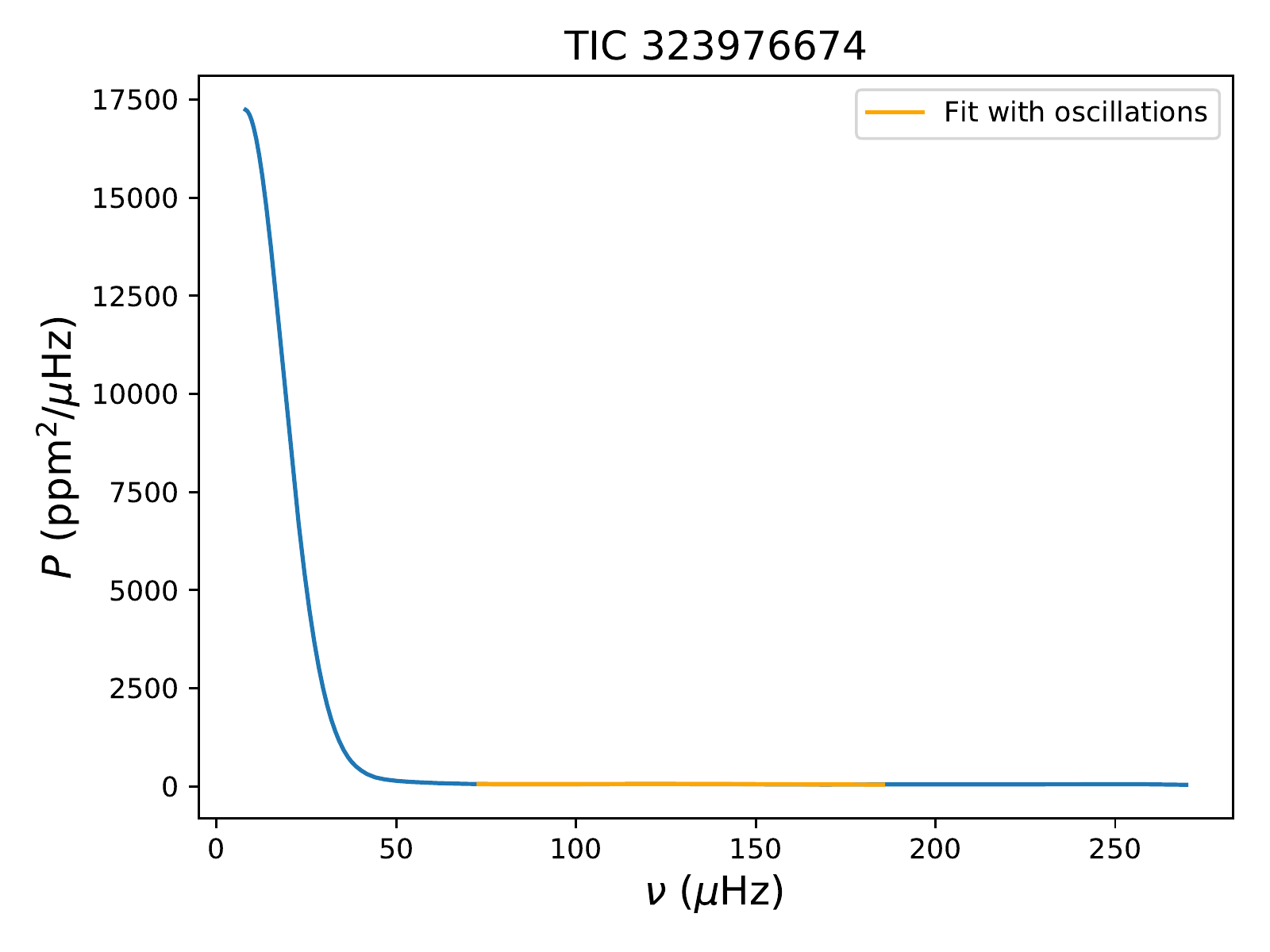}
\caption{Same as Fig.~\ref{fig-Kepler-high-rel-dev-1} for TIC 323976674.}
\label{fig-TESS-numax-14}
\end{figure*}

There are 9 TESS red giants for which we have a relative deviation of at least 10\% between our measurements obtained with our FRA pipeline and the measurements from \cite{Mackereth} once we apply an extra extra validation step for magnitudes $G > 9.5$, for which our analysis provides a largely overestimated $\numax$:
\begin{itemize}
\item TIC 220556666 (Fig.~\ref{fig-TESS-numax-6});
\item TIC 38510718 (Fig.~\ref{fig-TESS-numax-7});
\item TIC 237935915 (Fig.~\ref{fig-TESS-numax-8});
\item TIC 140527427 (Fig.~\ref{fig-TESS-numax-9});
\item TIC 323242564 (Fig.~\ref{fig-TESS-numax-10});
\item TIC 141911311 (Fig.~\ref{fig-TESS-numax-11});
\item TIC 452521951 (Fig.~\ref{fig-TESS-numax-12};
\item TIC 309657663 (Fig.~\ref{fig-TESS-numax-13});
\item TIC 323976674 (Fig.~\ref{fig-TESS-numax-14}).
\end{itemize}

\clearpage


\section{\textit{Kepler} red giants for which we have a relative deviation of at least 10\% compared to existing $\Dnu$ measurements}\label{appendix-5}

\begin{figure*}
\centering
\includegraphics[width=8.8cm]{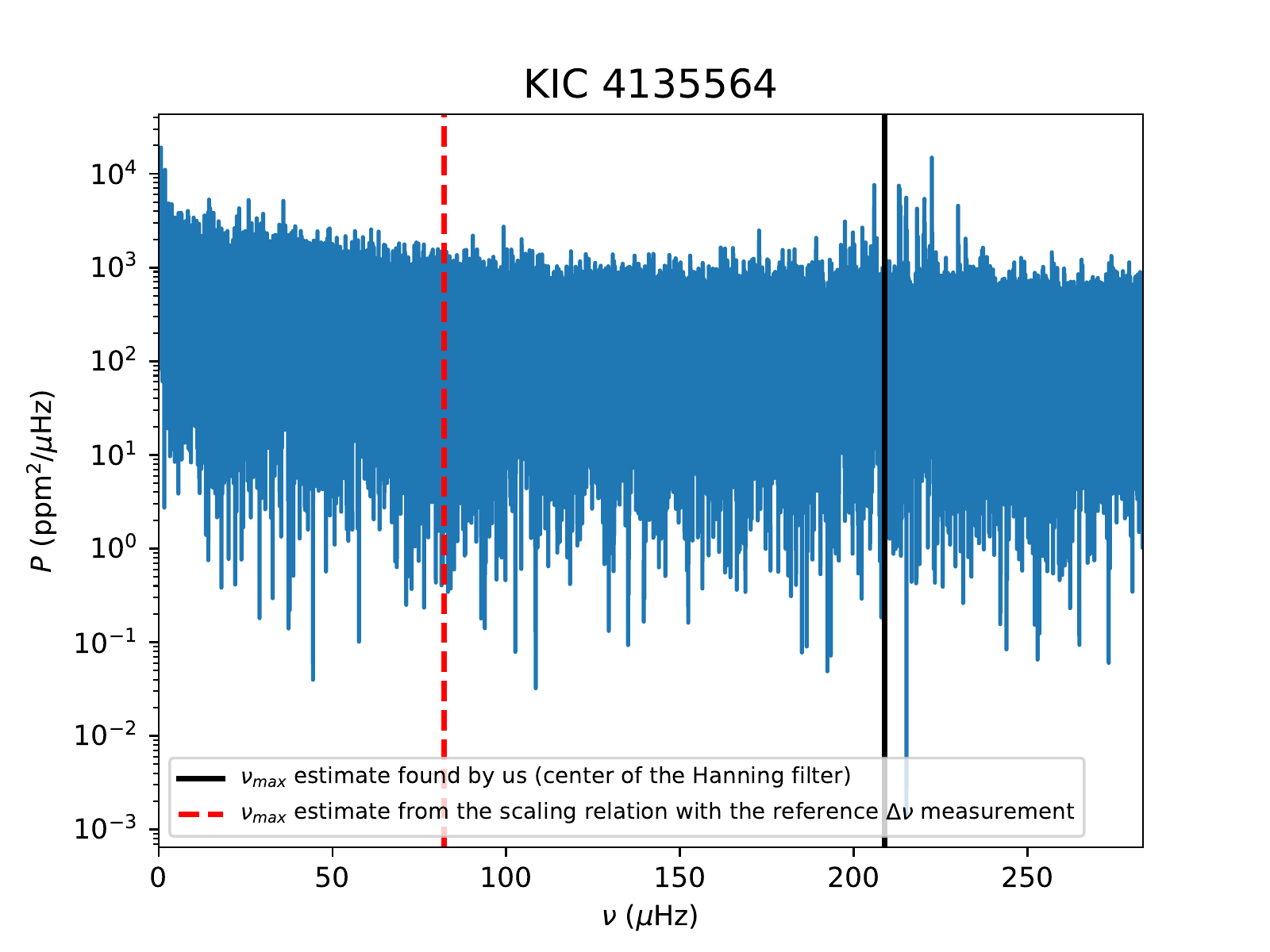}
\includegraphics[width=8.8cm]{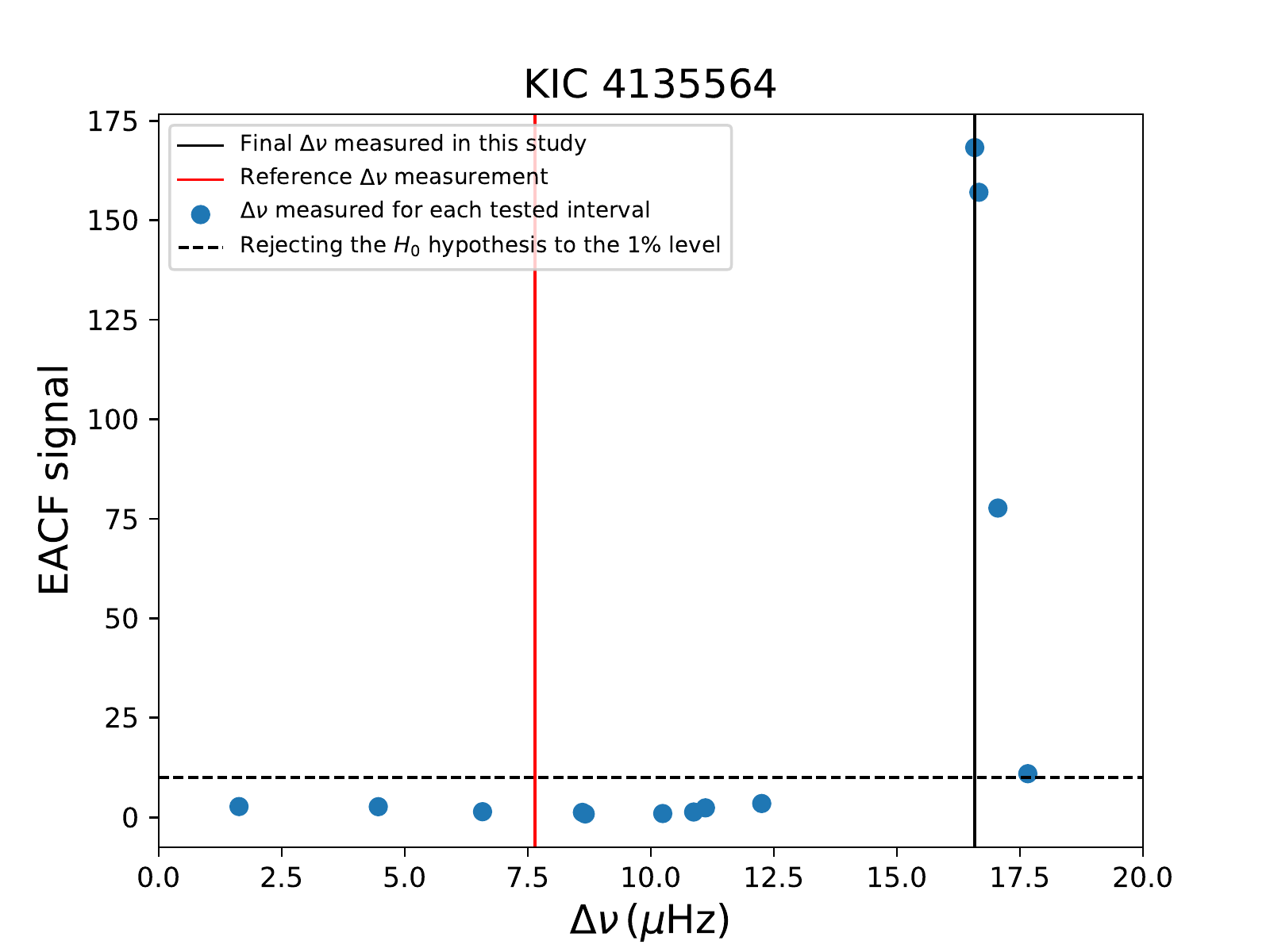}
\caption{\textit{Left:} Raw power spectrum of KIC 2021216. The vertical black line represents our $\numax$ estimate coresponding to the center of the Hanning filter used to window the spectrum, while the red dashed lines represents the $\numax$ estimate derived from the scaling relation with the reference $\Dnu$ measurement. \textit{Right:} Same as Fig.~\ref{fig-EACF} for KIC 2021216.}
\label{fig-Kepler-Dnu-1}
\end{figure*}

\begin{figure*}
\centering
\includegraphics[width=8.8cm]{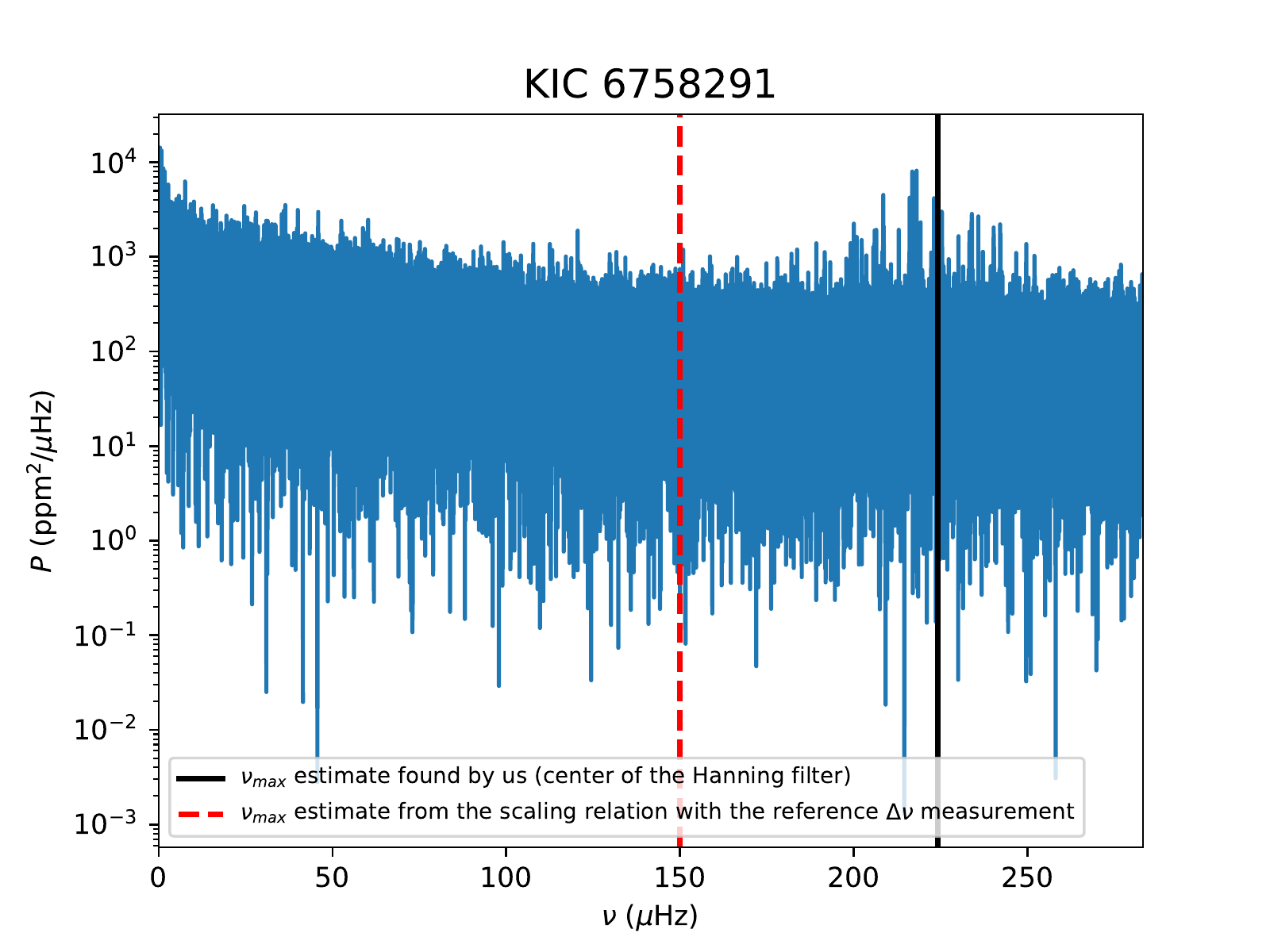}
\includegraphics[width=8.8cm]{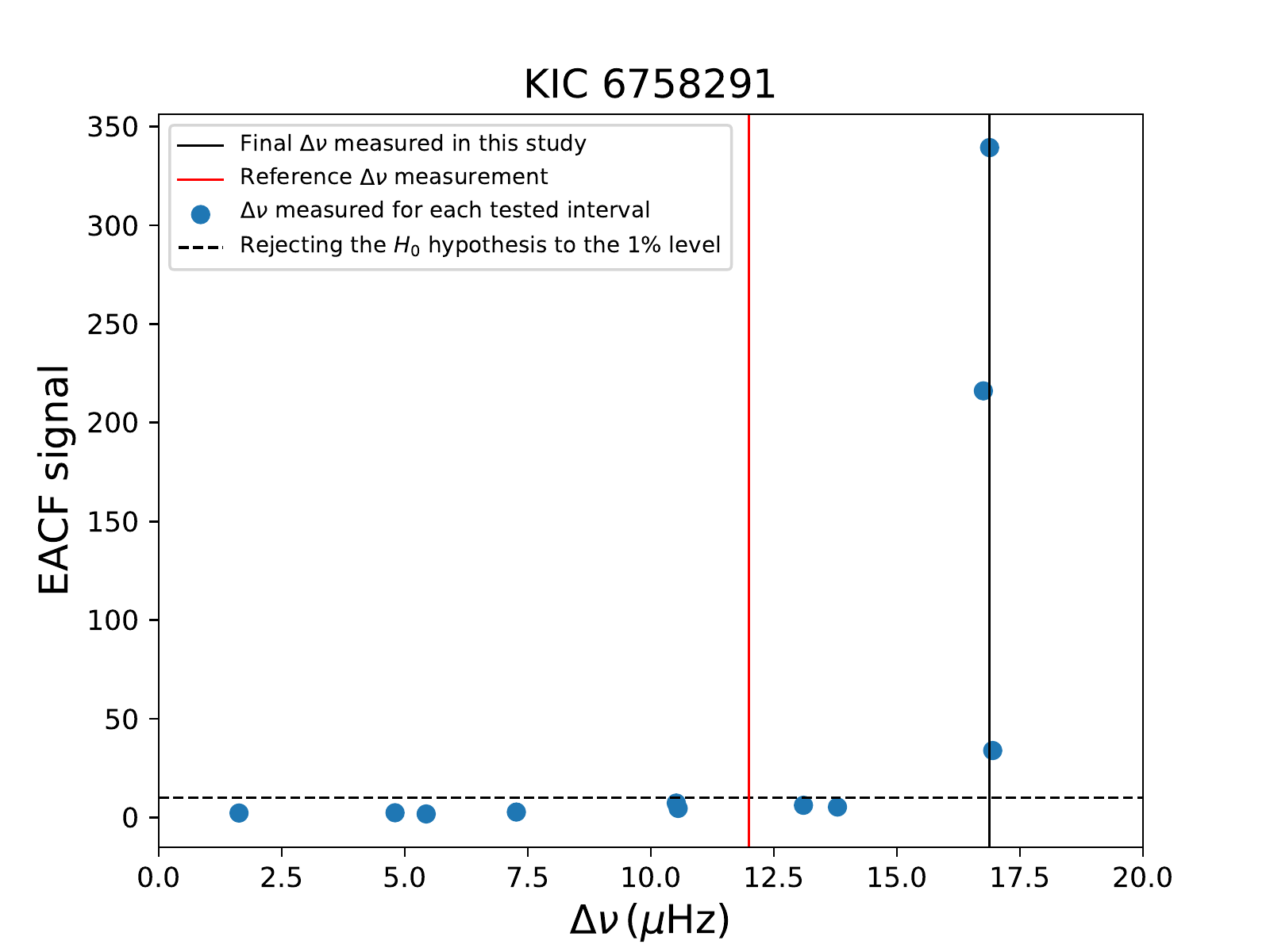}
\caption{Same as Fig.~\ref{fig-Kepler-Dnu-1} for KIC 6758291.}
\label{fig-Kepler-Dnu-2}
\end{figure*}

\begin{figure*}
\centering
\includegraphics[width=8.8cm]{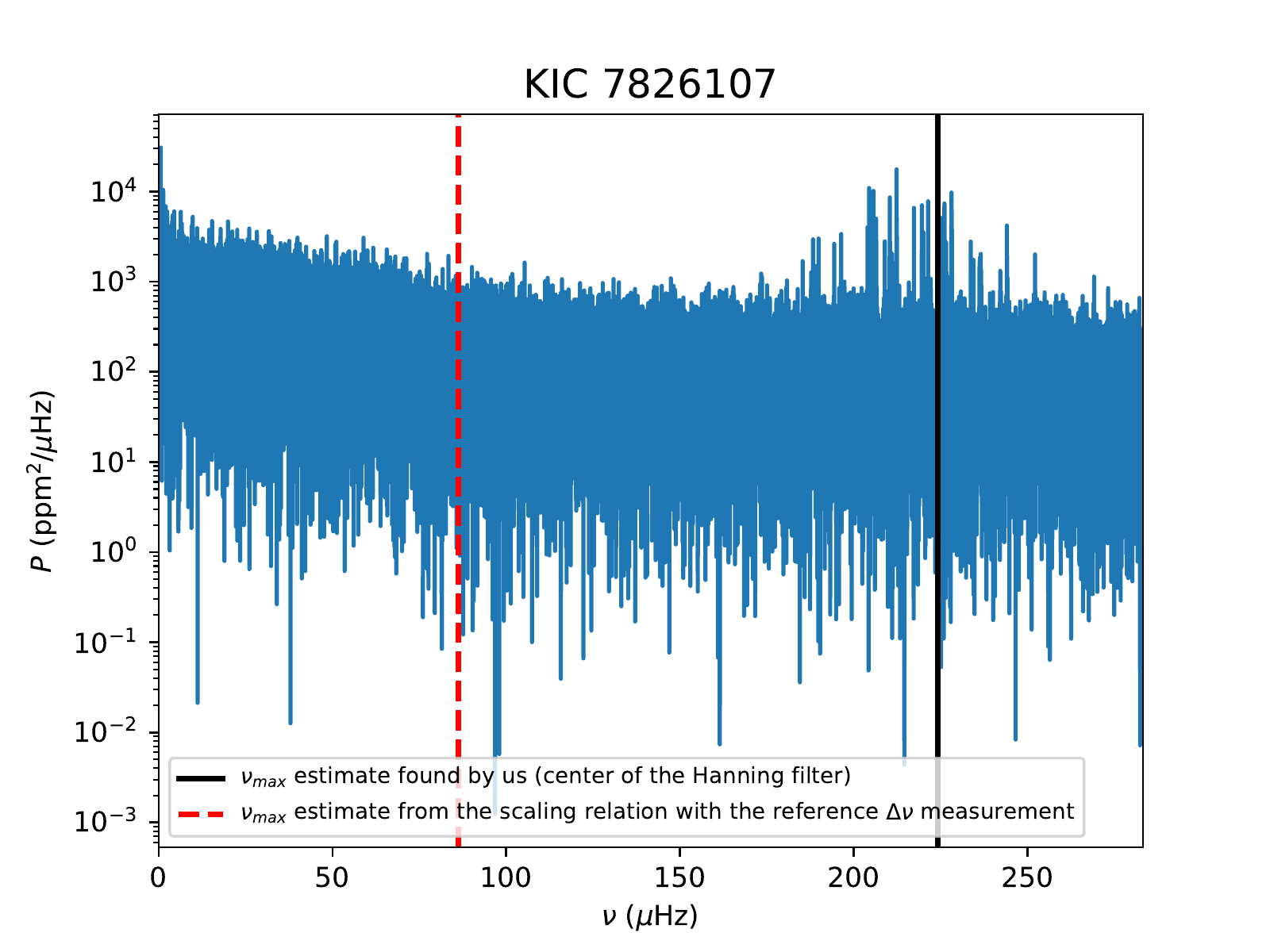}
\includegraphics[width=8.8cm]{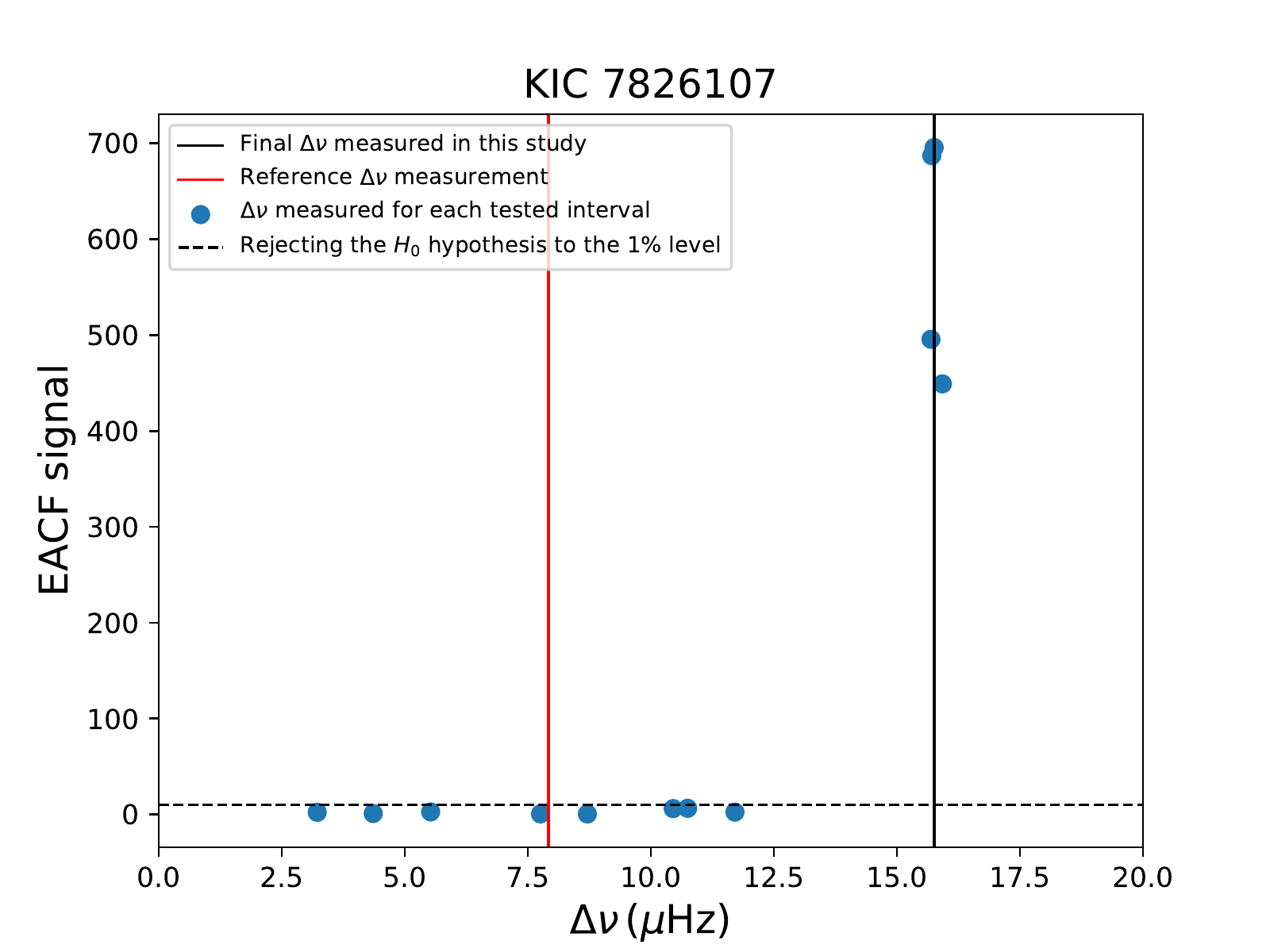}
\caption{Same as Fig.~\ref{fig-Kepler-Dnu-1} for KIC 7826107.}
\label{fig-Kepler-Dnu-3}
\end{figure*}

\begin{figure*}
\centering
\includegraphics[width=8.8cm]{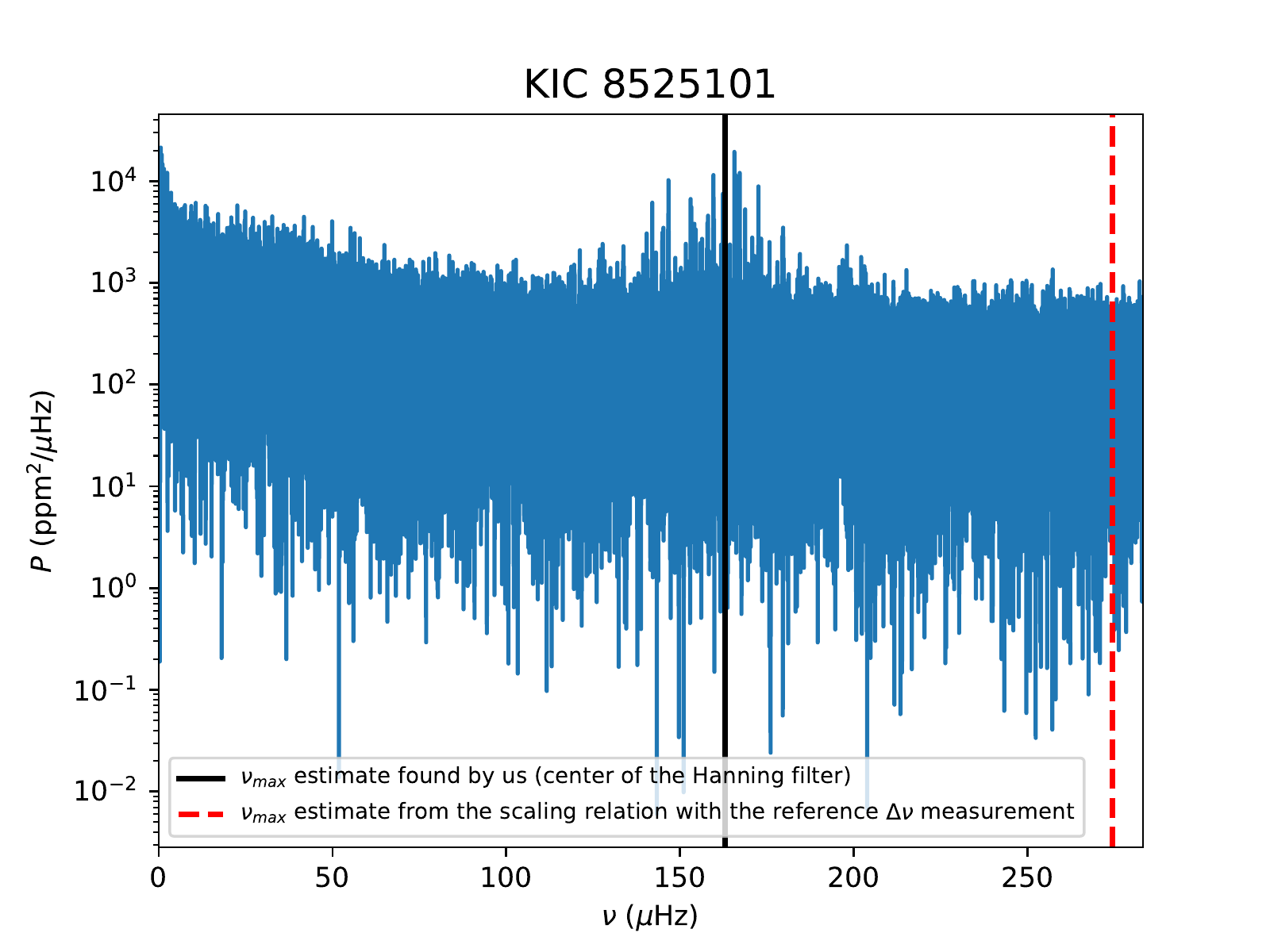}
\includegraphics[width=8.8cm]{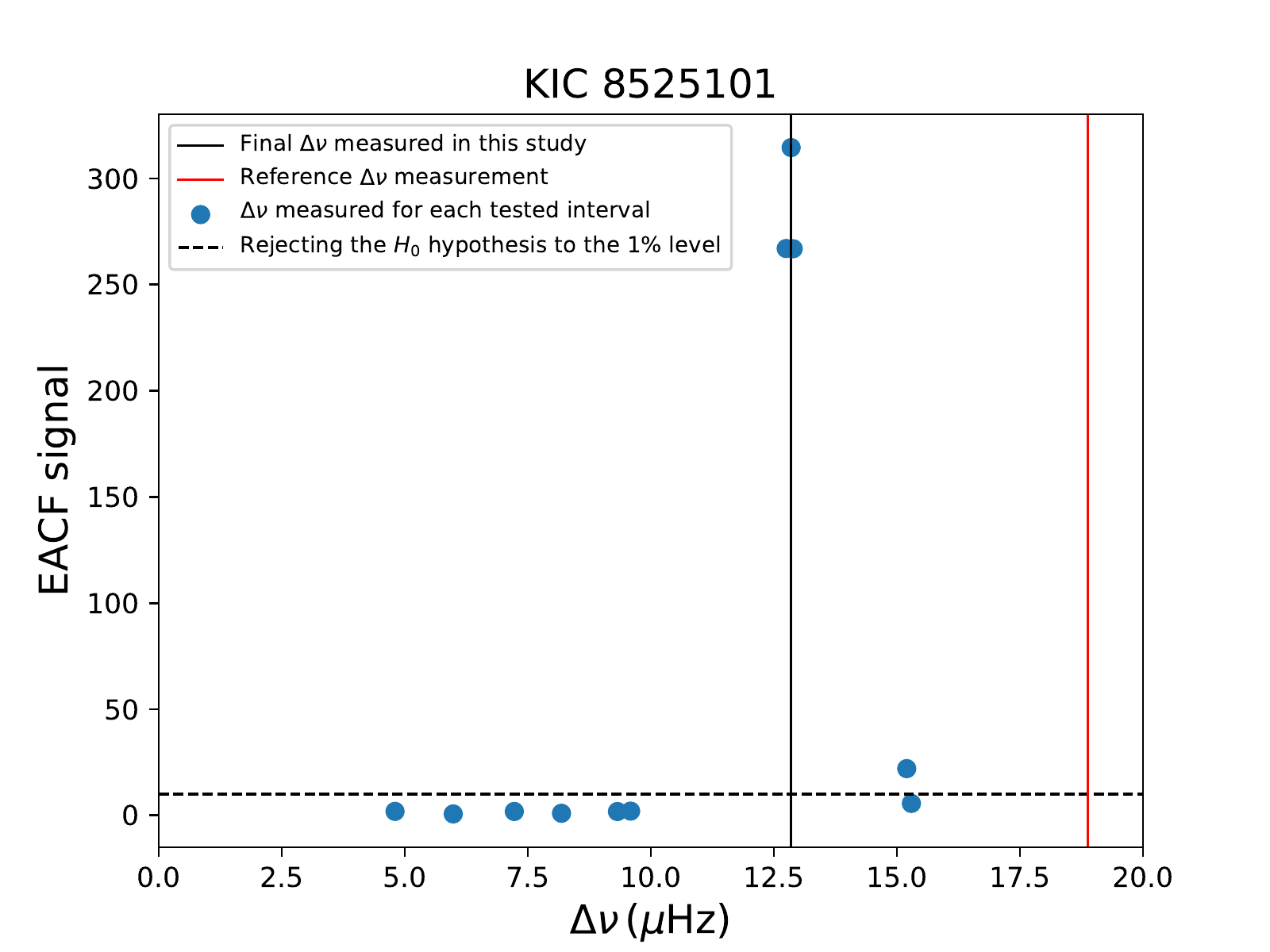}
\caption{Same as Fig.~\ref{fig-Kepler-Dnu-1} for KIC 8525101.}
\label{fig-Kepler-Dnu-4}
\end{figure*}

\begin{figure*}
\centering
\includegraphics[width=8.8cm]{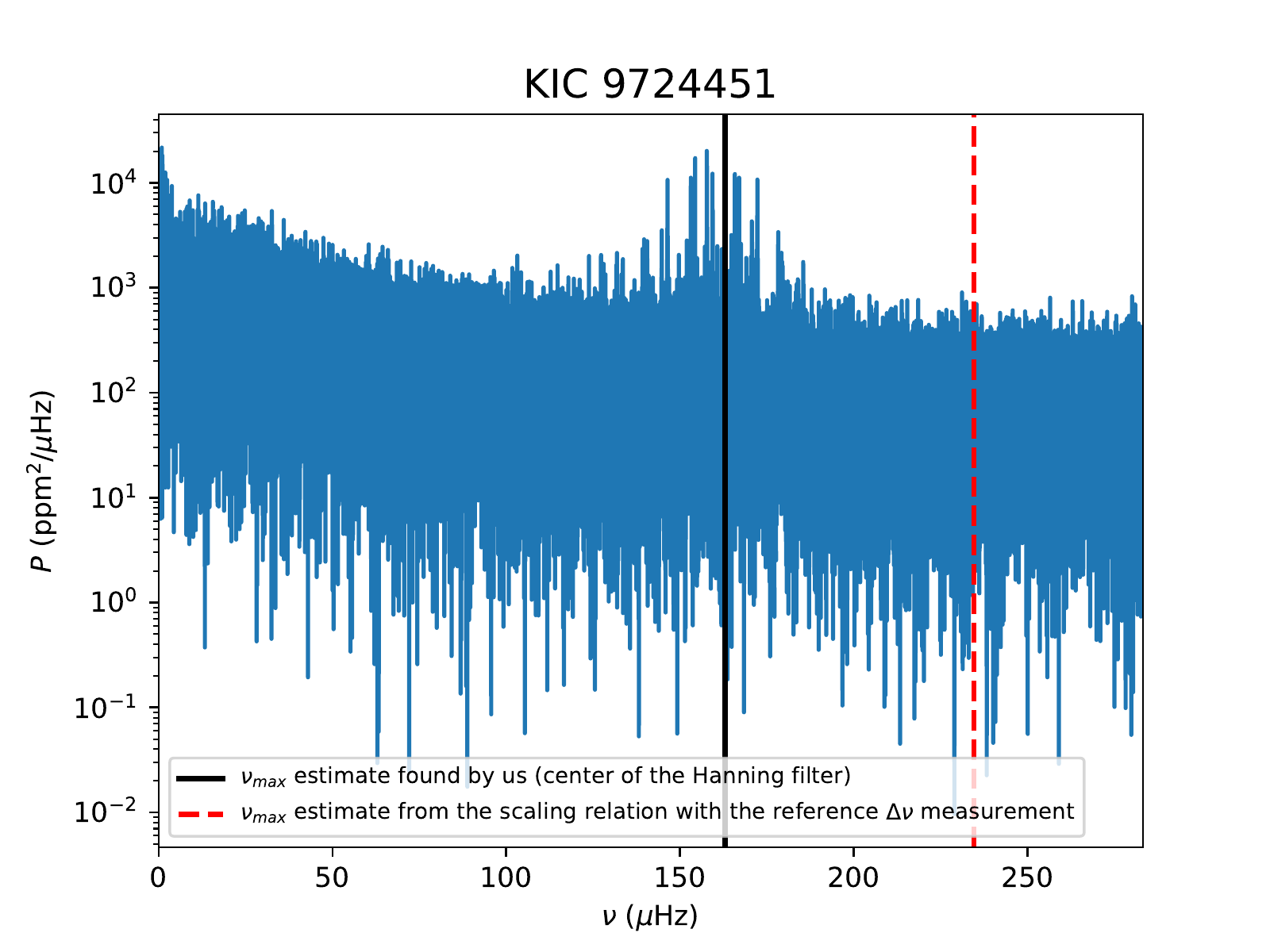}
\includegraphics[width=8.8cm]{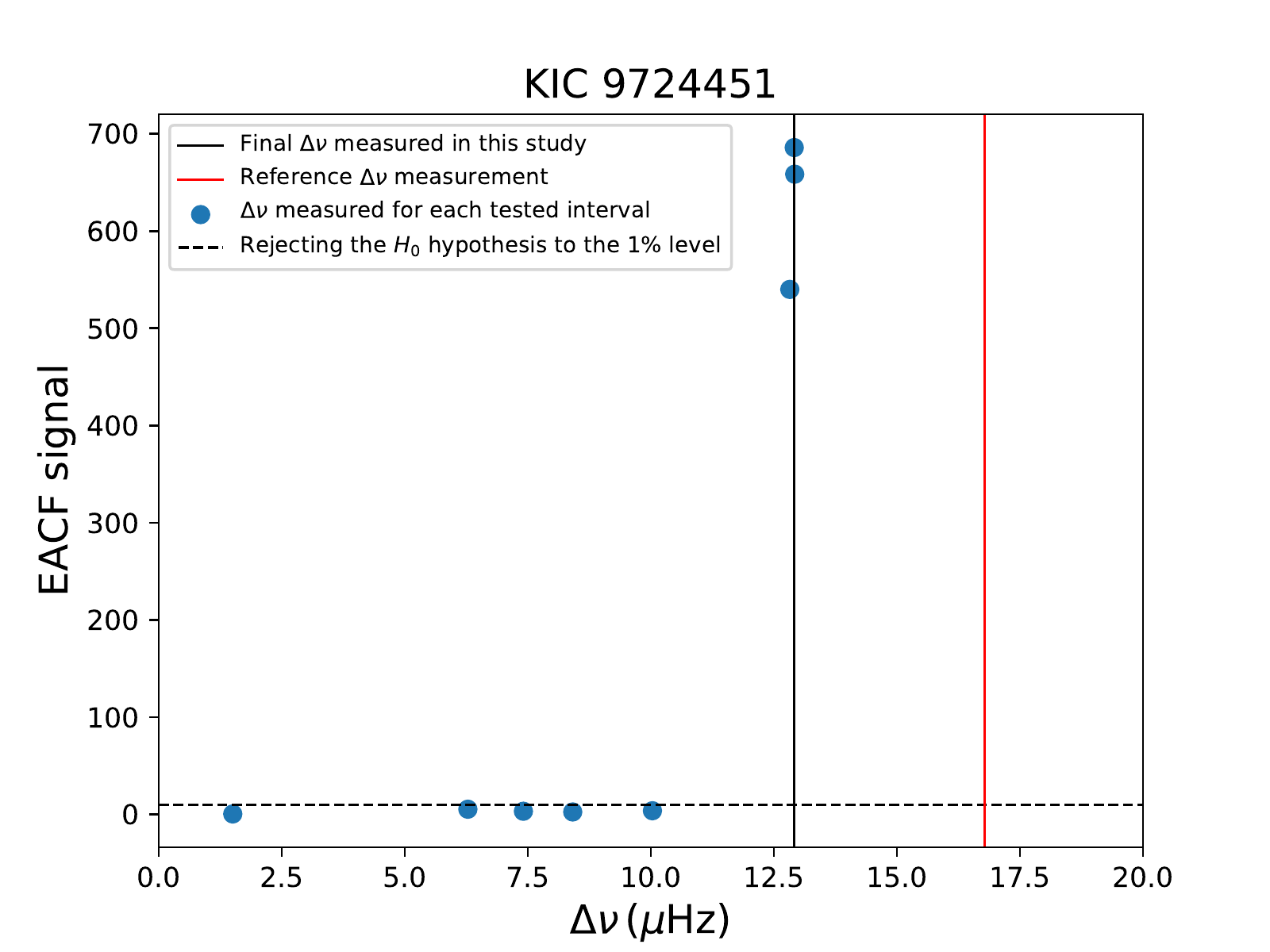}
\caption{Same as Fig.~\ref{fig-Kepler-Dnu-1} for KIC 9724451.}
\label{fig-Kepler-Dnu-5}
\end{figure*}

\begin{figure*}
\centering
\includegraphics[width=8.8cm]{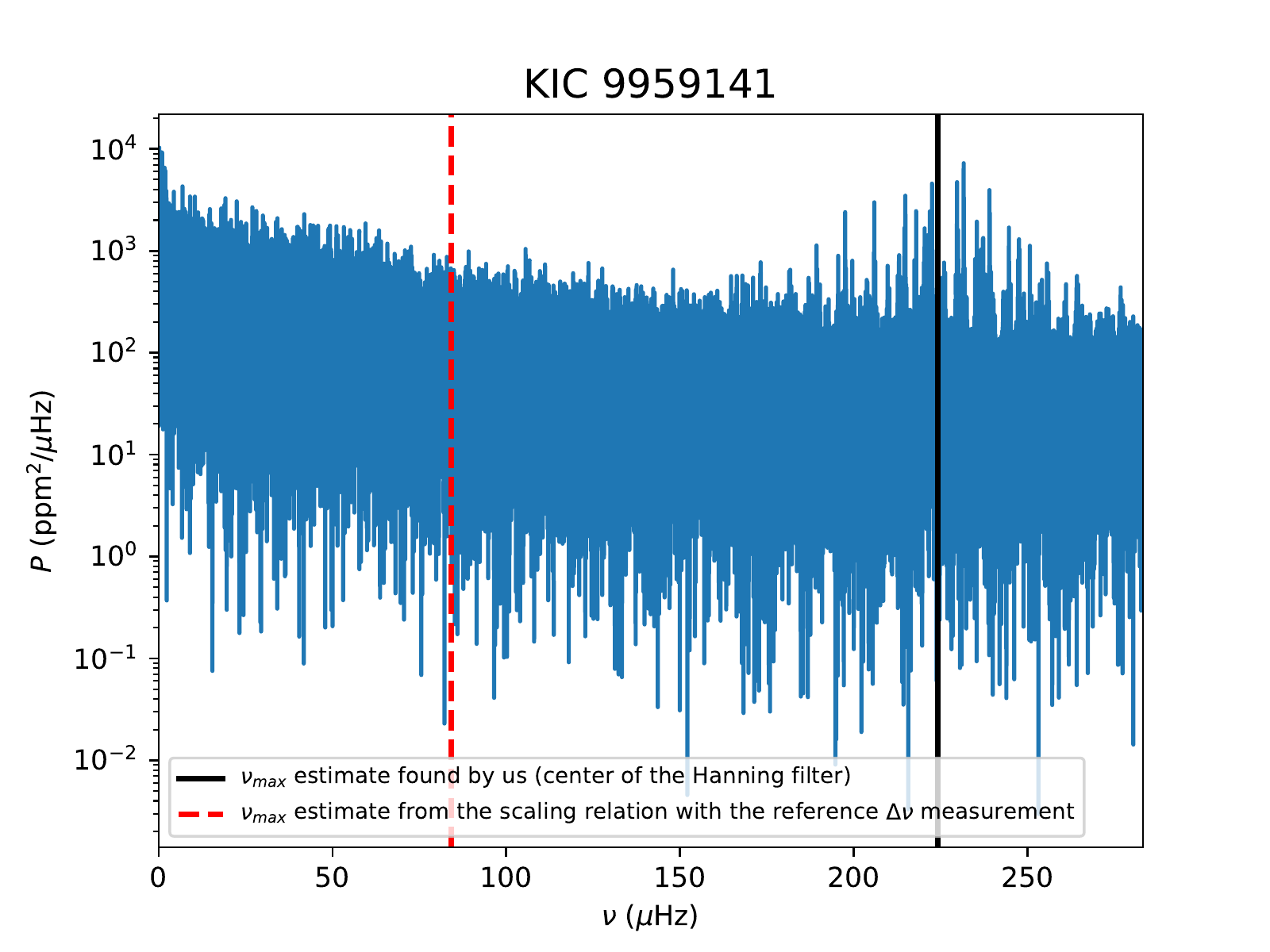}
\includegraphics[width=8.8cm]{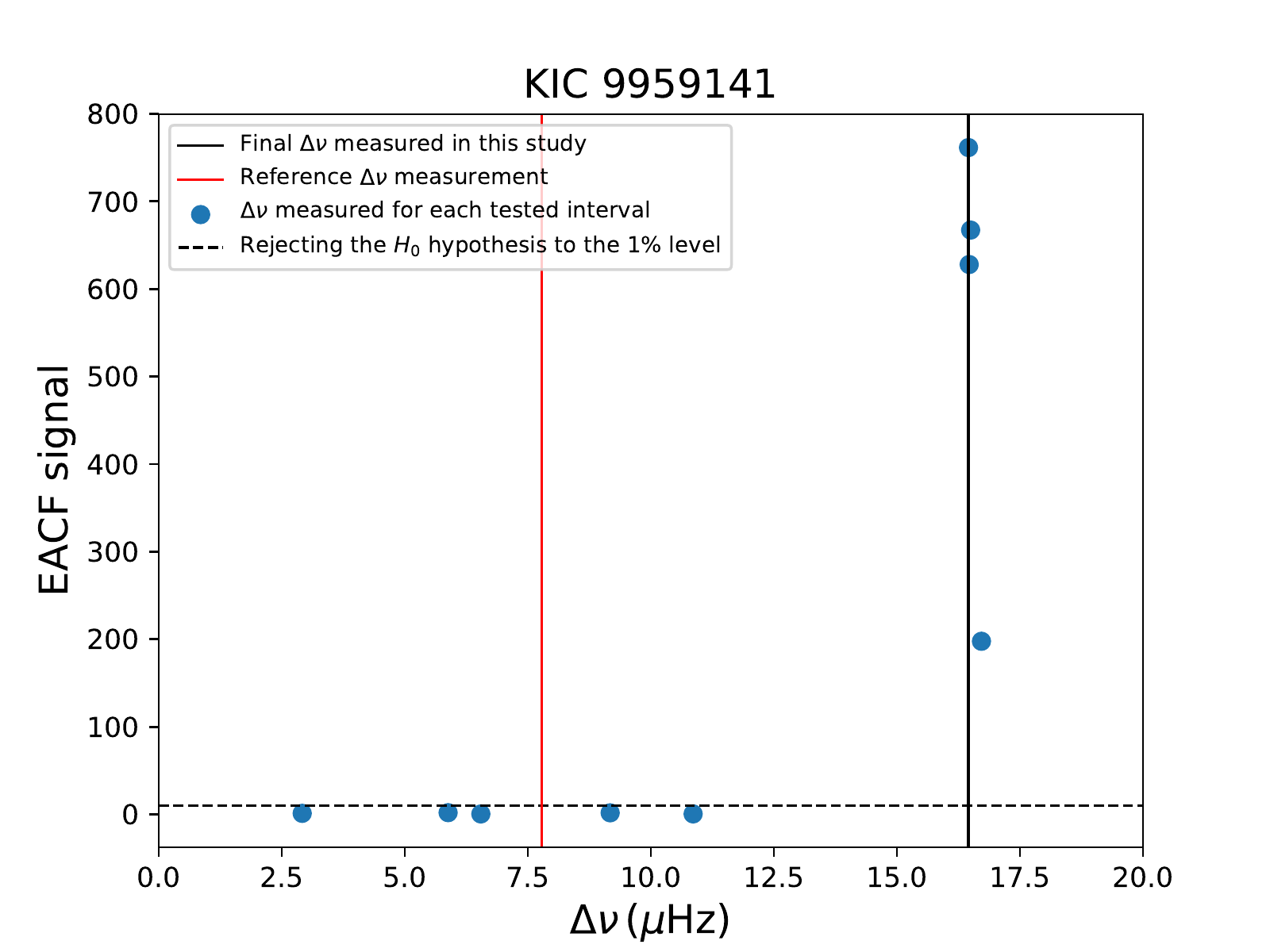}
\caption{Same as Fig.~\ref{fig-Kepler-Dnu-1} for KIC 9959141.}
\label{fig-Kepler-Dnu-6}
\end{figure*}

\begin{figure*}
\centering
\includegraphics[width=8.8cm]{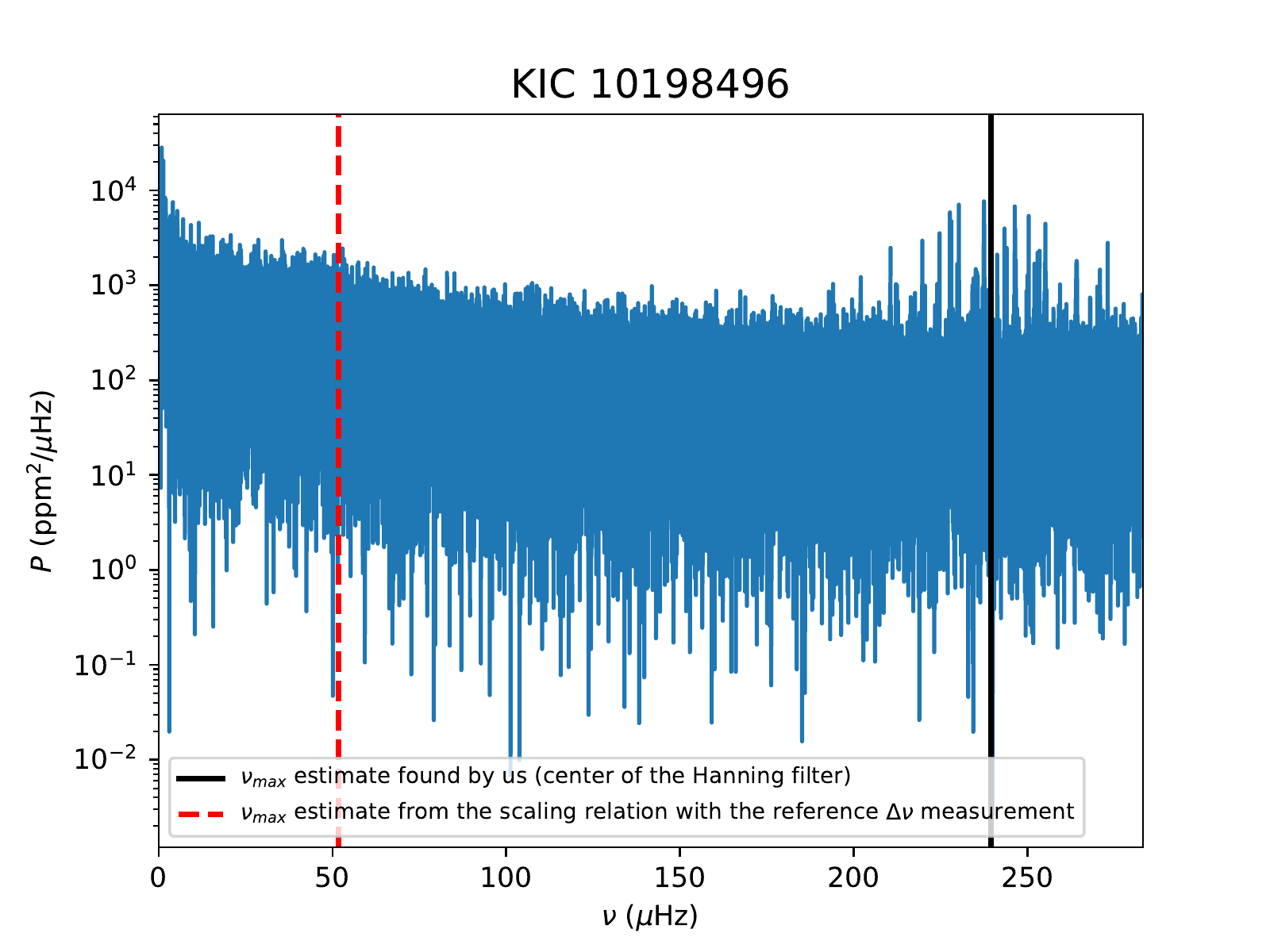}
\includegraphics[width=8.8cm]{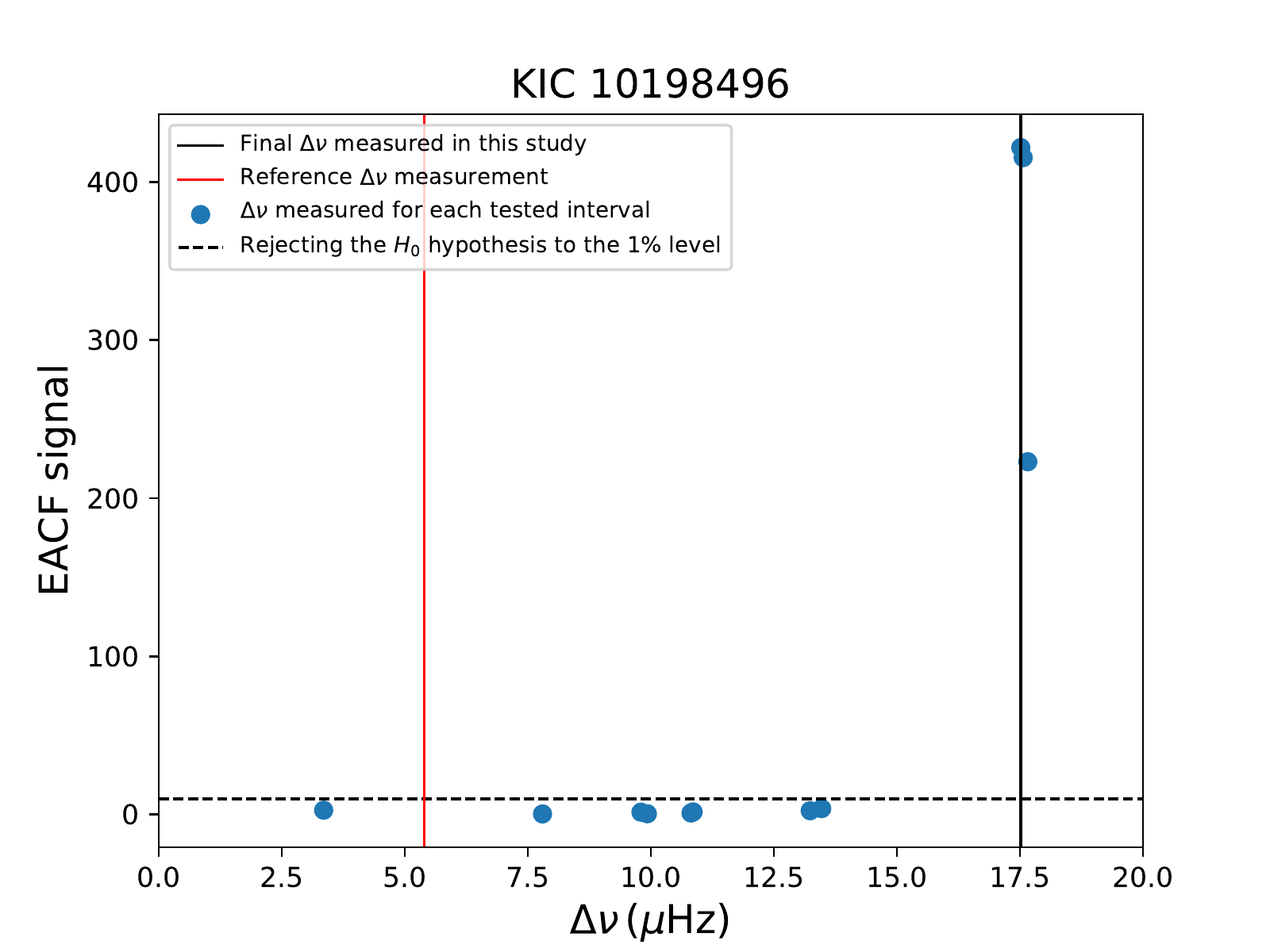}
\caption{Same as Fig.~\ref{fig-Kepler-Dnu-1} for KIC 10198496.}
\label{fig-Kepler-Dnu-7}
\end{figure*}

\begin{figure*}
\centering
\includegraphics[width=8.8cm]{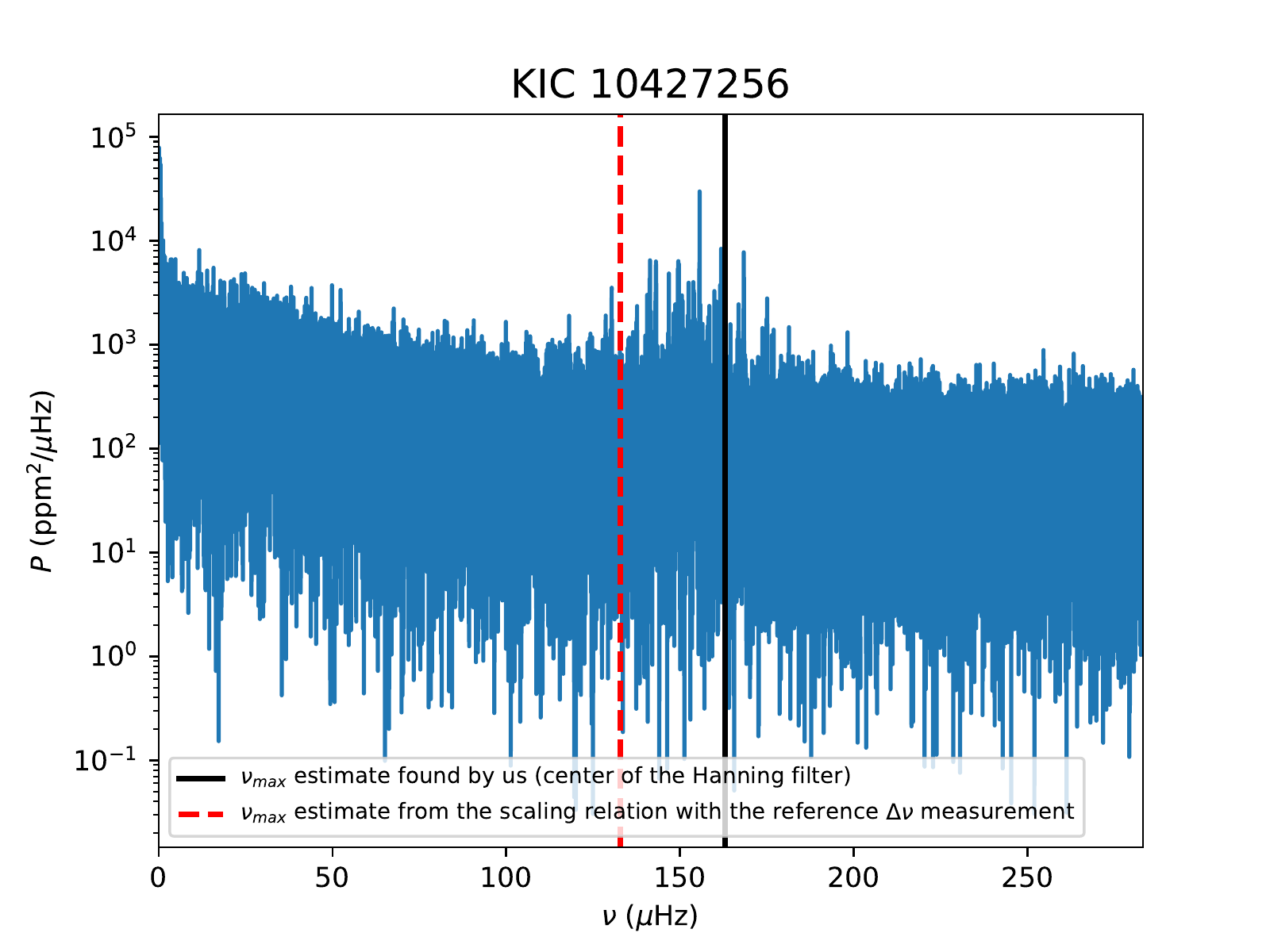}
\includegraphics[width=8.8cm]{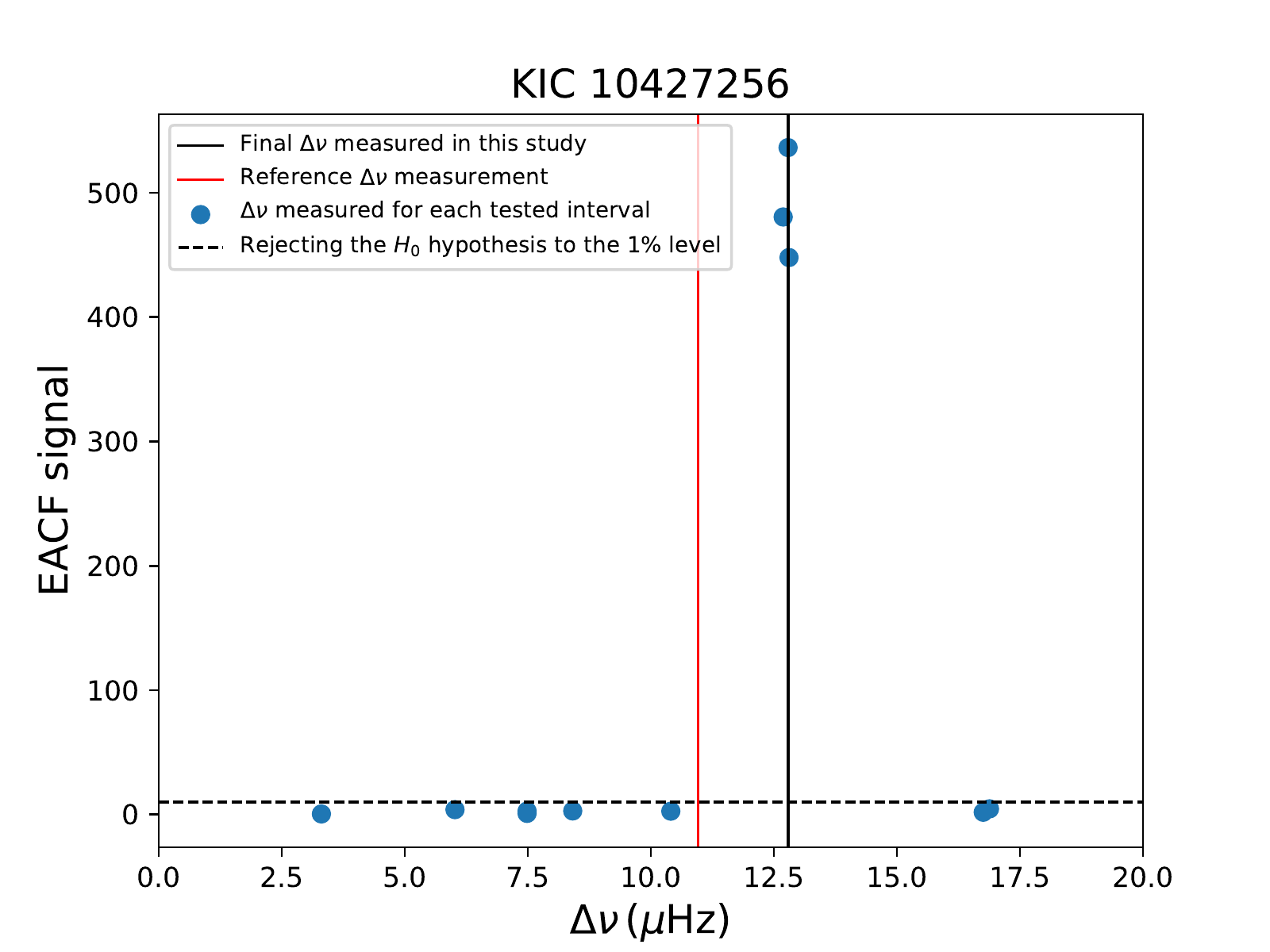}
\caption{Same as Fig.~\ref{fig-Kepler-Dnu-1} for KIC 10427256.}
\label{fig-Kepler-Dnu-8}
\end{figure*}

\begin{figure*}
\centering
\includegraphics[width=8.8cm]{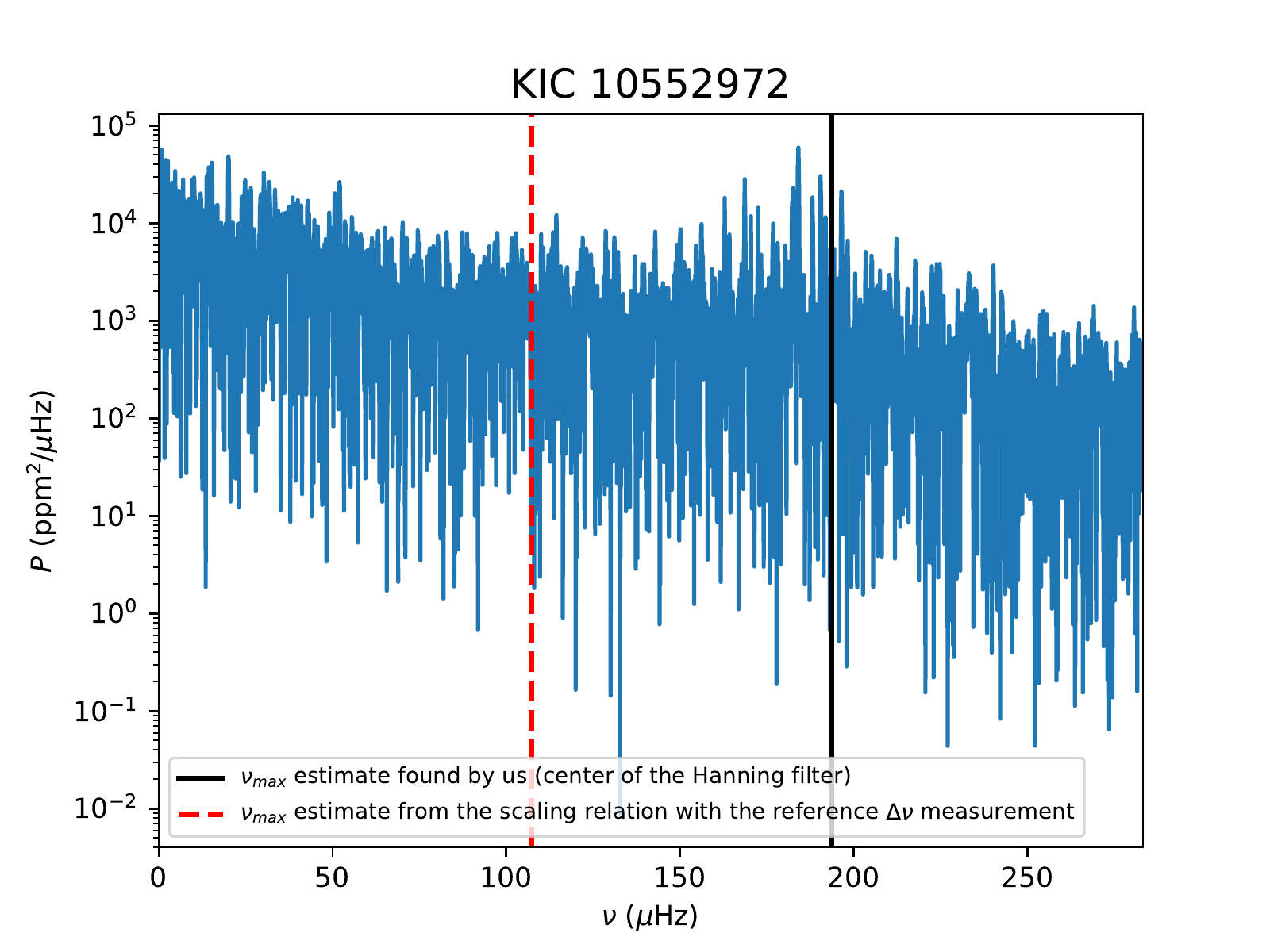}
\includegraphics[width=8.8cm]{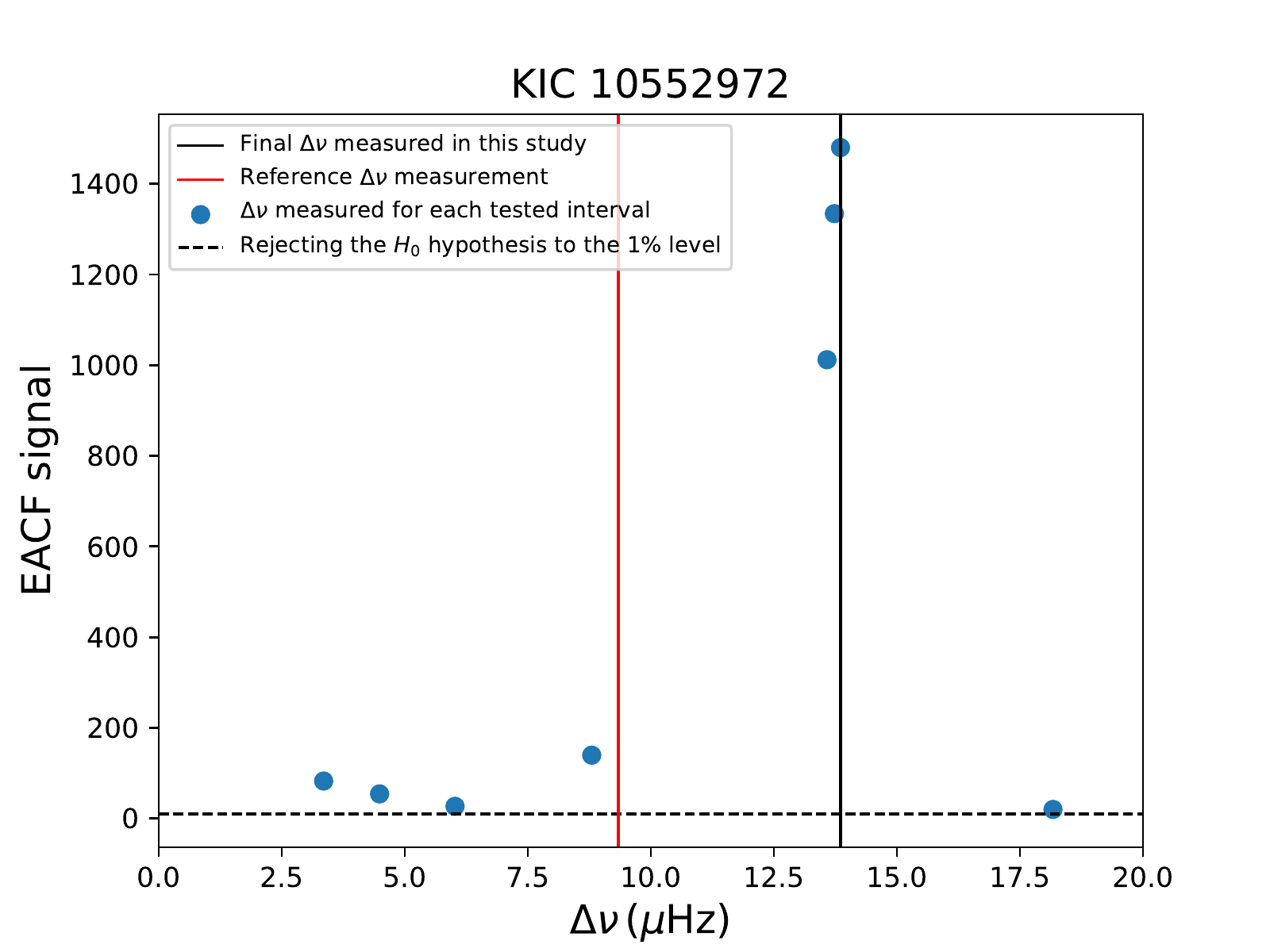}
\caption{Same as Fig.~\ref{fig-Kepler-Dnu-1} for KIC 10552972.}
\label{fig-Kepler-Dnu-9}
\end{figure*}

\begin{figure*}
\centering
\includegraphics[width=8.8cm]{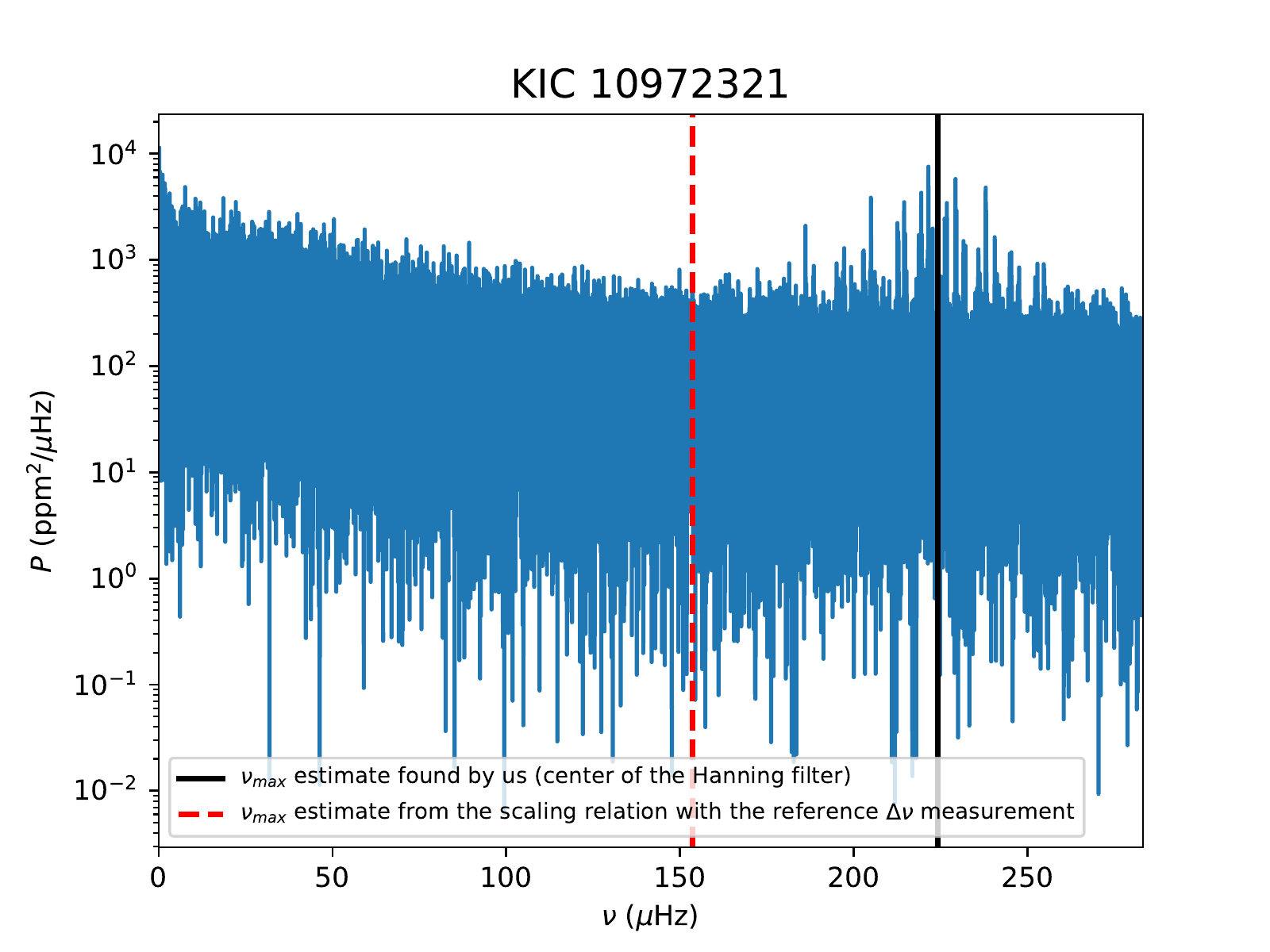}
\includegraphics[width=8.8cm]{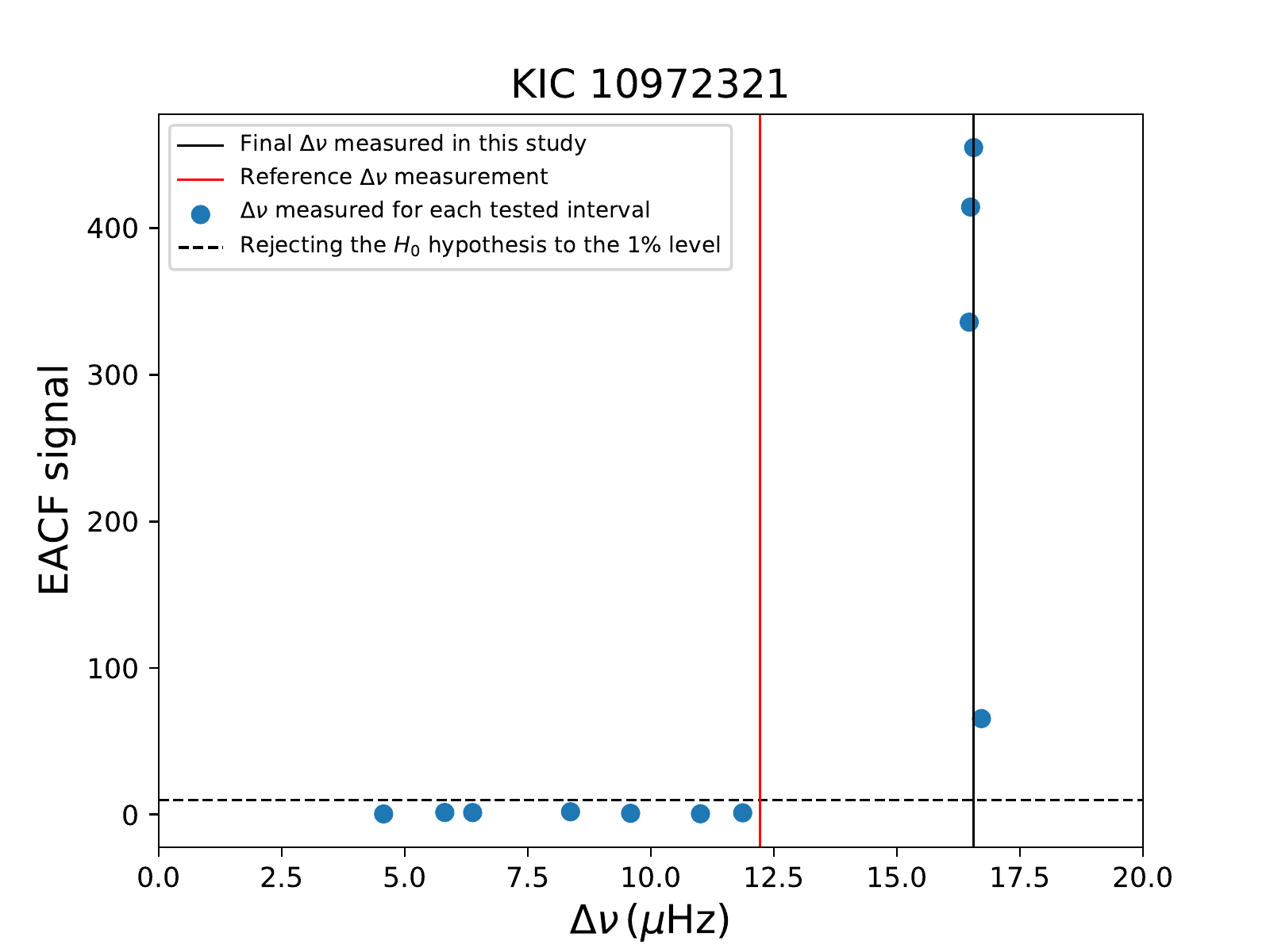}
\caption{Same as Fig.~\ref{fig-Kepler-Dnu-1} for KIC 10972321.}
\label{fig-Kepler-Dnu-10}
\end{figure*}

There are 10 \textit{Kepler} red giants for which the EACF method from \cite{Mosser_2009} provided an inaccurate $\Dnu$ measurement, while our version of the EACF method gives an accurate measurement:
\begin{itemize}
\item KIC 4135564 (Fig.~\ref{fig-Kepler-Dnu-1});
\item KIC 6758291 (Fig.~\ref{fig-Kepler-Dnu-2});
\item KIC 7826107 (Fig.~\ref{fig-Kepler-Dnu-3});
\item KIC 8525101 (Fig.~\ref{fig-Kepler-Dnu-4});
\item KIC 9724451 (Fig.~\ref{fig-Kepler-Dnu-5});
\item KIC 9959141 (Fig.~\ref{fig-Kepler-Dnu-6});
\item KIC 10198496 (Fig.~\ref{fig-Kepler-Dnu-7});
\item KIC 10427256 (Fig.~\ref{fig-Kepler-Dnu-8});
\item KIC 10552972 (Fig.~\ref{fig-Kepler-Dnu-9});
\item KIC 10972321 (Fig.~\ref{fig-Kepler-Dnu-10}).
\end{itemize}

\section{\textit{TESS} red giants for which we have a relative deviation of at least 10\% compared to existing $\Dnu$ measurements}\label{appendix-6}

\begin{figure*}
\centering
\includegraphics[width=8.8cm]{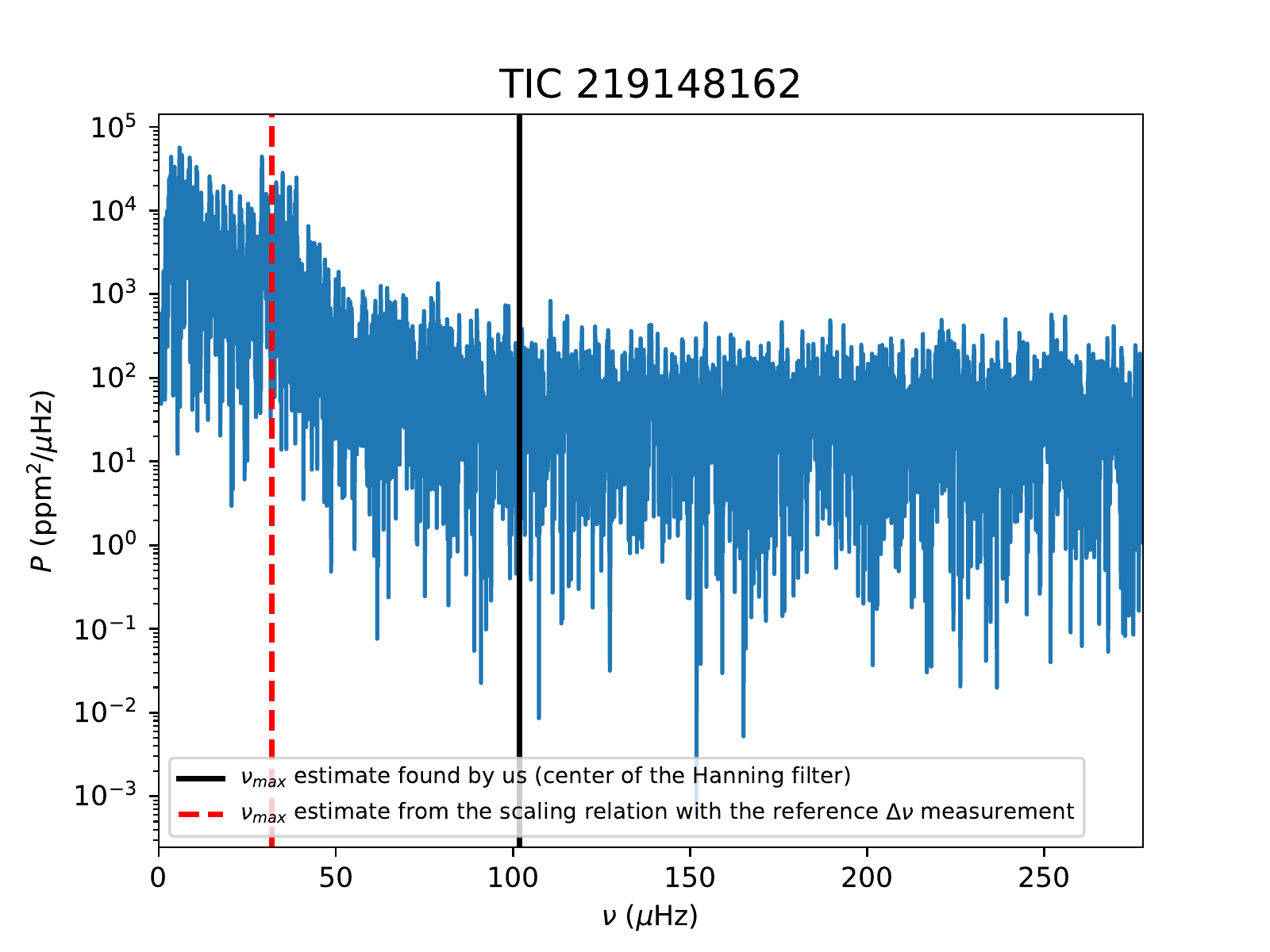}
\includegraphics[width=8.8cm]{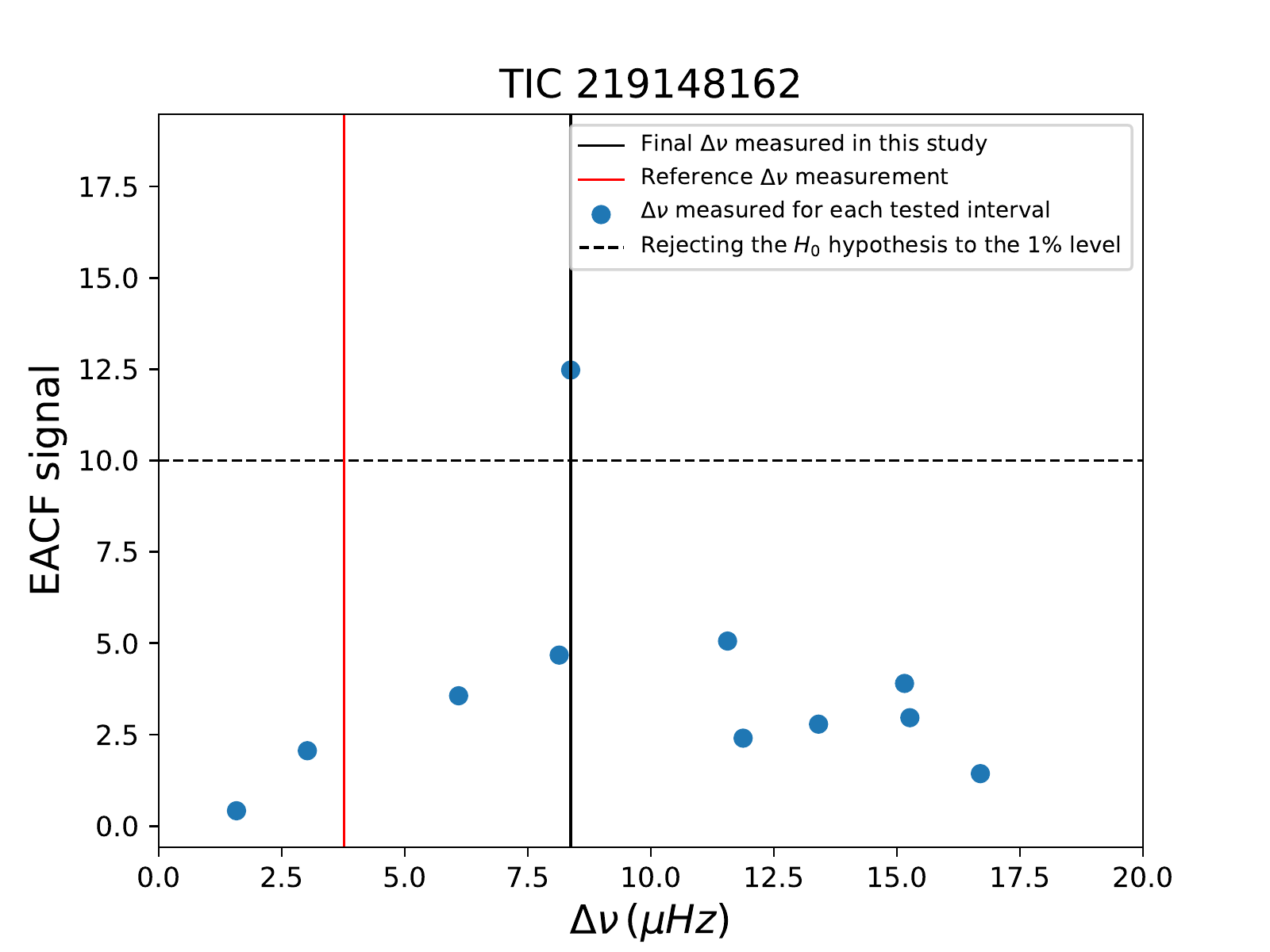}
\caption{Same as Fig.~\ref{fig-Kepler-Dnu-1} for TIC 219148162.}
\label{fig-TESS-Dnu-1}
\end{figure*}

\begin{figure*}
\centering
\includegraphics[width=8.8cm]{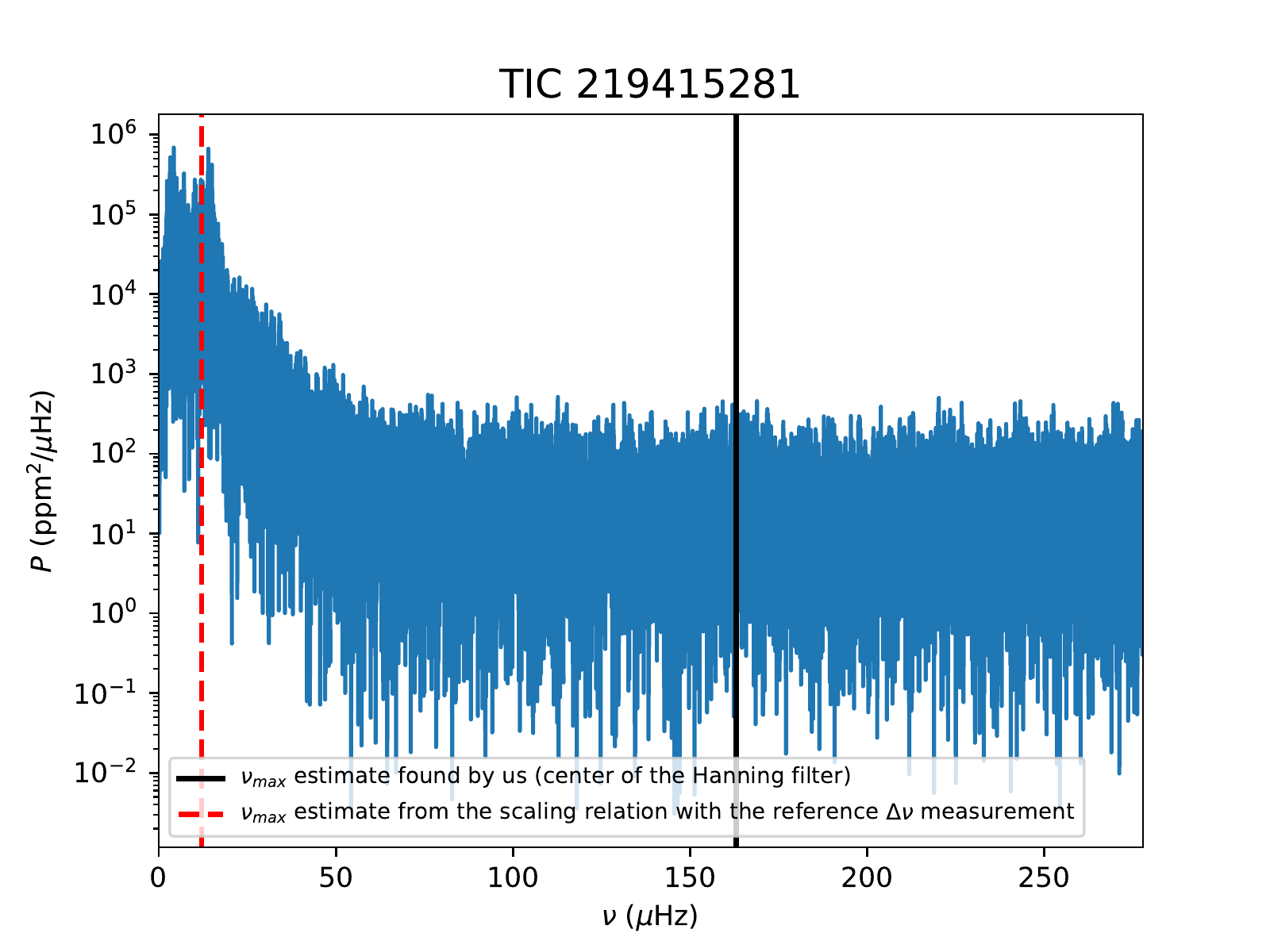}
\includegraphics[width=8.8cm]{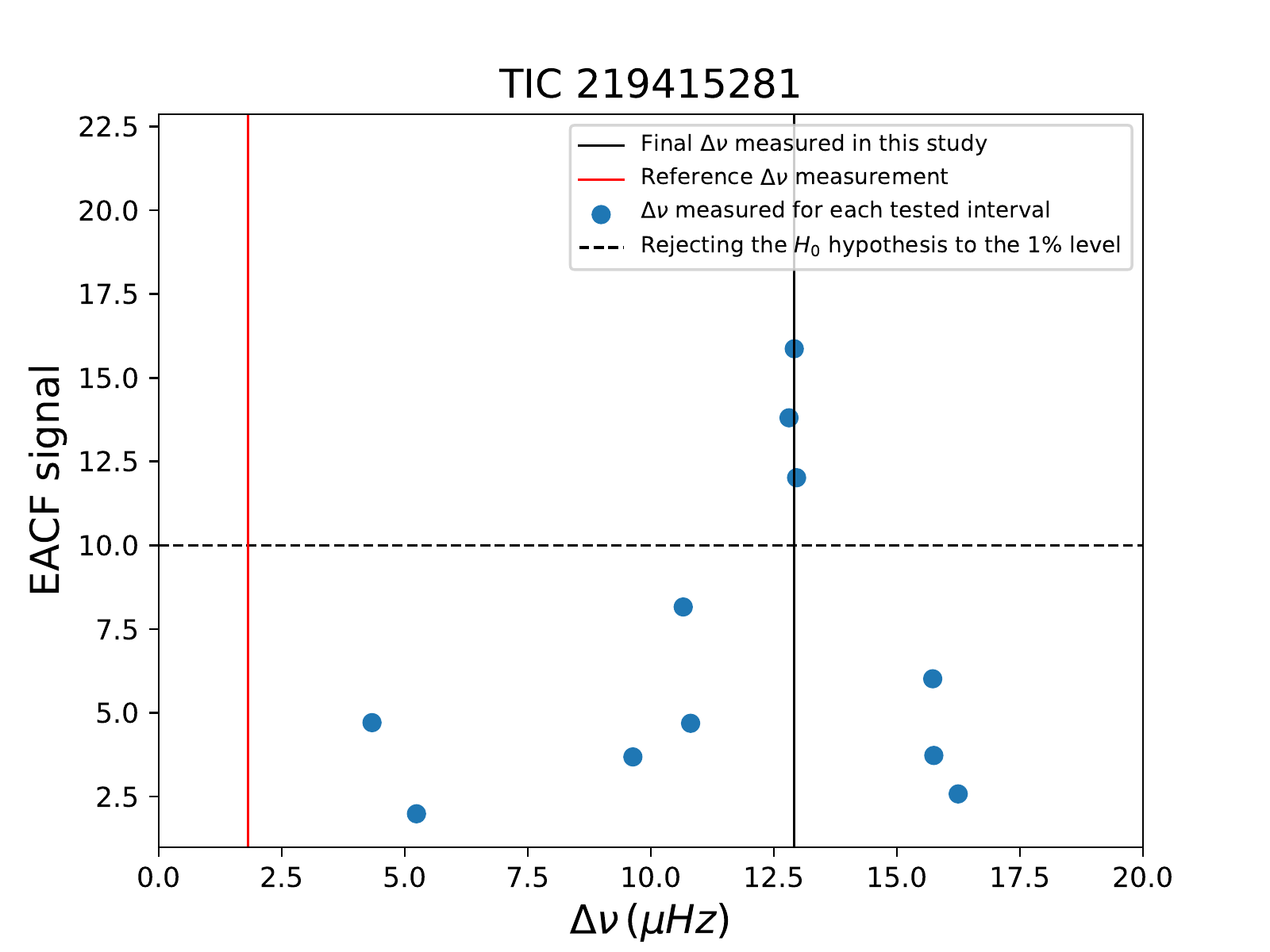}
\caption{Same as Fig.~\ref{fig-Kepler-Dnu-1} for TIC 219415281.}
\label{fig-TESS-Dnu-2}
\end{figure*}

\begin{figure*}
\centering
\includegraphics[width=8.8cm]{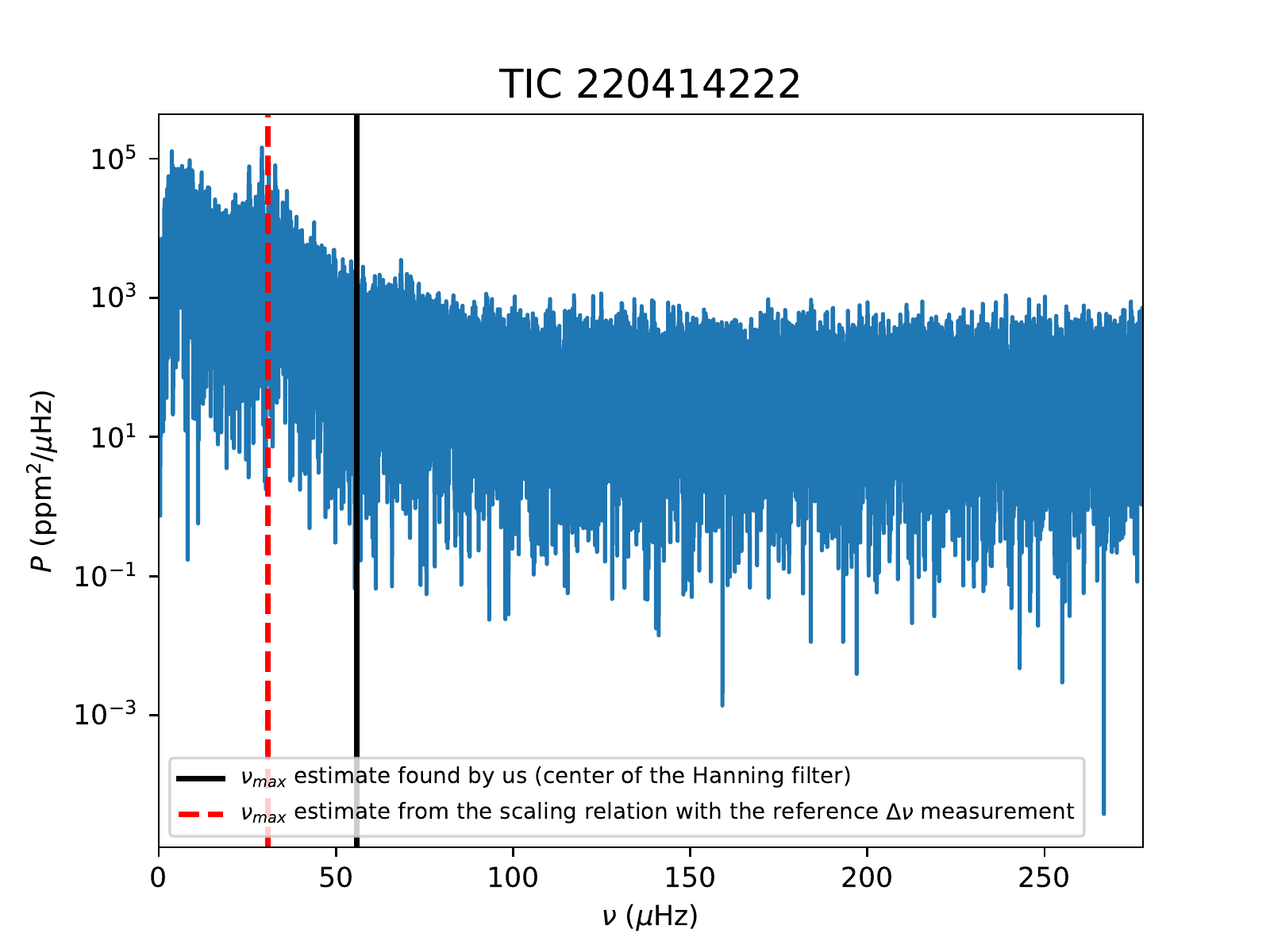}
\includegraphics[width=8.8cm]{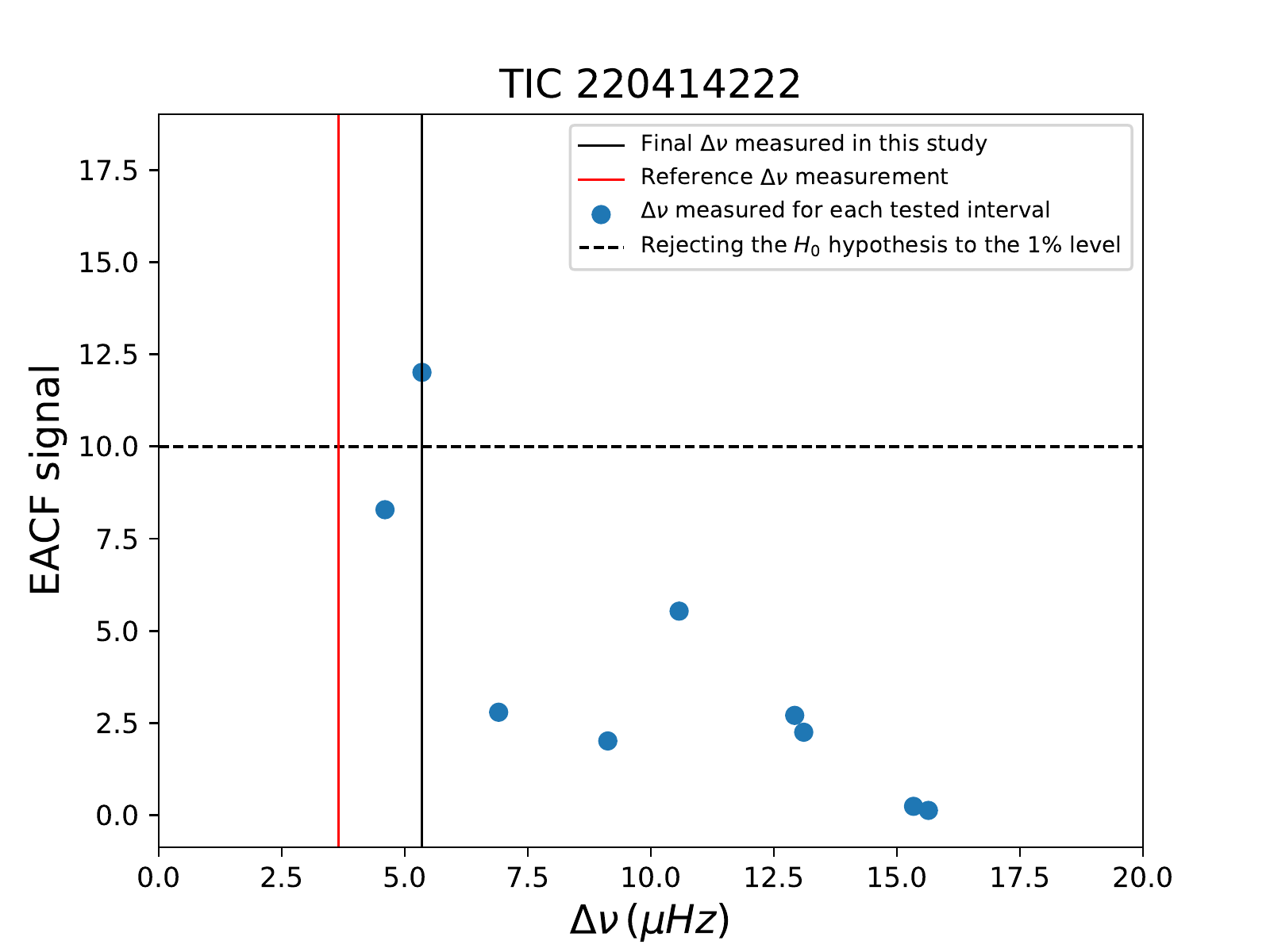}
\caption{Same as Fig.~\ref{fig-Kepler-Dnu-1} for TIC 220414222.}
\label{fig-TESS-Dnu-4}
\end{figure*}

\begin{figure*}
\centering
\includegraphics[width=8.8cm]{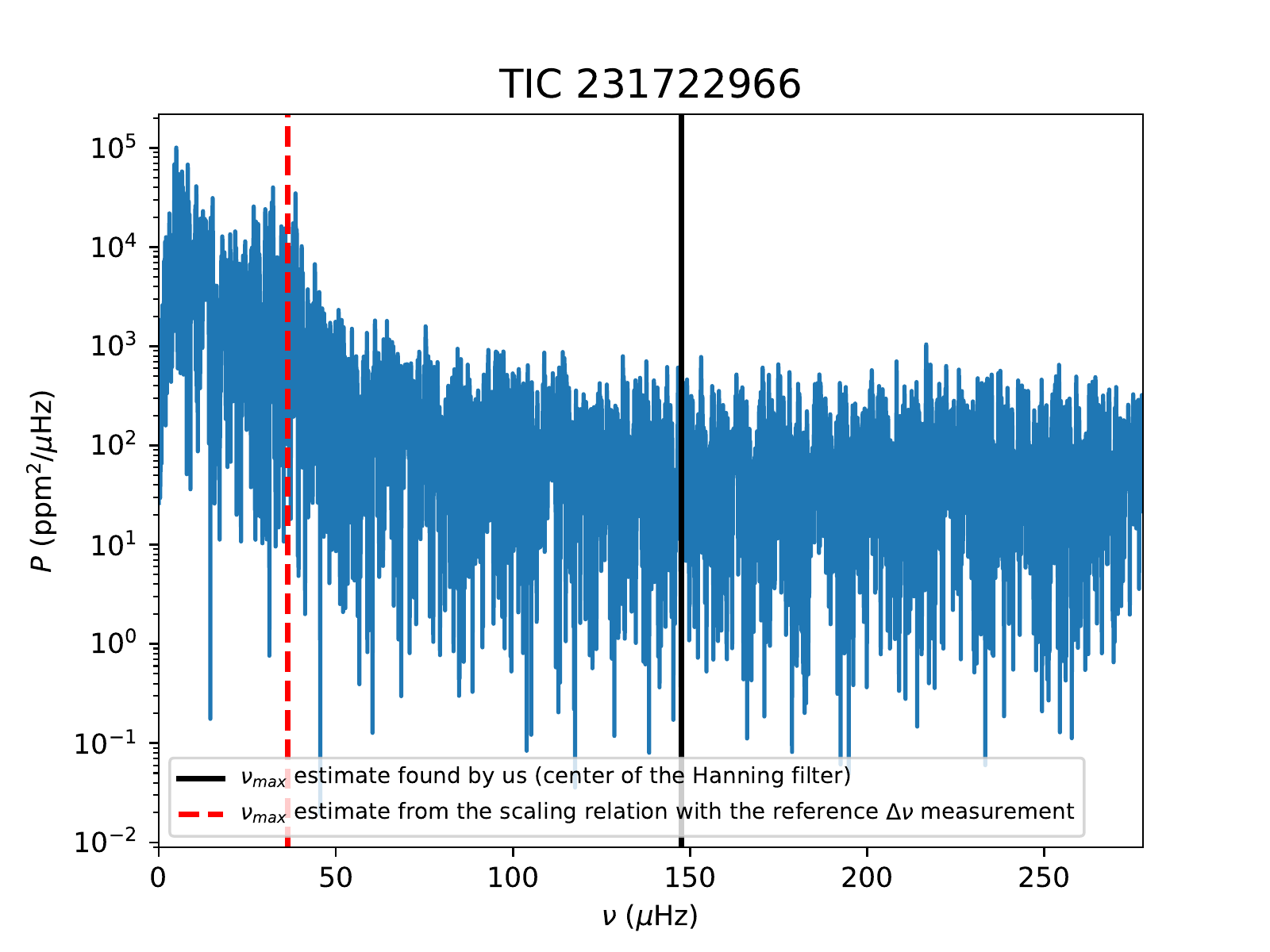}
\includegraphics[width=8.8cm]{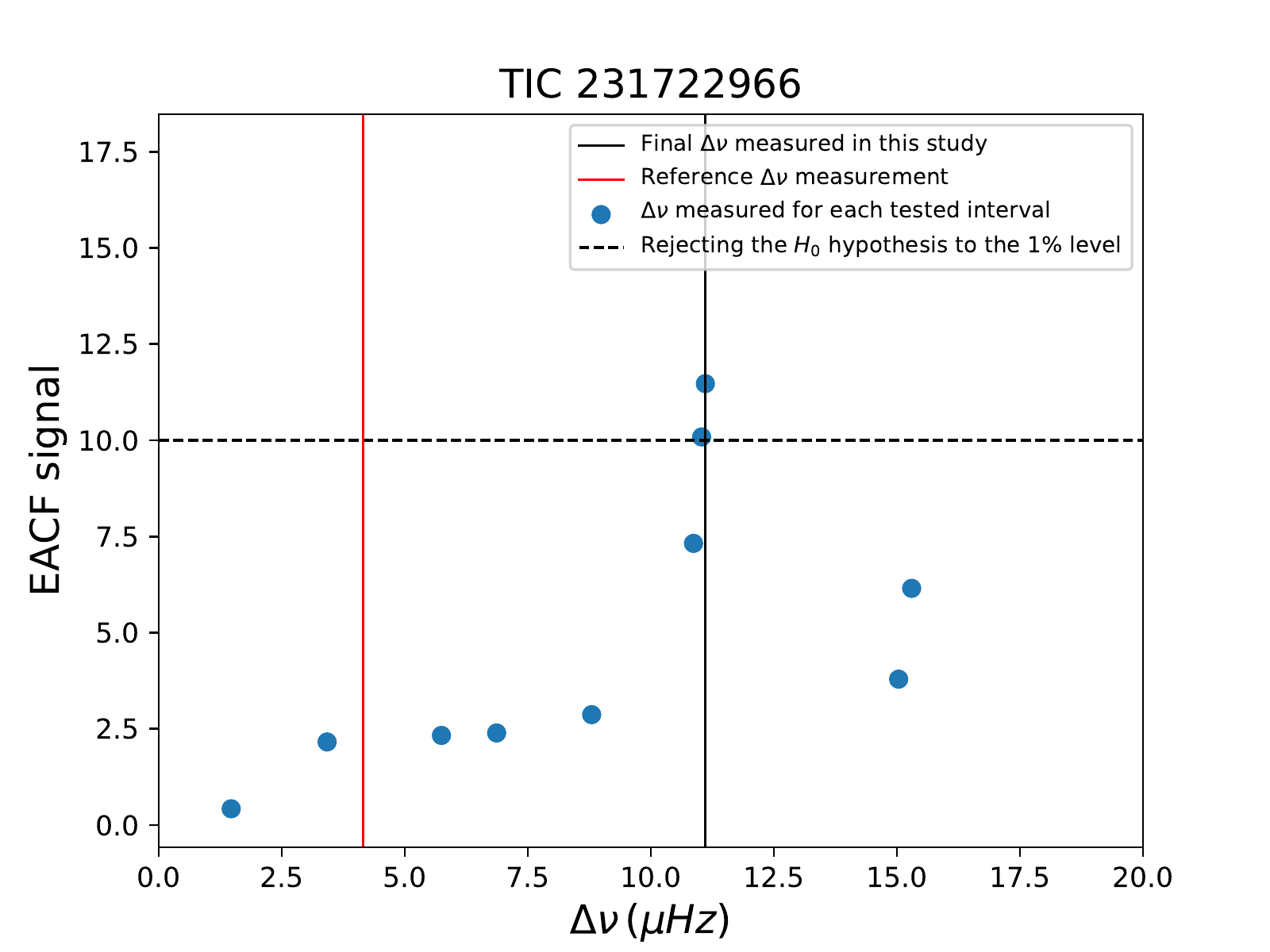}
\caption{Same as Fig.~\ref{fig-Kepler-Dnu-1} for TIC 231722966.}
\label{fig-TESS-Dnu-6}
\end{figure*}

\begin{figure*}
\centering
\includegraphics[width=8.8cm]{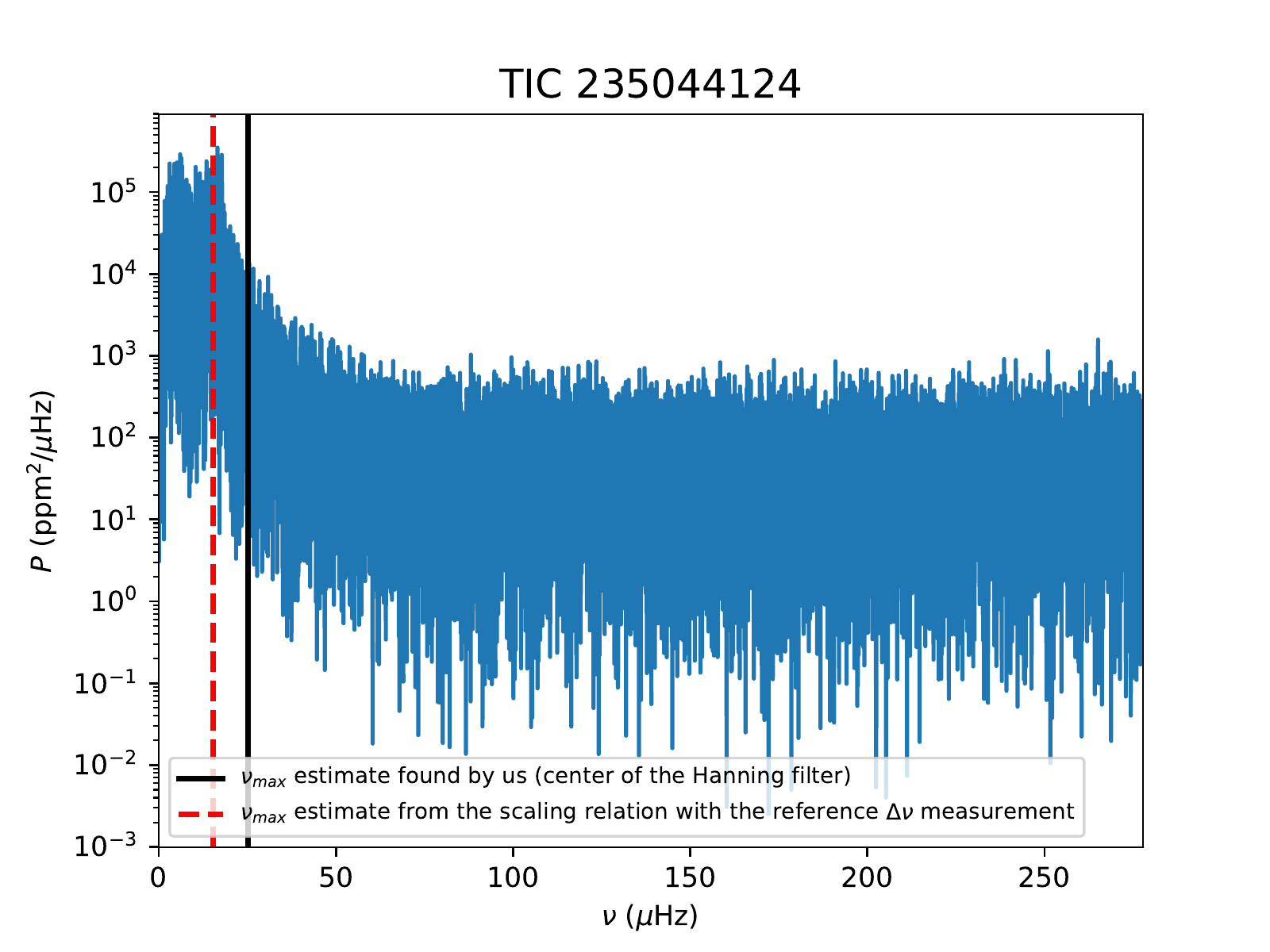}
\includegraphics[width=8.8cm]{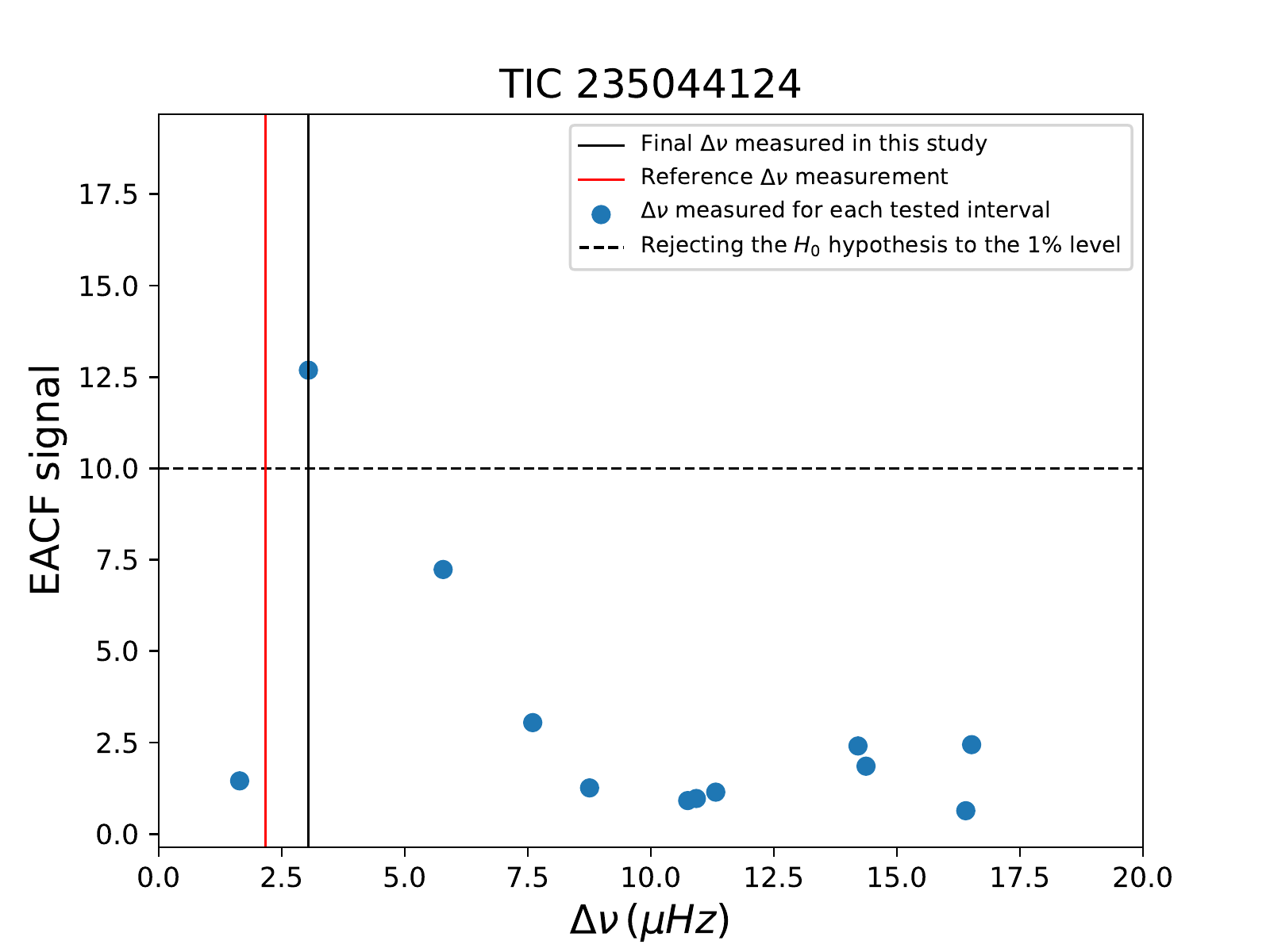}
\caption{Same as Fig.~\ref{fig-Kepler-Dnu-1} for TIC 235044124.}
\label{fig-TESS-Dnu-8}
\end{figure*}

\begin{figure*}
\centering
\includegraphics[width=8.8cm]{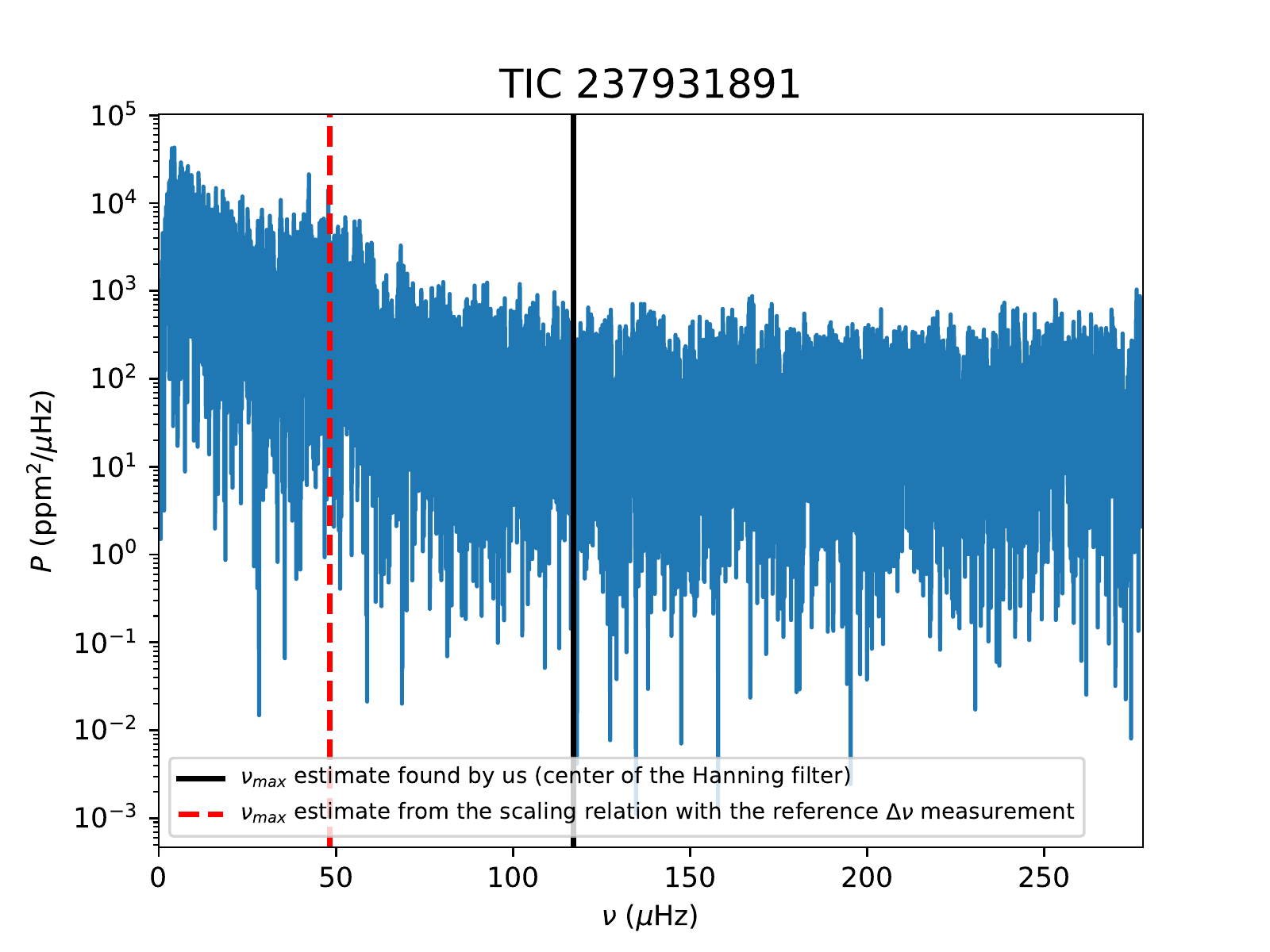}
\includegraphics[width=8.8cm]{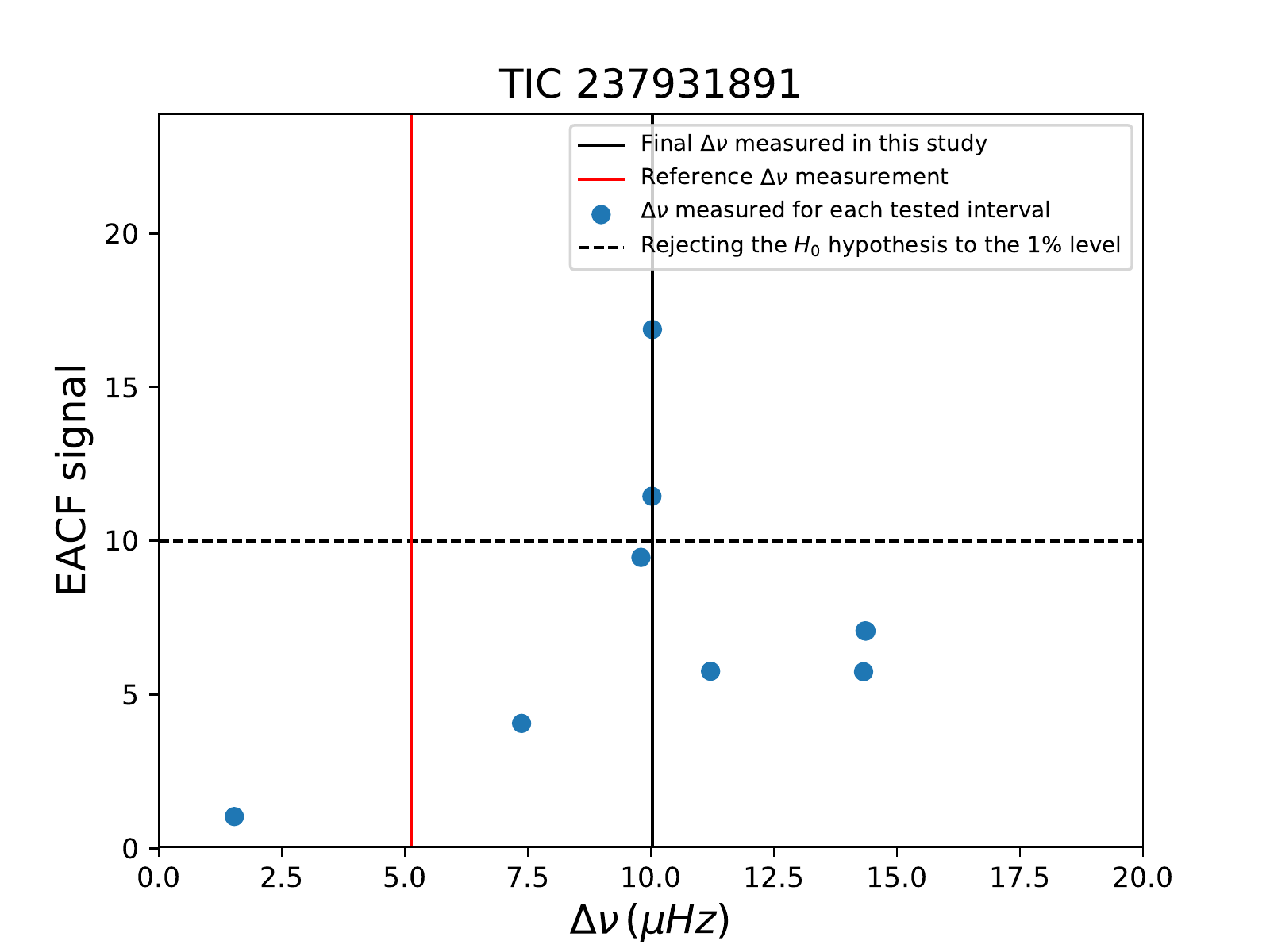}
\caption{Same as Fig.~\ref{fig-Kepler-Dnu-1} for TIC 237931891.}
\label{fig-TESS-Dnu-9}
\end{figure*}

\begin{figure*}
\centering
\includegraphics[width=8.8cm]{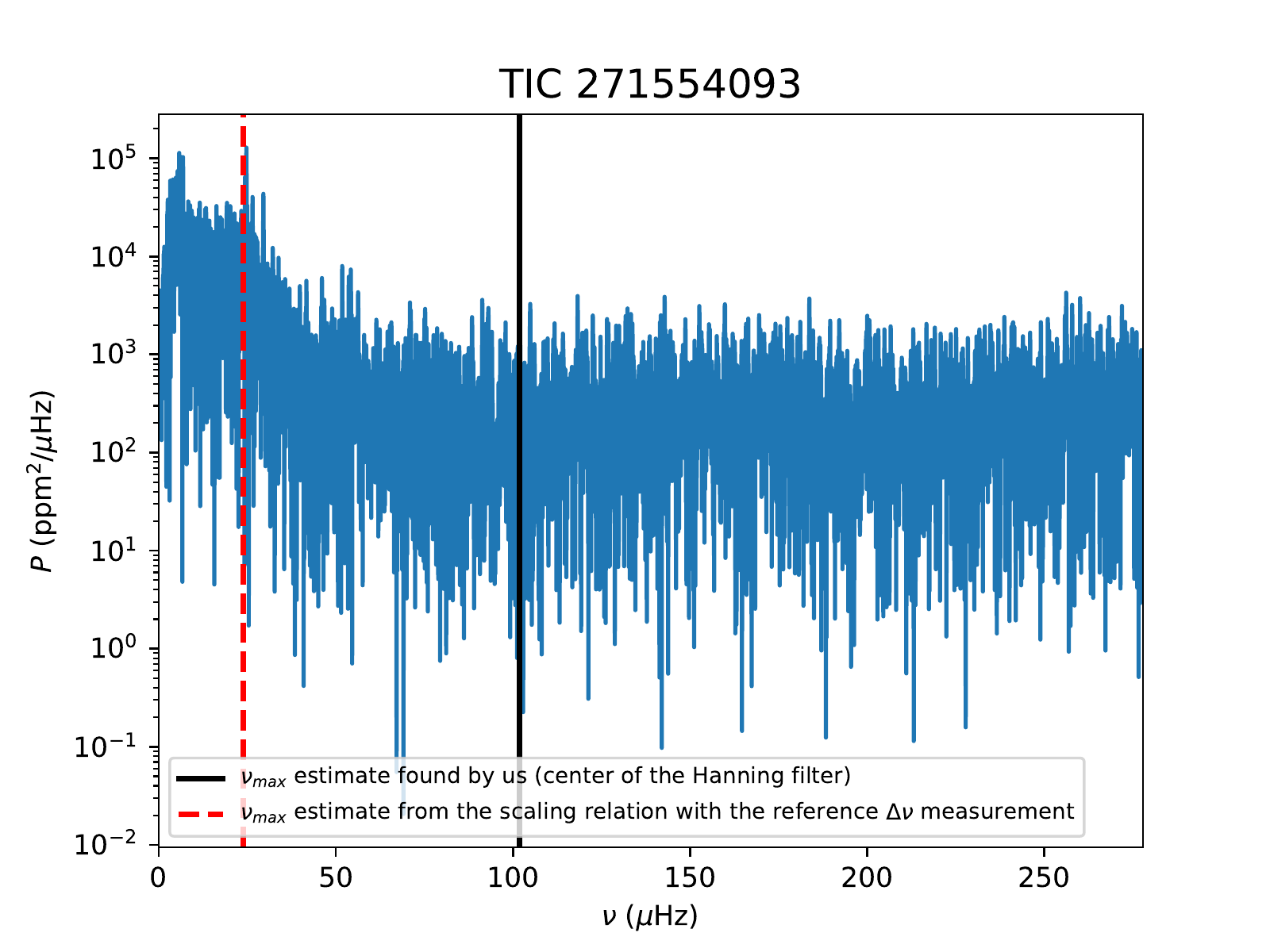}
\includegraphics[width=8.8cm]{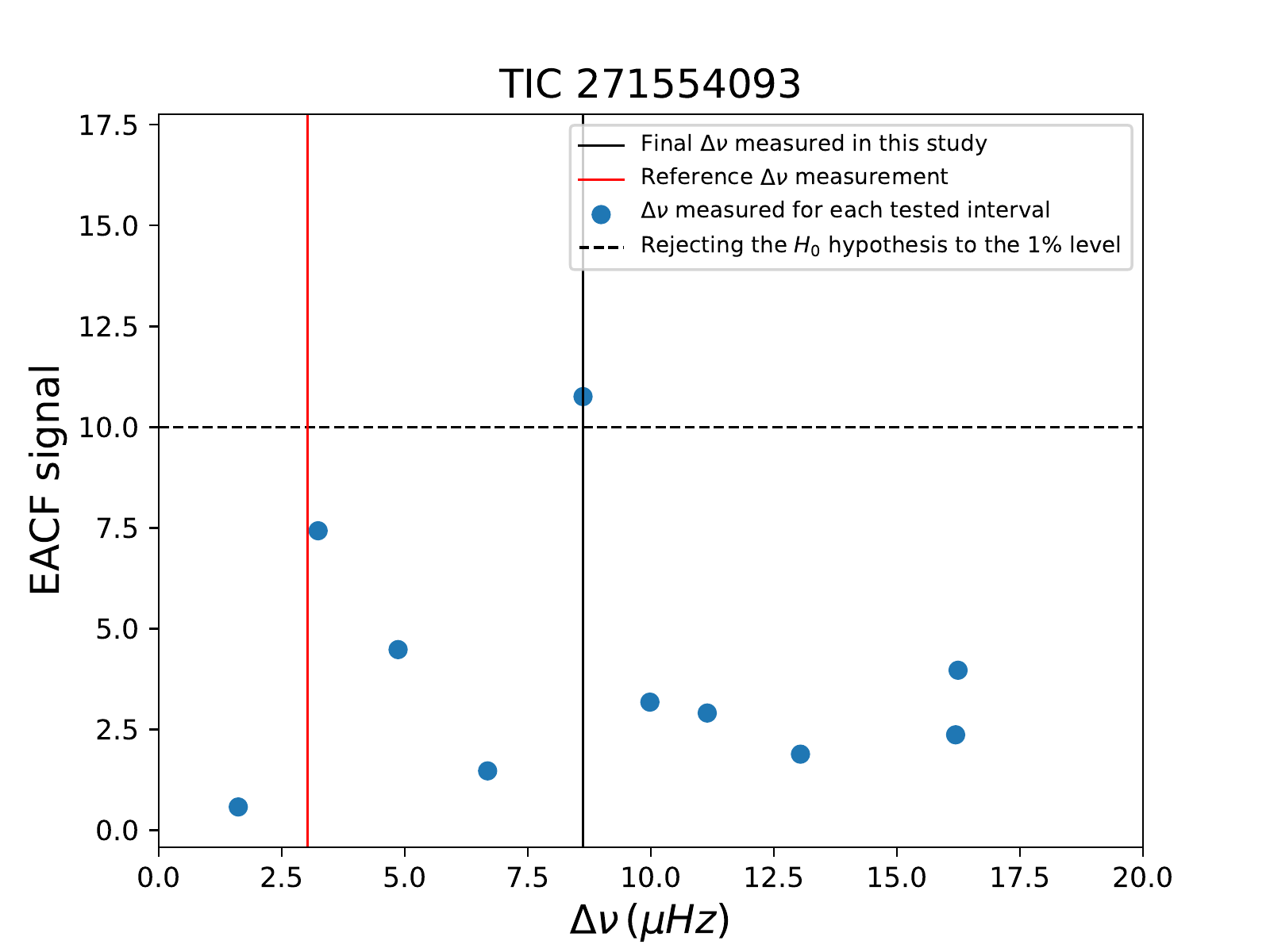}
\caption{Same as Fig.~\ref{fig-Kepler-Dnu-1} for TIC 271554093.}
\label{fig-TESS-Dnu-11}
\end{figure*}

\begin{figure*}
\centering
\includegraphics[width=8.8cm]{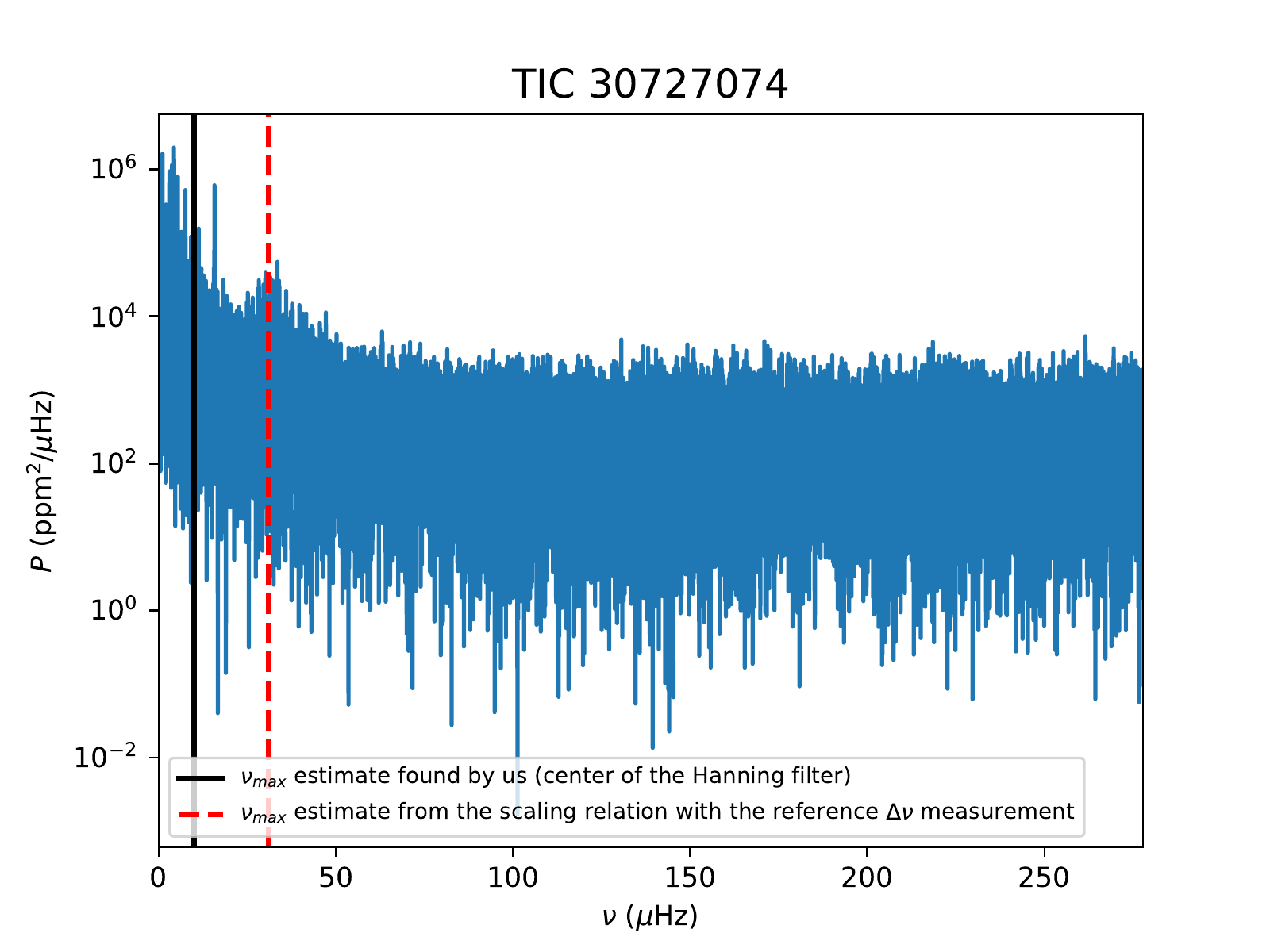}
\includegraphics[width=8.8cm]{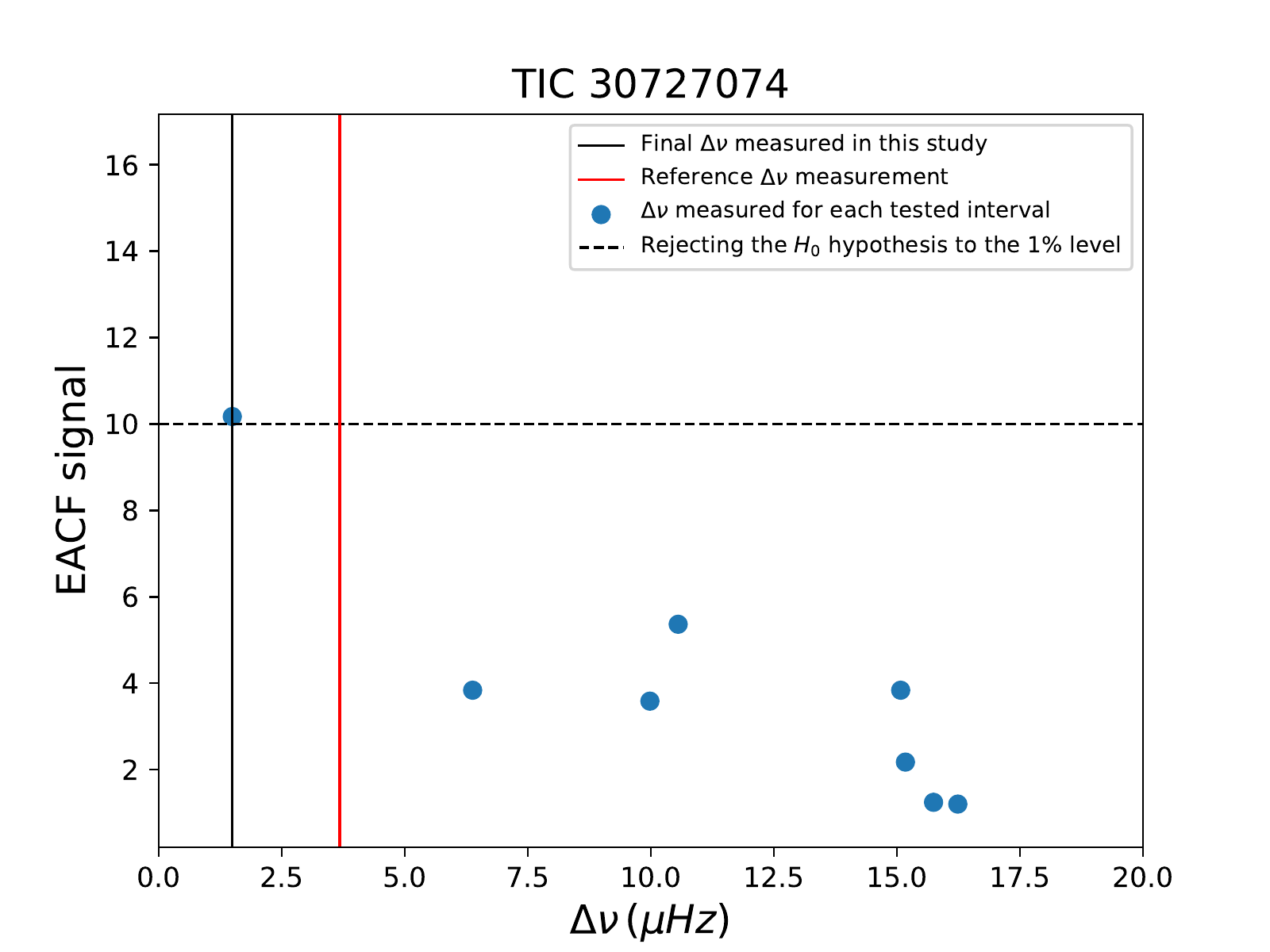}
\caption{Same as Fig.~\ref{fig-Kepler-Dnu-1} for TIC 30727074.}
\label{fig-TESS-Dnu-15}
\end{figure*}

\begin{figure*}
\centering
\includegraphics[width=8.8cm]{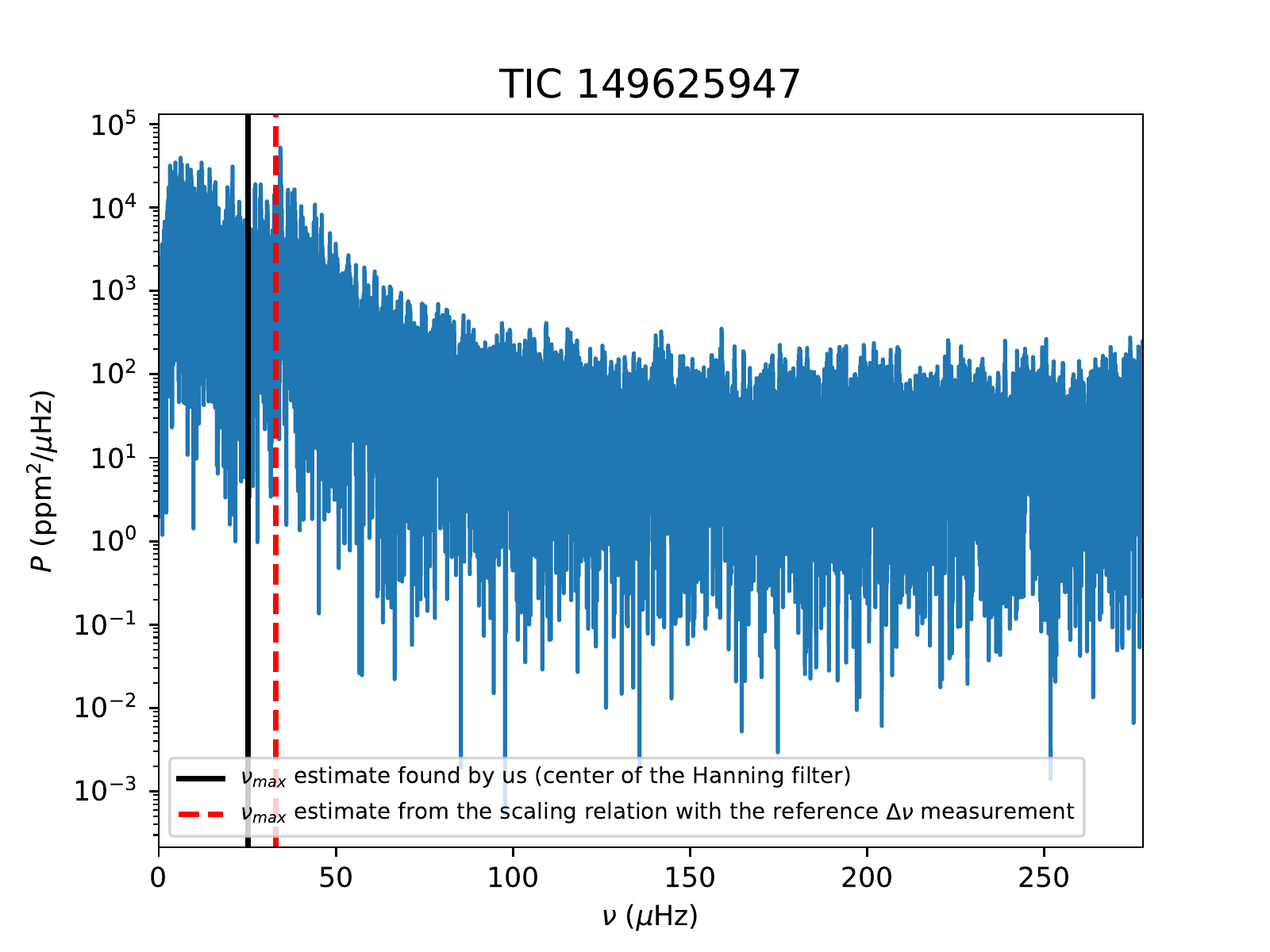}
\includegraphics[width=8.8cm]{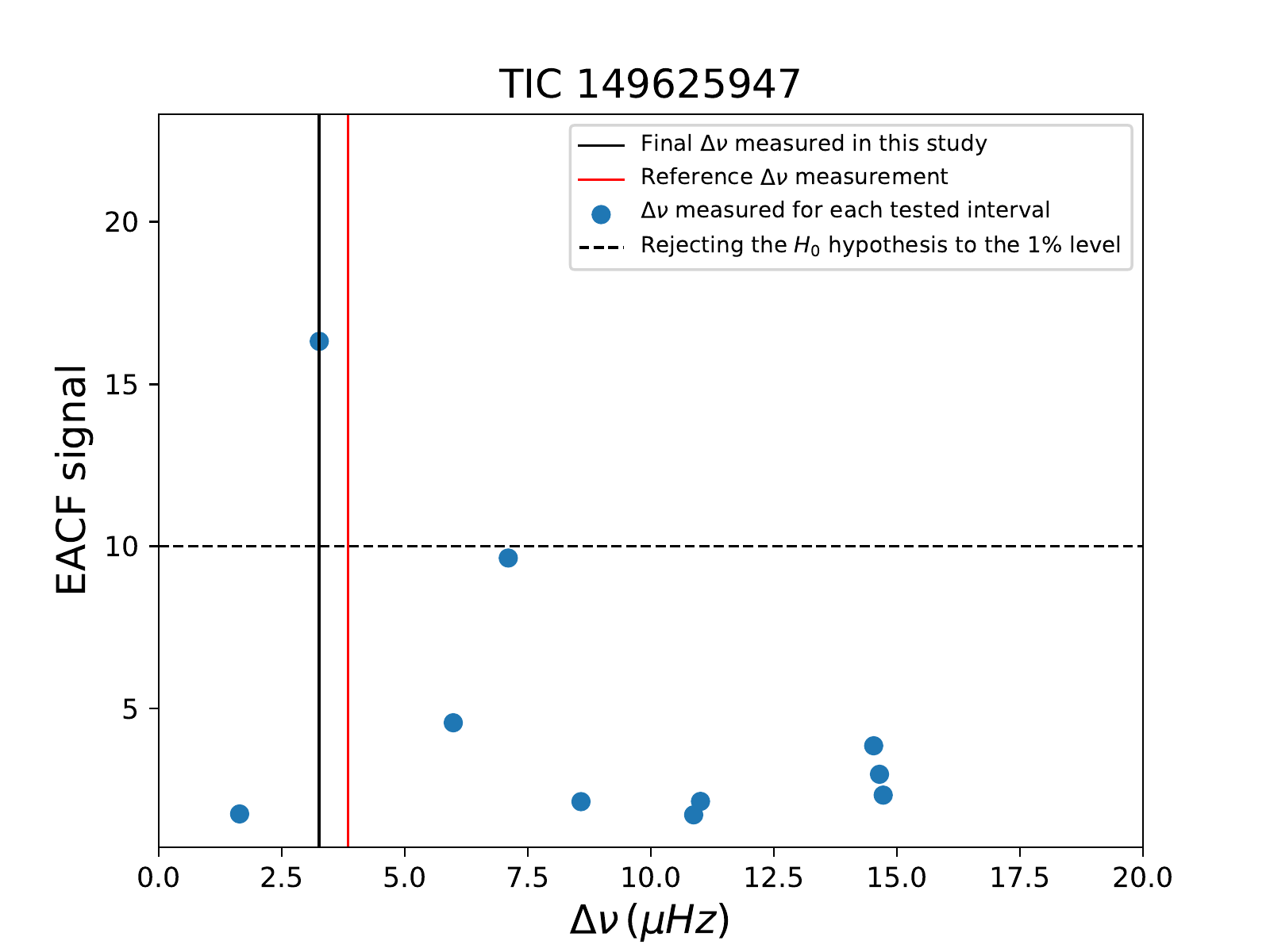}
\caption{Same as Fig.~\ref{fig-Kepler-Dnu-1} for TIC 149625947.}
\label{fig-TESS-Dnu-18}
\end{figure*}

\begin{figure*}
\centering
\includegraphics[width=8.8cm]{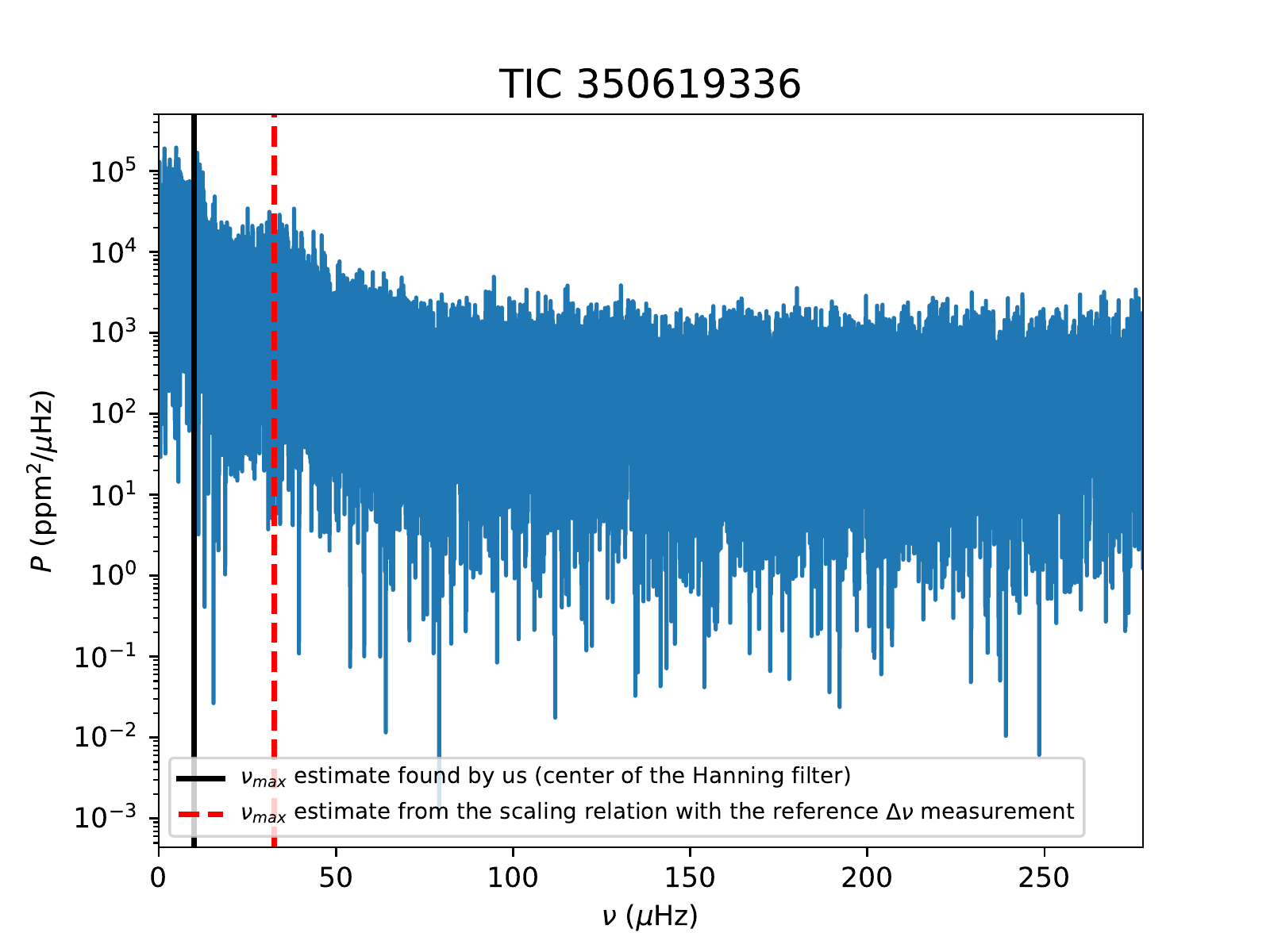}
\includegraphics[width=8.8cm]{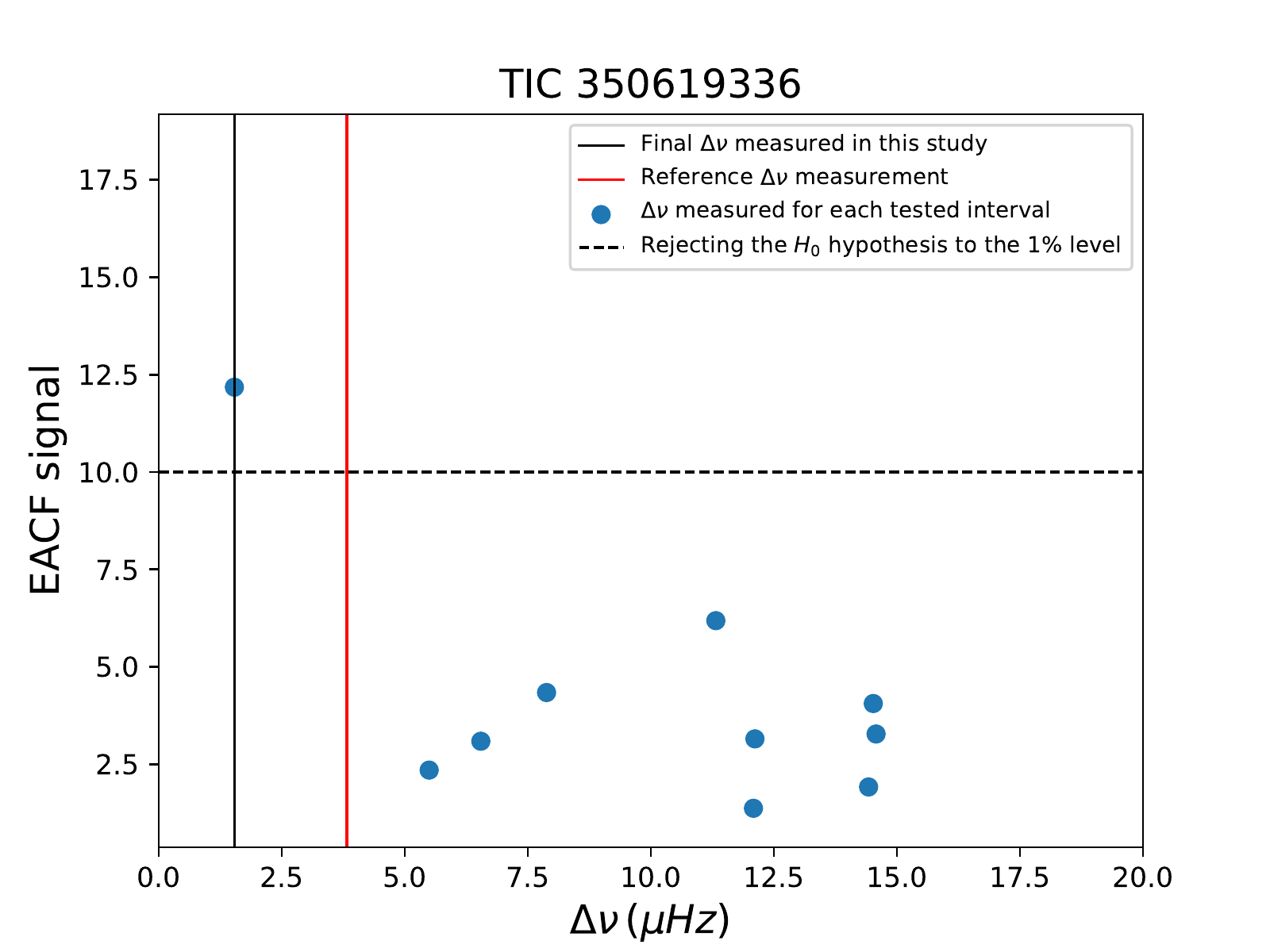}
\caption{Same as Fig.~\ref{fig-Kepler-Dnu-1} for TIC 350619336.}
\label{fig-TESS-Dnu-20}
\end{figure*}

There are 150 \textit{TESS} red giants for which we have a relative deviation of at least 10\% between our measurements obtained with our FRA pipeline and the measurements from \cite{Mackereth}, when looking for $\Dnu$ through a blind search. Here are some examples for which our analysis fails to provide an accurate $\numax$:
\begin{itemize}
\item TIC 219148162 (Fig.~\ref{fig-TESS-Dnu-1});
\item TIC 219415281 (Fig.~\ref{fig-TESS-Dnu-2});
\item TIC 220414222 (Fig.~\ref{fig-TESS-Dnu-4});
\item TIC 231722966 (Fig.~\ref{fig-TESS-Dnu-6});
\item TIC 235044124 (Fig.~\ref{fig-TESS-Dnu-8});
\item TIC 237931891 (Fig.~\ref{fig-TESS-Dnu-9});
\item TIC 271554093 (Fig.~\ref{fig-TESS-Dnu-11});
\item TIC 30727074 (Fig.~\ref{fig-TESS-Dnu-15});
\item TIC 149625947 (Fig.~\ref{fig-TESS-Dnu-18});
\item TIC 350619336 (Fig.~\ref{fig-TESS-Dnu-20}).
\end{itemize}

\end{document}